\documentclass[a4paper,10pt]{scrartcl}
\usepackage{microtype}
\usepackage{lmodern} 
\usepackage[utf8]{inputenc}
\setkomafont{disposition}{\normalcolor\bfseries}
\usepackage{amsmath, amssymb, enumitem, verbatim, color, tikzsymbols, tabularx, appendix}

\usepackage[style=authoryear, citetracker=true, maxcitenames=2, hyperref=true, url=true, isbn=false, doi=true, natbib=true, 
backend=bibtex,
bibstyle=authoryear, maxbibnames=30, sorting=nyt, uniquename=full, uniquelist=false]{biblatex}
\usepackage[colorlinks=true, allcolors=blue]{hyperref}
\usepackage{setspace}
\usepackage{floatrow}
\usepackage{placeins}
\usepackage{longtable, array, pdflscape, soul, bbm, subcaption, rotating}
\usepackage{ragged2e}
\setlist[description]{%
font={\bfseries\rmfamily}, 
}

\usepackage[affil-it]{authblk}

\usepackage{verbatimbox} 
\usepackage{etoolbox}
\usepackage{environ}
\NewEnviron{myresizeenv}{\resizebox{\linewidth}{!}{\BODY}}

\usepackage{tocloft}
\makeatletter
\renewcommand{\numberline}[1]{{\@cftbsnum #1\@cftasnum~}\@cftasnumb}
\makeatother

\setkomafont{caption}{\small}
\setkomafont{captionlabel}{\usekomafont{caption}}

\newcommand{\possessivecite}[1]{\citeauthor{#1}'s \citeyearpar{#1}}

\title{Demand-pull, technology-push, and the direction of technological change}
\author{\large Kerstin H\"otte\footnote{
Mail: khotte@turing.ac.uk. Postal address: Institute for New Economic Thinking, Manor Road Building, Manor Road, OX13UQ Oxford}
\\ 

\footnotesize{The Alan Turing Institute, London}\\
\footnotesize{Institute for New Economic Thinking, University of Oxford
}\\
\footnotesize{Faculty of Business Administration and Economics, University of Bielefeld
}\\
}
\date{\today}
\begin{document}
\pagenumbering{roman}
\maketitle

\begin{abstract}
This paper studies the impact of Demand-pull (DP) and Technology-push (TP) on growth, innovation, and the factor bias of technological change in a two-layer network of input-output (market) and patent citation (innovation) links among 307 6-digit US manufacturing industries in 1977-2012. Two types of TP and DP are distinguished: 
(1) DP and TP are between-layer spillovers when market demand shocks pull innovation and innovation pushes market growth. 
(2) Within-layer DP arises if downstream users trigger upstream innovation and growth, while TP effects spill over from up- to downstream industries. 	
The results support between- and within-layer TP: Innovation spillovers from upstream industries drive market growth and innovation. 
Within the market, upstream supply shocks stimulate growth, but this effect differs across industries. 
DP is not supported but shows a factor bias favoring labor, while TP comes with a shift towards non-production work. 
The results are strongest after the 2000s and shed light on the drivers of recent technological change and its factor bias.

\end{abstract}
\vspace{0.5cm}

\newpage


\onehalfspacing

\newpage
\pagenumbering{arabic}

\section{Introduction}
\label{sec:intro}
Shaping technological change is high on the political agenda to cope with the challenges of the twenty-first century, such as climate change or digitization \citep{ipcc2018special, brynjolfsson2012race}. 
Demand-pull (DP) and technology-push (TP) are drivers of technological change \citep{schmookler1966invention, meyer1969successful, mowery1979influence, von1976dominant, di2012technology} that can be stimulated by different policies \citep{rosenberg1982inside, nemet2009demand}. Effective policymaking requires an understanding of how these mechanisms interact and differ by impact. 

TP arises when technological breakthroughs enable the development and commercialization of novelties. 
DP emerges from market needs if inventors adapt R\&D efforts in response to the perceived market potential \citep{di2012technology, von1976dominant}. 
TP and DP are interdependent because R\&D objectives may be demand-selected and market needs may arise in response to innovation \citep{kline1986overview, cohen2010fifty, nemet2009demand, mowery1979influence, nelson1994co, saviotti2013co}. 

This paper studies the impact of TP and DP on qualitative and quantitative indicators of technological change in US manufacturing since the late 1970s. This period was characterized by a qualitative change of industrial production with rising capital intensity and shifts from production to non-production labor with changing skill-requirements \citep{acemoglu2002directed, karabarbounis2014global, elsby2013decline}. This raised concerns about the decline of the labor share and job market polarization with undesired effects on distribution and employment when low-skill and routine-intensive occupations become obsolete \citep{baltagi2005skill, autor2018automation, goldin2007long}. 
Here, it is analyzed how TP and DP contributed to industry growth, innovation, productivity, and changing production factor requirements.

\begin{figure}[ht]
{\centering
	\caption{Stylized representation of the two types of demand-pull and technology-push}
	\label{fig:schematic_network}
	\includegraphics[width=0.75\textwidth]{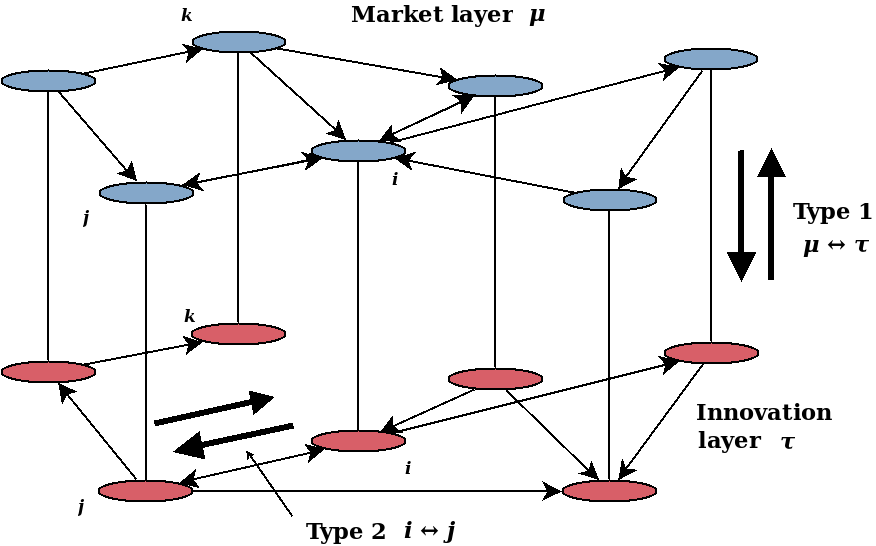}
	
}
\vspace{0.5cm}
\justifying
\footnotesize \noindent
Notes: This figure shows the two-layer network of a coupled input-output and patent citation network. Nodes in the network are industries $i, j, k$, and arrows connecting the nodes are flows of intermediate goods in the market layer $\mu$ (blue color) and patent citations in the innovation layer $\tau$ (red color). Demand-pull (DP) and technology-push (TP) Type 1 are between-layer spillovers, while Type 2 are within-layer effects arising from down- and upstream industries. 

\end{figure}

Previous studies of TP and DP mostly relied on market shares and sizes as proxies for DP and innovation outputs such as patents for TP to study their impact on growth, innovation, and productivity \citep[e.g.][]{jaffe1988demand, cohen2010fifty}.
This study relies on similar empirical proxies and analyzes TP and DP in a two-layer network that captures interactions between markets and innovation. 

The network is schematically illustrated in Fig. \ref{fig:schematic_network}. Industries $i,j,k$ are connected in the top layer, called ``market layer'' $\mu$ (red nodes), by cross-industrial flows of intermediate goods, and in the bottom layer, called ``innovation layer'' $\tau$ (blue nodes), by patent citations. 
The thick arrows indicate the two conceptually different mechanisms of TP and DP that are studied in this paper: 

\textbf{Type 1---Between-layer effects:} TP1 arises from technological breakthroughs reflected in surges of patenting in the innovation layer ($\tau \rightarrow \mu$) and DP1 arises from output shocks in the market ($\mu \rightarrow \tau$). 
TP1 is present if past innovation in an industry $i$ pushes growth in $i$'s market size, and DP1 is present if $i$'s market growth pulls innovation. 

\textbf{Type 2---Within-layer effects:} TP2 and DP2 arise from the supply (upstream) and demand (downstream) side of an industry. 
Within the market, DP2 effects ($j \rightarrow i$) are present when a rising demand from downstream customers $j$ for $i$'s outputs stimulates growth in the upstream industry $i$.
Within innovation, DP2 exists if an increased use of $i$'s inventions by downstream industry $j$ induces innovation in $i$. 
TP2 effects are the opposite ($i \rightarrow j$) when changes in the availability of inputs from suppliers $i$ induce downstream technological change in industry $j$. In the market, TP2 effects are present if an improved availability of production inputs drives growth in the downstream industry. In innovation, TP2 effects are given if upstream technological advance stimulates downstream innovation.

These effects are operationalized by network metrics that are empirically calculated using a panel of 307 NAICS 6-digit US manufacturing industries during 1977-2012. 
The market layer is inferred from empirical input-output (IO) tables provided by the Bureau of Economic Analysis (BEA). The patent citation network is compiled using US Patent and Trademark Office (USPTO) data and mapped to NAICS 6-digit industries through their technology classes \citep{goldschlag2020tracking}. 
The data is supplemented with data on productivity and production factor use \citep{bartlesman1996nber, becker2013nber}.\footnote{The data and code are available under a CC-BY 4.0 licsense \citep{hotte2023data}.} 

Using dynamic panel regressions, this paper analyzes the impact of both types of TP and DP on technological change measured by changing industry sizes in both layers (output and patent counts), productivity, and changing use of production factors. 
Industry sizes inform about the rise and decline of industries, productivity measures reflect the production efficiency \citep{oecd2001measuring}, and a changing factor input use (employment, wages, capital intensity, share of production labor) informs about the bias of technological change \citep{acemoglu2002directed}. 
To address cross-industrial heterogeneity in patterns of innovation \citep{pavitt1984sectoral}, the analysis of TP and DP is repeated for different subgroups of industries and subperiods. 

The results support within- and between-layer TP: Industries experiencing an expansion of innovation opportunities grow faster in the market (between-layer TP1) and innovate more (within-innovation TP2). These effects mostly arise from innovation spillovers from upstream industries. 
The support for these two TP effects is consistent across industries and strongest in patent-intensives industries. 

The results also support within-market TP2 effects arising from upstream market spillovers, which indicate positive supply shocks in the availability of production inputs. However, within-market TP2 is heterogeneous across industries and upstream network centrality shows a negative effect in some industries. One explanation for the ambiguous role of upstream market linkages lies in their interaction with the position of an industry in the supply chain \citep[cf.][]{mcnerney2018production} and a larger process of structural change towards increasingly processed goods (e.g. electronics, machinery, ICT). 
The results are much stronger during the post-2000s period, indicating that TP as driver of change became more important over time. 

DP is not supported as a driver of growth and innovation, but downstream linkages in the market are associated with a factor bias in favor of labor. 
An increasing centrality in the downstream network is associated with a higher labor demand, more investment, higher productivity, and a lower use of capital.\footnote{Downstream centrality indicates that an industry supplies goods that are essential inputs in many other industries (who themselves are important suppliers to others). Upstream centrality indicates that an industry's upstream suppliers are more specialized on few customers.} 
TP effects exhibit the opposite bias and are associated with a reallocation of labor from production towards non-production activities. 

Within the innovation layer, downstream spillovers (DP2) show a weak but statistically significant negative association with innovation. An increasing downstream use of innovations may indicate a higher level of technological maturity and more applied research. The combined results of within-innovation TP and DP effects suggest that technological breakthroughs trickle down the innovation chain from up- to downstream inventive activity. 

Finally, the analyses reveal path-dependence of market growth and innovation, which is stronger in innovation with an autocorrelation coefficient larger one. This indicates increasing returns to innovation, which may explain an increasing concentration of innovative activities.  

Four major limitations exist. First, patents are an indicator of radical innovation but are less suited to capture patterns of technology diffusion and use. They also suffer from a series of additional well-known limitations \citep[e.g.][]{jaffe2019patent, fontana2013reassessing, kogan2017technological}. 
Second, classification systems change over time, which hampers the study of long-term technology-industry links \citep{lafond2019long, yuskavage2007converting}. Third, drivers of technological change differ across firms, industries, time, and industrial maturity \citep{walsh1984invention, pavitt1984sectoral}. 
Fourth, this research is limited to US manufacturing and neglects other structural changes in the US economy. 

Nevertheless, this study offers relevant contributions and insights for research and policy. 
The two-layer network and the data offer a rich basis for future research that aims to understand how markets and technology interact (see also Sec. \ref{sec:disc}).
This understanding is essential as the societal challenges of today directly interact with technological change: Large-scale technological and economic change is necessary to cope with climate change and can be steered by adequate policy \citep{ipcc2018special}. Digitalization as a process of technological change needs to be steered to alleviate undesirable distributional effects and disruptions in the labor market. 
These challenges require an understanding of how radical innovation can be used to influence the evolution of markets for goods and labor, and how market forces can be mobilized to shape technological change. 

The results suggest that TP policies stimulating the creation of technological knowledge may influence market growth and innovation. The results also show that the distributional effects may depend on the driver of change. This is important for policy design. 
For example, it is a long-standing debate whether technology replaces or complements human labor and more recent research raised concerns about job market polarization and an unequal distribution of gains from technological change \citep[see][for an overview]{hotte2022technology}. 
This research suggests that the driver of change matters: TP shows a bias in favor of capital, while DP dynamics exhibit an opposite relationship. 
This is informative for innovation policy design when making the choice between DP or TP instruments. 

The study further shows that innovation is subject to increasing returns. This is important as increasing returns are drivers of endogenous growth \citep{romer1990endogenous}, which may leverage the effectiveness of innovation policy. But increasing returns may also cause technological lock-ins \citep{arthur1989competing}. 

The remainder of the paper is structured as follows: 
The next section provides an overview of the related literature. The theoretical framework is explained in Sec. \ref{sec:model}. Sec. \ref{sec:data} introduces the data. Sec. \ref{sec:results} summarizes the results, Sec. \ref{sec:disc} offers a discussion, and Sec. \ref{sec:concl} concludes. 

\section{Background}
\label{sec:lit}
TP and DP as drivers of technological and economic change are the subject of a long-lived debate \citep{schmookler1966invention, mowery1979influence, pakes1984exploration, cohen2010fifty, saviotti2013co}. 
DP suggests that R\&D follows the market: The perceived commercial potential of innovations offers an incentive for targeted R\&D. TP arises from technological opportunities that enable the development of new products and subsequent innovation. 
The two concepts differ by the assumptions made about the incentives that influence R\&D and production decisions, and about the sources of ideas for technological improvements. While DP emphasizes the role of users and customers, TP builds on external and internal research \citep{cohen1989innovation, kline1986overview, di2012technology}. 

Previous studies found support for both effects \citep[see][for an overview]{cohen2010fifty, di2012technology}. For example, using aggregate time series over the business cycle, \citet{geroski1995innovative} studied interactions between manufacturing outputs and innovative activity. They observed causal effects from outputs to innovation, but no support for the other way round. They highlighted the critical role of stochastic determinants, which may indicate supply shocks that point to TP from radical innovation. 
\citet{walsh1984invention} showed for the chemical industry that TP from radical breakthroughs drives market growth, which creates DP that induces incremental innovation.

The analysis in this paper is conceptually close to the seminal work by \citet{jaffe1988demand} who operationalized DP through the market shares of a firm in different industries and TP as the effect that arises from innovations in technologically related fields. The author found that push and pull effects cannot be distinguished empirically when explaining TFP growth. 

This paper builds on an empirical two-layer network approach composed of an IO and a patent citation layer. The two layers capture the co-evolution of markets and technology \citep[][]{nelson1994co}, which occurs when market dynamics correlate with innovation \citep[][]{saviotti2013co}. 

IO and patent citation networks can be used to qualitatively describe technology by the network position of an industry, firm, or patent. The position is determined by the bundle of input links pointing to the physical production inputs or existing technological knowledge embodied in patents that enables subsequent innovation. 
Two firms, industries, or patents are technologically similar if they have many overlapping up- and downstream links, indicating the capability to make use of similar inputs and to serve the needs of similar users \citep[e.g.][]{carvalho2014input, antony2012technology, acemoglu2016innovation, cai2017knowledge, huang2021network, atalay2011network, jaffe2019patent, cohen2000absorptive}. 

Technological similarity and direct network links enable cross-industrial spillovers of knowledge and market shocks. Other research based on patent data has shown that innovation spillovers may stimulate productivity growth \citep{antony2012technology}, innovation \citep{acemoglu2016innovation}, employment growth and R\&D investments \citep{buerger2012regional}, and may be a source of path-dependence of technological trajectories \citep{kay2014patent, taalbi2020evolution, huang2021network}. 

In IO networks, the characteristics of an industry's production technology are reflected in the bundle of input used and outputs produced. 
Other researchers observed that patterns of production input adoption \citep{carvalho2014input} and product market diversification \citep{boehm2022comparative} depend on pre-existing IO links. 
\citet{carvalho2014micro} showed that relatedness through IO links can be a moderating factor of output fluctuations. 

This study combines both types of networks. Existing concordance mappings of patents to industries often rely on the industrial classification of the firms that own patents in specific technological fields \citep[e.g.][]{kortum1997assigning, schmoch2003linking, dorner2018novel, van2014patent, lybbert2014getting, goldschlag2020tracking}. 
Concordances make it possible to study interactions between the evolution of patented technology and industries. Proving the economic validity of their concordance, \citet{goldschlag2020tracking} showed that industry-technology relationships are relatively stable. 

This study is not the first that considers the market and innovation network positions simultaneously. 
Next to \possessivecite{jaffe1988demand} seminal paper, \citet{bloom2013identifying} build on a similar framework but study the role of market rivalry and knowledge spillovers on firms' performance. 
\citeauthor{bloom2013identifying} used output linkages to capture rivalry in product markets and patent links to compute knowledge spillovers. They found that market rivalry has a negative on firm performance, while the impact of knowledge spillovers is positive. 

\section{Technological and economic change in a two-layered network}
\label{sec:model}
This section introduces the conceptual framework of the two-layer network, explains how the network is used to identify TP and DP, and introduces the measures of technological change. It concludes by a short outline of the steps that are undertaken in the empirical analysis.

\subsection{The economy as a two-layer network}
\label{sec:model_concept}
Technology is the capability to transform a bundle of inputs into outputs. Technological change occurs when the quality and/or quantity of inputs or outputs changes \citep{saviotti1997black, ruttan1959usher}. 
Industries use intermediate goods as production inputs and build on pre-existing knowledge encoded in cited patents to innovate, in addition to production factors such as labor and capital. 

Patent citations do not necessarily represent a direct flow of knowledge. 
Citations are a legal requirement to describe prior art and to limit the scope of the new patent. However, a patent citation still serves as an indicator of technological relatedness revealing that the knowledge encoded in the cited patent contributed to the technological basis onto which a patent builds \citep{jaffe2019patent, oecd2009use}.

In this paper, IO flows represent the market and patents are interpreted as innovations. The IO and patent citation links span a weighted, directed two-layer network consisting of a market layer $\mu$ and an innovation layer $\tau$. A node in the network represents an industry $i \in N$, which is connected with other industries $j \in N$ through IO links in the market and patent citations in the innovation layer, where $N$ is the set of industries in the economy. 
The layers $\alpha = \mu, \tau $ are linked as a duplex network where each industry has a representation in each layer. 

The links in the layers are given by the flow of goods $flow^{\mu}_{ij,t}$ and patent citations $flow^{\tau}_{ij,t}$ from an industry $j$ to $i$ with $i,j \in N$ in time $t$. 
These flows reveal two types of information about the technology used by $i$ and $j$: An input flow from $j$ to $i$ indicates that $i$ has the capability to use the outputs produced by $j$. 
It also reveals that $j$ has the capabilities to produce outputs that are valuable for $i$. 
Hence, the bundle of upstream (input) links and downstream (output) links contains qualitative information about the technology that is used in these industries. 

Here, the flows of goods and citations are transformed into input shares $w^{\alpha,up}_{ij,t}$ dividing the monetary flow (patent citation count) $flow^{\alpha}_{ij,t}$ by the sum of inputs $\sum_{j} flow^{\alpha}_{ij,t}$. Analogously, output shares $w^{\alpha,dw}_{ij,t}$ are obtained by dividing $flow^{\alpha}_{ij,t}$ by the sum of all outputs produced by industry $i$, i.e. $\sum_{k} flow^{\alpha}_{ki,t}$. Note that $w^{\alpha,up}_{ij,t} \neq w^{\alpha,dw}_{ji,t}$ due to the different weighting. They reflect different concepts: The input share $w^{\alpha,up}_{ij,t}$ reflects $j$'s relevance as an input provider for $i$, while $w^{\alpha,dw}_{ij,t}$ reflect $j$'s relevance as a customer or knowledge user of $i$. 
The normalization to shares improves the comparability of different data types (monetary flows, patents) and of industries that are heterogeneous by size. 

Each network layer can be represented as a quadratic, asymmetric $|N|\times |N|$ matrix $W_t^{\alpha, d} = \{ w_{ij, t}^{\alpha, d} \}_{i,j \in N}$ with positive non-zero entries if a link from $i$ to $j$ exists in time $t$. The superscript $d = up,dw$ indicates the direction of the links, i.e. $w^{\alpha,up}_{ij,t}$ ($w^{\alpha,dw}_{ij,t}$) indicates an upstream (downstream) link. 
The two-layer network is given by the set of both matrices: one representing the ``market layer'' and the other presenting the ``innovation layer'' as illustrated in Fig. \ref{fig:schematic_network}. 

\subsection{Network spillovers and centrality}
\label{sec:model_simil} 
The network data is used to derive indicators to capture up- and downstream TP2 and DP2 effects. 

\subsubsection{Spillovers}
Industries are connected in both layers and shocks in one industry may spill over to industries that are sufficiently close by their up- and downstream linkages \citep[see e.g.][]{carvalho2014input, acemoglu2016innovation, bloom2013identifying}.\footnote{For a short discussion of the theoretical foundations of spillovers and the underlying metrics, see \citet{bloom2013identifying}.} 

Closeness in networks and spillovers can be empirically captured in different ways. This paper focuses on spillovers from direct up- or downstream links $l^{\alpha,d}_{ij,t} = \mathbbm{1}(w^{\alpha, d}_{ij,t} \geq 0.05)$ that equal one if the weight of the link $w^{\alpha, d}_{ij,t}$ exceeds a threshold of five percent and to zero otherwise. Hence, it is assumed that upstream spillovers do only arise from those industries $j$ that are important for $i$ as a supplier and the goods produced by $j$ account for five or more percent in $i$'s input mix. 
Analogously, downstream spillovers arise only from those industries from which $i$ receives at least five percent of its market revenue or patent citations.\footnote{The threshold level of 5\% restricts the network to a maximal number of 20 most important up- and downstream linkages. Robustness checks confirm that the main regression results are qualitatively consistent for alternative threshold levels (2.5\%, 10\%, 20\%) and for spillovers calculated by using the weights $w^{\alpha, d}_{ij,t}$ instead of binary links. Spillovers could also be calculated on the basis of technological similarities. This paper focuses on direct links as this most directly captures the impact of an industry's neighbors in the network and as they show sufficient variation in the innovation layer to disentangle up- and downstream effects (see Sec. \ref{subsec:results_regression}). Restricting the analysis to spillovers from direct neighbors is a common approach in economic studies of the transmission of shocks from up- and downstream industries \citep[see e.g.][]{carvalho2019production, di2021stock, frohm2021spillovers}. This procedure also helps reduce noise in the data arising from the application of concordances when mapping patents to industries and harmonizing different vintages of the NAICS.}

Spillovers are calculated as
\begin{align}
Spill(A)^{\alpha, d}_{i,t} = \sum_{j \neq i}^N l^{\alpha, d}_{ij,t} \cdot A^{\alpha}_{j,t} 
\end{align}
with $\alpha = \mu, \tau$, $d = up, dw$. $A^{\tau}_{j,t}$ is the number of patents and $A^{\mu}_{j,t}$ is the amount of goods produced by $j$ in $t$. 
S
The level of $Spill(A)^{\alpha, d}_{i,t}$ changes either by an output shock in related industries $A^{\alpha}_{j,t}$ or by a change in the $i$'s up- or downstream network $l^{\alpha, d}_{ij,t}$. 

Increasing upstream spillovers in the market suggest that either existing suppliers grew by market size $A^{\mu}_{j,t}$ or that $i$ connected to a new and potentially larger supplier. 
Analogously, increasing upstream spillovers in the patent layer indicate that the knowledge pool ---proxied by patents $A^{\tau}_{j,t}$--- of the upstream industry increased or that new sources of patented knowledge were acquired through new citation links. 
Upstream spillovers are related to TP2 effects as $Spill(A)^{\alpha,up}_{i,t}$ reflects a change in the availability of production and innovation inputs. 
Generally, a high level of upstream spillovers suggests that an industry receives a high share of its inputs from large industries.

In contrast, increasing downstream spillovers suggest that an industry's customers grew by market size or new customers were acquired. In the patent layer, $Spill(A)^{\tau,dw}_{i,t}$ indicates that patents by $i$ are either cited by industries with a growing number of patents or by novel groups of knowledge users. 
Downstream spillovers indicate DP2 if they drive technological change in the upstream industry.

\subsubsection{Centrality}
\label{subsubsec:methods_centrality}
An industry is central in the network if it is well-connected with other industries. 
Network centrality is an indicator for the relevance of an industry as input provider or output user \citep[cf.][]{jackson2008social, carvalho2014micro}. 
Different approaches exist to measure network centrality \citep{jackson2008social}. 
Here, the PageRank $PR^{\alpha, d}_{i,t}$ is used. It is calculated by a recursive algorithm that ranks industries by the number and quality of link, whereby the quality is higher when the link connects to an industry that is itself ranked high by the PageRank \citep[][]{brin1998anatomy}. This study relies on a version of the PageRank that accounts for the weighted and directed nature of the links \citep{csardi2006igraph}.\footnote{Originally, the PageRank was developed by \citet{brin1998anatomy} and used in the early versions of the Google search engine to rank websites by their relevance to the users based on upstream links that are weighted by the relevance of the websites from which the links are coming. 
The PageRank is used because it can be computed on the basis of up- and downstream links and shows a sufficiently high variation. 
Exploratory analyses have shown that the results across different measures are robust, and correlation statistics show a high correlation between the PageRank and other measures of network centrality (see \ref{app:regression_data}).}

An industry can be central in two ways, and changes in the centrality are indicative of within- and between-layer DP and TP effects: 
(1) Downstream centrality indicates that an industry is a critical supplier. 
An increasing downstream centrality in the market (innovation) layer indicates that an industry $i$ became more important as a provider of goods (source of knowledge) in the network. 
This is associated with DP2 because it reflects an increasing importance of $i$'s outputs that are demanded by downstream industries. 
Support for DP2 is found if an increasing downstream centrality is associated with subsequent growth and innovation.

(2) Upstream centrality indicates that an industry is a critical customer. It uses goods or patents supplied by a great number of other industries and/or that account for a large share in their output bundle. 
If $PR^{\alpha,up}_{i,t}$ increases, the upstream market power of $i$ increases. This may imply a higher diversity of input sources, but it may also reflect a pattern of vertical specialization when upstream industries become more specialized to supply goods to $i$. 
An increasing upstream centrality can indicate TP2 because it reflects an improved input availability for industry $i$. TP2 is supported if this stimulates growth and innovation in the same layer.

Changes in the centrality may also inform about between-layer effects because centrality adds explanatory power to the pure size effect of an industry by controlling for the quality of $i$'s outputs. For example, an increasing number of patents of $i$ indicates a pure increase of innovation outputs $A^{\tau}_{i,t}$, while an increasing centrality $PR^{\tau, d}_{i,t}$ suggests that the innovations of the industry became more relevant. As we shall see in the empirical analysis in Sec. \ref{subsubsec:DP_and_TP}, the number of patents is not significantly related to market growth, but centrality is, which offers support for TP1.\footnote{Between-layer effects (DP1/TP1) as conceptualized in this paper include both: effects that come directly from an industry itself and effects that arise from the network. For example, increasing patents, spillovers, and centrality in the innovation layer suggest rising technological opportunities in an industry $i$. These opportunities may come from the industry itself, but also from its position in the innovation network. TP1 is supported if any of these effects is a driver of market growth. Analogously, DP1 is supported if rising output or rising market opportunities and improved input availability in the market positively correlate with subsequent innovation.}

\subsection{Technological change}
\label{sec:model_tech_change}
At the industry level, technological change is reflected in a changing composition of in- and outputs used to produce goods and for innovation. This leads to changes in industry sizes: Some industries grow, others shrink in relative and absolute terms. 

Technological change also manifests in a changing use of production factors, which is measured by productivity, labor requirements, capital and investment intensity. 
This \emph{qualitative} dimension of technological change is referred to as non-neutral or directed technological change. 

Patterns of innovation and technological change are heterogeneous across industries \citep{pavitt1984sectoral, kline1986overview, cohen2010fifty} and may differ over time. These sectoral patterns are taken into account by analyzing the role of TP and DP separately for different subsets of industries and separately for the first and second half of the period covered. The second subperiod was characterized by a major trade shock caused by Chinese import competition, which triggered larger disruptions in the US economy \citep{pierce2016surprisingly}. Splitting the sample by the pre- and post-2000s helps understand whether the drivers of technological change have changed over time. 

Wrapping up, this paper studies technological change in the following ways: 
\begin{enumerate}
\item The analysis begins with a description of the rise and decline of industries and shows how the characteristics of the networks evolved over time. 
\item Next, regressions are used to identify the impact of both types of TP and DP on industry growth $A^{\mu}_{i,t}$ and innovation $A^{\tau}_{i,t}$. 
\item Then, the direction of technological change is analyzed through a series of regressions of productivity indicators (value added per employee $(VA/L)_{i,t}$ as measure of labor productivity, total factor productivity $TFP_{i,t}$), and indicators about the use of different production factors (employment $L_{i,t}$, wages $W_{i,t}$, capital intensity $(K/L)_{i,t}$, investment per capita $(I/L)_{i,t}$, the share of production labor $(L^P/L)_{i,t}$, and the relative wage paid for production labor $(W^P/W)_{i,t}$).
\item Finally, the analyses are repeated for different groups of industries that differ by innovation intensity $(A^{\tau}/A^{\mu})_{i}$, market size $A^{\mu}_{i}$, centrality in the patent network $PR^{\tau,dw}_{i}$, up- and downstream centrality in the market network $PR^{\mu,d}_{i}$, by broad industry group (defined by their 2-digit NAICS code), and by Pavitt industry group using the coding proposed by \citet{bogliacino2016pavitt}. 
\end{enumerate}


\section{Data}
\label{sec:data}
The two-layer network is inferred from two different data sets covering the US manufacturing sector during the period 1977-2012. The data is available in five-year snapshots. 
The market layer is compiled with national account data provided by the Bureau of Economic Analysis (BEA). 
The data is combined with data on patents granted by the US Patent and Trademark Office (USPTO), which are classified by the Cooperative Patent Classification (CPC) system. Using the concordance tables by \citet{goldschlag2020tracking}, 4-digit CPC codes are mapped to industries and aggregated into five-year windows in accordance with the IO data (see \ref{app:data_patent} for more detail).
This enables the compilation of a cross-industry patent citation network for different periods. 

The networks are given by symmetric matrices, where the entries are flows of goods $flow^{\mu}_{ij,t}$ and patent citations $flow^{\tau}_{ij,t}$. 
The data on cross-industrial flows is harmonized to shares. 
The networks are used to compile the industry level citation-weighted patent stocks $A^{\tau}_{i,t}$, output $A^{\mu}_{i,t}$, centrality $PR^{\alpha, d}_{i,t}$, and spillovers $Spill(A)^{\alpha, d}_{i,t}$. 

The network data is complemented with the Manufacturing Productivity Database from the National Bureau of Economic Research (NBER) and US Census Bureau's Center for Economic Studies (CES) \citep{becker2013nber, bartlesman1996nber}. The industry level variables extracted from this data include five factor productivity $TFP_{i,t}$, value added per employee $(VA/L)_{i,t}$ which is used as a proxy of labor productivity, employment $L_{i,t}$, labor costs per employee (hereafter called ``Wage'') $W_{i,t}$, investment per employee $(I/L)_{i,t}$, capital intensity $(K/L)_{i,t}$, the share of production workers $(L^P/L)_{i,t}$, and the relative wage for production labor $(W^P/W)_{i,t}$.\footnote{Production labor includes jobs that are related to tasks like fabricating, processing, maintenance, repair, product development, and similar activities. Non-production labor covers tasks associated with supervision, sales, delivery, advertising, finance, legal issues, and similar services (see \url{https://www2.census.gov/programs-surveys/asm/technical-documentation/questionnaire/2021/instructions/MA_10000_Instructions.pdf} [accessed on 14/11/2022]).}  
The data in monetary terms ($A^{\mu}_{i,t}$, $VA_{i,t}$, $W_{i,t}$, $W^P_{i,t}$, $K_{i,t}$, $I_{i,t}$) is deflated using the industry level price deflator for the value of shipment from the NBER database. 

The final data consists of a balanced panel of 307 6-digit manufacturing industries. 
More aggregate data is used for additional robustness checks. 
The most important steps of the data compilation are summarized in \ref{app:data}. Additional detail is provided in \ref{supp:data_processing}.  

\section{Results}
\label{sec:results}
This section begins with a descriptive analysis of the two network layers. 
In Sec. \ref{subsec:results_regression}, it follows a series of regressions studying the impact of TP and DP. 

\FloatBarrier
\subsection{Descriptive analysis}
\label{sec:results_descr}

\begin{figure}[h!]
\caption{Upstream networks at the 4-digit level for different periods}
\label{fig:fourDcomp_netw_flow_up}
\hspace{0cm}\textbf{1977-1992}\vspace{0.25cm}

\centering
\begin{subfigure}{0.45\textwidth}		
	{\centering
		\caption{Input-output}			
		\label{fig:fourDcomp_netw_flow_up_noOV_io92}			
		\includegraphics[width=\textwidth]{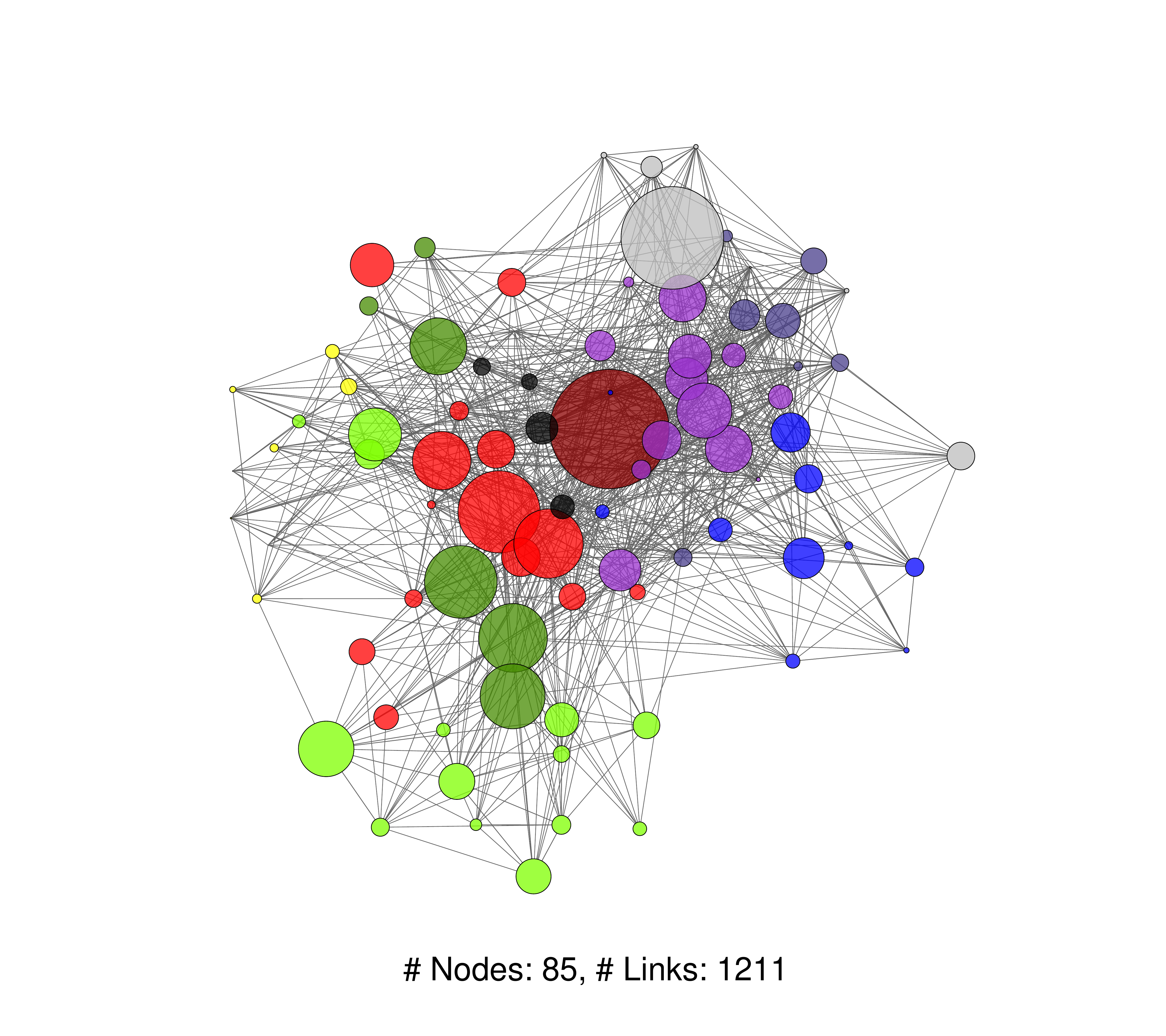}	
	}
\end{subfigure}
\begin{subfigure}{0.45\textwidth}		
	{\centering
		\caption{Patent citations}			
		\label{fig:fourDcomp_netw_flow_up_noOV_pat92}			
		\includegraphics[width=\textwidth]{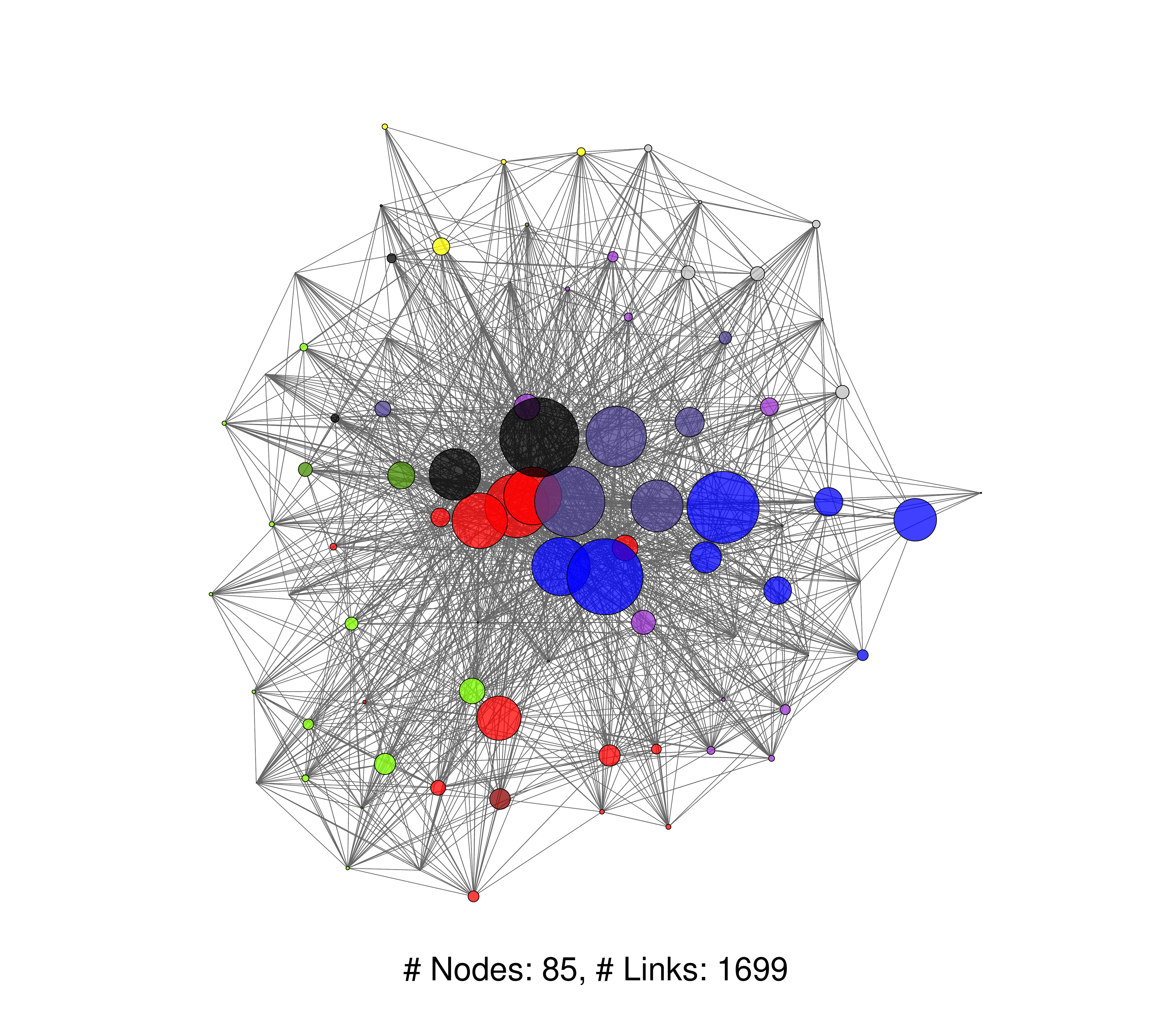}	
	}
\end{subfigure}

\hspace{0cm}\textbf{1997-2012}\vspace{0.25cm}

\centering
\begin{subfigure}{0.45\textwidth}		
	{\centering
		\caption{Input-output}			
		\label{fig:fourDcomp_netw_flow_up_noOV_io12}			
		\includegraphics[width=\textwidth]{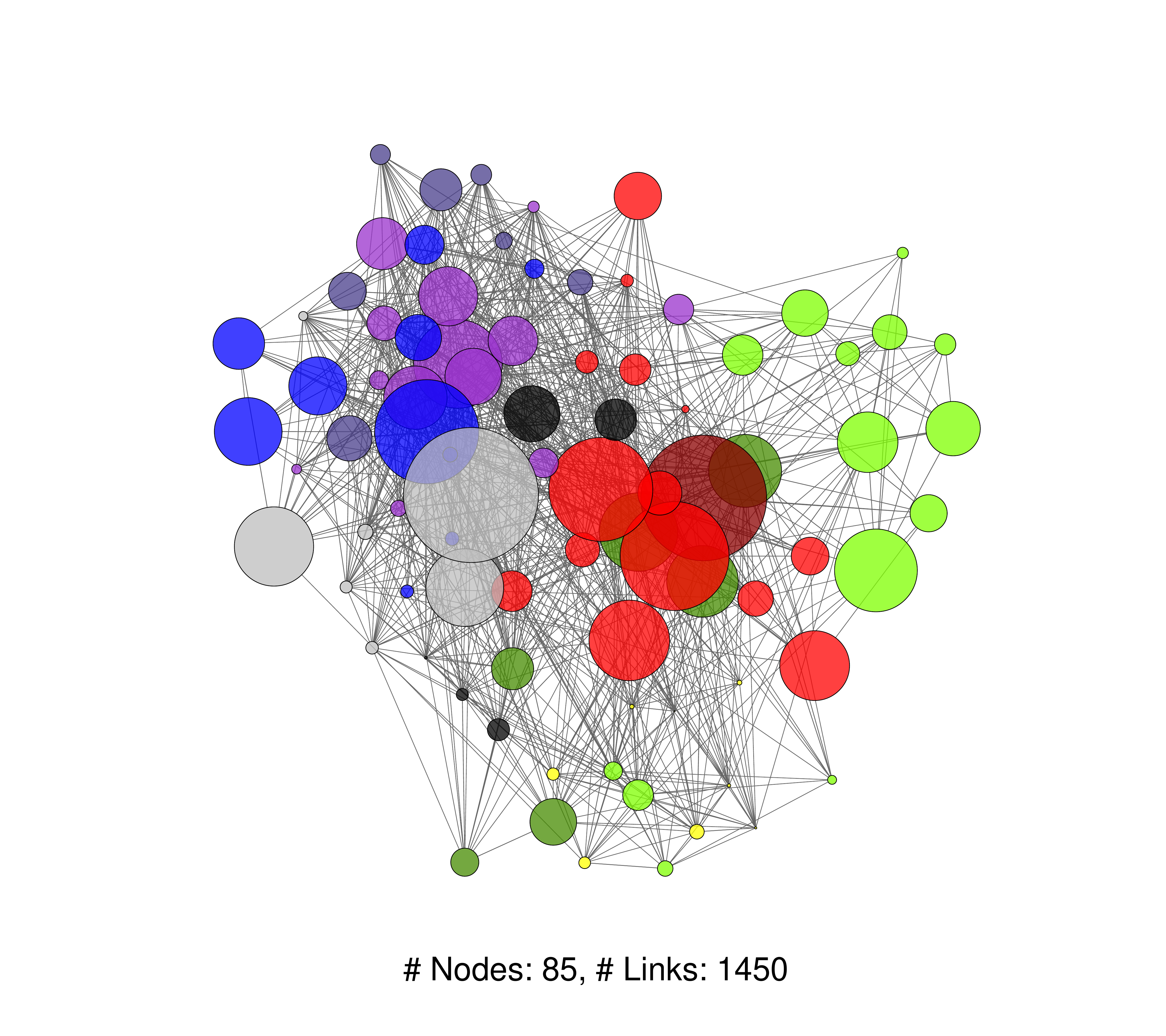}	
	}
\end{subfigure}
\begin{subfigure}{0.45\textwidth}		
	{\centering
		\caption{Patent citations}			
		\label{fig:fourDcomp_netw_flow_up_noOV_pat12}			
		\includegraphics[width=\textwidth]{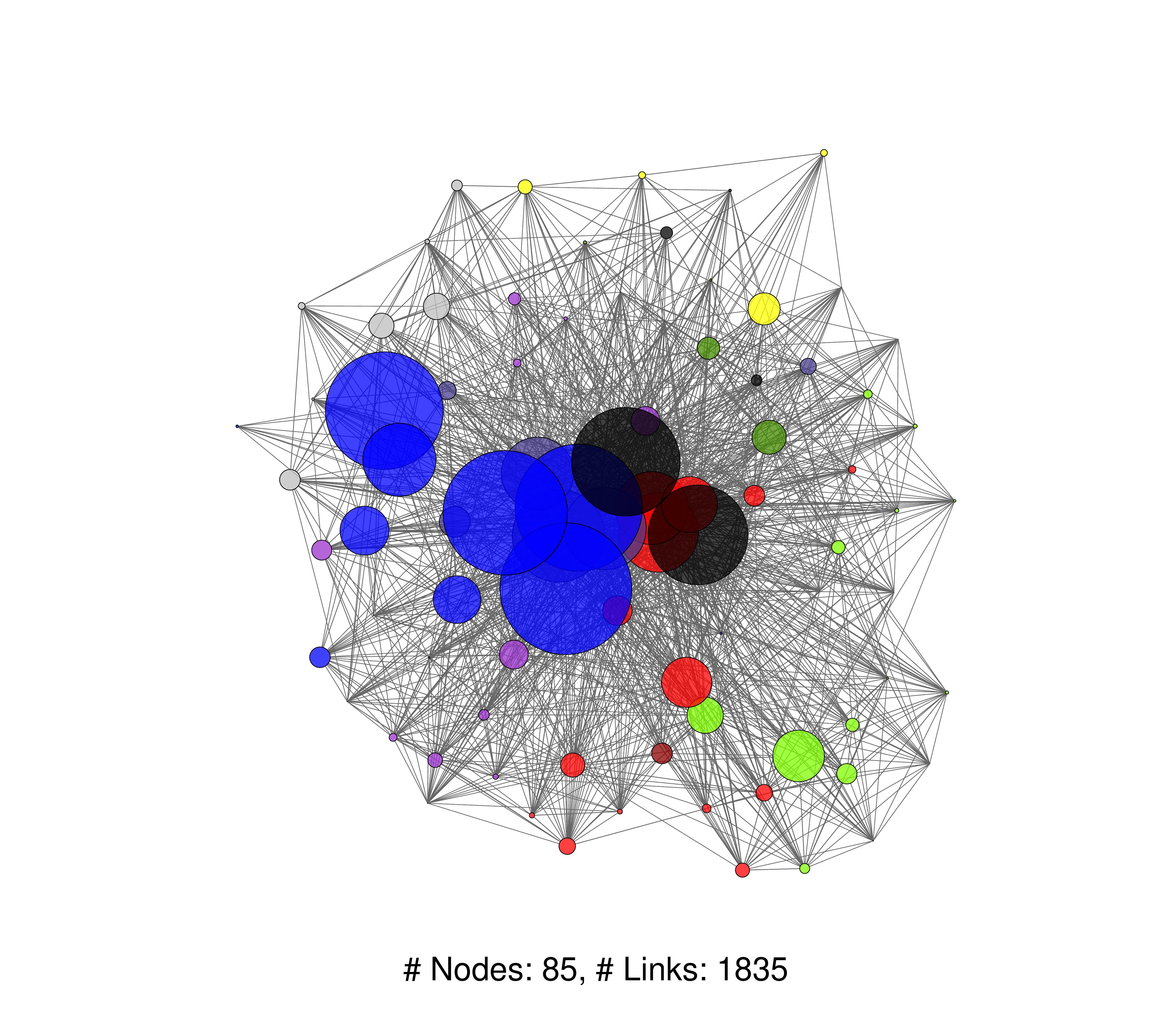}	
	}
\end{subfigure}

\includegraphics[width=1.05\textwidth]{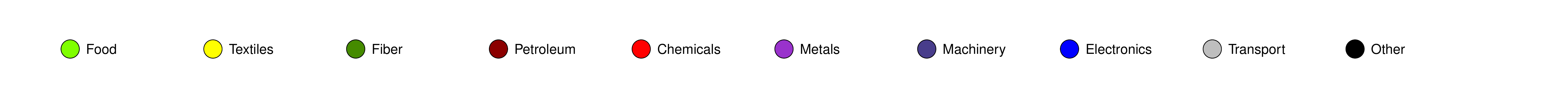}
\footnotesize 
\justifying \noindent
Notes: These figures show the network of upstream links (suppliers) at the 4-digit level for two different periods. Self-citations and within-sector IO flows are not shown. The colors indicate broad industrial categories given by groups of 3-digit level industries, i.e. Food (311-312), Textiles (314-316), Fiber (321-323), Petroleum (324), Chemicals (325-327), Metals (331,332), Machinery (333), Electronics (334-335), Transport (336), Other (337-339). A link between industries $i$ and $j$ is shown if $j$ is an important supplier to $i$, i.e. if the average of the weight $w^{in,\alpha}_{ij,t}$ during a subperiod exceeds a threshold level given by the average weight over all industry pairs and both subperiods plus one standard deviation ($\text{mean}(w^{in, \alpha}_{ij,t}) + \text{sd}(w^{in, \alpha}_{ij,t})$). 
The size of the nodes is proportional to the log size of an industry $A^{\alpha}_{i,t}$. 
Before taking logs, $A^{\alpha}_{i,t}$ had been scaled by its standard deviation over all industries and periods. This enables a comparison of sector sizes and their distribution over time and across layers. 
The figure is generated using the R-package \emph{igraph}, which makes use of the Fruchtermann-Reingold algorithm to allocate the nodes. The algorithm aims to minimize the crossing of links while keeping the length of the links similar. Technical detail is provided in \ref{app:network_plots}. 
\end{figure}

Fig. \ref{fig:fourDcomp_netw_flow_up} shows a series of upstream network plots at the 4-digit level for the first and the second half of the period. 
A link between two industries $i$ and $j$ is shown if the connecting weight $w^{\alpha,up}_{ij,t}$ is strong. 
The node sizes are proportional to the industry size $A^{\alpha}_{i,t}$. The algorithm that creates the plots tends to group industries with similar linking patterns together (see Notes in Fig. \ref{fig:fourDcomp_netw_flow_up} or \ref{app:network_plots} for technical details). 
The node colors indicate the broad industry group.  

This visual analysis shows a more balanced size distribution of industries in the market compared to the innovation layer, which is strongly dominated by electronics (blue), machinery (dark violet), and chemical manufacturing (red). 
The black color (``Other'') captures residual economic activities. The large size of black colored nodes in the innovation layer is an artifact of the concordance table, as this category captures many patents with CPC codes that are not straightforward to attribute to more specific industries.\footnote{``Other'' includes sectors with the 3-digit NAICS codes 337 and 339, which cover a wide range of residual manufacturing sectors ranging from special types of furniture, medical supplies, buttons, brushes, and miscellaneous other categories that do not fit well into the other 3-digit groups. Many of these industries map to a relatively high number of 4-digit CPC codes, which leads to a high number of patents attributed to these industries. The weighting scheme does not fully offset this effect. This does also affect the ranking of industries by patent count (see below in Table \ref{tab:sixDman_top10}). An illustrative example offers the sector ``Fastener, Button \& Pin'', which ranks very high by the number of patents. Its NAICS 6-digit code maps to almost 50 different 4-digit CPC codes, which is about five times more than the average number of CPC links per 6-digit industry. This industry falls into one of the residual categories starting with 33999.\label{footnote:concordance_issue_residual_category}
} 
The comparison over time suggests an increasing concentration in both layers. 

In the market layer, groups of industries with similar color tend to cluster together. 
The position of the clusters interacts with the position in the supply chain (``trophic level" \citep{mcnerney2018production}). Industries that are close to more primary resources (food processing (greenish), textiles (yellow)) or close to final consumers (food, textiles, electronics (blue), transportation (gray)) are located at the margins, while chemicals (red), metals (violet) and petroleum products (brown) with intermediate positions in the supply chain between primary input providers and end users take central positions. 

This pattern is different in the innovation layer where the largest nodes (electronics (blue), chemistry (red), and machinery (dark violet)) take the most central position, suggesting that they are important providers and users of innovations. 
It is also clearly visible that the electronics sector (including computer industries) grew most strongly over time. 

Statistics of the properties of the two network layers (see Table \ref{tab:sixDman_netw_stats_summary}) and the industry size rankings (Table \ref{tab:sixDman_top10}) confirm the visual impressions. 
The connectivity in the innovation layer is higher: An industry is connected to roughly 40-50 other industries by patent citations, but only 20-30 industries by IO links. 
While the characteristics of the IO network fluctuate without any clear trend, the innovation layer became increasingly connected and industries became more similar by their patent citation patterns. 
Both layers show a negatively valued, but stable assortativity: Larger and more connected industries tend to be linked to smaller and less connected industries (see \ref{tab:sixDman_netw_stats_summary}). 

\begin{landscape}

\begin{table}
	\caption{Top-10 ranking of industries by output and patent stock at the 6-digit level.} 
	\label{tab:sixDman_top10}
	\begin{myresizeenv}
		\begingroup \scriptsize
		\begin{tabular}{|p{0.01cm}p{0.7cm}p{3cm}p{0.7cm}p{0.7cm}p{3cm}p{0.7cm}p{0.7cm}p{3cm}p{0.7cm}p{0.7cm}p{3cm}p{0.8cm}|}
			\hline \hline \rule{0pt}{1.1\normalbaselineskip}
			& \multicolumn{ 12 }{l|}{\emph{Top-10 industries by Aggr. output ($A^{\mu}_{i,t}$) }} \\
			& \multicolumn{3}{c}{ 1977-1982 }&\multicolumn{3}{c}{ 1987-1992 } &\multicolumn{3}{c}{1997-2002} &\multicolumn{3}{c|}{2007-2012}\\
			\hline  \rule{0pt}{1.1\normalbaselineskip}			
			&324110 & Petroleum Refineries & 30.55 & 324110 & Petroleum Refineries & 17.24 & 324110 & Petroleum Refineries & 19.13 & 334413 & Semiconductor & 39.65 \\ 
			&324199 & Petroleum \& Coal Prod & 8.98 & 325211 & Plastics Material, Resin & 8.76 & 334413 & Semiconductor & 15.48 & 324110 & Petroleum Refineries & 20.15 \\ 
			&331111 & Iron$\text{ \&}$ Steel Mills & 8.28 & 311611 & Animal Slaughter & 7.57 & 331111 & Iron$\text{ \&}$ Steel Mills & 11.24 & 334111 & Electr Computer & 18.98 \\ 
			&311611 & Animal Slaughter & 8.13 & 324199 & Petroleum \& Coal Prod & 6.43 & 325211 & Plastics Material, Resin & 8.13 & 336112 & Light Utility Vhcl & 8.03 \\ 
			&325211 & Plastics Material, Resin & 7.36 & 331111 & Iron$\text{ \&}$ Steel Mills & 5.56 & 326191 & Plastics Plumb Fixture & 4.78 & 331111 & Iron$\text{ \&}$ Steel Mills & 7.44 \\ 
			&322121 & Paper Mills & 5.20 & 322130 & Paperboard Mills & 5.43 & 321113 & Sawmills & 4.73 & 325211 & Plastics Material, Resin & 5.36 \\ 
			&322122 & Newsprint Mills & 5.19 & 322122 & Newsprint Mills & 5.39 & 332710 & Machine Shops & 4.68 & 336391 & Motor Vhcl Air-Cond & 4.75 \\ 
			&325510 & Paint$\text{ \&}$ Coating & 4.54 & 322121 & Paper Mills & 5.38 & 334418 & Print Circuit Assembly & 4.09 & 336411 & Aircraft Mnft & 4.27 \\ 
			&322130 & Paperboard Mills & 4.41 & 325510 & Paint$\text{ \&}$ Coating & 4.82 & 311119 & Other Animal Food & 3.68 & 336111 & Automobile Mnft & 4.09 \\ 
			&327320 & Ready-Mix Concrete & 3.82 & 327320 & Ready-Mix Concrete & 3.98 & 322121 & Paper Mills & 3.61 & 336350 & Vhcl Power Train Parts & 3.38 \\ 				
	\hline \rule{0pt}{1.1\normalbaselineskip}  & \multicolumn{12}{l|}{Quartiles:}\\
			& \multicolumn{3}{c}{ 0.21, 0.54, 1.11 }& \multicolumn{3}{c}{ 0.21, 0.65, 1.3675 } & \multicolumn{3}{c}{0.22, 0.545, 1.19}& \multicolumn{3}{c|}{0.28, 0.46, 0.79}  \\
			\hline \hline \rule{0pt}{1.1\normalbaselineskip}
			& \multicolumn{ 12 }{l|}{\emph{Top-10 industries by Patent stock ($A^{\tau}_{i,t}$) }} \\
			& \multicolumn{3}{c}{ 1977-1982 }&\multicolumn{3}{c}{ 1987-1992 } &\multicolumn{3}{c}{1997-2002} &\multicolumn{3}{c|}{2007-2012}\\
			\hline  \rule{0pt}{1.1\normalbaselineskip}			
			&325520 & Adhesive Mnft & 16.23 & 334413 & Semiconductor & 19.33 & 334413 & Semiconductor & 25.97 & 334111 & Electr Computer & 30.12 \\ 
			&325998 & Misc Chem Prod & 15.31 & 325520 & Adhesive Mnft & 15.52 & 334111 & Electr Computer & 23.77 & 334413 & Semiconductor & 29.65 \\ 
			&334413 & Semiconductor & 13.73 & 334111 & Electr Computer & 14.87 & 339112 & Surgical \& Medical Instr & 14.09 & 339112 & Surgical \& Medical Instr & 15.08 \\ 
			&339993 & Fastener, Button, Pin & 11.62 & 325998 & Misc Chem Prod & 14.60 & 339993 & Fastener, Button, Pin & 13.60 & 333314 & Optical Instr$\text{ \&}$ Lens & 13.34 \\ 
			&334111 & Electr Computer & 9.47 & 339993 & Fastener, Button, Pin & 14.00 & 333314 & Optical Instr$\text{ \&}$ Lens & 12.48 & 334510 & Electromedical Instr & 11.79 \\ 
			&333314 & Optical Instr$\text{ \&}$ Lens & 9.05 & 333314 & Optical Instr$\text{ \&}$ Lens & 11.59 & 325520 & Adhesive Mnft & 12.33 & 339993 & Fastener, Button, Pin & 11.27 \\ 
			&333613 & Power Transmission Equ & 9.01 & 339112 & Surgical \& Medical Instr & 9.07 & 325998 & Misc Chem Prod & 11.48 & 334518 & Watch, Clock,$\text{ \&}$ Part & 11.14 \\ 
			&333911 & Pump$\text{ \&}$ Pumping Equ & 6.08 & 333613 & Power Transmission Equ & 8.04 & 334510 & Electromedical Instr & 11.02 & 334220 & Radio, TV, Communic & 11.04 \\ 
			&333612 & Speed Changer \& Gear & 5.99 & 334510 & Electromedical Instr & 7.61 & 334518 & Watch, Clock,$\text{ \&}$ Part & 9.47 & 325520 & Adhesive Mnft & 9.98 \\ 
			&334518 & Watch, Clock,$\text{ \&}$ Part & 5.67 & 334518 & Watch, Clock,$\text{ \&}$ Part & 6.57 & 334210 & Telephone Apparatus & 8.02 & 325998 & Misc Chem Prod & 9.23 \\ 
			
	\hline \rule{0pt}{1.1\normalbaselineskip}  & \multicolumn{12}{l|}{Quartiles:}\\
			& \multicolumn{3}{c}{ 0.0575, 0.33, 0.99 }& \multicolumn{3}{c}{ 0.06, 0.28, 0.8625 } & \multicolumn{3}{c}{0.06, 0.22, 0.7375}& \multicolumn{3}{c|}{0.05, 0.19, 0.6925} \\
			\hline
			\hline
		\end{tabular}
		\endgroup	
	\end{myresizeenv}
\vspace{0.25cm}
	\scriptsize \justifying
	
	\noindent
	Notes: Industries are ranked by deflated output (citation-weighted patent stock) $A^{\alpha}_{i,t}$ averaged across the time window indicated in the column header in decreasing order, i.e. showing the largest industries on top. The 6-digit number in the second column of each block shows the NAICS code of the corresponding industry, and the third column shows the value of $A^{\alpha}_{i,t}$.
	The values $A^{\alpha}_{i,t}$ were normalized before through division by the economy-wide average output (patent stock) in $t$, i.e. the mean value for each period equals one. 
	The bottom lines of each sub-table show the quartile values as indicators for the skewness of the distribution. 
	Deviations of the median from the average indicate skewness. 
\end{table}

\end{landscape}

Table \ref{tab:sixDman_top10} shows the Top-10 ranking of industries by output and patents. 
Petroleum Refineries (dark red color in the plots) rank persistently at the top position in the market until the 2007 when the semiconductor industry took over the first rank. We can also observe persistently high ranks for Iron \& Steel, Plastics Material \& Resin, and Semiconductors. Over time, industries associated with natural product processing (Paper Mills, Newsprint, Paperboard, Slaughtering) gradually disappeared from the top ranks. This was accompanied by the rise of machinery and electronics (Machine Shops, Aircraft, Vehicle Air-Conditioning, Automobile). 

The top ranks in the innovation layer are dominated by metal, machinery and electronics manufacturing, as indicated by the leading 2-digit code 33. Only two chemical industries rank high, but declined during the period covered (2-digit code 32). 
The patent ranks show an increasing dominance of ICT-related industries and specialist instruments (Semiconductors, Electronic Computers, Medical \& Optical Instruments, Watch \& Clock, Wireless Communication). 
The time trends in the patent ranking are more monotonous over time, while the rise of Semiconductors (during the 1990s) and Electronic Computers (2000s) came more abrupt in the market, as these two sectors did not occupy any of the Top-10 ranks during the decades before. 
The high rank of the ``Fastener, Button \& Pin'' industry is likely an artifact of the concordance table (see Footnote \ref{footnote:concordance_issue_residual_category}).

The bottom lines in the tables show the quartile distribution of the industry sizes. The data is normalized and the average size in each period equals one. A median value that deviates from one indicates a skewed distribution. 
Both layers show a skewed distribution with concentration at the top ranks. 
The table reveals a higher and increasing concentration in the innovation layer, while market concentration does not show any clear trend. 
These observations are consistent across different levels of aggregation (see \ref{app:add_descr_results} and \ref{supp:results_descr}).

\FloatBarrier
\subsection{Demand-pull, technology-push, and technological change}
\label{subsec:results_regression}
This section presents three sets of results showing first how different types of TP and DP affect industry growth and innovation, and second how these drivers interact with the direction of technological change reflected in the use of labor and capital. Finally, the results for different subgroups of industries are summarized.

\subsubsection{Demand-pull and technology-push}
\label{subsubsec:DP_and_TP}
To test the presence of DP and TP, the following regressions are run: 
\begin{align}
Y_{i,t} = \sum_{\alpha = \mu, \tau} \Bigl[ \beta^{\alpha}_{A} A^{\alpha}_{i,t-1} + \beta^{\alpha}_{PR} PR^{\alpha, d}_{i,t-1} + \beta^{\alpha, d}_{S} Spill(A)^{\alpha, d}_{i,t-1} \Bigr] + \mathbf{\beta}' \mathbf{X}_{i,t-1}
\end{align}
where $Y_{i,t}$ is a placeholder for output $A^{\mu}_{i,t}$ and patents $A^{\tau}_{i,t}$. $\mathbf{X}_{i,t-1}$ is a vector of industry level controls explained below. 
This specification allows testing for TP and DP in the following ways: 
\begin{description}
\item[Type 1---Between-layer effects:] Support for TP1 is found if lagged innovation is positively related to market growth $A^{\mu}_{i,t}$ reflected in significant, positive coefficients of $\beta^{\tau}_A$, $\beta^{\tau,d}_{PR}$, and $\beta^{\tau,d}_{S}$ in the regression of $A^{\mu}_{i,t}$. 

Analogously, support for DP1 is found if lagged market dynamics stimulate innovation $A^{\tau}_{i,t}$ reflected in significant, positive coefficients $\beta^{\mu}_A$, $\beta^{\mu,d}_{PR}$, and $\beta^{\mu,d}_{S}$ in the regression of $A^{\tau}_{i,t}$. 

\item[Type 2---Within-layer effects:] TP2 is supported if upstream supply shocks positively affect downstream growth, as reflected in significant, positive coefficients of $\beta^{\alpha,up}_{PR}$ and $\beta^{\alpha,up}_{S}$ in the regression of $A^{\alpha}_{i.t}$ within the same layer $\alpha$. 

Support for DP2 is found if downstream changes in demand patterns stimulate upstream growth and $\beta^{\alpha,dw}_{PR}$ and $\beta^{\alpha,dw}_{S}$ show positive and significant coefficients in the regression of $A^{\alpha}_{i,t}$.

\end{description}

A few technical details need to be mentioned. 
The up- and downstream PageRank in the innovation layer are highly correlated, with a correlation coefficient of 98.4\% (see Fig. \ref{fig:sixDcomp_correl_plots}). To avoid multicollinearity, the analysis includes only $PR^{\tau,dw}_{i,t-1}$, which is not informative about within-innovation DP2. To detect within-innovation TP2 and DP2, this study relies exclusively on up- and downstream spillovers $Spill(A)^{\tau, d}_{i,t-1}$. 
However, the innovation centrality $PR^{\tau,dw}_{i,t}$ is informative about TP1 effects as it indicates the importance of $i$'s innovation outputs for other industries, and it can be considered as an indicator for the quality of $i$'s innovations.

To deal with unobserved cross-industrial heterogeneity and uniform time trends, the regressions include two-ways time and industry fixed effects (FE). The lagged dependent variables $A^{\alpha}_{i,t-1}$ capture path dependence and industry level time trends in market growth and innovation, that are not driven by the two types of TP and DP under study.\footnote{Note that the impact of lagged changes of market size $A^{\mu}_{i,t-1}$ on current market size could be interpreted as an alternative type of within-industry DP effect as an increasing $A^{\mu}_{i,t}$ suggests a higher demand for the products of $i$. In the innovation layer, the impact of past on current innovation could be seen as an alternative type of TP when own technological breakthroughs stimulate subsequent innovation. This effect is not included in the conceptual framework to limit the scope of analysis and discussion of concepts. Instead, the impact of the autoregressive terms is referred to as path dependence. Between-layer DP and TP effects could be also conceptualized contemporaneously without a time lag interpreting DP1 as impact of current innovation on current market growth and TP1 as impact of current innovation on current market size. Here, this approach is not followed, and technological change is considered as a dynamic process, with past changes having an impact on future development. In a contemporaneous setting, the causal mechanisms would operate through expectations, as both processes (innovation and diffusion) are sluggish.}

Combining lagged dependent variables with individual FE violates the assumption of strict exogeneity as the demeaning process of the FE causes a correlation of the lagged regressor with the error term \citep{nickell1981biases}. This bias may be non-negligible, as the number of periods is small while the number of industries is large. 
The endogeneity can be overcome by using an instrumental variable (IV) approach (\citet{arellano1991some} (AB), \citet{blundell1998initial} (BB)). The main analyses presented in this section rely on the BB estimator with robust standard errors.\footnote{The AB and BB estimator rely on deeper lags as instruments for the lagged dependent variable to overcome the endogeneity. The BB specification is chosen because the BB model builds on stronger instruments compared to an AB estimator. It estimates the outcome variable in levels, includes time dummies, and it is estimated using a one-step procedure. The one-step method is preferred over a two-step routine for computational reasons. The maximal number of lags and all exogenous variables are used as IVs. To avoid concerns about instrument proliferation, the model is re-estimated with collapsed IVs as a robustness check. Additional robustness checks include, for example, deeper lags, one- and two-step estimation procedures and different specifications of the regression equation and empirical variables. The results are qualitatively robust, but sometimes with changing significance levels. Additional results from an AB estimator and a weighted FE regression are available in \ref{app:results_DP_TP}. The regressions are run using the \emph{pgmm} function of the R-package \emph{plm} \citep{croissant2008panel}.}

Market growth and innovation may also be driven by industry-specific structural changes and price shocks that are unrelated to TP and DP. The control variables $\mathbf{X}_{i,t-1}$ aim to capture these effects. The controls include employment $L_{i,t-1}$, wages $W_{i,t-1}$, capital intensity $(K/L)_{i,t-1}$, the employment share of production labor $(L^P/L)_{i,t-1}$, relative wages for production labor $(W^P/W)_{i,t-1}$, energy intensity $(E/L)_{i,t-1}$, and material costs per employee $(M/L)_{i,t-1}$. These variables control for the industry-specific and time variant exposure to changes in factor markets, which are not captured by industry and time FE. 
The controls also include lagged investment per capita $(I/L)_{i,t-1}$, as it might be the source of an increase in outputs and innovation. 
To cope with skewness, all variables (except for $(L^P/L)_{i,t-1}$) were log-linearized and outliers were removed. 
The network variables (node size $A^{\alpha}_{i,t}$, centrality $PR^{\alpha,d}_{i,t}$, spillovers $Spill(A)^{\alpha,d}_{i,t}$) are scaled by their standard deviation before logs were taken to make them quantitatively comparable. A detailed description of the data transformations is available in \ref{app:regression_data}. 

\begin{table}[H]
\begin{myresizeenv}
	\begingroup
	\begin{tabular}{|l|cccc|cccc|cccc|}
		\hline \hline  \rule{0pt}{1.075\normalbaselineskip}  
		& \multicolumn{4}{c|}{ \ul{Type 1}} & \multicolumn{4}{c|}{ \ul{Type 2}} & \multicolumn{4}{c|}{ \ul{Both}}\\  & \multicolumn{2}{c}{ $\tau \rightarrow \mu$} & \multicolumn{2}{c|}{ $\mu \rightarrow \tau$}
		& \multicolumn{2}{c}{ Market} & \multicolumn{2}{c|}{ Innovation}
		& \multicolumn{4}{c|}{}\\
		\hline  \rule{0pt}{1.075\normalbaselineskip}    & $A^{\mu}_{i,t}$ & $A^{\mu}_{i,t}$ & $A^{\tau}_{i,t}$ & $A^{\tau}_{i,t}$ & $A^{\mu}_{i,t}$ & $A^{\mu}_{i,t}$ & $A^{\tau}_{i,t}$ & $A^{\tau}_{i,t}$ & $A^{\mu}_{i,t}$ & $A^{\mu}_{i,t}$ & $A^{\tau}_{i,t}$ & $A^{\tau}_{i,t}$ \\ 
		& (1) & (2) & (3) & (4) & (5) & (6) & (7) & (8) & (9) & (10) & (11) & (12) \\ 
		\hline \rule{0pt}{1.075\normalbaselineskip}   $A^{\mu}_{i,t-1}$ & 0.6255*** & 0.6073*** & 0.0035 & -0.0058 & 0.7321*** & 0.6514*** &  &  & 0.6867*** & 0.6603*** & 0.0021 & -0.0058 \\ 
		& (0.0291) & (0.0242) & (0.0136) & (0.0127) & (0.0385) & (0.034) &  &  & (0.039) & (0.0344) & (0.0128) & (0.0117) \\ 
		$A^{\tau}_{i,t-1}$ & -0.0637 & -0.0246 & 1.034*** & 1.016*** &  &  & 1.044*** & 1.034*** & -0.0197 & -0.0351 & 1.061*** & 1.049*** \\ 
		& (0.0947) & (0.0849) & (0.0076) & (0.0068) &  &  & (0.0403) & (0.0314) & (0.0862) & (0.0778) & (0.0359) & (0.0303) \\ 
		$PR^{\mu,up}_{i,t-1}$ &  &  & -0.0055 & -1e-04 & 0.0066 & 0.01 &  &  & 0.0088 & 0.0047 & -0.0062 & -0.0021 \\ 
		&  &  & (0.0046) & (0.0041) & (0.015) & (0.0131) &  &  & (0.0145) & (0.0127) & (0.0046) & (0.0039) \\ 
		$PR^{\mu,dw}_{i,t-1}$ &  &  & 0.0021 & 0.0041 & -0.001 & 0.0036 &  &  & -0.0039 & 0.005 & 9e-04 & 0.0039 \\ 
		&  &  & (0.0033) & (0.0031) & (0.0145) & (0.0121) &  &  & (0.0137) & (0.0118) & (0.0033) & (0.003) \\ 
		$PR^{\tau,dw}_{i,t-1}$ & 0.1536** & 0.0928* &  &  &  &  & 0.0482** & 0.0314* & 0.1119** & 0.0924** & 0.031* & 0.0183. \\ 
		& (0.0478) & (0.0391) &  &  &  &  & (0.0167) & (0.0136) & (0.0376) & (0.0322) & (0.013) & (0.0109) \\ 
		$Spill(A)^{\mu,up}_{i,t-1}$ &  &  & -0.0126* & -0.0081 & 0.1054*** & 0.1043*** &  &  & 0.0975*** & 0.0868*** & -0.0133* & -0.0088 \\ 
		&  &  & (0.006) & (0.006) & (0.02) & (0.0189) &  &  & (0.0207) & (0.0195) & (0.0066) & (0.0063) \\ 
		$Spill(A)^{\mu,dw}_{i,t-1}$ &  &  & 0.0044 & 0.006 & -0.0394. & -0.007 &  &  & -0.0429* & -0.0178 & 0.002 & 0.0068 \\ 
		&  &  & (0.0058) & (0.005) & (0.0228) & (0.0192) &  &  & (0.02) & (0.0181) & (0.0053) & (0.0048) \\ 
		$Spill(A)^{\tau,up}_{i,t-1}$ & 2.435*** & 2.572*** &  &  &  &  & 0.7004** & 0.3476* & 2.363*** & 2.642*** & 0.4757** & 0.2599. \\ 
		& (0.7017) & (0.5889) &  &  &  &  & (0.2131) & (0.1672) & (0.6273) & (0.5391) & (0.156) & (0.1388) \\ 
		$Spill(A)^{\tau,dw}_{i,t-1}$ & -0.0606 & -0.0573 &  &  &  &  & -0.0539*** & -0.0369** & -0.0695. & -0.059. & -0.0514*** & -0.0378** \\ 
		& (0.0417) & (0.035) &  &  &  &  & (0.016) & (0.013) & (0.0369) & (0.0337) & (0.0144) & (0.0126) \\ 
		\hline \rule{0pt}{1.075\normalbaselineskip}  AR(1) & 0 & 0 & 1e-04 & 2e-04 & 0 & 0 & 7e-04 & 5e-04 & 0 & 0 & 5e-04 & 7e-04 \\ 
		AR(2) & 0.9373 & 0.9079 & 0.899 & 0.8057 & 0.9237 & 0.9583 & 0.6394 & 0.7465 & 0.9656 & 0.978 & 0.7307 & 0.7026 \\ 
		Sargan & 0 & 0 & 0 & 0.001 & 0 & 0 & 0 & 0 & 0 & 1e-04 & 0 & 0.0056 \\ 

		Controls &  & Y &  & Y &  & Y &  & Y &  & Y &  & Y \\ 
		$R^2$ & 0.9284 & 0.9354 & 0.996 & 0.9962 & 0.9363 & 0.9402 & 0.9952 & 0.9958 & 0.9332 & 0.9374 & 0.9957 & 0.996 \\ 
	\hline
		\hline
	\end{tabular}
	\endgroup
	\caption{Demand-pull and technology-push effects.}
	\label{tab:real_sum_output|citation_weighted_patent_stock_all_sectors_all_years_sys_1step_wControls}
\end{myresizeenv}

\vspace{0.25cm}

\justifying \footnotesize

\noindent
Notes: The table shows the regression results of output $A_{i,t}^{\mu}$ and patents $A_{i,t}^{\tau}$ on demand-pull and technology-push effects. The estimation is based on a two-ways Blundell-Bond (BB) system GMM model using a one-step estimation procedure. 
The controls included in all regressions are wages $W_{i,t}$, capital intensity $(K/L)_{i,t}$, investment per capita $(I/L)_{i,t}$, relative wages for production labor $(W^P/W)_{i,t}$, energy intensity $(E/L)^{}_{i,t-1}$, and material inputs per capita $(M/L)^{}_{i,t-1}$.   
Spillovers are calculated on the basis of first-order links. Variables measured in monetary terms are deflated using the industry level price deflators for the value of shipment obtained from the NBER-productivity database \citep{becker2013nber}. To cope with skewness and to obtain tractable coefficients, most variables are pre-processed (taking logs, removing outliers, scaling). Data in logs are patents and output $A^{\alpha }_{i,t}$, centrality $PR^{\alpha, d }_{i,t}$, spillovers $Spill(A)^{\alpha, d }_{i,t}$, employment $L_{i,t}$,  $(K/L)_{i,t}$, $(I/L)_{i,t}$, $W_{i,t}$, $(E/L)^{}_{i,t-1}$, $(M/L)^{}_{i,t-1}$ with $\alpha = \mu, \tau$ and $d = up, dw$. $A^{\alpha }_{i,t}$, $PR^{\alpha, d }_{i,t}$, and $Spill(A)^{\alpha, d }_{i,t}$ are scaled by division by their standard deviation to obtain comparable coefficients across the different network effects. A detailed description of the transformations and descriptive statistics of the regression data before and after the transformations are provided in \ref{app:data}. The rows AR(1), AR(2), and Sargan show the test statistics of the specification tests, i.e. testing for first- and second-order autocorrelation and the results of a Sargan test for validity of instruments \citep[see][]{roodman2009xtabond2}. 
The $R^2$ is proxied by the squared correlation between the fitted and observed values of the dependent variable. 

\end{table}

The regression results are presented in Table \ref{tab:real_sum_output|citation_weighted_patent_stock_all_sectors_all_years_sys_1step_wControls}. Even numbered columns show the results when industry level controls $\mathbf{X}_{i,t-1}$ are included. Their coefficients are not shown here but can be found in \ref{app:results_DP_TP}.\footnote{Among the controls, only wages, investment per capita, and energy intensity are significant. Wages exhibit a negative association with market growth, energy intensity a negative one with innovation, and investment per capita shows a positive relationship with both market growth and innovation.}

The first two columns show the isolated impact of TP1 on market growth. The coefficients of the patent centrality $PR^{\tau,dw}_{i,t-1}$ and upstream innovation spillovers $Spill(A)^{\tau,dw}_{i,t-1}$ enter both with positive and significant coefficients. This observation persists when within layer effects are included (column (9)-(10)). An increase in the normalized centrality by one percent is associated with an 0.1 percent increase in the normalized market size. Innovation spillovers exhibit a quantitatively stronger effect: A one percent increase is associated with 2.6 percent increase in market size. 
Note that the network variables $A^{\alpha}_{i,t}$, $PR^{\alpha,d}_{i,t}$, $Spill(A)^{\alpha,d}_{i,t}$ are all scaled and similar by their means (see Table \ref{app:regression_data}). This improves the comparability of the coefficients. However, the variation of the spillover measure is very low and small deviations may have large effects, which needs to be taken into consideration when comparing the coefficients.

The results show that there is no direct effect of within-sector innovation outputs $A^{\tau}_{i,t-1}$ on market growth when controlling for other innovation network effects. However, exploratory analyses have shown that $A^{\tau}_{i,t-1}$ exhibits a positive effect when other innovation layer effects are excluded. This may be explained by the high correlation between $A^{\tau}_{i,t-1}$ and $PR^{\tau,dw}_{i,t-1}$, while the centrality is more powerful in explaining market growth. 

A rising patent centrality $PR^{\tau, dw}_{i,t-1}$ indicates that $i$'s innovations became increasingly relevant in the innovation network, which may be considered as a quality indicator of $i$'s innovations. Rising spillovers $Spill(A)^{\tau,d}_{i,t-1}$ suggest an increase in technological opportunities arising from neighboring industries. Both measures can be interpreted as an increase of technological opportunities, and thus the positive coefficients in column (1)-(2) and (9)-(10) offer support for TP1. This effect is mostly driven by the network effects, but less from the innovations produced by the industry itself. Note that the industry that produces innovations is not necessarily the main user of these inventions. This may explain why TP1 mainly arises from network effects and less from within-sector innovation effects.
Downstream innovation spillovers $Spill(A)^{\tau,dw}_{i,t-1}$ enter with a negative coefficient, which is only weakly significant and becomes insignificant when within-market effects are added (column (10)). 

Column (3)-(4) show the isolated effect of DP1 on innovation. None of the market variables ($A^{\mu}_{i,t-1}$, $PR^{\mu,d}_{i,t-1}$, $Spill(A)^{\mu,d}_{i,t-1}$) shows any significant interaction with innovation $A^{\tau}_{i,t}$, suggesting the absence of DP1 effects. This does not change when simultaneously controlling for within-innovation layer dynamics (column (11)-(12)) or altering the model specification (see \ref{app:results_DP_TP}). 

The next two columns (5)-(6) show the results for within-market DP2 and TP2 effects. 
They offer support for TP2 in the market layer: Spillovers from upstream suppliers $Spill(A)^{\mu,up}_{i,t-1}$ enter with a significant positive coefficient. 
Market growth in upstream industries can be interpreted as a positive supply shock, which is positively correlated with downstream growth. This finding persists when between-layer effects are added (column (12)-(13)). 
The effect is small: Compared to TP1 arising from innovation in upstream industries, it plays an economically less important role. 
Within the market, there is no evidence of DP2 effects arising from downstream customers.\footnote{Type 2 effects are not robust if a weighted FE instead of an AB or BB estimator is used. In an FE specification (see e.g. Table \ref{tab:real_sum_output|citation_weighted_patent_stock_all_sectors_all_years_FE_weighted_wControls} in \ref{app:results_DP_TP}) market up- and downstream centrality enter with significant positive coefficients while the effect of spillovers diminishes.}
Column (7)-(8) inform about the existence of within-innovation DP2 and TP2. 
The results offer support for TP2: Upstream spillovers $Spill(A)^{\tau,up}_{i,t-1}$ enter with a small but positive coefficient. 
This observation is robust across different estimation methods (AB, FE) with even more significant effects.

The empirical findings contradict the existence of within-innovation DP2: Downstream innovation spillovers $Spill(A)^{\tau,dw}_{i,t-1}$ show a negative coefficient, which is not significant in an AB and weighted FE model (see \ref{app:results_DP_TP}). 
The positive effect of upstream spillovers quantitatively dominates and is about six times larger than the negative effect of $Spill(A)^{\tau,dw}_{i,t-1}$. 

Beyond TP and DP, the results in Table \ref{tab:real_sum_output|citation_weighted_patent_stock_all_sectors_all_years_sys_1step_wControls} show that growth in both layers is path dependent: The lagged dependent variable $A^{\alpha}_{i,t-1}$ enters with a positive and strongly significant coefficient. Path dependence is stronger in innovation compared to the market and enters with an autocorrelation coefficient, which is larger than one. 
One percent more patents $A^{\tau}_{i,t-1}$ is associated with approximately 1.05 percent more innovation five years later. The coefficient of output $A^{\mu}_{i,t-1}$ scores persistently below one, suggesting a decreasing explanatory power of the autoregressive term over time. This indicates increasing returns to innovation, which is in line with the observation that concentration in the innovation layer grew stronger compared to the market (see Sec. \ref{sec:results_descr}). 

\FloatBarrier
\subsubsection{The direction of technological change}
\label{subsubsec:direction_of_change}

This section studies whether and how TP and DP interact with the direction of technological change. 
This is analyzed through a series of regression analyses showing how productivity, labor demand, capital use, and production labor interact with TP and DP effects arising from both layers. 

The dependent variables are not directly associated with one of the two layers, which inhibits the distinction of within- and between-layer effects. Instead, it is referred to DP (TP) whenever an effect is driven by the market (innovation) layer. 

The regression models look similar as above and are given by 
\begin{align}
Y_{i,t} = \beta_{Y} Y_{i,t-1} + \sum_{\alpha = \mu, \tau} \Bigl[ \beta^{\alpha}_{A} A^{\alpha}_{i,t-1} + \beta^{\alpha}_{PR} PR^{\alpha, d}_{i,t-1} + \beta^{\alpha, d}_{S} Spill(A)^{\alpha, d}_{i,t-1} \Bigr] + \mathbf{\beta}' \mathbf{X}_{i,t-1}
\end{align}
where $Y_{i,t}$ is a placeholder for productivity measured as $TFP_{i,t}$, and value added per worker $(VA/L)_{i,t}$, labor $L_{i,t}$, wages $W_{i,t}$, capital intensity $(K/L)_{i,t}$, per capita investments $(I/L)_{i,t}$, the share of production labor $(L^P/L)_{i,t}$, and the relative wage for production labor $(W^P/W)_{i,t}$. $\mathbf{X}_{i,t-1}$ is a vector of industry level controls explained above. 

	\begin{table}[]
		\begin{myresizeenv}
			\begingroup
			\begin{tabular}{|l|cc|cc|cc|cc|}
				\hline \hline  \rule{0pt}{1.075\normalbaselineskip}  
				& \multicolumn{2}{c|}{ {Productivity}} 
				& \multicolumn{2}{c|}{ {Labor \& Wages}} 
				& \multicolumn{2}{c|}{ {Capital \& Investment}} 
				& \multicolumn{2}{c|}{ {Production labor}} \\
				\hline  \rule{0pt}{1.075\normalbaselineskip}   & $TFP_{i,t}$ &  $(VA/L)_{i,t}$ & $L_{i,t}$  & $W_{i,t}$  & $(K/L)_{i,t}$ & $(I/L)_{i,t}$ & $(L^P/L)_{i,t}$ & $(W^P/W)_{i,t}$ \\ 
				\rule{0pt}{1.075\normalbaselineskip}  & (1) & (2) & (3) & (4) & (5) & (6) & (7) & (8)  \\ 
				\hline \rule{0pt}{1.075\normalbaselineskip}   
				$A^{\mu}_{i,t-1}$ &  0.0275. &  -0.1469*  & -0.0895* & -0.2082***  & 0.0492* & -0.1578. & -0.0149*** &  -0.0186* \\ 
				 & (0.0159) & (0.0629) & (0.0438) & (0.0527) & (0.0213) & (0.0923) & (0.0043) & (0.0075) \\ 
				$A^{\tau}_{i,t-1}$ & 0.0096 & -0.0482 &  -0.019 & -0.0632 & 0.0924 & -0.1016 & -0.0265* & -0.0306 \\ 
				& (0.0446) & (0.1675) & (0.1171) & (0.1402) & (0.0568) & (0.2458) & (0.0115) & (0.0201) \\ 
				$PR^{\mu,up}_{i,t-1}$ & 0.0082. & 0.047* & 0.0223. & 0.0596*** & -0.0048 & 0.0602* & 0.0013 & 0.0069** \\ 
				& (0.0048) & (0.0196) & (0.0136) & (0.0164) & (0.0067) & (0.0288) & (0.0013) & (0.0024) \\ 
				$PR^{\mu,dw}_{i,t-1}$ & 0.0123** & 0.057*** & 0.0265* & 0.0398** & -0.0096. & 0.0619** & -1e-04 & 0 \\ 
				& (0.0041) & (0.0163) & (0.0112) & (0.0136) & (0.0055) & (0.0239) & (0.0011) & (0.0019) \\ 
				$PR^{\tau,dw}_{i,t-1}$ & -0.0166 & -0.0463 & 0.0136 & 0.0841 & 0.0384 & 0.1945. & -0.0064 & 0.0265** \\ 
				& (0.0192) & (0.0745) & (0.0514) & (0.0623) & (0.0252) & (0.1093) & (0.0051) & (0.0089) \\ 
				$Spill(A)^{\mu,up}_{i,t-1}$ & -0.0117 & -0.0815* & -0.0555. & -0.0672. & 0.0319* & -0.1051. & -0.0025 & 0.0088. \\ 
				& (0.0097) & (0.0411) & (0.0284) & (0.0344) & (0.0139) & (0.0604) & (0.0028) & (0.0049) \\ 
				$Spill(A)^{\mu,dw}_{i,t-1}$ & -0.0135* & -0.0176 & -0.0079 & -0.0085 & -0.0035 & -0.0165 & 7e-04 & 0.0033 \\ 
				& (0.0062) & (0.0252) & (0.0175) & (0.0212) & (0.0086) & (0.0371) & (0.0017) & (0.003) \\ 
				$Spill(A)^{\tau,up}_{i,t-1}$ & -0.7479*** & -1.207 & -0.4783 & -1.336* & 0.0448 & 1.124 & -0.201*** & -0.2156* \\ 
				& (0.2002) & (0.7875) & (0.5451) & (0.661) & (0.2676) & (1.159) & (0.0543) & (0.0946) \\ 
				$Spill(A)^{\tau,dw}_{i,t-1}$ & -0.0225 & 0.0382 & 0.0322 & 0.0451 & -0.0286 & 0.1815. & 0.0043 & -0.0041 \\ 
				& (0.0167) & (0.066) & (0.0459) & (0.0554) & (0.0224) & (0.0971) & (0.0045) & (0.0079) \\ 
				\hline \hline \rule{0pt}{1.075\normalbaselineskip}   
				Controls &  Y & Y & Y & Y & Y & Y & Y &  Y \\ 
				$R^2$ & 0.7746 & 0.9024 & 0.9359 & 0.927 & 0.927 & 0.8339 & 0.8832 & 0.6776 \\ 
				\hline
				\hline
			\end{tabular}
			\endgroup
			\caption{Productivity, labor, capital, and production labor} 
			\label{tab:TFP|real_vadd|emp|real_pay|cap|real_invest|share_prode|real_relative_prodw_all_sectors_all_years_FE_weighted_wControls}
		\end{myresizeenv}
		\vspace{0.25cm}
		
		\justifying \footnotesize
		
		\noindent
		Notes: The table shows the regression results of productivity, labor demand, capital use, and production labor on demand-pull and technology-push effects. The estimation is based on a two-ways weighted fixed-effects (FE) model. The weights used in the regressions are $A^{\mu}_{i,t}$ in the TFP regression and $L_{i,t}$ in all other regressions. Each regression include the respective lagged dependent variable as a control. The controls included in all regressions are wages $W_{i,t}$, capital intensity $(K/L)_{i,t}$, investment per capita $(I/L)_{i,t}$, relative wages for production labor $(W^P/W)_{i,t}$, energy intensity $(E/L)^{}_{i,t-1}$, and material inputs per capita $(M/L)^{}_{i,t-1}$.   
		Spillovers are calculated on the basis of first-order links. Variables measured in monetary terms are deflated using the industry level price deflators for the value of shipment obtained from the NBER-productivity database \citep{becker2013nber}. To cope with skewness and to obtain tractable coefficients, most variables are pre-processed (taking logs, removing outliers, scaling). Data in logs are patents and output $A^{\alpha }_{i,t}$, centrality $PR^{\alpha, d }_{i,t}$, spillovers $Spill(A)^{\alpha, d }_{i,t}$, employment $L_{i,t}$,  $(K/L)_{i,t}$, $(I/L)_{i,t}$, $W_{i,t}$, $(E/L)^{}_{i,t-1}$, $(M/L)^{}_{i,t-1}$ with $\alpha = \mu, \tau$ and $d = up, dw$. $A^{\alpha }_{i,t}$, $PR^{\alpha, d }_{i,t}$, and $Spill(A)^{\alpha, d }_{i,t}$ are scaled by division by their standard deviation to obtain comparable coefficients across the different network effects. A detailed description of the transformations and descriptive statistics of the regression data before and after the transformations are provided in \ref{app:data}. 
	\end{table}
	
The regression models again include lagged dependent variables and may be subject to a dynamic panel bias \citep{nickell1981biases}. 
This bias may be addressed using a BB or AB estimator as before, but these estimators suffer from weak instruments. Hence, a weighted FE approach with industry and time FE and two-ways clustered standard errors are used. The weights in the regression of $TFP_{i,t}$ are the market size $A^{\mu}_{i,t}$ to capture the impact on productivity for an average good produced. All other regressions rely on employment $L_{i,t}$ as weights to capture the different relevance of industries for the US labor market. Hence, the results inform on how the labor productivity, wages, or capital intensity of an average employee are affected instead of capturing the effect on an average industry. 
Results using alternative estimators are reported in \ref{app:results_direction}. The findings are qualitatively consistent across the different specifications, sometimes with variation in their statistical significance. 

The results are shown in Table \ref{tab:TFP|real_vadd|emp|real_pay|cap|real_invest|share_prode|real_relative_prodw_all_sectors_all_years_FE_weighted_wControls}. To keep it short, only the results of the full models that include market and patent layer effects simultaneously are presented.\footnote{Additional results with separate layers can be found in the \ref{supp:regression_direction_separate_layers}.}  
All regressions include industry level controls $\mathbf{X}_{i,t-1}$, which are not shown here but can be found in \ref{app:results_direction}. 

Columns (1)-(2) show the results for productivity measured as five factor TFP and value added per employee. Columns (3)-(4) inform about labor demand as reflected by employment and wages. Columns (5)-(6) show the effects on capital use and investment intensity. The last two columns (7)-(8) show to which extent an industry relies on production and non-production labor.

An increasing downstream centrality in the market $PR^{\mu,dw}_{i,t-1}$ shows a positive association with productivity growth and a factor bias in favor of labor. It is positively correlated with labor demand and wages (column (3)-(4)), negatively with capital use (column (5)), and positively with investments per capita (column (6)). 
The positive impact on labor demand appears to dominate the investment effect, leading to a net negative relationship between market downstream centrality and capital use per employee. 
Market upstream centrality $PR^{\mu,up}_{i,t-1}$ shows similar effects, namely a positive correlation with wages and employment, relative wages for production labor, labor productivity, and investment. 

Further, the results suggest that upstream spillovers $Spill(A)^{\alpha,up}_{i,t-1}$ from both layers $\alpha = \mu, \tau$ are associated with a decline in productivity (column (1)-(2)), wages (column (4)), and a decreasing demand for production labor (column (7)-(8)). These spillovers are positively correlated with a higher capital intensity. This finding is qualitatively robust across different estimators, but the significance varies across estimators.\footnote{It may seem inconsistent that innovation spillovers show a positive relationship with industry output (Table \ref{tab:real_sum_output|citation_weighted_patent_stock_all_sectors_all_years_sys_1step_wControls}) but a negative association with productivity and labor, while the effect on capital is insignificant (Table \ref{tab:TFP|real_vadd|emp|real_pay|cap|real_invest|share_prode|real_relative_prodw_all_sectors_all_years_FE_weighted_wControls}). The results in these two tables rely on different estimation methods, and the effects in Table \ref{tab:TFP|real_vadd|emp|real_pay|cap|real_invest|share_prode|real_relative_prodw_all_sectors_all_years_FE_weighted_wControls} are weighted. As mentioned in the main text, the findings are qualitatively consistent across different estimation methods, but the statistical significance of the observations on productivity, labor, and capital varies (see Sec. \ref{app:add_regression_results}). For example, the positive correlation between spillovers and capital is statistically stronger when using an AB estimator. The qualitative consistency supports the explanation that the increase of output is associated with a shift from labor to capital. Moreover, TFP is five factor productivity, and energy and material use are not considered here.}
Moreover, TP arising from more innovation $A^{\tau}_{i,t-1}$, higher upstream innovation spillovers $Spill(A)^{\tau,up}_{i,t-1}$ and centrality $PR^{\tau,dw}_{i.t-1}$ is associated with a shift from production to non-production labor (columns (7)-(8)) as they both show a positive correlation with the share of production labor $(L^P/L)_{i,t}$ but no significant effect on labor $L_{i,t}$. 

Market growth $A^{\mu}_{i,t-1}$ is associated with decreasing employment and wages, a lower share of production labor, and less significantly, with an increasing capital intensity, higher TFP, and lower labor productivity. 
These findings are qualitatively robust across different estimators, but less or not significant in other model specifications. 
The previous section (\ref{subsubsec:DP_and_TP}) has shown that TP1 was a key driver of market growth, and Table \ref{tab:TFP|real_vadd|emp|real_pay|cap|real_invest|share_prode|real_relative_prodw_all_sectors_all_years_FE_weighted_wControls} indicates that market growth is associated with a factor bias in favor of capital.
The combined results suggest that TP-driven technological change is biased in favor of capital and associated with a labor demand shift towards non-production labor.

\subsubsection{Sectoral patterns of innovation}
Patterns of innovation, technological change, and sources of knowledge differ across industries and may change over time \citep{pavitt1984sectoral, kline1986overview}. To capture this, the regressions are repeated for different subsets of industries and subperiods (1987-1997, 2002-2012).\footnote{Technically, the two subperiods cover the periods 1977-1997 and 1992-2012. They overlap in two 5-year snapshots. The overlap only matters for the first stage of the IV regressions, as the BB estimator uses the second lag as in instrument for the first lag. Hence, the first two periods are lost due to the estimation procedure.\label{footnote:subperiods}}   
To simplify the representation, the main observations are only verbally summarized. The regression results for the two subperiods are provided in \ref{app:results_DP_TP}-\ref{app:results_direction} and those for the industry subgroups in \ref{supp:results_sectoral_patterns_of_inno}. 

Table \ref{tab:Overview_industry_subsamples} gives an overview of the industry subsets and shows whether TP and DP are supported. The mapping of NAICS 6-digit codes to subgroups can be found in Table \ref{tab:NAICStoSubsample_map}. The plus symbols (+) indicate support, minus symbols (-) indicate that opposite, i.e. negative effects were found, and tilde symbols ($\sim$) suggest that the results were ambiguous across variables and/or model specifications.\footnote{In Table \ref{tab:Overview_industry_subsamples}, it is refrained from showing exact coefficients for two reasons: (1) Using symbols instead of coefficients simplifies the representation as it summarizes multiple variables that track DP and TP effects. For example, TP1 is tracked through $A^{\tau}_{i,t-1}$, $PR^{\tau,d}_{i,t-1}$ and $Spill(A)^{\tau,d}_{i,t-1}$. Hence, it would be necessary to show all three coefficients. (2) Showing coefficients could suggest more precision than present, as the significance and strength of the effects varies across different estimation methods. A discussion of the limitations and consistency across methods can be found in the previous sections (\ref{subsubsec:DP_and_TP}, \ref{subsubsec:direction_of_change}). For brevity, an analogous discussion is omitted here.} Next, the results are summarized stepwise for each effect (TP1, DP1, TP2, DP2). 
\begin{table}[h!]
	\centering
\begin{myresizeenv}
	\small
	\begin{tabular}{|p{3.45cm}p{2.2cm}|p{0.55cm}|p{0.575cm}|p{0.575cm}|p{0.575cm}|p{0.575cm}|p{0.575cm}|p{0.575cm}|}
	\hline 
	\hline \rule{0pt}{1.075\normalbaselineskip}
	Name&Definition&N&TP1&DP1&TP2$^\mu$&TP2$^\tau$&DP2$^\mu$&DP2$^\tau$\\
	\hline \rule{0pt}{1.075\normalbaselineskip}
	All sectors&&2456&+++&&+++&+&&-\\
	\hline \rule{0pt}{1.075\normalbaselineskip}
	First subperiod&1987-1997&1535&+& & & & & \\
	Second subperiod&2002-2012&1535&+++& &+++&+++&-&- - -\\
	\hline \rule{0pt}{1.075\normalbaselineskip}
	Food Processing&2-digit 31&520&++& & & & & \\
	Non-Metallic&2-digit 32&656& & & &++& & \\
	Metallic and Machinery&2-digit 33&1280&+++&$\sim$&+++& &-&- -\\
	\hline \rule{0pt}{1.075\normalbaselineskip}
	Innovation-intensive &$>$ md($A^{\tau}/A^{\mu})$&1224&+++&+& &++&$\sim$&- - -\\
	Non-Inno.-intensive &$\leq$ md($A^{\tau}/A^{\mu})$&1232&+&-&- -&+&-&- \\
	\hline \rule{0pt}{1.075\normalbaselineskip}
	Big&$>$ md($A^{\mu})$&1224& &+& &+& &- -\\
	Small&$\leq$ md($A^{\mu})$&1232&++& & &++&-&- -\\	
	\hline \rule{0pt}{1.075\normalbaselineskip}
	Science-intensive&BP (2016)&360& & & & & &- - -\\
	Supplier-dominated&BP (2016)&840&++& & & &-&- - -\\
	Production-intensive&BP (2016)&1208&+++&$\sim$& & &-&-\\					
	\hline \rule{0pt}{1.075\normalbaselineskip}
	Patent-central&$>$ md($PR^{\tau,dw})$&1224&+++&+& &+++&-&- - -\\
	Not Patent-central&$\leq$ md($PR^{\tau,dw})$&1232&++&- - -&- - -& +&-&-\\				
	\hline \rule{0pt}{1.075\normalbaselineskip}
	Market-central (up)&$>$ md($PR^{\mu,up})$&1224& &+& & &-&- -\\
	Not Market-central (up)&$\leq$ md($PR^{\mu,up})$&1232&++& & &+&- -&-\\
	\hline \rule{0pt}{1.075\normalbaselineskip}
	Market-central (dw)&$>$ md($PR^{\mu,dw})$&1224&++&$\sim$&++& &-&-\\
	Not Market-central (dw)&$\leq$ md($PR^{\mu,dw})$&1232& &-& & & &-\\
	\hline 
	\hline
\end{tabular}
\caption{Overview - Sectoral patterns of innovation}
\label{tab:Overview_industry_subsamples}
\end{myresizeenv}
\footnotesize
\vspace{0.5cm}
\justifying

\noindent
Notes: This table shows how the industry subsets are named, defined, and it summarizes the results for different effects. md is short for median. 
The correspondence list of how 6-digit NAICS industries are grouped can be found in \ref{tab:NAICStoSubsample_map}. The column N shows the number of observations per sample. 
The plus (minus) symbols +++, ++, + (- - -, - -, -) indicate strong, moderate, weak positive (negative) effects by statistical significance and consistency across different model settings. A tilde $\sim$ is shown whenever the effects are ambiguous for different variables that capture the same concept (e.g. qualitative ambiguity between $A^{\alpha}_{i,t-1}$ and $PR^{\alpha,d}_{i,t-1}$). 
The summary of the results is based on a large sample of regression analyses of different industry subsets and different model specifications. Positive and negative effects are only indicated, if the results are consistent across specifications. These results are published along with the research data and statistical outputs \citep{hotte2023data}. An extract of these results is available in \ref{supp:results_sectoral_patterns_of_inno}. 
The two subperiods (1987-1997 and 2002-2012) technically cover the periods 1977-1997 and 1992-2012, which is due to the estimation procedure (see Footnote \ref{footnote:subperiods} and \ref{app:results_DP_TP}). 
Food, Non-Metal, Metal are short for the Food processing sector identified by 2-digit NAICS codes 31, Non-metallic (Wood, Fiber, Chemical) manufacturing (2-digit NAICS code 32), Metallic and Machinery manufacturing (2-digit NAICS code 33). 
Innovation-intensive, Big, Patent and Market central industries are defined by whether the corresponding variable taken from the network data and averaged over time ranges above or below the cross-industry median value. 
The classes Science-intensive, Supplier-dominated, Production-intensive are categories motivated by the taxonomy introduced by \citet{pavitt1984sectoral}. The mapping to 6-digit NAICS industries is based on the tables provided by \citet{bogliacino2016pavitt} (short: BP (2016)). 
\end{table}

\begin{description}
\item[TP1:] This effect is consistently supported in all subsamples with varying levels of significance. 
TP1 arises from both innovation centrality $PR^{\tau,dw}_{i,t-1}$ and spillovers $Spill(A)^{\tau,up}_{i,t-1}$, while the effect from upstream spillovers is larger and more significant in most groups. 
The effect is statistically and economically strongest during the second subperiod, and in Metallic and Machinery manufacturing, Innovation-intensive, Small, Production-intensive sectors, those with a high innovation centrality, and those with a low upstream but high downstream centrality in the market. 

\item[DP1:] Support for DP1 effects is ---if at all--- only weakly significant, ambiguous, or even negative; hence market growth, centrality, and spillovers do not show a clearly positive effect on innovation. 
Generally, but only poorly significant, upstream market centrality tends to have a negative effect while downstream centrality rather shows positive effects. The opposite holds for spillovers: Downstream (upstream) spillovers are rather negatively (positively) correlated with innovation. 

\item[TP2 in the market:] Upstream spillovers $Spill(A)^{\mu,up}_{i,t-1}$ show a positive association with market growth during the second subperiod, in Metallic and Machinery manufacturing and industries with high market downstream centrality. 
Upstream market centrality $PR^{\mu,up}_{i,t-1}$ exhibits opposite effects: It shows negative coefficients in sectors with a low innovation-intensity and low innovation centrality. 

\item[TP2 in innovation:] Upstream innovation spillovers display an unambiguously positive association with subsequent innovation, with varying significance. It is significant during the second subperiod, in Non-Metallic manufacturing, Innovation-intensive, Big, Production-intensive sectors, and sectors with a high and low innovation centrality, and with a low upstream and high downstream market centrality.  		
The coefficients are small in Big industries and those with a low innovation centrality, and high downstream market centrality. 

\item[DP2 in the market:] DP2 from downstream links is not supported. Downstream spillovers enter unambiguously with negative coefficients. Only in Innovation-intensive industries the effect is ambiguous as downstream centrality enters with a weakly significant positive coefficient. 

\item[DP2 in innovation:] This effect is absent and, whenever a significant effect is observed, the effect is negative, contradicting DP2 to be a driver of innovation. Almost all industry subsamples show a statistically significant but quantitatively small negative correlation between downstream spillovers $Spill(A)^{\tau,dw}_{i,t-1}$ and innovation outputs. This effect is strongest in Metallic manufacturing, Innovation-intensive, Big, Small, Science-intensive, Suppliers-dominated, sectors with a high innovation centrality, and a high downstream market centrality. 
\end{description}

Summing up, the results further confirm that the effects from innovation on the market are stronger than vice versa (TP1). 
The positive effects of innovation mostly arise from upstream spillovers. In the market, up- and downstream spillovers are ambiguous: Downstream spillovers show rather positive interactions with innovation, while the opposite holds for upstream effects. 

There is clear support for TP2 in the innovation layer, while the effect of TP2 in the market is ambiguous. The ambiguity largely comes from the qualitatively different impact of upstream centrality and upstream spillovers. 
Generally, the results suggest that DP2 is not supported. 
Most of the results are much stronger for the second subperiod of the sample, and sometimes not or only weakly significant in the years before.

The qualitative results are largely consistent across industry subgroups, but the relative importance of TP and DP as drivers of innovation and market growth differs. For example, TP1 and TP2 effects are qualitatively consistent across all subsets of industries and subperiods. But their impact is strongest in industries that are most reliant on patents, and belong to Metallic and Machinery Manufacturing. This coincides with the rise of the ICT sector, which grew strongly by market size and innovation activity (see Fig. \ref{fig:threeDcomp_netw_flow_up}). Many of the industries belonging to this broad industrial group are classified as Small industries, even though they have become large ones over time. The rise of industries that are driven by TP may explain why the TP effects became stronger in the post-2000s.

The results suggest a negative correlation between DP and TP forces, as in many industries that are positively driven by TP show a negative effect of DP. However, innovation-intensive and patent-central industries are an exception and both DP in the market and TP in innovation simultaneously drive technological change, suggesting a co-evolutionary dynamic between innovation and market growth in these sectors.

\FloatBarrier
\section{Discussion}
\label{sec:disc}
The key results can be wrapped up as follows: 
\begin{enumerate}
\item Both network layers show different dynamics: The innovation layer became increasingly connected, technologically similar, and skewed, while connectivity, similarity, and size rankings in the market are sluggish without a clear trend. The size distribution in the market shares is more balanced compared to innovation. 
Both layers show the rise of ICT-related and other electronics industries. 

\item The results reveal path-dependence in both layers, which stronger in innovation and indicates increasing returns. This may explain why innovation became more concentrated than the market. 

\item The results support TP1: Past innovation shows a strong positive association with subsequent market growth. TP1 mostly arises from upstream innovation spillovers, which indicate an expansion of the technological knowledge base upon which an industry relies. 

Also the centrality in the innovation network is positively associated with market growth but less significant. Innovation centrality shows how well an industry's own R\&D activity is integrated in the innovation network and informs about an industry's access and exchange of technological knowledge. 

\item The results support within-innovation TP2: Upstream spillovers stimulate subsequent innovation. This observation is consistent with previous research \citep[e.g.][]{acemoglu2016innovation, antony2012technology, jaffe1986technological}.

\item Within-market TP2 effects are ambiguous. There is support for a positive effect of upstream market spillovers, which is strongest during the second subperiod and in Metallic manufacturing sectors with a high downstream centrality in the market. These spillovers indicate an increased availability of production inputs. 

Upstream effects in the market are heterogeneous across industries: Upstream market centrality shows a negative effect in industries with a low reliance on innovation. 

\item The results do not support DP effects. While DP1 effects are weak and ambiguous across industries, the lack of support for DP2 effects is clear: Across almost all industry groups, downstream effects enter with negative coefficients. 

\item Up- and downstream innovation spillovers enter with opposite coefficients, which might be explained by their different meaning. 
Upstream spillovers inform about the pool of technological knowledge upon which the downstream industry relies. 
Downstream spillovers suggest that an industry's own innovations are increasingly used (cited). This may reflect a higher level of technological maturity, which may associated with a lower potential to grow. 
However, the negative coefficients are small and only weakly significant. 

\item The analysis in Sec. \ref{subsubsec:direction_of_change} indicates that the factor bias of technological change may depend on its driver. 
DP arising from a higher centrality in the customer network in the market is associated with higher productivity, more labor demand, and less capital use. This indicates a factor bias in favor of labor, albeit DP does not show any positive effect on output or innovation. 
In contrast, TP and pure size effects are associated with lower wages and a reallocation from production to non-production labor, and show a statistically weak positive relationship with capital use. 

Further, TP from innovation spillovers and productivity show a negative correlation, which might be attributed to the shift from production to non-production labor. Previous research has shown that the increasing service share in manufacturing coincided with lower productivity growth, which may be an issue of mismeasurement as intangible capital is difficult to measure \citep{corrado2009intangible, lafond2022productivity, baily2016us, baumol2012cost}. 
\end{enumerate}

\subsection{Demand-pull, technology-push, and the direction of technological change}
The analysis suggests that TP from between-layer effects and within-layer upstream linkages is a key driver of industry growth and innovation in US manufacturing since the late 1970s, especially since the 2000s. A major source of TP are innovation spillovers from upstream industries.
TP from innovation is largest in those industries that rely on patents (Innovation-intensive, high centrality in the innovation layer, Metallic and Machinery manufacturing), and in Small industries, and those with a low upstream, but high downstream centrality in the market. 
These industry groups strongly overlap (see Table \ref{tab:NAICStoSubsample_map}) and many of them belong to the group of Metallic and Machinery manufacturing industries.  
It is not surprising to find TP driven by innovation to be strongest in industries, that use patents as means of IP protection. Patents may perform worse as an innovation indicator in other industries and TP effects from non-patented innovations cannot be captured by this framework. 

However, the analytical results are well in line with the descriptives. The post-1980s in the US were characterized by the rise of ICT-related and electronics industries, which largely belong to those industries that show the strongest impact of TP. 
The rise of these industries accelerated in recent years, coinciding with the observation that TP is most significant in the post-2000s. Hence, the increasing relevance of TP as a driver of innovation and market growth may be attributed to industrial change and the take-off of ICT-related industries, whose growth is more strongly driven by TP compared to other industries (especially Non-Metallic Manufacturing such as Textiles, Wood, Fiber, and Paper).

The analyses suggest that innovation-induced TP comes with a shift from production to non-production labor. This is consistent with previous research that documented patterns of capital deepening driven by ICT \citep{corrado2009intangible} and heterogeneous employment effects that are positive for high-skill, non-routine, and service jobs but negative for low-skilled production labor \citep[see][for an overview]{hotte2022technology}.  
However, this paper does not inform about the net impact of TP from innovation on labor markets at the aggregate level, as it only analyzes within-manufacturing effects and ignores the rise of services \citep{gallipoli2018structural}. 

In contrast to TP, DP does not show any significant positive impact on market growth or innovation, but the results indicate that DP arising from customer relationships in the market is associated with productivity growth and more employment. This observation points to the importance of the demand side for the creation of new jobs \citep{bessen2019automation, hotte2022technology}. 

Upstream linkages that reflect supply conditions and the availability of technological knowledge and physical production inputs show positive effects on industrial growth, while downstream spillovers show the opposite effect. 
Despite not directly comparable, this finding has conceptual overlaps with the observations made by  \citet{bloom2013identifying}. At the firm level, the authors identified positive effects of upstream knowledge spillovers in R\&D, and negative effects of downstream spillovers, which they interpret as market rivalry and cannibalization effects. 

Another interesting observation is the role of up- and downstream centrality in the market. 
While upstream centrality tends to show negative correlations with innovation and growth, the opposite holds for downstream centrality. The effects are only weakly significant and matter only in certain subgroups of industries. Nevertheless, this observation is interesting as it points to the relevance of an industry's position in the supply chain. Other research has shown that industries with a deep embeddedness in the supply network tend to grow faster as the positive effects of productivity growth accumulate along the supply chain \citep{mcnerney2018production}. 
Here, data limitations prevented the calculation of output multipliers, which would be an adequate indicator to capture this. 
However, the market up- and downstream centrality exhibit similar properties, as industries with a high downstream centrality often produce highly processed goods (transportation equipment, electronics, chemicals, processed food). Those with a high upstream centrality are more frequently found in sectors that process raw material (mills, raw plastics, primary material processing) (cf. Table \ref{tab:NAICStoSubsample_map} and \ref{supp:add_descriptives_industry_rankings}). The statistical analyses control for sector FE. Hence, changes in the market up- and downstream centrality reflect whether an industry moved up- or down in the supply chain. 
It seems that a move down the supply chain is rather positively associated with innovation, market growth, and labor. 

The results of this study are consistent with \citet{acemoglu2016innovation} who observed a strong positive effect of within-innovation TP2 arising from upstream innovation spillovers and a weak negative effect of downstream linkages, contradicting within-innovation DP2 effects. This is at odds with \citet{taalbi2020evolution} who found support for both DP2 and TP2 as drivers of innovation. \possessivecite{taalbi2020evolution} analysis relies on an innovation output network that captures innovations defined as significantly improved and commercialized products and their actual use in downstream industries. This indicates potential reasons for the lack of support for DP2 effects in this and \possessivecite{acemoglu2016innovation} study: Patent citations capture the flow of knowledge between industries, which does not necessarily coincide with technology supply and demand.
The contradictory results indicate that whether technology demand can pull upstream innovation may be sensitive to the way of measurement. 

Further, the dominance of TP may be specific for the period under study. The strong support of TP appears to be driven by the rise of the ICT sector, which characterized the four decades covered by this study. The dominance of TP could be a feature of emergent technologies. \citet{walsh1984invention} showed for the chemical industry that TP from radical breakthroughs may drive growth in the market, which in turn creates DP effects that induce incremental innovation. The analysis in this paper has shown that the rapid and radical technological change that happened in the ICT sector during the past decades was primarily driven by TP. However, this should not lead to the conclusion that the TP dominance will persist in future when incrementally enhanced ICT applications diffuse.

This study investigated innovation captured by patents. Patents are an indicator of technological inventions, which are only the first step in the chain of invention, innovation, and diffusion \citep{dosi2010technical}. Other research has shown that DP policy can effectively stimulate the diffusion and incremental innovation in existing technologies, while proving less effective as a stimulus of radical inventions \citep{nemet2009demand, hoppmann2013two}. Thus, the lack of support for DP may be due to the dominance of the ICT sector in its infancy and the choice of patents as means of measurement. Innovation policy goals (invention, innovation, diffusion) need to be carefully defined when it comes to the choice of the appropriate instruments.

\subsection{Limitations and research implications}
\label{sec:limit_outlook}

So far, this study has ignored the role of trade. The exposure to import competition from low wage countries can be a driver of the reallocation from low- to high-tech industries \citep{bernard2006survival}. 
The accession of China to the WTO in 2001 was a large shock to the US manufacturing sector and associated with a sharp drop in US manufacturing employment \citep{pierce2016surprisingly}. For European countries, \citet{bloom2016trade} have shown that Chinese import competition was a driver of technological change, inducing the reallocation of production towards more productive firms and spurring innovation within trade exposed firms. 

The increased trade exposure of US manufacturing after 2001 may be another explanation for the take-off of TP during the second subperiod. 
However, the rise of the ICT sector in innovation began already before the 2000s, even though the impact of TP was weak and non-existent in innovation. Whether the trade-shock caused by the Chinese WTO accession ``activated'' TP and accelerated the market uptake of the ICT sector and decline of more primary manufacturing activities is beyond the scope of this study. 

It should be also noted that this study is limited to manufacturing. Various studies documented the decline of US manufacturing since the 1980s \citep[e.g.][]{elsby2013decline, fort2018new}. 
It would be important to verify whether the observed patterns are general or specific for the US manufacturing sector during the period under study. 

Further, the analysis is subject to three more technical limitations. 
First, patents as a measure of innovation are imperfect. The use of patents to protect IP varies across industries \citep{fontana2013reassessing, arundel1998percentage, cohen2000protecting}, patents vary by value, and not every patent indicates a technological breakthrough \citep{trajtenberg1990penny, kogan2017technological}. Sometimes patents are only filed for defensive purposes to protect a pre-existing, but not a new invention \citep{granstrand1999economics}. Over time, citation practices may have changed \citep{hall2005market, marmor1980approach}. 
These limitations are partly addressed by restricting the sample to manufacturing where patents are a common means of intellectual property protection \citep{blank2012intellectual}, by controlling for industry and time FE, and by using citation-weighted patents. 

Second, studying innovation and industrial evolution over time is challenging because of non-static classification systems \citep{marmor1980approach, yuskavage2007converting, lafond2019long}. This analysis relies on NAICS industry codes that are purposely designed to describe industries by their production processes. The codes are regularly (quasi-endogenously) updated to meet this purpose. 
This can be one explanation for the less skewed sector-size distribution, and possibly also for the higher stability of the IO network. 

The aim of this paper is the economic analysis of technological change, which justifies the choice of NAICS codes rather than patent classes. Inferring from patents to industrial dynamics is a challenging endeavor \citep{antonelli2014economics, dosi2010technical}, not least because the industry where a patent is filed is not necessarily the same industry where the patented invention is used. 
Various concordances exist \citep{lybbert2014getting, van2014patent, goldschlag2020tracking, dorner2018novel}, and a systematic, dynamic comparison between these concordances and their implications for economic studies would be a valuable research avenue. 
It would be also interesting to compare the results of this study with an approach using patent classes as means of description: Classifying IO flows by their correspondence in patent classes can be insightful to understand the impact of DP on the dynamics of patented innovations. But this is beyond the scope of this paper. 

Third, this paper studies TP and DP at the aggregate level. 
But patterns of innovation, knowledge sources, and IP practices differ across firms, industries, and technology fields \citep{pavitt1984sectoral, carlsson1991nature, blank2012intellectual}. 
\citet{walsh1984invention} documented that whether DP or TP dominates may be a matter of industry maturity. 
The static dimension of sector heterogeneity is captured by the FE approach and by the analysis of subsamples. Exploring the dynamic dimension of industrial heterogeneity could be an interesting avenue for future work. 
For example, future research may take account of the distinction between radical and incremental innovation, and the chronology of the technology cycle. 
This might also confirm whether the results of this study are driven by the ICT sector, which was nascent during the early periods under study. 

\section{Conclusions}
\label{sec:concl}
Understanding and shaping technological change is a key task for policymakers in the 21st century. Climate change mitigation requires a dramatic and fast transition to climate-friendly technology, \citep{ipcc2018special} and digitalization may disrupt labor markets with undesirable effects on income distribution \citep{brynjolfsson2012race, autor2018automation}. 
Understanding the impact of TP and DP as drivers of technological change can help develop effective policies to steer the process and to mitigate undesirable side effects. 

This paper conceptualized TP and DP as between- and within-layer effects and studied their impact on market growth, innovation, and the direction of technological change in an empirical two-layer network of IO and patent citation links between US manufacturing industries. 

The results strongly support TP, suggesting that an increase of the available technological knowledge base can stimulate growth in the market and subsequent innovation. Increasing returns to innovation further suggest that this may be a self-reinforcing process. 
This provides evidence that R\&D policy, that stimulates the creation of relevant technological knowledge, may be an effective instrument to speed up industry level technological change. 
However, the results also show that technological change driven by TP is factor-biased, which may require supplementary policy to alleviate undesirable distributional effects. 

It was also seen that DP tends to show an opposite factor bias, which may help develop political instruments that offset the labor-saving effect of TP. 
The results bear important insights for the debate on whether and how technological change is labor-saving: The results indicate that the factor bias of technological change may be dependent on its driver. 

In contrast to previous research relying on other measures of innovation, the results do not support DP as a driver of innovation and market growth. Innovation was measured by patents, which are indicative for radical inventions but may be less suited to capture actual patterns of technology diffusion and use. Innovation policymakers need to be clear about their goals when it comes to the choice of DP or TP instruments. This study offers support that TP is a driver of technological change pushed by patented novelties, but it should not question existing evidence on the effectiveness of DP as a driver of diffusion and incremental change.

\section*{Acknowledgements}
I want to thank Angelo Secchi and Antoine Mandel who facilitated this work at an early stage. Further, many thanks to Herbert Dawid, Markus Trunschke, Angelos Theodorakopoulos, Peter Persoon, Francois Lafond, and the participants of the Annual GENED meeting 2020, the INET Complexity group, and the OMPTEC-FoW seminar for helpful comments. Further gratitude is owed to the Competence Center for Research Data at Bielefeld University. 
I am indebted to  three anonymous reviewers whose valuable feedback helped to significantly improve this research. 
Moreover, I gratefully acknowledge the financial support by the German Academic Foundation, Deutsch-Französische Hochschule, Bielefeld Graduate School of Economics and Management, and from the EU Horizon 2020 Research and Innovation programme under grant agreement No 822330 TECHNEQUALITY.

\newpage

\printbibliography

@article{hoppmann2013two,
	title={The two faces of market support—How deployment policies affect technological exploration and exploitation in the solar photovoltaic industry},
	author={Hoppmann, Joern and Peters, Michael and Schneider, Malte and Hoffmann, Volker H},
	journal={Research Policy},
	volume={42},
	number={4},
	pages={989--1003},
	year={2013},
	publisher={Elsevier}, 
	doi={10.1016/j.respol.2013.01.002}
}

@article{frohm2021spillovers,
	title={Spillovers in global production networks},
	author={Frohm, Erik and Gunnella, Vanessa},
	journal={Review of International Economics},
	volume={29},
	number={3},
	pages={663--680},
	year={2021},
	publisher={Wiley Online Library}, 
	doi={10.1111/roie.12526}
}

@article{bessen2019automation,
	title={Automation and jobs: When technology boosts employment},
	author={Bessen, James},
	journal={Economic Policy},
	volume={34},
	number={100},
	pages={589--626},
	year={2019},
	publisher={Oxford University Press}, 
	doi={10.1093/epolic/eiaa001}
}

@techreport{hotte2022technology,
	title={Technology and jobs: A systematic literature review},
	author={H{\"o}tte, Kerstin and Somers, Melline and Theodorakopoulos, Angelos},
	institution={Cornell University},
	type={arXiv preprint arXiv:2204.01296},
	year={2022}
}

@article{elsby2013decline,
	title={{The decline of the US labor share}},
	author={Elsby, Michael WL and Hobijn, Bart and {\c{S}}ahin, Ay{\c{s}}eg{\"u}l},
	journal={Brookings Papers on Economic Activity},
	volume={2013},
	number={2},
	pages={1--63},
	year={2013},
	publisher={Brookings Institution Press}, 
	doi={10.1353/eca.2013.0016}
}

@article{karabarbounis2014global,
	title={The global decline of the labor share},
	author={Karabarbounis, Loukas and Neiman, Brent},
	journal={The Quarterly Journal of Economics},
	volume={129},
	number={1},
	pages={61--103},
	year={2014},
	publisher={Oxford University Press}, 
	doi={10.1353/eca.2013.0016}
}

@article{baltagi2005skill,
	title={Skill-biased technical change in US manufacturing: a general index approach},
	author={Baltagi, Badi H and Rich, Daniel P},
	journal={Journal of Econometrics},
	volume={126},
	number={2},
	pages={549--570},
	year={2005},
	publisher={Elsevier}, 
	doi={10.1016/j.jeconom.2004.05.013}
}

@techreport{autor2018automation,
	title={Is automation labor-displacing? Productivity growth, employment, and the labor share},
	author={Autor, David and Salomons, Anna},
	year={2018},
	type={National Bureau of Economic Research Working Paper No. 24871}, 
	doi={10.3386/w24871}
}

@techreport{lafond2022productivity,
	title={Why is productivity slowing down?},
	author={Goldin, Ian and Koutroumpis, Pantelis and Lafond, Fran{\c{c}}ois and Winkler, Julian},
	year={2022},
    type={INET Oxford Working Paper No. 2022-08},
	institution={Institute for New Economic Thinking at the Oxford Martin School, University of Oxford}, 
    note={Forthcoming in the Journal of Economic Literature}, 
    url={https://www.inet.ox.ac.uk/files/2022-7-WP-Paper-2-Why-is-Productivity-Slowing-Down-OMPTEC-FoW-FoD-Ian-copy_2022-06-09-133241_hfcj.pdf}
}

@techreport{baily2016us,
	title={{Why is US productivity growth so slow? Possible explanations and policy responses}},
	author={Baily, Martin Neil and Montalbano, Nicholas},
	institution={Brookings Institution},
	type={Hutchins Center Working Paper 22},  
	year={2016}
}

@book{baumol2012cost,
	title={The cost disease: Why computers get cheaper and health care doesn't},
	author={Baumol, William J},
	year={2012},
	publisher={Yale university press}
}

@article{blundell1998initial,
	title={Initial conditions and moment restrictions in dynamic panel data models},
	author={Blundell, Richard and Bond, Stephen},
	journal={{Journal of Econometrics}},
	volume={87},
	number={1},
	pages={115--143},
	year={1998},
	publisher={Elsevier}, 
	doi={10.1016/S0304-4076(98)00009-8}
}

@article{arellano1991some,
	title={{Some tests of specification for panel data: Monte Carlo evidence and an application to employment equations}},
	author={Arellano, Manuel and Bond, Stephen},
	journal={{The Review of Economic Studies}},
	volume={58},
	number={2},
	pages={277--297},
	year={1991},
	publisher={Wiley-Blackwell}, 
	doi={10.2307/2297968}
}

@article{nickell1981biases,
	title={Biases in dynamic models with fixed effects},
	author={Nickell, Stephen},
	journal={Econometrica},
	pages={1417--1426},
	year={1981},
	publisher={JSTOR}, 
	doi={10.2307/1911408}
}

@article{bogliacino2016pavitt,
	title={The Pavitt Taxonomy, revisited: patterns of innovation in manufacturing and services},
	author={Bogliacino, Francesco and Pianta, Mario},
	journal={Economia Politica},
	volume={33},
	number={2},
	pages={153--180},
	year={2016},
	publisher={Springer}, 
	doi={10.1007/s40888-016-0035-1}
}

@article{fort2018new,
	title={{New perspectives on the decline of US manufacturing employment}},
	author={Fort, Teresa C and Pierce, Justin R and Schott, Peter K},
	journal={Journal of Economic Perspectives},
	volume={32},
	number={2},
	pages={47--72},
	year={2018}, 
	doi={10.1257/jep.32.2.47}
}

@article{mcnerney2018production,
	title={How production networks amplify economic growth},
	author={McNerney, James and Savoie, Charles and Caravelli, Francesco and Carvalho, Vasco M and Farmer, J Doyne},
	journal={{Proceedings of the National Academy of Sciences}},
	volume={119},
	number={1},
	pages={e2106031118},
	year={2022},
	publisher={{National Academy of Sciences}}, 
	doi={10.1073/pnas.2106031118}
}

@book{antonelli2014economics,
	title={The economics of innovation, new technologies and structural change},
	author={Antonelli, Cristiano},
	year={2014},
	publisher={Routledge}
}

@article{lafond2019long,
	title={Long-run dynamics of the US patent classification system},
	author={Lafond, Fran{\c{c}}ois and Kim, Daniel},
	journal={Journal of Evolutionary Economics},
	volume={29},
	number={2},
	pages={631--664},
	year={2019},
	publisher={Springer}, 
	doi={10.1007/s00191-018-0603-3}
}

@article{marmor1980approach,
	title={{The approach of the United States Patent and Trademark Office to finding prior art}},
	author={Marmor, Alfred C},
	journal={Journal of Chemical Information and Computer Sciences},
	volume={20},
	number={1},
	pages={6--9},
	year={1980},
	publisher={ACS Publications}, 
	doi={10.1021/ci60021a003}
}

@article{hall2005market,
	title={Market value and patent citations},
	author={Hall, Bronwyn H and Jaffe, Adam and Trajtenberg, Manuel},
	journal={{RAND Journal of Economics}},
	volume={36}, 
	number={1}, 
	pages={16--38},
	year={2005},
	publisher={JSTOR}, 
	url={https://www.jstor.org/stable/1593752}
}

@article{kogan2017technological,
	title={Technological innovation, resource allocation, and growth},
	author={Kogan, Leonid and Papanikolaou, Dimitris and Seru, Amit and Stoffman, Noah},
	journal={The Quarterly Journal of Economics},
	volume={132},
	number={2},
	pages={665--712},
	year={2017},
	publisher={Oxford University Press}, 
	doi={10.1093/qje/qjw040}
}

@article{walsh1984invention,
	title={Invention and innovation in the chemical industry: Demand-pull or discovery-push?},
	author={Walsh, Vivien},
	journal={Research Policy},
	volume={13},
	number={4},
	pages={211--234},
	year={1984},
	publisher={Elsevier}, 
	doi={10.1016/0048-7333(84)90015-5}
}

@article{geroski1995innovative,
	title={Innovative activity over the business cycle},
	author={Geroski, Paul A and Walters, Chris F},
	journal={The Economic Journal},
	volume={105},
	number={431},
	pages={916--928},
	year={1995},
	publisher={Oxford University Press Oxford, UK}, 
	doi={10.2307/2235158}
}

@article{jaffe1986technological,
	title={Technological opportunity and spillovers of R\&D: Evidence from firms' patents, profits, and market value},
	author={Jaffe, Adam B},
	journal={The American Economic Review},
	volume={76},
	number={5},
	pages={984--1001},
	year={1986}, 
	url={http://www.jstor.org/stable/1816464}
}

@article{jaffe1988demand,
	title={Demand and supply influences in R\&D intensity and productivity growth},
	author={Jaffe, Adam B},
	journal={The Review of Economics and Statistics},
	pages={431--437},
	year={1988},
	publisher={JSTOR}, 
	doi={10.2307/1926781}
}

@article{nelson1994co,
	title={The co-evolution of technology, industrial structure, and supporting institutions},
	author={Nelson, Richard R},
	journal={Industrial and Corporate Change},
	volume={3},
	number={1},
	pages={47--63},
	year={1994},
	publisher={Oxford University Press}, 
	doi={10.1093/icc/3.1.47}
}

@article{buerger2012regional,
	title={Regional dynamics of innovation: Investigating the co-evolution of patents, research and development (R\&D), and employment},
	author={Buerger, Matthias and Broekel, Tom and Coad, Alex},
	journal={Regional Studies},
	volume={46},
	number={5},
	pages={565--582},
	year={2012},
	publisher={Taylor \& Francis}, 
	doi={10.1080/00343404.2010.520693}
}

@article{taalbi2020evolution,
	title={Evolution and structure of technological systems -- An innovation output network},
	author={Taalbi, Josef},
	journal={Research Policy},
	volume={49},
	number={8},
	pages={104010},
	year={2020},
	publisher={Elsevier}, 
	doi={10.1016/j.respol.2020.104010}
}

@article{arundel1998percentage,
	title={What percentage of innovations are patented? Empirical estimates for European firms},
	author={Arundel, Anthony and Kabla, Isabelle},
	journal={Research Policy},
	volume={27},
	number={2},
	pages={127--141},
	year={1998},
	publisher={Elsevier}, 
	doi={10.1016/S0048-7333(98)00033-X}
}

@article{fontana2013reassessing,
	title={{Reassessing patent propensity: Evidence from a dataset of R\&D awards, 1977--2004}},
	author={Fontana, Roberto and Nuvolari, Alessandro and Shimizu, Hiroshi and Vezzulli, Andrea},
	journal={Research Policy},
	volume={42},
	number={10},
	pages={1780--1792},
	year={2013},
	publisher={Elsevier}, 
	doi={10.1016/j.respol.2012.05.014}
}

@techreport{cohen2000protecting,
	title={{Protecting their intellectual assets: Appropriability conditions and why US manufacturing firms patent (or not)}},
	author={Cohen, Wesley M and Nelson, Richard and Walsh, John P},
	year={2000},
	type={{National Bureau of Economic Research Working Paper No. 7552}},
	doi={10.3386/w7552}
}

@techreport{cai2017knowledge,
	title={{Knowledge Diffusion, Trade and Innovation across Countries and 
	Sectors}},
	author={Cai, Jie and Li, Nan and Santacreu, Ana Maria},
	type={{Working Paper 2017-029A}}, 
	institution={{Federal Reserve Bank of St. Louis Research Division}}, 
	year={2017}, 
	doi={10.20955/wp.2017.029}
}

@article{trajtenberg1990penny,
	title={A penny for your quotes: patent citations and the value of innovations},
	author={Trajtenberg, Manuel},
	journal={The Rand Journal of Economics},
	pages={172--187},
	year={1990},
	publisher={JSTOR}, 
	doi={10.2307/2555502}
}

@article{saviotti2013co,
	title={The co-evolution of innovation, demand and growth},
	author={Saviotti, Pier Paolo and Pyka, Andreas},
	journal={Economics of Innovation and New Technology},
	volume={22},
	number={5},
	pages={461--482},
	year={2013},
	publisher={Taylor \& Francis}, 
	doi={10.1080/10438599.2013.768492}
}

@incollection{pakes1984exploration,
	title={{9. An exploration into the determinants of research intensity}},
	booktitle={{R\&D, Patents and Productivity}}, 
	author={Pakes, Ariel and Schankerman, Mark},
	editor={Griliches, Zvi},
	year={1984},
	publisher={University of Chicago Press}, 
	doi={10.7208/9780226308920-012}
}

@article{mowery1979influence,
	title={The influence of market demand upon innovation: a critical review of some recent empirical studies},
	author={Mowery, David and Rosenberg, Nathan},
	journal={Research Policy},
	volume={8},
	number={2},
	pages={102--153},
	year={1979},
	publisher={Elsevier}, 
	doi={10.1016/0048-7333(79)90019-2}
}

@book{rosenberg1982inside,
	title={Inside the black box: technology and economics},
	author={Rosenberg, Nathan},
	year={1982},
	publisher={{Cambridge University Press}}
}

@book{granstrand1999economics,
	title={The economics and management of intellectual property},
	author={Granstrand, Ove},
	year={1999},
	publisher={Edward Elgar Publishing}
}

@article{von1976dominant,
	title={The dominant role of users in the scientific instrument innovation process},
	author={Von Hippel, Eric},
	journal={Research Policy},
	volume={5},
	number={3},
	pages={212--239},
	year={1976},
	publisher={Elsevier}, 
	doi={10.1016/0048-7333(76)90028-7}
}

@book{meyer1969successful,
	title={Successful industrial innovation},
	author={Myers, Sumner and Marquis, Donald G},
	journal={National Science Foundation, Washington DC},
	pages={17--69},
	year={1969}
}

@book{schmookler1966invention,
	title={Invention and economic growth},
	author={Schmookler, Jacob},
	year={1966},
	publisher={Harvard University Press}, 
	doi={10.4159/harvard.9780674432833}
}

@article{brin1998anatomy,
	title={The anatomy of a large-scale hypertextual web search engine},
	author={Brin, Sergey and Page, Lawrence},
	journal={Computer Networks and ISDN Systems},
	volume={30},
	number={1-7},
	pages={107--117},
	year={1998},
	publisher={Elsevier}, 
	doi={10.1016/S0169-7552(98)00110-X}
}

@article{bloom2013identifying,
	title={Identifying technology spillovers and product market rivalry},
	author={Bloom, Nicholas and Schankerman, Mark and Van Reenen, John},
	journal={Econometrica},
	volume={81},
	number={4},
	pages={1347--1393},
	year={2013},
	publisher={Wiley Online Library}, 
	doi={10.3982/ECTA9466}
}

@incollection{saviotti1997black,
	title={Black boxes and variety in the evolution of technologies},
	author={Saviotti, Pier Paolo},
	booktitle={Economics of Structural and Technological Change},
	pages={195--223},
	year={1997},
	publisher={Routledge}
}

@article{kymn1990aggregation,
	title={Aggregation in input--output models: a comprehensive review, 1946--71},
	author={Kymn, Kern O},
	journal={Economic Systems Research},
	volume={2},
	number={1},
	pages={65--93},
	year={1990},
	publisher={Taylor \& Francis}, 
	doi={10.1080/09535319000000008}
}

@techreport{bartlesman1996nber,
	title={{The NBER manufacturing productivity database}},
	author={Bartlesman, Eric and Gray, Wayne B},
	year={1996},
	type={{National Bureau of Economic Research Technical Working Paper No. 0205}},
	doi={10.3386/t0205}
}

@techreport{becker2013nber,
	title={{NBER-CES manufacturing industry database: Technical notes}},
	author={Becker, Randy and Gray, Wayne and Marvakov, Jordan},
	type={{National Bureau of Economic Research Working Paper No. 5809}},
	institution={National Bureau of Economic Research}, 
	year={2013}, 
	url={https://citeseerx.ist.psu.edu/viewdoc/download?doi=10.1.1.368.671&rep=rep1&type=pdf}
}

@techreport{di2021stock,
	title={Stock market spillovers via the global production network: transmission of US monetary policy},
	author={Di Giovanni, Julian and Hale, Galina},
	year={2021},
	type={{National Bureau of Economic Research Working Paper No. 28827}},
	institution={National Bureau of Economic Research}, 
	doi={10.3386/w28827}
}

@techreport{hotte2023data,
	author = {Kerstin H\"otte},
	title = {{Data publication: Demand-pull, technology-push and the direction of technological change}},
	year = {2023},
	institution={Bielefeld University}, 
	doi={10.4119/unibi/2967659}, 
	type={{Data Publication}}
}

@article{atalay2011network,
	title={Network structure of production},
	author={Atalay, Enghin and Hortacsu, Ali and Roberts, James and Syverson, Chad},
	journal={Proceedings of the National Academy of Sciences},
	volume={108},
	number={13},
	pages={5199--5202},
	year={2011},
	publisher={National Academy Sciences}, 
	doi={10.1073/pnas.1015564108}
}

@article{csardi2006igraph,
	title={The igraph software package for complex network research},
	author={Csardi, Gabor and Nepusz, Tamas},
	journal={InterJournal, Complex Systems},
	volume={1695},
	number={5},
	pages={1--9},
	year={2006}, 
	url = {https://igraph.org},
}

@article{carlsson1991nature,
	title={{On the nature, function and composition of technological systems}},
	author={Carlsson, Benny and Stankiewicz, Rikard},
	journal={Journal of Evolutionary Economics},
	volume={1},
	number={2},
	pages={93--118},
	year={1991},
	publisher={Springer}, 
	doi = {10.1007/BF01224915}
}

@incollection{ipcc2018special, 
	title = {{Summary for Policymakers}}, 
	booktitle = {{An IPCC Special Report on the impacts of global warming of 
	1.5 degree C above pre-industrial levels and related global greenhouse gas emission 
	pathways, in the context of strengthening the global response to the threat 
	of climate change, sustainable development, and efforts to eradicate poverty, 
	Summary for Policymakers}}, 
	editor = {Masson-Delmotte, V and Zhai, P and Pörtner, HO and Roberts, D 
	and Skea, J and Shukla, P R and Pirani, A and Moufouma-Okia, W and Péan,  C 
	and Pidcock, R and Connors, S and Matthews, J B R. and Chen, Y and Zhou, X and 
	Gomis, M I and Lonnoy, E and Maycock, T and Tignor, M and Waterfield, T}, 
	author = {IPCC}, 
	publisher = {{World Meteorological Organization, Geneva, Switzerland}}, 
	year = {2018}
}

@article{arthur1989competing,
	title={Competing technologies, increasing returns, and lock-in by historical 
	events},
	author={Arthur, W Brian},
	journal={{The Economic Journal}},
	volume={99},
	number={394},
	pages={116--131},
	year={1989},
	publisher={JSTOR}, 
	doi = {10.2307/2234208}
}

@book{brynjolfsson2012race,
	title={{Race against the machine: How the digital revolution is accelerating 
	innovation, driving productivity, and irreversibly transforming employment and 
	the economy}},
	author={Brynjolfsson, Erik and McAfee, Andrew},
	year={2012},
	publisher={Digital Frontier Press}
}

@article{bernard2006survival,
	title={Survival of the best fit: Exposure to low-wage countries and the (uneven) growth of US manufacturing plants},
	author={Bernard, Andrew B and Jensen, J Bradford and Schott, Peter K},
	journal={Journal of International Economics},
	volume={68},
	number={1},
	pages={219--237},
	year={2006},
	publisher={Elsevier}, 
	doi={10.1016/j.jinteco.2005.06.002}
}

@article{pierce2016surprisingly,
	title={The surprisingly swift decline of US manufacturing employment},
	author={Pierce, Justin R and Schott, Peter K},
	journal={American Economic Review},
	volume={106},
	number={7},
	pages={1632--62},
	year={2016}, 
	doi={10.1257/aer.20131578}
}

@article{bloom2016trade,
	title={Trade induced technical change? The impact of Chinese imports on innovation, IT and productivity},
	author={Bloom, Nicholas and Draca, Mirko and Van Reenen, John},
	journal={The Review of Economic Studies},
	volume={83},
	number={1},
	pages={87--117},
	year={2016},
	publisher={Oxford University Press}, 
	doi={10.1093/restud/rdv039}
}

@article{pavitt1984sectoral,
	title={{Sectoral patterns of technical change: Towards a taxonomy and a theory}},
	author={Pavitt, Keith},
	journal={Research Policy},
	volume={13},
	pages={343--373},
	year={1984}, 
	doi={10.1016/0048-7333(84)90018-0}
}

@inproceedings{yuskavage2007converting,
	title={{Converting historical industry time series data from SIC to NAICS}},
	author={Yuskavage, Robert E and others},
	booktitle={The Federal Committee on Statistical Methodology 2007 Research Conference. 5-7 November 2007},
	year={2007}, 
	url={https://www.bea.gov/system/files/papers/P2007-7.pdf}
}

@article{croissant2008panel,
	title={Panel data econometrics in R: The plm package},
	author={Croissant, Yves and Millo, Giovanni},
	journal={Journal of Statistical Software},
	volume={27},
	number={2},
	year={2008}, 
	doi={10.18637/jss.v027.i02}
}

@article{nemet2009demand,
	title={{Demand-pull, technology-push, and government-led incentives for non-incremental technical change}},
	author={Nemet, Gregory F},
	journal={Research Policy},
	volume={38},
	number={5},
	pages={700--709},
	year={2009},
	publisher={Elsevier}, 
	doi={10.1016/j.respol.2009.01.004}
}

@article{cohen2000absorptive,
	title={{Absorptive capacity: A new perspective on learning and innovation}},
	author={Cohen, Wesley M and Levinthal, Daniel A},
	journal={{Administrative Science Quarterly, Special Issue: Strategic Learning in a Knowledge economy}},
	volume={35},
	number={1},
	pages={39--67},
	year={2000},
	publisher={Elsevier}, 
	doi={10.2307/2393553}
}

@incollection{kline1986overview,
	title={{An overview of innovation}},
	author={Kline, Stephen J and Rosenberg, Nathan},
	booktitle={{The positive sum strategy: Harnessing technology for economic growth}},
	editor={Ralph Landau and Nathan Rosenberg},
	pages={275--306},
	year={1986},
	publisher={National Academies Press}, 
	address={Washington, DC}, 
	doi={10.17226/612}
}

@article{dorner2018novel,
	title={{A novel technology-industry concordance table based on linked inventor-establishment data}},
	author={Dorner, Matthias and Harhoff, Dietmar},
	journal={Research Policy},
	volume={47},
	number={4},
	pages={768--781},
	year={2018},
	publisher={Elsevier}, 
	doi={10.1016/j.respol.2018.02.005}
}

@techreport{schmoch2003linking,
	title={{Linking technology areas to industrial sectors}},
	author={Schmoch, Ulrich and Laville, Francoise and Patel, Pari and Frietsch, Rainer},
	type={{Final Report to the European Commission, DG Research}},
	institution={European Commission},
	year={2003}
}

@book{oecd2001measuring,
	title={{Measuring productivity-OECD manual: Measurement of aggregate and industry-level productivity growth}},
	author={OECD},
	year={2001},
	institution={{Organisation for Economic Co-operation and Development (OECD)}}, 
	publisher={{OECD Publishing, Paris}},
	doi={10.1787/9789264194519-en}
}

@article{goldschlag2020tracking,
	title={Tracking the technological composition of industries with algorithmic patent concordances},
	author={Goldschlag, Nathan and Lybbert, Travis J and Zolas, Nikolas J},
	journal={Economics of Innovation and New Technology},
	volume={29},
	number={6},
	pages={582--602},
	year={2020},
	publisher={Taylor \& Francis}, 
	doi={10.1080/10438599.2019.1648014}
}

@article{hotte2021rise,
	title={The rise of science in low-carbon energy technologies},
	author={H{\"o}tte, Kerstin and Pichler, Anton and Lafond, Fran{\c{c}}ois},
	journal={Renewable and Sustainable Energy Reviews},
	volume={139},
	pages={110654},
	year={2021},
	publisher={Elsevier}, 
	doi={10.1016/j.rser.2020.110654}
}

@techreport{hotte2021data_rise,
	author = {H\"otte, Kerstin and Pichler, Anton and Lafond, Fran\c{c}ois},
	title = {{Data Publication: The scientific knowledge base of low carbon energy technologies}},
	year = {2021},
	institution={Bielefeld University}, 
	doi={10.4119/unibi/2950291}, 
	type={{Data Publication}}
}

@article{lybbert2014getting,
	title={{Getting patents and economic data to speak to each other: An ‘algorithmic links with probabilities’ approach for joint analyses of patenting and economic activity}},
	author={Lybbert, Travis J and Zolas, Nikolas J},
	journal={Research Policy},
	volume={43},
	number={3},
	pages={530--542},
	year={2014},
	publisher={Elsevier}, 
	doi={10.1016/j.respol.2013.09.001}
}

@article{kortum1997assigning,
	title={{Assigning patents to industries: tests of the Yale technology concordance}},
	author={Kortum, Samuel and Putnam, Jonathan},
	journal={Economic Systems Research},
	volume={9},
	number={2},
	pages={161--176},
	year={1997},
	publisher={Taylor \& Francis}, 
	doi={10.1080/09535319700000011}
}

@article{carvalho2019production,
	title={Production networks: A primer},
	author={Carvalho, Vasco M and Tahbaz-Salehi, Alireza},
	journal={Annual Review of Economics},
	volume={11},
	pages={635--663},
	year={2019},
	publisher={Annual Reviews}, 
	doi={10.1146/annurev-economics-080218-030212}
}

@techreport{BEA2009IO, 
	title={{{Concepts and Methods of the U.S. Input-Output Accounts. Measuring the Nation's Economy}}}, 
	author={Horrowitz, Karen and Planting, Mark},
	year={2006}, 
	type={{Bureau of Economic Analysis (BEA), U.S. Department of Commerce: Input-Output Manual}}, 
	url={https://apps.bea.gov/papers/pdf/IOmanual_092906.pdf}
	
}

@article{carvalho2014micro,
	title={{From micro to macro via production networks}},
	author={Carvalho, Vasco M},
	journal={Journal of Economic Perspectives},
	volume={28},
	number={4},
	pages={23--48},
	year={2014}, 
	doi={10.1257/jep.28.4.23}
}

@book{jackson2008social,
	title={{Social and economic networks}},
	author={Jackson, Matthew O},
	year={2008},
	publisher={Princeton University Press}
}

@techreport{blank2012intellectual,
	title={{Intellectual property and the US economy: Industries in focus}},
	author={Blank, Rebecca M and Kappos, David J},
	type={{Economics and Statistics Administration \& US Patent and Trademark Office}},
	year={2012}, 
	url={https://www.uspto.gov/sites/default/files/news/publications/IP_Report_March_2012.pdf}
}

@article{ruttan1959usher,
	title={{Usher and Schumpeter on invention, innovation, and technological change}},
	author={Ruttan, Vernon W},
	journal={{The Quarterly Journal of Economics}},
	pages={596--606},
	year={1959},
	publisher={JSTOR}, 
	doi={10.2307/1884305}
}

@incollection{oecd2009use,
	title={{Chapter 6: The use and analysis of citations in patents}},
	author={OECD},
	booktitle={{OECD Patent Statistics Manual}},
	pages={105--123},
	year={2009},
	publisher={OECD}, 
	doi={10.1787/9789264056442-en}
}

@article{huang2021network,
	title={Network structure and economic growth},
	author={Huang, Jingong},
	journal={Economics Letters},
	volume={207},
	pages={110022},
	year={2021},
	publisher={Elsevier}, 
	doi={10.1016/j.econlet.2021.110022}
}

@article{acemoglu2016innovation,
	title={{Innovation network}},
	author={Acemoglu, Daron and Akcigit, Ufuk and Kerr, William R},
	journal={Proceedings of the National Academy of Sciences},
	pages={201613559},
	year={2016},
	publisher={National Academy of Sciences}, 
	doi={10.1073/pnas.1613559113}
}

@article{kay2014patent,
	title={{Patent overlay mapping: Visualizing technological distance}},
	author={Kay, Luciano and Newman, Nils and Youtie, Jan and Porter, Alan L and 
	Rafols, Ismael},
	journal={{Journal of the Association for Information Science and Technology}},
	volume={65},
	number={12},
	pages={2432--2443},
	year={2014},
	publisher={Wiley Online Library}, 
	doi={10.1002/asi.23146}
}

@article{boehm2022comparative,
  title={The comparative advantage of firms},
  author={Boehm, Johannes and Dhingra, Swati and Morrow, John},
  journal={Journal of Political Economy},
  volume={130},
  number={12},
  pages={3025--3100},
  year={2022},
  publisher={The University of Chicago Press Chicago, IL}, 
  doi={10.1086/720630}
}

@Manual{qlcMatrix2018,
	title = {{qlcMatrix: Utility Sparse Matrix Functions for Quantitative Language
	Comparison}},
	author = {Michael Cysouw},
	year = {2018},
	note = {R package version 0.9.7},
	url = {https://CRAN.R-project.org/package=qlcMatrix},
}

@techreport{van2014patent,
	title={{Patent Statistics: Concordance IPC V8--NACE Rev. 2}},
	author={Van Looy, Bart and Vereyen, Caro and Schmoch, Ulrich},
	type={Technical report}, 
	institution={Eurostat, European Commission},
	year={2014}
}

@incollection{dosi2010technical,
	title={{Technical change and industrial dynamics as evolutionary processes}},
	author={Dosi, Giovanni and Nelson, Richard R},
	booktitle={{Handbook of the Economics of Innovation}},
	volume={1},
	pages={51--127},
	year={2010},
	publisher={Elsevier}, 
	doi={10.1016/S0169-7218(10)01003-8}
}

@incollection{cohen2010fifty,
	title={{Fifty years of empirical studies of innovative activity and 
	performance}},
	author={Cohen, Wesley M},
	booktitle={{Handbook of the Economics of Innovation}},
	volume={1},
	pages={129--213},
	year={2010},
	publisher={Elsevier}
}

@article{antony2012technology,
	title={{Technology flows between sectors and their impact on large-scale 
	firms}},
	author={Antony, J{\"u}rgen and Grebel, Thomas},
	journal={Applied Economics},
	volume={44},
	number={20},
	pages={2637--2651},
	year={2012},
	publisher={Taylor \& Francis}, 
	doi={10.1080/00036846.2011.566191}
}

@article{di2012technology,
	title={{Technology push and demand pull perspectives in innovation studies: 
	Current findings and future research directions}},
	author={Di Stefano, Giada and Gambardella, Alfonso and Verona, Gianmario},
	journal={Research Policy},
	volume={41},
	number={8},
	pages={1283--1295},
	year={2012},
	publisher={Elsevier}, 
	doi={10.1016/j.respol.2012.03.021}
}

@incollection{jaffe2019patent,
	title={{Patent citation data in social science research: Overview and best practices}},
	author={Jaffe, Adam B and De Rassenfosse, Ga{\'e}tan},
	booktitle={{Research Handbook on the Economics of Intellectual Property Law}},
	year={2019},
	publisher={Edward Elgar Publishing},  
	doi={10.4337/9781789903997.00043}
}

@article{romer1990endogenous,
	title={{Endogenous technological change}},
	author={Romer, Paul M},
	journal={Journal of Political Economy},
	volume={98},
	number={5, Part 2},
	pages={71--102},
	year={1990},
	publisher={The University of Chicago Press}, 
	doi={10.1086/261725}
}

@article{cohen1989innovation,
	title={{Innovation and learning: the two faces of R \& D}},
	author={Cohen, Wesley M and Levinthal, Daniel A},
	journal={The Economic Journal},
	volume={99},
	number={397},
	pages={569--596},
	year={1989},
	publisher={Oxford University Press Oxford, UK}, 
	doi={10.2307/2233763}
}

@article{acemoglu2002directed,
	title={{Directed technical change}},
	author={Acemoglu, Daron},
	journal={The Review of Economic Studies},
	volume={69},
	number={4},
	pages={781--809},
	year={2002},
	publisher={Wiley-Blackwell}, 
	doi={10.1111/1467-937X.00226}
}

@article{roodman2009xtabond2,
	title={{How to do xtabond2: An introduction to difference and system GMM in Stata}},
	author={Roodman, David},
	journal={The STATA Journal},
	volume={9},
	number={1},
	pages={86--136},
	year={2009},
	publisher={SAGE Publications Sage CA: Los Angeles, CA}, 
	doi={10.1177/1536867X0900900106}
}

@article{corrado2009intangible,
	title={Intangible capital and US economic growth},
	author={Corrado, Carol and Hulten, Charles and Sichel, Daniel},
	journal={Review of Income and Wealth},
	volume={55},
	number={3},
	pages={661--685},
	year={2009},
	publisher={Wiley Online Library}, 
	doi={10.1111/j.1475-4991.2009.00343.x}
}

@techreport{goldin2007long,
	title={Long-run changes in the US wage structure: Narrowing, widening, polarizing},
	author={Goldin, Claudia and Katz, Lawrence F},
	year={2007},
	institution={National Bureau of Economic Research}, 
	type={Working Paper No. w13568}, 
	doi={10.3386/w13568}
}

@techreport{carvalho2014input,
	title={{Input diffusion and the evolution of production networks}},
	author={Carvalho, Vasco M and Voigtl{\"a}nder, Nico},
	year={2014},
	institution={{National Bureau of Economic Research}}, 
	type={Working Paper No. w20025}, 
	doi={10.3386/w20025}
}

@article{lerner1994importance,
	title={The importance of patent scope: an empirical analysis},
	author={Lerner, Joshua},
	journal={The Rand Journal of Economics},
	pages={319--333},
	year={1994},
	publisher={JSTOR}, 
	doi={10.2307/2555833}
}

@article{gallipoli2018structural,
	title={Structural transformation and the rise of information technology},
	author={Gallipoli, Giovanni and Makridis, Christos A},
	journal={Journal of Monetary Economics},
	volume={97},
	pages={91--110},
	year={2018},
	publisher={Elsevier}, 
	doi={10.1016/j.jmoneco.2018.05.005}
}

\newpage
\renewcommand{\appendixname}{Appendix}
\renewcommand{\thesection}{\Alph{section}} \setcounter{section}{0}
\renewcommand{\thefigure}{\Alph{section}.\arabic{figure}} \setcounter{figure}{0}
\renewcommand{\thetable}{\Alph{section}.\arabic{table}} \setcounter{table}{0}
\renewcommand{\theequation}{\Alph{section}.\arabic{table}} \setcounter{equation}{0}
\part*{Appendix}
\appendix

\label{appendix}

\section{Methods}
\label{app:methods}
\subsection{Data compilation}
\label{app:data}
\subsubsection{Input-output data}
\label{app:data_io}
BEA provides detailed current and historical benchmark IO tables in quinquennial frequency dating back to 1947.\footnote{The data was downloaded from \url{https://www.bea.gov/industry/benchmark-input-output-data} and \url{https://www.bea.gov/industry/historical-benchmark-input-output-tables} [Both accessed in Oct 2021].} The most disaggregate data at the 6-digit level is used. The data is accounting data, which show monetary flows between industries including final demand, and dummy positions that ensure the financial closure. Accounting positions are largely but not perfectly compatible with NAICS or Standard Industrial Classification (SIC) codes. 
The data stepwise converted into a time-consistent and convenient format. First, the data was transformed from accounting positions into industry codes, i.e. SIC codes for the 1977-1987 data and into NAICS for later periods. 
The industry codes are harmonized to the NAICS 2002 version using multiple concordances provided by BEA.\footnote{Detailed explanations of conceptual and technical issues (e.g. changing classification systems, ambiguous mappings) that arose during the compilation are available in \ref{supp:data_processing_IO}.} 

After harmonizing the data, a matrix of monetary flows between 1179 distinct 6-digit NAICS industries has been obtained for each period. The entries of the matrix are input flows $flow^{\mu}_{ij,t}$ indicating the monetary value of the inputs that $i$ buys from $j$ in time $t$. 
Division of the flows by the row sums $\sum_j flow^{\mu}_{ij,t}$ gives the input shares $w^{\mu,up}_{i,t}$. The output shares $w^{\mu,dw}_{i,t}$ are obtained by division by column sums. 
Note that some rows and columns are empty for some $t$. This results from the harmonization procedure to uniform NAICS codes and can happen when the classification changes. Industries can emerge or disappear over time. For example, industries associated with computer technologies were less granular in the 70s compared to the 90s. This is often associated with a split (merge) of pre-existing industries.

\subsubsection{Patent data}
\label{app:data_patent}
The innovation layer is taken from a patent data set compiled for an earlier project \citep{hotte2021data_rise, hotte2021rise}. 
The data contains a list of USPTO patents including the grant year and CPC technology classes, relying on the USPTO Master Classification File from January 2020.\footnote{\url{https://bulkdata.uspto.gov/data/patent/classification/}} 
This file offers a list that maps individual patents to one or more CPC classes at the most disaggregate level. 

To obtain NAICS level patent data, the patents-to-CPC and the CPC-to-NAICS provided by \citet{goldschlag2020tracking} were merged. The latter is a probabilistic mapping and comes along with probability weights whenever the one CPC class maps to multiple NAICS. 
These weights are used when compiling industry level patent stocks and industry-to-industry citation counts. 
To compile the 5-year snapshot of patent stocks, weighted patents per NAICS class were aggregated for each time window. 
To obtain the number of patents of an industry $i$ and in time $t$, all patents of a given CPC 4-digit that show a link to NAICS 6-digit sector $i$ are summed up after having been multiplied with the corresponding weight. For example, patents classified into CPC class A01B map to NAICS 6-digit sector 115112 with a weight of 0.9996819 and to sector 237010 with a weight 0.0003181. Hence, all patents that were granted in the time window $t$ and classified by A01B were summed up to get the patent count, then multiplied by 0.99996819, and assigned to sector 115112. Industry 115112 does also map to other CPC 4-digit classes (A01C, A01G, E01H). The patent counts from these classes are summed up in the same way and added to the industry level patent count.\footnote{The data and the R-code for the data compilation are published under a CC-BY 4.0 license as \citet{hotte2023data}.} 
Note that the number of CPC codes per patent are very heterogeneous, which may lead to double counts. Here, the heterogeneity is ignored, as a higher number of co-classifications is often coinciding with a higher value of the patent \citep{lerner1994importance}. This may justify that the patent enters the aggregate data with a higher weight compared to patents with fewer CPC codes.\footnote{It should be noted that the final analyses rely on citation-weighted patents, which may lead to a repeated double counting when patents map to multiple CPC classes and are frequently cited. However, the non-citation-weighted and the weighted patent stock are very high (see Fig. \ref{fig:sixDcomp_correl_plots}).}

Patent stocks rely on the time windows prior to the benchmark year. For example, for the patent stock in 1977, all patents granted in 1973-1977 are summed up. However, one could also argue to use the subsequent time window 1977-1981 to compile patents for 1977. The IO data is a time snapshot of the last year in the time window. The preceding time window is used for four major reasons: (1) This paper uses granted patents, and the time lag between patent application and grant often accounts for a few years. (2) Innovation is a dynamic concept comprising the process of invention, innovation and commercialization, and diffusion. Using the earlier time window takes account of the diffusion lag. (3) Patents are seen as a proxy for the stock of available technological knowledge, and patents that will be granted in future are not yet available as knowledge for current use. (4) This approach is consistent with other research where discounted patent stocks were used as proxies of innovation and technological knowledge \citep[e.g.][]{antony2012technology, huang2021network}.

The same procedure is applied to the citation data, where both the citing and the cited patent both are mapped to NAICS codes. 
In numbers, more than 37.66 M citation links between 3.75 M individual patents are first expanded to the CPC 4-digit level and then aggregated into citation counts for each NAICS-to-NAICS pair in the relevant period. 

These NAICS-to-NAICS citation counts are transformed into a symmetric matrix where the entries $flow^{\tau}_{ij,t}$ correspond to the flow of citations from $i$ to $j$, i.e. the number of times that $i$ cites patents from industry $j$. 
As above, the entries of $flow^{\tau}_t$ are transformed into input shares $w_{ij,t}^{\tau, d}$ through division by the row sum $\sum_j flow^{\tau, d}_{ij,t}$. Output shares $w^{\tau,dw}_{i,t}$ are obtained by division by column sums. Additional detail on the data processing is provided in \ref{supp:data_processing_patents}. 

\subsubsection{Supplementary data and processing}
\label{app:data_suppl_processing}
The data is supplemented with data from the NBER-CES Manufacturing Productivity Database \citep{becker2013nber, bartlesman1996nber}.\footnote{\url{https://www.nber.org/research/data/nber-ces-manufacturing-industry-database}}
For the main analyses, 6-digit level data is used and the subset of manufacturing industries. More aggregate level and additional data on non-manufacturing sectors have been used in earlier exploratory analyses and validation tests.\footnote{The data is available in the accompanying research data publication \citep{hotte2023data}.} 
Robustness checks with more aggregate data aim to cope with concerns about the reliability of the classification approach, as classification systems change over time and many sequential transformations were necessary. 

The data is unbalanced panel data, i.e. some industries have no data entry for output flows or patent counts in some periods. 
For the main analysis, industries with incomplete coverage were removed. The final data is characterized by $A^{\alpha}_{i,t}>0 \; \; \forall \; \; t, \alpha$. 
This reduces the sample size from 473 to 307 6-digit manufacturing industries. 
All variables that are measured in monetary terms are deflated using the price deflator for the value of shipment (\emph{piship}) from the NBER-CES productivity database. 

Both networks (cross-industrial flow and share matrices) and the raw patent data are used to construct industry level variables. 
Using the raw patent data, aggregate citation-weighted patent stocks $A^{\tau}_{i,t}$ at the industry level are compiled. The citation weights are used to control for the heterogeneity of patents by value \citep[e.g.][]{jaffe2019patent}. 

Using the IO flow network, the sum of output given by the column sum $A^{\mu}_{i,t} = \sum_k flow^{\mu}_{ki,t}$ is used as a measure for the market size. 
To cope with potential inconsistency across time, additional robustness checks are made with normalized data, dividing all entries by the cross-industry average $\tfrac{1}{|N|}\sum_j {A}^{\alpha}_{j,t}$ for each $t$.  
The normalized size measures the size relative to other industries and is used to illustrate the evolution of size ranking of industries over time (see Sec. \ref{sec:results_descr}). In the normalized data, the cross-industry average equals one. 

The networks $W^{\alpha, d}_{t}$ are used to compile the network centrality $PR^{\alpha, d}_{i,t}$. Other centrality measures (degree, strength) are used for robustness checks.  

The weight matrices $W^{\alpha, d}_{t}$ are further used to compute the cosine similarity matrices $\Sigma^{\alpha, d}_{t} = \{ \sigma_{ij, t}^{\alpha, d} \}_{i,j \in N}$ using a sparsity-robust version by \citet{qlcMatrix2018}. The similarity matrix is used to show how technological similarities in the network evolved (see below and Table \ref{tab:sixDman_netw_stats_summary}). 
The matrices $W^{\alpha, d}_{i,t}$ and industry sizes $A^{\alpha}_{i,t}$ are used to compute cross-industry spillovers $Spill(A)^{\alpha, d}_{i,t}$. 

In the robustness checks at other aggregation levels, all measures are re-compiled from the network data at the respective aggregation level as network properties (e.g. centrality, density, clustering) may change unsystematically when the aggregation level changes \citep{kymn1990aggregation}.

\subsection{Network plots}
\label{app:network_plots}
In the network plots shown in Sec. \ref{sec:results_descr}, links between two industries $i$ and $j$ if $j$ is a sufficiently important input supplier to $i$ and the weight $w^{\alpha,up}_{ij,t}$ exceeds a threshold level defined by the average of weights across all industry-pairs and periods plus one standard deviation. 

The node size is scaled proportionally to $A^{\alpha}_{i,t}$, which is the average value over the time window. To make the sizes in the market and innovation layer comparable, the average value is normalized by division through the cross-industry standard deviation over the full time horizon. After this normalization, the log is taken using a $log(1+x)$ formula to deal with $<1$ values.  

The plots are generated using the plotting function of the \emph{igraph} package in R using the Fruchtermann-Reingold algorithm for the layout. This algorithm bases on the principle to minimize the number of crossing edges while keeping them at roughly equal lengths. This comes with the side effect that nodes with similar linking patterns tend to group together.
\FloatBarrier

\subsection{Data transformations}
\label{app:regression_data}
Before performing the statistical analyses, a series of data transformations was undertaken to harmonize the data by scale and to cope with outliers and skewness. 
\begin{table}[h]
	
	\centering
	\caption{Overview statistics of regression variables}
	\label{tab:overview_stats_regr_vars_6}
	\begin{myresizeenv}
		
		\begingroup
		\begin{tabular}{|l|cccc|cccc|}
			\hline \hline \rule{0pt}{1.075\normalbaselineskip} 
			&\multicolumn{4}{c|}{Before transformation}&\multicolumn{4}{c|}{After transformation} \\
			\hline \rule{0pt}{1.075\normalbaselineskip} & Mean & Min & Max & Median & Mean & Min & Max & Median \\ 
			\hline 	\rule{0pt}{1.075\normalbaselineskip} $A^{\tau}_{i}$ & 2557 & 0.05 & 141771 & 504.30 & 5.873 & 0.05 & 11.86 & 6.22 \\ 
			$A^{\tau*}_{i,t}$ & 47145 & 0.02 & 2741069 & 6722.00 & 8.445 & 0.02 & 14.82 & 8.81 \\ 
			$A^{\mu}_{i,t}$ & 5447 & 0.30 & 480728 & 2498.00 & 7.594 & 0.26 & 13.08 & 7.82 \\ 
			$A^{\mu*}_{i,t}$ & 5862 & 0.36 & 698224 & 2870.00 & 7.726 & 0.31 & 13.46 & 7.96 \\ 
			\hline \rule{0pt}{1.075\normalbaselineskip}
			$PR^{\tau,up}_{i,t}$ & 0.0033 & 0.00 & 0.0699 & 0.00 & 1.056 & 0.40 & 4.261 & 0.79 \\ 
			$PR^{\tau,dw}_{i,t}$ & 0.0033 & 0.00 & 0.0697 & 0.00 & 1.058 & 0.40 & 4.258 & 0.82 \\ 
			$PR^{\mu,up}_{i,t}$ & 0.0033 & 0.00 & 0.2095 & 0.00 & 1.016 & 0.40 & 5.349 & 0.74 \\ 
			$PR^{\mu,dw}_{i,t}$ & 0.0033 & 0.00 & 0.1344 & 0.00 & 1.084 & 0.40 & 4.908 & 0.83 \\ 
			\hline \rule{0pt}{1.075\normalbaselineskip}
			$Spill(A)^{\tau,up}_{i,t}$ & 646738 & 197324.00 & 1321936 & 543613.00 & 4.064 & 3.03 & 4.892 & 4.01 \\ 
			$Spill(A)^{\tau*,up}_{i,t}$ & 12140991 & 3280630.00 & 28901043 & 9762547.00 & 6.941 & 5.80 & 7.969 & 6.88 \\ 
			$Spill(A)^{\mu,up}_{i,t}$ & 477533 & -129887.00 & 2505736 & 350955.00 & 3.587 & -2.64 & 5.528 & 3.59 \\ 
			$Spill(A)^{\mu*,up}_{i,t}$ & 537552 & -134443.00 & 2505413 & 435077.00 & 3.753 & -2.67 & 5.528 & 3.80 \\ 
			$Spill(A)^{\tau,dw}_{i,t}$ & 271075 & -9.07 & 1367711 & 144102.00 & 2.352 & -0.00 & 4.926 & 2.73 \\ 
			$Spill(A)^{\tau*,dw}_{i,t}$ & 4989650 & -124.90 & 29603733 & 2218575.00 & 4.574 & -0.01 & 7.993 & 5.41 \\ 
			$Spill(A)^{\mu,dw}_{i,t}$ & 232683 & -30240.00 & 3479138 & 88113.00 & 2.217 & -1.39 & 5.855 & 2.28 \\ 
			$Spill(A)^{\mu*,dw}_{i,t}$ & 258969 & -113022.00 & 3175549 & 106248.00 & 2.343 & -2.51 & 5.764 & 2.45 \\ 
			\hline \rule{0pt}{1.075\normalbaselineskip}
			$TFP^{}_{i,t}$ & 4.136 & 0.04 & 326.2 & 0.96 & 0.6717 & 0.04 & 2.428 & 0.67 \\ 
			$(VA/L)^{}_{i,t}$ & 105.1 & 10.46 & 2404 & 75.76 & 4.353 & 2.44 & 7.785 & 4.34 \\ 
			$(VA/L)^{*}_{i,t}$ & 4053 & 19.62 & 366784 & 1995.00 & 7.615 & 3.03 & 12.81 & 7.60 \\ 
			
			$L^{}_{i,t}$ & 36.43 & 0.74 & 469.5 & 22.67 & 3.197 & 0.55 & 6.154 & 3.16 \\ 
			$W^{}_{i,t}$ & 29.85 & 5.42 & 101.4 & 27.85 & 3.309 & 1.86 & 4.629 & 3.36 \\ 
			$W^{*}_{i,t}$ & 1278 & 9.38 & 84494 & 687.90 & 6.548 & 2.34 & 11.34 & 6.54 \\ 
			
			$(K/L)^{}_{i,t}$ & 117.2 & 5.15 & 1958 & 75.10 & 0.6649 & 0.05 & 3.025 & 0.56 \\ 
			$(K/L)^{*}_{i,t}$ & 4506 & 33.45 & 222749 & 2110.00 & 3.15 & 0.29 & 7.709 & 3.10 \\ 
			
			$(I/L)^{}_{i,t}$ & 7.125 & 0.20 & 221.6 & 4.42 & 1.774 & 0.18 & 5.405 & 1.69 \\ 
			$(I/L)^{*}_{i,t}$ & 276.4 & 1.10 & 12254 & 123.00 & 4.829 & 0.74 & 9.414 & 4.82 \\ 
			
			$(L^P/L)_{i,t}$ & 0.716 & 0.29 & 0.931 & 0.74 & 0.5378 & 0.26 & 0.658 & 0.56 \\ 
			$(W^P/W)_{i,t}$ & 1.226 & 0.86 & 2.581 & 1.17 & 1.226 & 0.86 & 2.581 & 1.17 \\ 
			$(W^P/W)^{*}_{i,t}$ & 0.858 & 0.48 & 1.195 & 0.87 & 0.858 & 0.48 & 1.195 & 0.87 \\ 
			
			\hline \rule{0pt}{1.075\normalbaselineskip} 
			$Vship^{}_{i,t}$ & 240.2 & 20.16 & 10509 & 158.50 & 5.097 & 3.05 & 9.26 & 5.07 \\ 
			$Vship^{*}_{i,t}$ & 8939 & 33.80 & 565041 & 4149.00 & 8.366 & 3.55 & 13.24 & 8.33 \\ 
			$(M/L)^{}_{i,t}$ & 135.8 & 5.59 & 9122 & 77.90 & 4.412 & 1.89 & 9.119 & 4.37 \\ 
			$(M/L)^{*}_{i,t}$ & 4575 & 37.19 & 322902 & 2025.00 & 7.649 & 3.64 & 12.69 & 7.61 \\ 
			$(E/L)^{}_{i,t}$ & 5.183 & 0.12 & 160.6 & 2.06 & 1.323 & 0.12 & 5.085 & 1.12 \\ 
			$(E/L)^{*}_{i,t}$ & 152.9 & 0.47 & 7031 & 58.73 & 4.143 & 0.38 & 8.858 & 4.09 \\
			\hline
			\hline
		\end{tabular}
		\endgroup
		
	\end{myresizeenv}
	
	\vspace{0.25cm}
	
	\justifying \footnotesize
	
	\noindent
	Notes: This table shows the overview statistics of the variables included in the regression equations before and after data transformation. The last block of rows shows the data entries of the patent counts when the data is not weighted by patent citations. * indicates that the row entry refers to the citation-weighted value for patents and deflated data for market data in monetary terms. 
\end{table}
Table \ref{tab:overview_stats_regr_vars_6} shows an overview of the variables included in the regression analyses. 
The columns at the left-hand side show the raw data values, and the right-hand side columns show the values after a series of data transformations that are done to make the data more comparable and to cope with outliers and highly skewed distributions. 
The transformation steps include in sequential order: 
\begin{enumerate}
	\item Network variables are scaled by their standard deviation to make the size $A^{\alpha}_{i,t}$, centrality $PR^{\alpha,d}_{i,t}$, and spillovers $Spill(A)^{\alpha,d}_{i,t}$ comparable across the three variables and across the two layers. 
	\item All variables except for the share of production labor ($(L^P/L)_{i,t}$ transformed to log values using the formula $\log(1+x)$ to cope with $<1$ values. The log-linearization is done to cope with highly skewed data. 
	\item Outliers are removed according to an interquartile range (IQR) based formula. Those values are treated as outliers that are beyond the 25/75\% quantile values minus/plus $(a \cdot IQR)$ with $a = 30$ in the baseline models. Robustness checks are made with more restrictive (i.e. $a=5$ and $a=10$) removal rules. The regression results are qualitatively consistent with the baseline. 	
\end{enumerate}
The full code that was used to compile and process the data will be made available upon publication of this paper. 

\FloatBarrier

\section{Additional descriptive information}
\label{app:add_descr_results}

\begin{table}
	\centering
	\caption{Aggregate network statistics over time at the 6-digit level.} 
	\label{tab:sixDman_netw_stats_summary}
	
	\begin{myresizeenv}
		\begingroup
		
		\begin{tabular}{|l|cccc|cccc|}
			\hline \hline  \rule{0pt}{1.075\normalbaselineskip}  
			& \multicolumn{4}{c|}{\ul{Input-output}}&\multicolumn{4}{c|}{\ul{Patent}}  \\ \rule{0pt}{1.075\normalbaselineskip}
			&1977-1982&1987-1992&1997-2002&2007-2012&1977-1982&1987-1992&1997-2002&2007-2012\\
			\hline 
			\multicolumn{9}{|l|}{\emph{Flow matrix - upstream network}}\\
			\hline \rule{0pt}{1.075\normalbaselineskip}
			Density & 0.07 & 0.07 & 0.07 & 0.10 & 0.14 & 0.15 & 0.16 & 0.17 \\ 
			Avg. degree & 22.02 & 21.86 & 20.08 & 29.02 & 44.43 & 46.39 & 49.84 & 51.33 \\ 
			Avg. weight & 0.87 & 0.83 & 0.88 & 0.75 & 0.36 & 0.35 & 0.34 & 0.34 \\ 
			Reciprocity & 0.15 & 0.14 & 0.12 & 0.33 & 0.36 & 0.34 & 0.32 & 0.29 \\ 
			Transitivity & 0.34 & 0.34 & 0.30 & 0.34 & 0.42 & 0.43 & 0.45 & 0.46 \\ 
			Diameter & 3.00 & 3.00 & 4.00 & 2.00 & 2.00 & 2.00 & 2.00 & 2.00 \\ 
			Mean dist. & 1.43 & 1.41 & 1.43 & 1.49 & 1.07 & 1.04 & 1.02 & 1.02 \\ 
			Assort. by degree & -0.25 & -0.24 & -0.19 & -0.33 & -0.12 & -0.10 & -0.07 & -0.06 \\ 
			Assort. by size & -0.00 & -0.02 & -0.03 & -0.02 & -0.02 & -0.01 & -0.00 & -0.00 \\ 
			\hline				
			\hline 
			\multicolumn{9}{|l|}{\emph{Flow matrix - downstream network}}\\
			\hline \rule{0pt}{1.075\normalbaselineskip}
			Density & 0.08 & 0.07 & 0.07 & 0.09 & 0.14 & 0.15 & 0.16 & 0.16 \\ 
			Avg. degree & 24.15 & 22.50 & 20.91 & 28.39 & 43.72 & 45.42 & 48.95 & 50.33 \\ 
			Reciprocity & 0.17 & 0.17 & 0.13 & 0.30 & 0.36 & 0.35 & 0.33 & 0.30 \\ 
			
			\hline				
			\hline 
			\multicolumn{9}{|l|}{\emph{Cosine similarity - upstream network}}\\
			\hline \rule{0pt}{1.075\normalbaselineskip}
			
			Density & 0.28 & 0.27 & 0.28 & 0.31 & 0.38 & 0.40 & 0.40 & 0.40 \\ 
			Avg. degree & 86.33 & 83.22 & 85.20 & 94.89 & 116.73 & 121.32 & 122.52 & 123.42 \\ 
			Avg. weight & 12.84 & 11.60 & 12.20 & 9.37 & 23.69 & 26.35 & 28.40 & 31.50 \\ 
			Transitivity & 0.67 & 0.68 & 0.75 & 0.63 & 0.72 & 0.74 & 0.75 & 0.74 \\ 
			
			\hline				
			\hline 
			\multicolumn{9}{|l|}{\emph{Cosine similarity - downstream network}}\\
			\hline \rule{0pt}{1.075\normalbaselineskip}
			
			Density & 0.25 & 0.24 & 0.23 & 0.29 & 0.39 & 0.40 & 0.41 & 0.41 \\ 
			Avg. degree & 76.85 & 73.10 & 71.95 & 88.92 & 118.84 & 121.95 & 124.19 & 125.52 \\ 
			Avg. weight & 9.59 & 8.82 & 7.72 & 9.47 & 25.35 & 26.52 & 29.41 & 30.93 \\ 
			Transitivity & 0.61 & 0.62 & 0.58 & 0.67 & 0.72 & 0.74 & 0.74 & 0.73 \\ 
			
			\hline \hline
			
		\end{tabular}

		\endgroup
	\end{myresizeenv}
	\vspace{0.25cm}
	
	\justifying \footnotesize
	\noindent
	Notes: The upper part of the table shows a series of network statistics compiled at the basis of the up- and downstream links in the market- and innovation layer for different time windows. The links in these time windows are averaged. 
	The lower parts of the table summarize the network characteristics of the cosine similarity network given by the symmetric $|N| \times |N|$ cosine similarity matrix $\Sigma^{\alpha, d}_t$, where the pairwise similarities $\sigma^{\alpha, d}_{ij,t}$ are the weights of a link connecting $i$ and $j$. 		
	The metrics are compiled using the R-package \emph{igraph} \citep{csardi2006igraph}. For an introduction to the use of these metrics see also \citet{jackson2008social}. 
	Those variables that are identical for the up- and downstream network are shown only once. They are identical because of the normalization of weights (Avg. weight) to shares or by the nature of the data (Diameter, Transitivity, Mean distance, Assortativity). 
\end{table}

Table \ref{tab:sixDman_netw_stats_summary} shows the properties of the two network layers, constructed on the basis of up- and downstream links. 
In addition, the bottom part of the table shows the network characteristics of the up- and downstream similarity network. This is a network that shows industries as being connected if they are very similar by their bundle of upstream (or downstream) links.

Upstream similarity in the market $\sigma^{\mu,up}_{ij,t}$ indicates that a pair of industries relies on a similar bundle of intermediate goods as production inputs.  An increase in $Spill(A)^{\mu,up}_{i,t}$ indicates a rise in the competition for these inputs as either, competitors that use the same inputs grew ($A^{\mu}_{j,t} \uparrow$) or the extent to which input requirements overlap increased ($\sigma^{\mu,up}_{ij,t} \uparrow$). 

Downstream similarity in the market $\sigma^{\mu,dw}_{ij,t}$ measures the overlap of $i$'s and $j$'s customer links, which indicates that the outputs of $i$ and $j$ serve similar customer needs. This can be also an indicator for competition if the outputs produced by $i$ and $j$ are substitutes, but it may also indicate demand synergies if the outputs are complements. 
An increasing similarity of two industries over time may be an indicator of technological and economic convergence. 

The first lines of each section of the table show the network density, which measures the connectivity. It shows the ratio of actual over potential network links. Both layers are sparsely connected. The innovation layer is denser and shows an increase in the up- and downstream density over time, while connectivity in the market does not exhibit any clear trend. 
The average degree (second line of each block in the table) indicates the average number of industries to which an industry is connected. 
On average, an industry is connected to 20-30 customers and suppliers in the market and cites patents from 44-51 other industries. 
Both network layers show a negatively valued assortativity: Larger and more connected industries tend to link more often to smaller and less connected industries. 

The density in the cosine similarity networks is an indicator of technological convergence, measuring whether industries became more similar on average. In line with the connectivity trends, similarity in the market fluctuates without any clear trend, while industries became increasingly similar by patent citations. 

\begin{figure}[h]
	\begin{subfigure}[]{0.45\textwidth}
		\includegraphics[width=\textwidth]{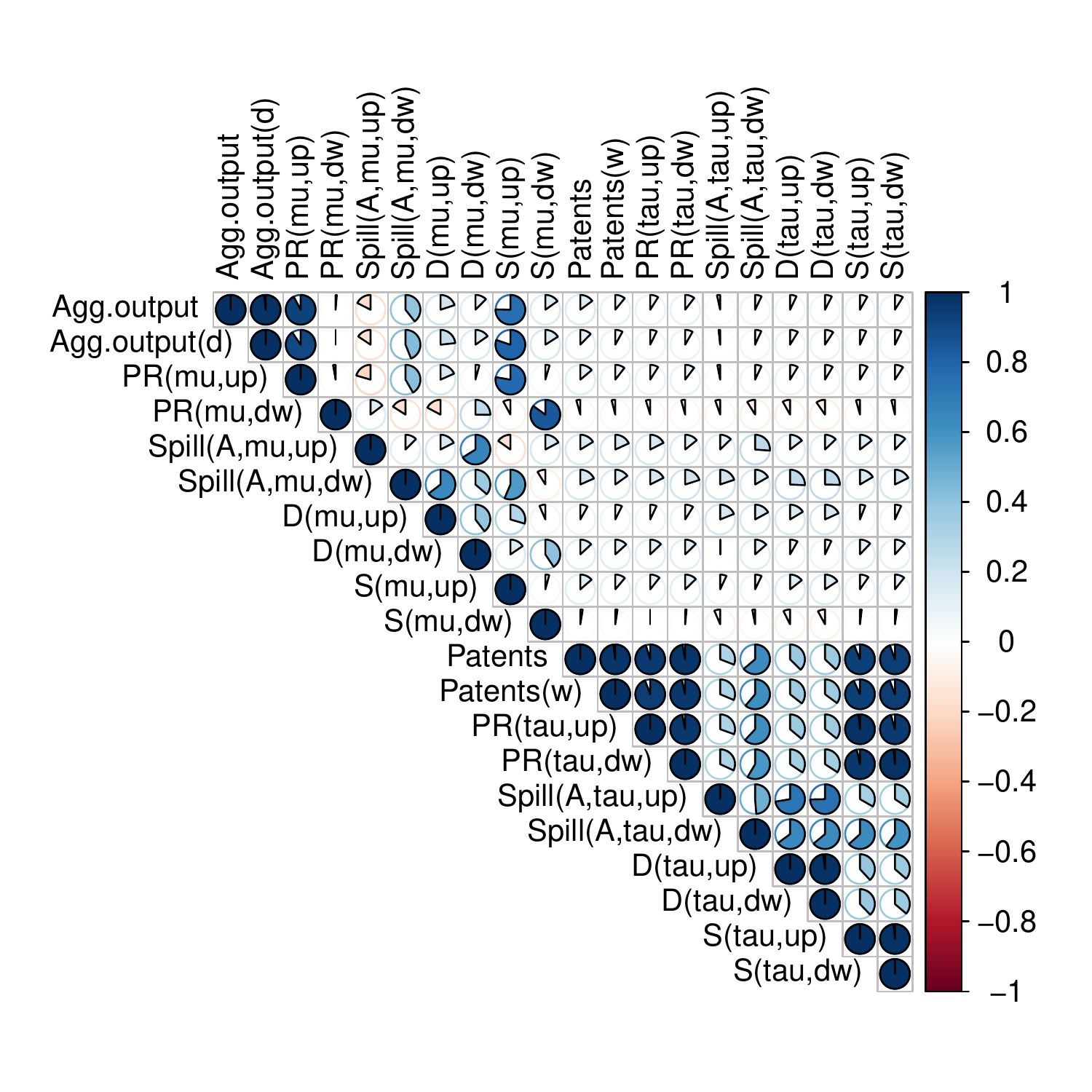}
		\caption{1977-1992}
		\label{fig:corrplot_1982}
	\end{subfigure}
	\begin{subfigure}[]{0.45\textwidth}
		\includegraphics[width=\textwidth]{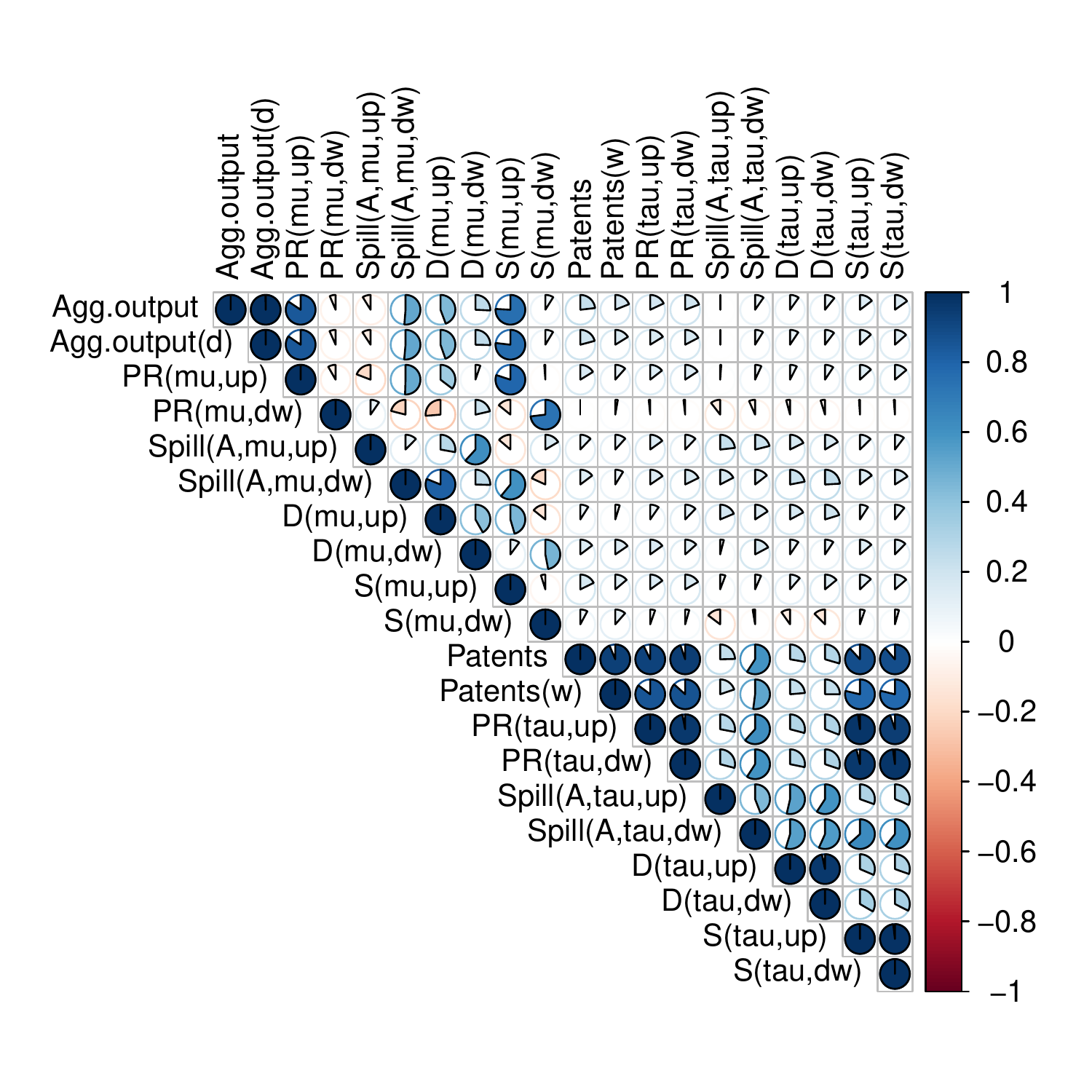}
		
		\caption{1997-2012}
		\label{fig:corrplot_2012}
	\end{subfigure}
	
	\caption{Pairwise correlations of network indicators in two subperiods}

	\scriptsize \justifying 
	\noindent 
	Notes: These figures show a correlation plot between different pairs of network variables at the 6-digit level. Figure \ref{fig:corrplot_1982} (\ref{fig:corrplot_2012}) shows the average correlations during the first (second) half of the period that is studied. To calculate the average correlations, the data is averaged over the respective periods before the correlation coefficient is computed. $PR$ is short for PageRank, $D$ for degree, $S$ for strength. $Patents (w)$ is the weighted patent stock. The correlation at the diagonal is by definition equal one. The colors and the shape of the ellipses indicate the strength of correlation. 
	\label{fig:sixDcomp_correl_plots}
\end{figure}

Figure \ref{fig:sixDcomp_correl_plots} shows the pairwise correlation of different indicators used to describe the network across and within layers. The figure illustrates a that patent up- and downstream network properties are by far higher correlated compared to the equivalent metrics in the market. The correlation among the network metrics in the innovation layer is always positive, while there are some negative correlations in the market, e.g. between the up- and downstream centrality or upstream spillovers and market size. The figure also shows that the correlation patterns are fairly robust over time when comparing the two subperiods.

\newpage

\section{Additional regression results}
\label{app:add_regression_results}
\subsection{Demand-pull and technology-push}
\FloatBarrier
\label{app:results_DP_TP}

Table \ref{tab:real_sum_output|citation_weighted_patent_stock_all_sectors_all_years_sys_1step_wControls_appendix} show the same results as discussed in Sec. \ref{subsubsec:DP_and_TP} but including industry level controls. 
Most of the coefficients of the controls are not significant, suggesting that these effects are likely already captured by the FE or the autoregressive term. Only wages, investment per capita, and energy intensity are significant. Wages exhibit a negative association with market growth, energy intensity shows a negative one with innovation, and investment per capita shows a positive relationship with both market growth and innovation.
\begin{table}[h]
\begin{myresizeenv}
\begingroup
\begin{tabular}{|l|cccc|cccc|cccc|}
	\hline \hline  \rule{0pt}{1.075\normalbaselineskip}  
	& \multicolumn{4}{c|}{ \ul{Type 1}} & \multicolumn{4}{c|}{ \ul{Type 2}} & \multicolumn{4}{c|}{ \ul{Both}}\\  & \multicolumn{2}{c}{ $\tau \rightarrow \mu$} & \multicolumn{2}{c|}{ $\mu \rightarrow \tau$}
	& \multicolumn{2}{c}{ Market} & \multicolumn{2}{c|}{ Innovation}
	& \multicolumn{4}{c|}{}\\
	\hline  \rule{0pt}{1.075\normalbaselineskip}    & $A^{\mu}_{i,t}$ & $A^{\mu}_{i,t}$ & $A^{\tau}_{i,t}$ & $A^{\tau}_{i,t}$ & $A^{\mu}_{i,t}$ & $A^{\mu}_{i,t}$ & $A^{\tau}_{i,t}$ & $A^{\tau}_{i,t}$ & $A^{\mu}_{i,t}$ & $A^{\mu}_{i,t}$ & $A^{\tau}_{i,t}$ & $A^{\tau}_{i,t}$ \\ 
	& (1) & (2) & (3) & (4) & (5) & (6) & (7) & (8) & (9) & (10) & (11) & (12) \\ 
	\hline \rule{0pt}{1.075\normalbaselineskip}   $A^{\mu}_{i,t-1}$ & 0.6255*** & 0.6073*** & 0.0035 & -0.0058 & 0.7321*** & 0.6514*** &  &  & 0.6867*** & 0.6603*** & 0.0021 & -0.0058 \\ 
	& (0.0291) & (0.0242) & (0.0136) & (0.0127) & (0.0385) & (0.034) &  &  & (0.039) & (0.0344) & (0.0128) & (0.0117) \\ 
	$A^{\tau}_{i,t-1}$ & -0.0637 & -0.0246 & 1.034*** & 1.016*** &  &  & 1.044*** & 1.034*** & -0.0197 & -0.0351 & 1.061*** & 1.049*** \\ 
	& (0.0947) & (0.0849) & (0.0076) & (0.0068) &  &  & (0.0403) & (0.0314) & (0.0862) & (0.0778) & (0.0359) & (0.0303) \\ 
	$PR^{\mu,up}_{i,t-1}$ &  &  & -0.0055 & -1e-04 & 0.0066 & 0.01 &  &  & 0.0088 & 0.0047 & -0.0062 & -0.0021 \\ 
	&  &  & (0.0046) & (0.0041) & (0.015) & (0.0131) &  &  & (0.0145) & (0.0127) & (0.0046) & (0.0039) \\ 
	$PR^{\mu,dw}_{i,t-1}$ &  &  & 0.0021 & 0.0041 & -0.001 & 0.0036 &  &  & -0.0039 & 0.005 & 9e-04 & 0.0039 \\ 
	&  &  & (0.0033) & (0.0031) & (0.0145) & (0.0121) &  &  & (0.0137) & (0.0118) & (0.0033) & (0.003) \\ 
	$PR^{\tau,dw}_{i,t-1}$ & 0.1536** & 0.0928* &  &  &  &  & 0.0482** & 0.0314* & 0.1119** & 0.0924** & 0.031* & 0.0183. \\ 
	& (0.0478) & (0.0391) &  &  &  &  & (0.0167) & (0.0136) & (0.0376) & (0.0322) & (0.013) & (0.0109) \\ 
	$Spill(A)^{\mu,up}_{i,t-1}$ &  &  & -0.0126* & -0.0081 & 0.1054*** & 0.1043*** &  &  & 0.0975*** & 0.0868*** & -0.0133* & -0.0088 \\ 
	&  &  & (0.006) & (0.006) & (0.02) & (0.0189) &  &  & (0.0207) & (0.0195) & (0.0066) & (0.0063) \\ 
	$Spill(A)^{\mu,dw}_{i,t-1}$ &  &  & 0.0044 & 0.006 & -0.0394. & -0.007 &  &  & -0.0429* & -0.0178 & 0.002 & 0.0068 \\ 
	&  &  & (0.0058) & (0.005) & (0.0228) & (0.0192) &  &  & (0.02) & (0.0181) & (0.0053) & (0.0048) \\ 
	$Spill(A)^{\tau,up}_{i,t-1}$ & 2.435*** & 2.572*** &  &  &  &  & 0.7004** & 0.3476* & 2.363*** & 2.642*** & 0.4757** & 0.2599. \\ 
	& (0.7017) & (0.5889) &  &  &  &  & (0.2131) & (0.1672) & (0.6273) & (0.5391) & (0.156) & (0.1388) \\ 
	$Spill(A)^{\tau,dw}_{i,t-1}$ & -0.0606 & -0.0573 &  &  &  &  & -0.0539*** & -0.0369** & -0.0695. & -0.059. & -0.0514*** & -0.0378** \\ 
	& (0.0417) & (0.035) &  &  &  &  & (0.016) & (0.013) & (0.0369) & (0.0337) & (0.0144) & (0.0126) \\ 
	$W^{}_{i,t-1}$ &  & -0.0173 &  & -0.0047. &  & -0.0276** &  & -0.0056 &  & -0.0333** &  & -0.0036 \\ 
	&  & (0.0139) &  & (0.0028) &  & (0.0104) &  & (0.0042) &  & (0.0121) &  & (0.0034) \\ 
	$(K/L)^{}_{i,t-1}$ &  & 0.0238 &  & -4e-04 &  & 0.0207 &  & -0.0065 &  & 0.0152 &  & -0.0029 \\ 
	&  & (0.0157) &  & (0.0034) &  & (0.0132) &  & (0.0044) &  & (0.0145) &  & (0.0037) \\ 
	$(L^P/L)_{i,t-1}$ &  & 0.2489. &  & -0.0135 &  & 0.1575 &  & -0.0215 &  & 0.1662 &  & -0.0247 \\ 
	&  & (0.1279) &  & (0.0323) &  & (0.1089) &  & (0.0424) &  & (0.1241) &  & (0.0341) \\ 
	$(I/L)^{}_{i,t-1}$ &  & 0.0289* &  & 0.0118*** &  & 0.0315** &  & 0.0108*** &  & 0.0216. &  & 0.0107*** \\ 
	&  & (0.0114) &  & (0.0029) &  & (0.0104) &  & (0.0032) &  & (0.0112) &  & (0.0029) \\ 
	$(E/L)^{}_{i,t-1}$ &  & -0.0032 &  & -0.0084** &  & 8e-04 &  & -0.0116** &  & 0.0048 &  & -0.0086** \\ 
	&  & (0.0143) &  & (0.003) &  & (0.0122) &  & (0.0037) &  & (0.013) &  & (0.003) \\ 
	$(M/L)^{}_{i,t-1}$ &  & 0.0011 &  & -9e-04 &  & 0.0024 &  & 0.0049 &  & 0.014 &  & 4e-04 \\ 
	&  & (0.0161) &  & (0.0037) &  & (0.0127) &  & (0.005) &  & (0.0134) &  & (0.0036) \\ 
	$(W^P/W)_{i,t-1}$ &  & -0.2207. &  & -0.0413 &  & -0.2831** &  & 0.0278 &  & -0.2376* &  & 0.011 \\ 
	&  & (0.1287) &  & (0.0251) &  & (0.1071) &  & (0.0252) &  & (0.116) &  & (0.0226) \\ 
	\hline \rule{0pt}{1.075\normalbaselineskip}  AR(1) & 0 & 0 & 1e-04 & 2e-04 & 0 & 0 & 7e-04 & 5e-04 & 0 & 0 & 5e-04 & 7e-04 \\ 
	AR(2) & 0.9373 & 0.9079 & 0.899 & 0.8057 & 0.9237 & 0.9583 & 0.6394 & 0.7465 & 0.9656 & 0.978 & 0.7307 & 0.7026 \\ 
	Sargan & 0 & 0 & 0 & 0.001 & 0 & 0 & 0 & 0 & 0 & 1e-04 & 0 & 0.0056 \\ 
	Controls &  & Y &  & Y &  & Y &  & Y &  & Y &  & Y \\ 
	$R^2$ & 0.9284 & 0.9354 & 0.996 & 0.9962 & 0.9363 & 0.9402 & 0.9952 & 0.9958 & 0.9332 & 0.9374 & 0.9957 & 0.996 \\ 
	\hline
	\hline
\end{tabular}
\endgroup
\caption{Demand-pull and technology-push effects including controls.}
\label{tab:real_sum_output|citation_weighted_patent_stock_all_sectors_all_years_sys_1step_wControls_appendix}
\end{myresizeenv}

\vspace{0.25cm}
\justifying \footnotesize
\noindent
Notes: The table shows the regression results of output $A_{i,t}^{\mu}$ and patents $A_{i,t}^{\tau}$ on demand-pull and technology-push effects. The estimation is based on a two-ways Blundell-Bond (BB) system GMM model using a one-step estimation procedure. The controls included in all regressions are wages $W_{i,t}$, capital intensity $(K/L)_{i,t}$, investment per capita $(I/L)_{i,t}$, relative wages for production labor $(W^P/W)_{i,t}$, energy intensity $(E/L)^{}_{i,t-1}$, and material inputs per capita $(M/L)^{}_{i,t-1}$.    Spillovers are calculated on the basis of first-order links. Variables measured in monetary terms are deflated using the industry level price deflators for the value of shipment obtained from the NBER-productivity database \citep{becker2013nber}. To cope with skewness and to obtain tractable coefficients, most variables are pre-processed (taking logs, removing outliers, scaling). Data in logs are patents and output $A^{\alpha }_{i,t}$, centrality $PR^{\alpha, d }_{i,t}$, spillovers $Spill(A)^{\alpha, d }_{i,t}$, employment $L_{i,t}$,  $(K/L)_{i,t}$, $(I/L)_{i,t}$, $W_{i,t}$, $(E/L)^{}_{i,t-1}$, $(M/L)^{}_{i,t-1}$ with $\alpha = \mu, \tau$ and $d = up, dw$. $A^{\alpha }_{i,t}$, $PR^{\alpha, d }_{i,t}$, and $Spill(A)^{\alpha, d }_{i,t}$ are scaled by division by their standard deviation to obtain comparable coefficients across the different network effects. A detailed description of the transformations and descriptive statistics of the regression data before and after the transformations are provided in \ref{app:data}. 
The rows AR(1), AR(2), and Sargan show the test statistics of the specification tests, i.e. testing for first- and second-order autocorrelation and the results of a Sargan test for validity of instruments \citep[see][]{roodman2009xtabond2}. 

\end{table}
\begin{table}[h]
	\begin{myresizeenv}
		\begingroup
		\begin{tabular}{|l|cccc|cccc|cccc|}
			\hline \hline  \rule{0pt}{1.075\normalbaselineskip}  
			& \multicolumn{4}{c|}{ \ul{Type 1}} & \multicolumn{4}{c|}{ \ul{Type 2}} & \multicolumn{4}{c|}{ \ul{Both}}\\  & \multicolumn{2}{c}{ $\tau \rightarrow \mu$} & \multicolumn{2}{c|}{ $\mu \rightarrow \tau$}
			& \multicolumn{2}{c}{ Market} & \multicolumn{2}{c|}{ Innovation}
			& \multicolumn{4}{c|}{}\\
			\hline  \rule{0pt}{1.075\normalbaselineskip}    & $A^{\mu}_{i,t}$ & $A^{\mu}_{i,t}$ & $A^{\tau}_{i,t}$ & $A^{\tau}_{i,t}$ & $A^{\mu}_{i,t}$ & $A^{\mu}_{i,t}$ & $A^{\tau}_{i,t}$ & $A^{\tau}_{i,t}$ & $A^{\mu}_{i,t}$ & $A^{\mu}_{i,t}$ & $A^{\tau}_{i,t}$ & $A^{\tau}_{i,t}$ \\ 
			\rule{0pt}{1.075\normalbaselineskip}  & (1) & (2) & (3) & (4) & (5) & (6) & (7) & (8) & (9) & (10) & (11) & (12) \\ 
			\hline \rule{0pt}{1.075\normalbaselineskip}   $A^{\mu}_{i,t-1}$ & 0.3431*** & 0.306*** & 0.0108 & 0.0314 & 0.8299*** & 0.6401*** &  &  & 0.5894*** & 0.5672*** & 0.0391 & 0.0401 \\ 
			& (0.082) & (0.0596) & (0.0243) & (0.0218) & (0.0918) & (0.0795) &  &  & (0.1084) & (0.0832) & (0.0261) & (0.027) \\ 
			$A^{\tau}_{i,t-1}$ & -0.2727 & -0.2537 & 0.8206*** & 0.6053*** &  &  & 0.7537*** & 0.3797** & -0.3969 & -0.1191 & 0.7526*** & 0.3715** \\ 
			& (0.4007) & (0.304) & (0.0567) & (0.0755) &  &  & (0.1279) & (0.1365) & (0.527) & (0.3246) & (0.126) & (0.1255) \\ 
			$PR^{\mu,up}_{i,t-1}$ &  &  & -0.0015 & -0.0018 & -0.0686* & -0.055. &  &  & -0.0806** & -0.0682* & -0.0081 & -0.0094 \\ 
			&  &  & (0.0052) & (0.0051) & (0.0304) & (0.0322) &  &  & (0.0308) & (0.0278) & (0.0056) & (0.0057) \\ 
			$PR^{\mu,dw}_{i,t-1}$ &  &  & -6e-04 & -7e-04 & -0.0916** & -0.0607* &  &  & -0.0762** & -0.052* & -0.0036 & -0.0019 \\ 
			&  &  & (0.0052) & (0.0044) & (0.0295) & (0.0289) &  &  & (0.0281) & (0.0244) & (0.005) & (0.0042) \\ 
			$PR^{\tau,dw}_{i,t-1}$ & 0.5732* & 0.377* &  &  &  &  & 0.0285 & 0.0095 & 0.7663** & 0.4663** & -0.02 & -0.0046 \\ 
			& (0.2659) & (0.173) &  &  &  &  & (0.045) & (0.0432) & (0.2921) & (0.1773) & (0.0419) & (0.0427) \\ 
			$Spill(A)^{\mu,up}_{i,t-1}$ &  &  & 0.0017 & 2e-04 & -0.059 & -0.0373 &  &  & -0.0132 & -0.0207 & 0.0074 & 0.0032 \\ 
			&  &  & (0.0086) & (0.008) & (0.0486) & (0.0457) &  &  & (0.0522) & (0.0411) & (0.009) & (0.0096) \\ 
			$Spill(A)^{\mu,dw}_{i,t-1}$ &  &  & 0.0064 & 0.0025 & -0.2074*** & -0.1591*** &  &  & -0.1184** & -0.12** & 0.0099 & 0.0115 \\ 
			&  &  & (0.0057) & (0.006) & (0.0402) & (0.0367) &  &  & (0.0452) & (0.0405) & (0.0061) & (0.0071) \\ 
			$Spill(A)^{\tau,up}_{i,t-1}$ & 10.37** & 6.151** &  &  &  &  & 1.853*** & 1.598*** & 11.62*** & 8.457*** & 1.517*** & 1.322*** \\ 
			& (3.25) & (2.199) &  &  &  &  & (0.4343) & (0.4336) & (2.676) & (1.961) & (0.355) & (0.3967) \\ 
			$Spill(A)^{\tau,dw}_{i,t-1}$ & 0.2045 & 0.0152 &  &  &  &  & -0.0145 & 0.0341 & 0.1705 & 0.0418 & 0.0013 & 0.0329 \\ 
			& (0.1595) & (0.1065) &  &  &  &  & (0.0326) & (0.0326) & (0.1544) & (0.1052) & (0.0327) & (0.0317) \\ 
			$W^{}_{i,t-1}$ &  & 0.112* &  & -0.0202. &  & 0.1671** &  & 0.0024 &  & 0.1328** &  & 1e-04 \\ 
			&  & (0.0568) &  & (0.0114) &  & (0.0567) &  & (0.0163) &  & (0.0471) &  & (0.0155) \\ 
			$(K/L)^{}_{i,t-1}$ &  & -0.0374 &  & 0.02 &  & 0.0069 &  & 0.0317. &  & -0.0112 &  & 0.0384* \\ 
			&  & (0.0767) &  & (0.0153) &  & (0.0717) &  & (0.0186) &  & (0.0763) &  & (0.0175) \\ 
			$(L^P/L)_{i,t-1}$ &  & -0.7078 &  & 0.2495. &  & -0.3813 &  & 0.1315 &  & -0.0839 &  & 0.2425 \\ 
			&  & (0.666) &  & (0.1496) &  & (0.5939) &  & (0.1788) &  & (0.6121) &  & (0.1692) \\ 
			$(I/L)^{}_{i,t-1}$ &  & 0.044 &  & 0.0055 &  & 0.0387 &  & 0.0053 &  & 0.0558. &  & 0.0015 \\ 
			&  & (0.04) &  & (0.0061) &  & (0.0338) &  & (0.0073) &  & (0.0326) &  & (0.0068) \\ 
			$(E/L)^{}_{i,t-1}$ &  & -0.1974*** &  & -0.028** &  & -0.2233*** &  & -0.0383** &  & -0.1512** &  & -0.0376** \\ 
			&  & (0.0536) &  & (0.0106) &  & (0.0476) &  & (0.014) &  & (0.0471) &  & (0.0122) \\ 
			$(M/L)^{}_{i,t-1}$ &  & 0.0141 &  & 0.0321* &  & -0.0345 &  & 0.0185 &  & -0.0803 &  & 0.0201 \\ 
			&  & (0.0746) &  & (0.0137) &  & (0.064) &  & (0.0201) &  & (0.0665) &  & (0.0172) \\ 
			$(W^P/W)_{i,t-1}$ &  & 0.0194 &  & -0.0642 &  & -0.247 &  & -0.0824 &  & -0.2412 &  & -0.0821 \\ 
			&  & (0.2929) &  & (0.0584) &  & (0.2868) &  & (0.0626) &  & (0.2703) &  & (0.0632) \\ 
			\hline \rule{0pt}{1.075\normalbaselineskip}  AR(1) & 0 & 0 & 2e-04 & 4e-04 & 0 & 0 & 3e-04 & 0.0136 & 0 & 0 & 5e-04 & 0.0125 \\ 
			AR(2) & 0.4028 & 0.3798 & 0.7767 & 0.5895 & 0.8789 & 0.8175 & 0.5217 & 0.3777 & 0.4567 & 0.523 & 0.5125 & 0.3298 \\ 
			Sargan & 1e-04 & 0 & 0 & 3e-04 & 0 & 0 & 0 & 1e-04 & 4e-04 & 6e-04 & 0 & 1e-04 \\ 
			Controls &  & Y &  & Y &  & Y &  & Y &  & Y &  & Y \\ 
			$R^2$ & 0.0034 & 0.0137 & 0.6304 & 0.6608 & 0.0027 & 0.0027 & 0.6421 & 0.6892 & 0.0054 & 0.0151 & 0.6462 & 0.6883 \\ 
			\hline
			\hline
		\end{tabular}
		\endgroup\caption{Demand-pull and technology-push effects --- Arellano-Bond estimator.}

		\label{tab:real_sum_output|citation_weighted_patent_stock_all_sectors_all_years_diff_2step_wControls_links}
	\end{myresizeenv}
	\vspace{0.25cm}
	
	\justifying \footnotesize
	
	\noindent
	Notes: The table shows the regression results of output $A_{i,t}^{\mu}$ and patents $A_{i,t}^{\tau}$ on demand-pull and technology-push effects. The estimation is based on a two-ways Arellano-Bond (AB) first-difference model. The controls included in all regressions are wages $W_{i,t}$, capital intensity $(K/L)_{i,t}$, investment per capita $(I/L)_{i,t}$, relative wages for production labor $(W^P/W)_{i,t}$, energy intensity $(E/L)^{}_{i,t-1}$, and material inputs per capita $(M/L)^{}_{i,t-1}$.    Spillovers are calculated on the basis of first-order links. Variables measured in monetary terms are deflated using the industry level price deflators for the value of shipment obtained from the NBER-productivity database \citep{becker2013nber}. To cope with skewness and to obtain tractable coefficients, most variables are pre-processed (taking logs, removing outliers, scaling). Data in logs are patents and output $A^{\alpha }_{i,t}$, centrality $PR^{\alpha, d }_{i,t}$, spillovers $Spill(A)^{\alpha, d }_{i,t}$, employment $L_{i,t}$,  $(K/L)_{i,t}$, $(I/L)_{i,t}$, $W_{i,t}$, $(E/L)^{}_{i,t-1}$, $(M/L)^{}_{i,t-1}$ with $\alpha = \mu, \tau$ and $d = up, dw$. $A^{\alpha }_{i,t}$, $PR^{\alpha, d }_{i,t}$, and $Spill(A)^{\alpha, d }_{i,t}$ are scaled by division by their standard deviation to obtain comparable coefficients across the different network effects. A detailed description of the transformations and descriptive statistics of the regression data before and after the transformations are provided in \ref{app:data}. 
\end{table}

\begin{table}[h]
	\begin{myresizeenv}
		\begingroup
		
		\begin{tabular}{|l|cccc|cccc|cccc|}
			\hline \hline  \rule{0pt}{1.075\normalbaselineskip}  
			& \multicolumn{4}{c|}{ \ul{Type 1}} & \multicolumn{4}{c|}{ \ul{Type 2}} & \multicolumn{4}{c|}{ \ul{Both}}\\  & \multicolumn{2}{c}{ $\tau \rightarrow \mu$} & \multicolumn{2}{c|}{ $\mu \rightarrow \tau$}
			& \multicolumn{2}{c}{ Market} & \multicolumn{2}{c|}{ Innovation}
			& \multicolumn{4}{c|}{}\\
			\hline  \rule{0pt}{1.075\normalbaselineskip}    & $A^{\mu}_{i,t}$ & $A^{\mu}_{i,t}$ & $A^{\tau}_{i,t}$ & $A^{\tau}_{i,t}$ & $A^{\mu}_{i,t}$ & $A^{\mu}_{i,t}$ & $A^{\tau}_{i,t}$ & $A^{\tau}_{i,t}$ & $A^{\mu}_{i,t}$ & $A^{\mu}_{i,t}$ & $A^{\tau}_{i,t}$ & $A^{\tau}_{i,t}$ \\ 
			\rule{0pt}{1.075\normalbaselineskip}  & (1) & (2) & (3) & (4) & (5) & (6) & (7) & (8) & (9) & (10) & (11) & (12) \\ 
			\hline \rule{0pt}{1.075\normalbaselineskip}   $A^{\mu}_{i,t-1}$ & 0.3838*** & 0.3538*** & 0.0111. & 0.0046 & 0.4142*** & 0.3616*** &  &  & 0.3997*** & 0.3601*** & 0.0128* & 0.0048 \\ 
			& (0.0213) & (0.0223) & (0.0061) & (0.0065) & (0.0259) & (0.0275) &  &  & (0.0259) & (0.0273) & (0.006) & (0.0064) \\ 
			$A^{\tau}_{i,t-1}$ & 0.2423*** & 0.1392. & 0.7802*** & 0.7295*** &  &  & 0.8407*** & 0.7835*** & 0.2542*** & 0.1491* & 0.8375*** & 0.7838*** \\ 
			& (0.0718) & (0.0753) & (0.0134) & (0.0152) &  &  & (0.0202) & (0.0212) & (0.0712) & (0.0745) & (0.0202) & (0.0212) \\ 
			$PR^{\mu,up}_{i,t-1}$ &  &  & -0.0019 & -0.0015 & 0.0228** & 0.0277*** &  &  & 0.0221** & 0.0267** & -0.0022 & -0.0017 \\ 
			&  &  & (0.0021) & (0.0021) & (0.0084) & (0.0084) &  &  & (0.0083) & (0.0083) & (0.0021) & (0.0021) \\ 
			$PR^{\mu,dw}_{i,t-1}$ &  &  & -0.0019 & -0.0014 & 0.0326*** & 0.0393*** &  &  & 0.0314*** & 0.0375*** & -0.0019 & -0.0013 \\ 
			&  &  & (0.0017) & (0.0017) & (0.007) & (0.007) &  &  & (0.0069) & (0.007) & (0.0017) & (0.0017) \\ 
			$PR^{\tau,dw}_{i,t-1}$ & 0.0775* & 0.0541. &  &  &  &  & -0.0094 & -0.0159* & 0.0727* & 0.0514 & -0.0124. & -0.0177* \\ 
			& (0.0316) & (0.0328) &  &  &  &  & (0.0074) & (0.0075) & (0.0314) & (0.0326) & (0.0074) & (0.0076) \\ 
			$Spill(A)^{\mu,up}_{i,t-1}$ &  &  & -0.0061 & -0.0058 & 8e-04 & -0.0016 &  &  & 0.0107 & 0.0048 & -0.0052 & -0.0053 \\ 
			&  &  & (0.0042) & (0.0041) & (0.0169) & (0.0167) &  &  & (0.0169) & (0.0168) & (0.0042) & (0.0041) \\ 
			$Spill(A)^{\mu,dw}_{i,t-1}$ &  &  & 0.0024 & 0.0026 & -0.0278** & -0.0242* &  &  & -0.0275** & -0.0259* & 0.0021 & 0.0027 \\ 
			&  &  & (0.0026) & (0.0026) & (0.0106) & (0.0106) &  &  & (0.0106) & (0.0106) & (0.0025) & (0.0025) \\ 
			$Spill(A)^{\tau,up}_{i,t-1}$ & 1.268*** & 1.216*** &  &  &  &  & 0.4938*** & 0.509*** & 1.343*** & 1.268*** & 0.4822*** & 0.4925*** \\ 
			& (0.3483) & (0.349) &  &  &  &  & (0.0924) & (0.0924) & (0.3466) & (0.3461) & (0.0928) & (0.0928) \\ 
			$Spill(A)^{\tau,dw}_{i,t-1}$ & -0.0915** & -0.0824** &  &  &  &  & -0.0222** & -0.0146. & -0.0887** & -0.0766** & -0.021** & -0.0143. \\ 
			& (0.0288) & (0.0289) &  &  &  &  & (0.0077) & (0.0077) & (0.0286) & (0.0286) & (0.0077) & (0.0077) \\ 
			$W^{}_{i,t-1}$ &  & 0.0246* &  & -0.0014 &  & 0.0237* &  & 9e-04 &  & 0.025* &  & 6e-04 \\ 
			&  & (0.0102) &  & (0.0024) &  & (0.0102) &  & (0.0023) &  & (0.0103) &  & (0.0024) \\ 
			$(K/L)^{}_{i,t-1}$ &  & 0.0033 &  & 0.0042 &  & 0.0053 &  & 0.005. &  & 2e-04 &  & 0.0049 \\ 
			&  & (0.0127) &  & (0.003) &  & (0.0123) &  & (0.003) &  & (0.0126) &  & (0.003) \\ 
			$(L^P/L)_{i,t-1}$ &  & -0.0915 &  & 0.0142 &  & -0.1232 &  & 0.0174 &  & -0.0646 &  & 0.0166 \\ 
			&  & (0.1251) &  & (0.0292) &  & (0.1231) &  & (0.0291) &  & (0.1237) &  & (0.0291) \\ 
			$(I/L)^{}_{i,t-1}$ &  & 0.0174* &  & 0.0057** &  & 0.0149. &  & 0.006** &  & 0.0174* &  & 0.0059** \\ 
			&  & (0.0088) &  & (0.0021) &  & (0.0087) &  & (0.0021) &  & (0.0087) &  & (0.0021) \\ 
			$(E/L)^{}_{i,t-1}$ &  & -0.0672*** &  & -0.0151*** &  & -0.0782*** &  & -0.0149*** &  & -0.0733*** &  & -0.0143*** \\ 
			&  & (0.0113) &  & (0.0027) &  & (0.0111) &  & (0.0027) &  & (0.0112) &  & (0.0027) \\ 
			$(M/L)^{}_{i,t-1}$ &  & 0.0223. &  & 0.0101*** &  & 0.0325** &  & 0.0087** &  & 0.0257* &  & 0.0084** \\ 
			&  & (0.0123) &  & (0.0029) &  & (0.012) &  & (0.0029) &  & (0.0122) &  & (0.0029) \\ 
			$(W^P/W)_{i,t-1}$ &  & -0.0021 &  & -0.0431* &  & 0.0035 &  & -0.041* &  & 0.0135 &  & -0.0417* \\ 
			&  & (0.0766) &  & (0.0182) &  & (0.076) &  & (0.018) &  & (0.0757) &  & (0.018) \\ 
			\hline \rule{0pt}{1.075\normalbaselineskip}   Controls &  & Y &  & Y &  & Y &  & Y &  & Y &  & Y \\ 
			$R^2$ & 0.8118 & 0.8175 & 0.9955 & 0.9957 & 0.8114 & 0.8197 & 0.9956 & 0.9958 & 0.816 & 0.8224 & 0.9956 & 0.9958 \\ 
			\hline
			\hline
		\end{tabular}
		
		\endgroup
		\caption{Demand-pull and technology-push effects --- weighted FE model} 
		\label{tab:real_sum_output|citation_weighted_patent_stock_all_sectors_all_years_FE_weighted_wControls}
	\end{myresizeenv}
	\vspace{0.25cm}
	
	\justifying \footnotesize
	
	\noindent
	Notes: The table shows the regression results of output $A_{i,t}^{\mu}$ and patents $A_{i,t}^{\tau}$ on demand-pull and technology-push effects. The estimation is based on a two-ways weighted fixed-effects (FE) model. The weights used in the regressions are $A^{\mu}_{i,t}$ in the TFP regression and $L_{i,t}$ in all other regressions.  Spillovers are calculated on the basis of first-order links. Variables measured in monetary terms are deflated using the industry level price deflators for the value of shipment obtained from the NBER-productivity database \citep{becker2013nber}. To cope with skewness and to obtain tractable coefficients, most variables are pre-processed (taking logs, removing outliers, scaling). Data in logs are $A^{\alpha }_{i,t}$, $PR^{\alpha, d }_{i,t}$, $Spill(A)^{\alpha, d }_{i,t}$, $L_{i,t}$,  $(K/L)_{i,t}$, $(I/L)_{i,t}$, $W_{i,t}$, $(K/L)^{}_{ij,t-1}$, $(E/L)^{}_{ij,t-1}$, $(M/L)^{}_{ij,t-1}$, $W^{P}_{ij,t-1}$ with $\alpha = \mu, \tau$ and $d = up, dw$. $A^{\alpha }_{i,t}$, $PR^{\alpha, d }_{i,t}$, and $Spill(A)^{\alpha, d }_{i,t}$ are scaled by division by their standard deviation to obtain comparable coefficients across the different network effects. A detailed description of the transformations and descriptive statistics of the regression data before and after the transformations are provided in \ref{app:data}.
\end{table}

Table \ref{tab:real_sum_output|citation_weighted_patent_stock_all_sectors_all_years_diff_2step_wControls_links} and \ref{tab:real_sum_output|citation_weighted_patent_stock_all_sectors_all_years_FE_weighted_wControls} show the results using an AB and a weighted FE regression. The weights in the FE regression are output for the regression testing the impact of TP and DP on market growth  $A^{\mu}_{i,t}$ and citation-weighted patents for the regression assessing the impact on innovation $A^{\tau}_{i,t}$.

Table \ref{tab:real_sum_output|citation_weighted_patent_stock_all_sectors_first_subperiod_sys_1step_wControls} and \ref{tab:real_sum_output|citation_weighted_patent_stock_all_sectors_second_subperiod_sys_1step_wControls} show the results for the two different subperiods. 
The data of the subperiods is technically taken the six 5-year snapshots covering 1977-1997 and 1992-2012. This leads to an overlap of two observations (1992 and 1997) in the estimation procedure. The BB estimator makes use of the second time lag of the dependent variable to build an instrument for the first lag of the dependent variable \citep{blundell1998initial, roodman2009xtabond2}. This leads to an exclusion of the first two observations. Hence, the second stage models for the two subperiods do not overlap in time and effectively cover the period from 1987-1992 and 1997-2012.  
\begin{table}[h]
\begin{myresizeenv}
\begingroup
\begin{tabular}{|l|cccc|cccc|cccc|}
	\hline \hline  \rule{0pt}{1.075\normalbaselineskip}  
	& \multicolumn{4}{c|}{ \ul{Type 1}} & \multicolumn{4}{c|}{ \ul{Type 2}} & \multicolumn{4}{c|}{ \ul{Both}}\\  & \multicolumn{2}{c}{ $\tau \rightarrow \mu$} & \multicolumn{2}{c|}{ $\mu \rightarrow \tau$}
	& \multicolumn{2}{c}{ Market} & \multicolumn{2}{c|}{ Innovation}
	& \multicolumn{4}{c|}{}\\
	\hline  \rule{0pt}{1.075\normalbaselineskip}    & $A^{\mu}_{i,t}$ & $A^{\mu}_{i,t}$ & $A^{\tau}_{i,t}$ & $A^{\tau}_{i,t}$ & $A^{\mu}_{i,t}$ & $A^{\mu}_{i,t}$ & $A^{\tau}_{i,t}$ & $A^{\tau}_{i,t}$ & $A^{\mu}_{i,t}$ & $A^{\mu}_{i,t}$ & $A^{\tau}_{i,t}$ & $A^{\tau}_{i,t}$ \\ 
	\rule{0pt}{1.075\normalbaselineskip}  & (1) & (2) & (3) & (4) & (5) & (6) & (7) & (8) & (9) & (10) & (11) & (12) \\ 
	\hline \rule{0pt}{1.075\normalbaselineskip}   $A^{\mu}_{i,t-1}$ & 0.5807*** & 0.5313*** & 0.0252 & 0 & 0.5993*** & 0.4622*** &  &  & 0.5934*** & 0.4794*** & 0.0281 & 0.0035 \\ 
	& (0.0854) & (0.0946) & (0.0208) & (0.0196) & (0.098) & (0.0932) &  &  & (0.0854) & (0.0926) & (0.0206) & (0.0202) \\ 
	$A^{\tau}_{i,t-1}$ & -0.0539 & -0.0557 & 1.092*** & 1.045*** &  &  & 0.9694*** & 0.9981*** & 0.0116 & 0.0648 & 1.027*** & 1.027*** \\ 
	& (0.1777) & (0.1536) & (0.0218) & (0.02) &  &  & (0.1289) & (0.0688) & (0.1535) & (0.1416) & (0.0844) & (0.056) \\ 
	$PR^{\mu,up}_{i,t-1}$ &  &  & -0.0128 & -0.0047 & 0.0072 & -0.0051 &  &  & 0.0091 & -0.013 & -0.0197 & -0.0068 \\ 
	&  &  & (0.0127) & (0.0075) & (0.0309) & (0.029) &  &  & (0.0314) & (0.0294) & (0.0159) & (0.0085) \\ 
	$PR^{\mu,dw}_{i,t-1}$ &  &  & -0.0035 & 0.0018 & 0.0492 & -0.0506 &  &  & 0.0037 & -0.05 & -0.0052 & 9e-04 \\ 
	&  &  & (0.0117) & (0.0077) & (0.0384) & (0.0363) &  &  & (0.0368) & (0.0352) & (0.0135) & (0.008) \\ 
	$PR^{\tau,dw}_{i,t-1}$ & 0.1011 & 0.0809 &  &  &  &  & 0.1253 & 0.0574 & 0.0759 & 0.0421 & 0.06 & 0.0267 \\ 
	& (0.0932) & (0.0742) &  &  &  &  & (0.0872) & (0.0373) & (0.0651) & (0.0595) & (0.0459) & (0.0255) \\ 
	$Spill(A)^{\mu,up}_{i,t-1}$ &  &  & 0.0022 & 0.0028 & 0.0757 & -0.0686 &  &  & 0.0173 & -0.092 & -0.0038 & -0.0136 \\ 
	&  &  & (0.0369) & (0.0262) & (0.0984) & (0.086) &  &  & (0.0998) & (0.0913) & (0.0381) & (0.0285) \\ 
	$Spill(A)^{\mu,dw}_{i,t-1}$ &  &  & 0.0036 & 0.003 & 0.0481 & 0.0641 &  &  & 0.0242 & 0.0569 & 5e-04 & 0 \\ 
	&  &  & (0.0156) & (0.0104) & (0.0427) & (0.044) &  &  & (0.0418) & (0.0462) & (0.0139) & (0.0111) \\ 
	$Spill(A)^{\tau,up}_{i,t-1}$ & 1.827 & 2.831* &  &  &  &  & -0.2011 & -0.6991 & 1.827. & 2.389* & -0.4824 & -0.6271* \\ 
	& (1.163) & (1.165) &  &  &  &  & (0.5279) & (0.4404) & (1.029) & (1.015) & (0.348) & (0.2954) \\ 
	$Spill(A)^{\tau,dw}_{i,t-1}$ & -0.0337 & -0.0037 &  &  &  &  & -0.0314 & -0.0066 & -0.0539 & -0.049 & -0.0194 & -0.0054 \\ 
	& (0.0679) & (0.0695) &  &  &  &  & (0.0308) & (0.0236) & (0.0641) & (0.0676) & (0.0251) & (0.0212) \\ 
	$W^{}_{i,t-1}$ &  & -0.0192 &  & -0.011* &  & -0.004 &  & -0.0085 &  & -0.0136 &  & -0.0053 \\ 
	&  & (0.0211) &  & (0.0049) &  & (0.019) &  & (0.0069) &  & (0.02) &  & (0.0059) \\ 
	$(K/L)^{}_{i,t-1}$ &  & 0.0791. &  & 0.0074 &  & 0.0611 &  & 0.0164 &  & 0.0553 &  & 0.0082 \\ 
	&  & (0.0473) &  & (0.009) &  & (0.038) &  & (0.014) &  & (0.0377) &  & (0.0099) \\ 
	$(L^P/L)_{i,t-1}$ &  & 0.535. &  & -0.0097 &  & 0.0544 &  & 0.0562 &  & 0.0925 &  & 0.0252 \\ 
	&  & (0.315) &  & (0.0721) &  & (0.2134) &  & (0.0893) &  & (0.2318) &  & (0.0676) \\ 
	$(I/L)^{}_{i,t-1}$ &  & 0.0493* &  & 0.0084. &  & 0.0422* &  & 0.0066 &  & 0.0375* &  & 0.0075 \\ 
	&  & (0.0204) &  & (0.0044) &  & (0.0173) &  & (0.0071) &  & (0.0178) &  & (0.0057) \\ 
	$(E/L)^{}_{i,t-1}$ &  & -3e-04 &  & -0.0055 &  & -0.027 &  & -0.028* &  & -0.0276 &  & -0.0075 \\ 
	&  & (0.035) &  & (0.0062) &  & (0.0238) &  & (0.0111) &  & (0.0242) &  & (0.0075) \\ 
	$(M/L)^{}_{i,t-1}$ &  & -0.0498 &  & 0.0023 &  & 1e-04 &  & 0.0133 &  & 0.0062 &  & 8e-04 \\ 
	&  & (0.0403) &  & (0.0074) &  & (0.026) &  & (0.0122) &  & (0.0279) &  & (0.0075) \\ 
	$(W^P/W)_{i,t-1}$ &  & -0.4005 &  & -0.1213* &  & -0.3473 &  & -0.1095 &  & -0.1988 &  & -0.1154. \\ 
	&  & (0.2947) &  & (0.0557) &  & (0.2158) &  & (0.0834) &  & (0.2344) &  & (0.0605) \\ 
	\hline \rule{0pt}{1.075\normalbaselineskip}  AR(1) & 0 & 0 & 0.0378 & 0.0384 & 0 & 2e-04 & 0.0227 & 0.0477 & 0 & 1e-04 & 0.0479 & 0.0534 \\ 
	AR(2) & 0.5155 & 0.6397 & 0.8436 & 0.8225 & 0.5133 & 0.4159 & 0.8396 & 0.8501 & 0.5298 & 0.4975 & 0.8122 & 0.818 \\ 
	Sargan & 0.0232 & 2e-04 & 0.0011 & 4e-04 & 1e-04 & 1e-04 & 0 & 3e-04 & 9e-04 & 0.0063 & 1e-04 & 0 \\ 
	\hline \hline \rule{0pt}{1.075\normalbaselineskip}   Controls &  & Y &  & Y &  & Y &  & Y &  & Y &  & Y \\ 
	$R^2$ & 0.9435 & 0.9398 & 0.995 & 0.9958 & 0.9476 & 0.949 & 0.9908 & 0.9942 & 0.9466 & 0.9467 & 0.9944 & 0.9954 \\ 
	\hline
	\hline
\end{tabular}
\endgroup

\caption{Demand-pull and technology-push effects --- first subperiod.}

\label{tab:real_sum_output|citation_weighted_patent_stock_all_sectors_first_subperiod_sys_1step_wControls}
\end{myresizeenv}

\vspace{0.25cm}

\justifying \footnotesize

\noindent	
Notes: The table shows the regression results of output $A_{i,t}^{\mu}$ and patents $A_{i,t}^{\tau}$ on demand-pull and technology-push effects. The estimation is based on a two-ways Blundell-Bond (BB) system GMM model. The controls included in all regressions are wages $W_{i,t}$, capital intensity $(K/L)_{i,t}$, investment per capita $(I/L)_{i,t}$, relative wages for production labor $(W^P/W)_{i,t}$, energy intensity $(E/L)^{}_{i,t-1}$, and material inputs per capita $(M/L)^{}_{i,t-1}$.    Spillovers are calculated on the basis of first-order links. Variables measured in monetary terms are deflated using the industry level price deflators for the value of shipment obtained from the NBER-productivity database \citep{becker2013nber}. To cope with skewness and to obtain tractable coefficients, most variables are pre-processed (taking logs, removing outliers, scaling). Data in logs are patents and output $A^{\alpha }_{i,t}$, centrality $PR^{\alpha, d }_{i,t}$, spillovers $Spill(A)^{\alpha, d }_{i,t}$, employment $L_{i,t}$,  $(K/L)_{i,t}$, $(I/L)_{i,t}$, $W_{i,t}$, $(E/L)^{}_{i,t-1}$, $(M/L)^{}_{i,t-1}$ with $\alpha = \mu, \tau$ and $d = up, dw$. $A^{\alpha }_{i,t}$, $PR^{\alpha, d }_{i,t}$, and $Spill(A)^{\alpha, d }_{i,t}$ are scaled by division by their standard deviation to obtain comparable coefficients across the different network effects. A detailed description of the transformations and descriptive statistics of the regression data before and after the transformations are provided in \ref{app:data}. 
The rows AR(1), AR(2), and Sargan show the test statistics of the specification tests, i.e. testing for first- and second-order autocorrelation and the results of a Sargan test for validity of instruments \citep[see][]{roodman2009xtabond2}. 

\end{table}

\begin{table}[h]
	\begin{myresizeenv}
		\begingroup
\begin{tabular}{|l|cccc|cccc|cccc|}
	\hline \hline  \rule{0pt}{1.075\normalbaselineskip}  
	& \multicolumn{4}{c|}{ \ul{Type 1}} & \multicolumn{4}{c|}{ \ul{Type 2}} & \multicolumn{4}{c|}{ \ul{Both}}\\  & \multicolumn{2}{c}{ $\tau \rightarrow \mu$} & \multicolumn{2}{c|}{ $\mu \rightarrow \tau$}
	& \multicolumn{2}{c}{ Market} & \multicolumn{2}{c|}{ Innovation}
	& \multicolumn{4}{c|}{}\\
	\hline  \rule{0pt}{1.075\normalbaselineskip}    & $A^{\mu}_{i,t}$ & $A^{\mu}_{i,t}$ & $A^{\tau}_{i,t}$ & $A^{\tau}_{i,t}$ & $A^{\mu}_{i,t}$ & $A^{\mu}_{i,t}$ & $A^{\tau}_{i,t}$ & $A^{\tau}_{i,t}$ & $A^{\mu}_{i,t}$ & $A^{\mu}_{i,t}$ & $A^{\tau}_{i,t}$ & $A^{\tau}_{i,t}$ \\ 
	\rule{0pt}{1.075\normalbaselineskip}  & (1) & (2) & (3) & (4) & (5) & (6) & (7) & (8) & (9) & (10) & (11) & (12) \\ 
	\hline \rule{0pt}{1.075\normalbaselineskip}   $A^{\mu}_{i,t-1}$ & 0.4469*** & 0.4156*** & -0.006 & -9e-04 & 0.522*** & 0.5089*** &  &  & 0.5242*** & 0.5032*** & -0.0074 & -0.0026 \\ 
	& (0.0409) & (0.0323) & (0.0213) & (0.0201) & (0.0459) & (0.0519) &  &  & (0.0514) & (0.049) & (0.0236) & (0.0193) \\ 
	$A^{\tau}_{i,t-1}$ & -0.1027 & 0.057 & 1.022*** & 1.007*** &  &  & 1.053*** & 1.077*** & -0.001 & 0.0045 & 1.084*** & 1.07*** \\ 
	& (0.1537) & (0.1308) & (0.0111) & (0.0124) &  &  & (0.0536) & (0.0482) & (0.1241) & (0.1255) & (0.0468) & (0.0388) \\ 
	$PR^{\mu,up}_{i,t-1}$ &  &  & -0.0014 & -8e-04 & 0.0076 & 0.0183 &  &  & 0.0115 & 0.0135 & 0.0014 & -0.0011 \\ 
	&  &  & (0.0057) & (0.0059) & (0.0162) & (0.0135) &  &  & (0.0171) & (0.0142) & (0.0058) & (0.0056) \\ 
	$PR^{\mu,dw}_{i,t-1}$ &  &  & 0.0017 & 0.004 & -0.0045 & 0.0034 &  &  & 0.0026 & 0.0033 & 0.004 & 0.0036 \\ 
	&  &  & (0.004) & (0.004) & (0.016) & (0.0133) &  &  & (0.0164) & (0.0134) & (0.0041) & (0.0037) \\ 
	$PR^{\tau,dw}_{i,t-1}$ & 0.3035*** & 0.1018. &  &  &  &  & 0.0681* & 0.0098 & 0.2048*** & 0.1362** & 0.0293 & 0.0047 \\ 
	& (0.0835) & (0.0567) &  &  &  &  & (0.0272) & (0.0169) & (0.0563) & (0.05) & (0.0195) & (0.014) \\ 
	$Spill(A)^{\mu,up}_{i,t-1}$ &  &  & -0.0118* & -0.0085 & 0.1106*** & 0.0956*** &  &  & 0.0989*** & 0.0831*** & -0.0148* & -0.0101 \\ 
	&  &  & (0.006) & (0.0062) & (0.0216) & (0.0217) &  &  & (0.0225) & (0.0209) & (0.0067) & (0.0065) \\ 
	$Spill(A)^{\mu,dw}_{i,t-1}$ &  &  & 0.0015 & 0.0058 & -0.0266 & -0.0373 &  &  & -0.0342 & -0.0442. & -0.0029 & 0.0032 \\ 
	&  &  & (0.0066) & (0.0061) & (0.0247) & (0.0245) &  &  & (0.0237) & (0.0227) & (0.0063) & (0.0058) \\ 
	$Spill(A)^{\tau,up}_{i,t-1}$ & 3.601** & 2.055* &  &  &  &  & 1.973*** & 1.071*** & 2.738** & 2.793*** & 1.366*** & 0.9312*** \\ 
	& (1.185) & (0.8783) &  &  &  &  & (0.3517) & (0.2817) & (0.882) & (0.8156) & (0.299) & (0.2517) \\ 
	$Spill(A)^{\tau,dw}_{i,t-1}$ & -0.1629. & -0.1361* &  &  &  &  & -0.0883*** & -0.0581** & -0.1768** & -0.1412* & -0.0737*** & -0.0517*** \\ 
	& (0.0874) & (0.0579) &  &  &  &  & (0.0255) & (0.0181) & (0.0636) & (0.0572) & (0.019) & (0.0151) \\ 
	$W^{}_{i,t-1}$ &  & -0.0125 &  & -0.0013 &  & -0.0389. &  & -0.0012 &  & -0.0382* &  & -0.0027 \\ 
	&  & (0.0261) &  & (0.0051) &  & (0.0214) &  & (0.0082) &  & (0.0191) &  & (0.0056) \\ 
	$(K/L)^{}_{i,t-1}$ &  & 0.0077 &  & 9e-04 &  & -0.0078 &  & -0.0038 &  & -0.0139 &  & -0.0015 \\ 
	&  & (0.0246) &  & (0.005) &  & (0.0203) &  & (0.0066) &  & (0.0217) &  & (0.0054) \\ 
	$(L^P/L)_{i,t-1}$ &  & 0.3218 &  & 0.0316 &  & 0.385 &  & -0.0255 &  & 0.2817 &  & -0.0374 \\ 
	&  & (0.2343) &  & (0.0499) &  & (0.2443) &  & (0.0715) &  & (0.2319) &  & (0.0523) \\ 
	$(I/L)^{}_{i,t-1}$ &  & 0.0249 &  & 0.0099* &  & 0.0177 &  & 0.0089. &  & 0.0162 &  & 0.01* \\ 
	&  & (0.0193) &  & (0.0048) &  & (0.0225) &  & (0.0046) &  & (0.0195) &  & (0.0045) \\ 
	$(E/L)^{}_{i,t-1}$ &  & 0.0293 &  & -0.0121** &  & 0.0456* &  & -0.0067 &  & 0.0472* &  & -0.0089. \\ 
	&  & (0.0211) &  & (0.0044) &  & (0.023) &  & (0.0062) &  & (0.0195) &  & (0.0048) \\ 
	$(M/L)^{}_{i,t-1}$ &  & 0.0011 &  & 0.001 &  & 0.0178 &  & 0.003 &  & 0.0216 &  & 0.0032 \\ 
	&  & (0.0276) &  & (0.0052) &  & (0.0211) &  & (0.0076) &  & (0.0213) &  & (0.0049) \\ 
	$(W^P/W)_{i,t-1}$ &  & -0.4345* &  & -0.0334 &  & -0.7856*** &  & 0.0127 &  & -0.5908** &  & 0.0176 \\ 
	&  & (0.1802) &  & (0.0409) &  & (0.2207) &  & (0.0399) &  & (0.1873) &  & (0.0341) \\ 
	\hline \rule{0pt}{1.075\normalbaselineskip}  AR(1) & 1e-04 & 1e-04 & 2e-04 & 3e-04 & 0 & 0 & 1e-04 & 0.0016 & 0 & 0 & 0.001 & 0.0019 \\ 
	AR(2) & 0.5792 & 0.3485 & 0.3267 & 0.2549 & 0.4284 & 0.2814 & 0.301 & 0.176 & 0.5271 & 0.4214 & 0.2331 & 0.1869 \\ 
	Sargan & 0 & 0 & 0 & 0 & 0 & 0 & 0.0234 & 4e-04 & 0 & 0 & 5e-04 & 4e-04 \\ 
	\hline \hline \rule{0pt}{1.075\normalbaselineskip}   Controls &  & Y &  & Y &  & Y &  & Y &  & Y &  & Y \\ 
	$R^2$ & 0.9009 & 0.929 & 0.9964 & 0.9964 & 0.9308 & 0.9316 & 0.9943 & 0.9963 & 0.9184 & 0.9275 & 0.9958 & 0.9964 \\ 
	\hline
	\hline
\end{tabular}
		\endgroup
		\caption{Demand-pull and technology-push effects --- second subperiod}
		\label{tab:real_sum_output|citation_weighted_patent_stock_all_sectors_second_subperiod_sys_1step_wControls}
	\end{myresizeenv}
\vspace{0.25cm}

\justifying \footnotesize

\noindent	
Notes: The table shows the regression results of output $A_{i,t}^{\mu}$ and patents $A_{i,t}^{\tau}$ on demand-pull and technology-push effects. The estimation is based on a two-ways Blundell-Bond (BB) system GMM model. The controls included in all regressions are wages $W_{i,t}$, capital intensity $(K/L)_{i,t}$, investment per capita $(I/L)_{i,t}$, relative wages for production labor $(W^P/W)_{i,t}$, energy intensity $(E/L)^{}_{i,t-1}$, and material inputs per capita $(M/L)^{}_{i,t-1}$.    Spillovers are calculated on the basis of first-order links. Variables measured in monetary terms are deflated using the industry level price deflators for the value of shipment obtained from the NBER-productivity database \citep{becker2013nber}. To cope with skewness and to obtain tractable coefficients, most variables are pre-processed (taking logs, removing outliers, scaling). Data in logs are patents and output $A^{\alpha }_{i,t}$, centrality $PR^{\alpha, d }_{i,t}$, spillovers $Spill(A)^{\alpha, d }_{i,t}$, employment $L_{i,t}$,  $(K/L)_{i,t}$, $(I/L)_{i,t}$, $W_{i,t}$, $(E/L)^{}_{i,t-1}$, $(M/L)^{}_{i,t-1}$ with $\alpha = \mu, \tau$ and $d = up, dw$. $A^{\alpha }_{i,t}$, $PR^{\alpha, d }_{i,t}$, and $Spill(A)^{\alpha, d }_{i,t}$ are scaled by division by their standard deviation to obtain comparable coefficients across the different network effects. A detailed description of the transformations and descriptive statistics of the regression data before and after the transformations are provided in \ref{app:data}.
The rows AR(1), AR(2), and Sargan show the test statistics of the specification tests, i.e. testing for first- and second-order autocorrelation and the results of a Sargan test for validity of instruments \citep[see][]{roodman2009xtabond2}. 
\end{table}

\FloatBarrier

\subsection{The direction of technological change}
\label{app:results_direction}
This section provides a sample of additional results and robustness checks about the impact of TP and DP on the direction of technological change. 
Table \ref{tab:TFP|real_vadd|emp|real_pay|cap|real_invest|share_prode|real_relative_prodw_all_sectors_all_years_FE_weighted_wControls_appendix} shows the same results as discussed in Sec. \ref{subsubsec:direction_of_change} including industry level controls and autoregressive terms. 

Table \ref{tab:TFP|real_vadd|emp|real_pay|cap|real_invest|share_prode|real_relative_prodw_all_sectors_all_years_sys_1step_wControls} and \ref{tab:TFP|real_vadd|emp|real_pay|cap|real_invest|share_prode|real_relative_prodw_all_sectors_all_years_diff_2step_wControls} the regression results from models relying on alternative estimators (one-step System-GMM à la \citet{blundell1998initial} and two-step difference GMM à la \citet{arellano1991some}). Both control for two-ways FE, i.e. include time and industry dummies. 
As the test statistics suggest, these models severely suffer from weak instruments. Additional tests, which are not presented here, suggest that these estimation difficulties cannot be resolved if instruments are collapsed or deeper lags are used as instruments. 

Notably, the performance of these models is much better if different subperiods or subgroups of industries are considered suggesting non-trivial patterns over time and industries, which are hard to identify in these linear models. However, a deeper search for suitable identification strategies is beyond the scope of this paper. Lastly, the results for the two subperiods are provided in \ref{tab:TFP|real_vadd|emp|real_pay|cap|real_invest|share_prode|real_relative_prodw_all_sectors_first_subperiod_FE_weighted_wControls} and \ref{tab:TFP|real_vadd|emp|real_pay|cap|real_invest|share_prode|real_relative_prodw_all_sectors_second_subperiod_FE_weighted_wControls}. 
\newpage

\begin{sidewaystable}[H]
\begin{myresizeenv}
\begingroup
\small
\begin{tabular}{|l|cccc|cccc|cccc|cccc|}
	\hline \hline  \rule{0pt}{1.075\normalbaselineskip}  
	& \multicolumn{4}{c|}{ {Productivity}} 
	& \multicolumn{4}{c|}{ {Labor \& Wages}} 
	& \multicolumn{4}{c|}{ {Capital \& Investment}} 
	& \multicolumn{4}{c|}{ {Production labor}} \\
	\hline  \rule{0pt}{1.075\normalbaselineskip}   & $TFP_{i,t}$ & $TFP_{i,t}$ & $(VA/L)_{i,t}$ & $(VA/L)_{i,t}$ & $L_{i,t}$ & $L_{i,t}$ & $W_{i,t}$ & $W_{i,t}$ & $(K/L)_{i,t}$ & $(K/L)_{i,t}$ & $(I/L)_{i,t}$ & $(I/L)_{i,t}$ & $(L^P/L)_{i,t}$ & $(L^P/L)_{i,t}$ & $(W^P/W)_{i,t}$ & $(W^P/W)_{i,t}$ \\ 
	\rule{0pt}{1.075\normalbaselineskip}  & (1) & (2) & (3) & (4) & (5) & (6) & (7) & (8) & (9) & (10) & (11) & (12) & (13) & (14) & (15) & (16) \\ 
	\hline \rule{0pt}{1.075\normalbaselineskip}   $A^{\mu}_{i,t-1}$ & 0.0459** & 0.0275. & -0.1465* & -0.1469* & -0.2349*** & -0.0895* & -0.2227*** & -0.2082*** & 0.1234*** & 0.0492* & -0.1769* & -0.1578. & -0.0099* & -0.0149*** & -0.0138. & -0.0186* \\ 
	& (0.0166) & (0.0159) & (0.0624) & (0.0629) & (0.0426) & (0.0438) & (0.0526) & (0.0527) & (0.0216) & (0.0213) & (0.0886) & (0.0923) & (0.0041) & (0.0043) & (0.0072) & (0.0075) \\ 
	$A^{\tau}_{i,t-1}$ & 0.0966* & 0.0096 & -0.1031 & -0.0482 & -0.4202*** & -0.019 & -0.0747 & -0.0632 & 0.2042*** & 0.0924 & -0.598* & -0.1016 & -0.038*** & -0.0265* & -0.0669*** & -0.0306 \\ 
	& (0.0452) & (0.0446) & (0.1587) & (0.1675) & (0.1152) & (0.1171) & (0.1337) & (0.1402) & (0.0587) & (0.0568) & (0.237) & (0.2458) & (0.0112) & (0.0115) & (0.0195) & (0.0201) \\ 
	$PR^{\mu,up}_{i,t-1}$ & 0.0034 & 0.0082. & 0.042* & 0.047* & 0.0258. & 0.0223. & 0.0585*** & 0.0596*** & -0.0017 & -0.0048 & 0.0577. & 0.0602* & 0.0013 & 0.0013 & 0.0068** & 0.0069** \\ 
	& (0.0051) & (0.0048) & (0.0196) & (0.0196) & (0.0142) & (0.0136) & (0.0166) & (0.0164) & (0.0072) & (0.0067) & (0.0294) & (0.0288) & (0.0014) & (0.0013) & (0.0024) & (0.0024) \\ 
	$PR^{\mu,dw}_{i,t-1}$ & 0.0063 & 0.0123** & 0.046** & 0.057*** & 0.024* & 0.0265* & 0.0371** & 0.0398** & -0.0027 & -0.0096. & 0.0629** & 0.0619** & -7e-04 & -1e-04 & -3e-04 & 0 \\ 
	& (0.0043) & (0.0041) & (0.0162) & (0.0163) & (0.0117) & (0.0112) & (0.0136) & (0.0136) & (0.0059) & (0.0055) & (0.0242) & (0.0239) & (0.0011) & (0.0011) & (0.002) & (0.0019) \\ 
	$PR^{\tau,dw}_{i,t-1}$ & -0.0041 & -0.0166 & -0.1013 & -0.0463 & -0.155** & 0.0136 & 0.0284 & 0.0841 & 0.1312*** & 0.0384 & 0.1126 & 0.1945. & -0.0026 & -0.0064 & 0.0248** & 0.0265** \\ 
	& (0.0196) & (0.0192) & (0.0734) & (0.0745) & (0.052) & (0.0514) & (0.0619) & (0.0623) & (0.0262) & (0.0252) & (0.1078) & (0.1093) & (0.005) & (0.0051) & (0.0087) & (0.0089) \\ 
	$Spill(A)^{\mu,up}_{i,t-1}$ & -0.0111 & -0.0117 & -0.0742. & -0.0815* & -0.0743* & -0.0555. & -0.0718* & -0.0672. & 0.0424** & 0.0319* & -0.0883 & -0.1051. & -9e-04 & -0.0025 & 0.0107* & 0.0088. \\ 
	& (0.0103) & (0.0097) & (0.0412) & (0.0411) & (0.0297) & (0.0284) & (0.0348) & (0.0344) & (0.0151) & (0.0139) & (0.0615) & (0.0604) & (0.0029) & (0.0028) & (0.005) & (0.0049) \\ 
	$Spill(A)^{\mu,dw}_{i,t-1}$ & -0.016* & -0.0135* & -0.0161 & -0.0176 & 0.0032 & -0.0079 & -0.004 & -0.0085 & -0.0046 & -0.0035 & -0.0041 & -0.0165 & -2e-04 & 7e-04 & 0.0032 & 0.0033 \\ 
	& (0.0065) & (0.0062) & (0.0252) & (0.0252) & (0.0181) & (0.0175) & (0.0212) & (0.0212) & (0.0092) & (0.0086) & (0.0376) & (0.0371) & (0.0018) & (0.0017) & (0.0031) & (0.003) \\ 
	$Spill(A)^{\tau,up}_{i,t-1}$ & -0.7711*** & -0.7479*** & -1.471. & -1.207 & -0.754 & -0.4783 & -1.533* & -1.336* & 0.1511 & 0.0448 & 0.3971 & 1.124 & -0.2311*** & -0.201*** & -0.3043** & -0.2156* \\ 
	& (0.2122) & (0.2002) & (0.7862) & (0.7875) & (0.5666) & (0.5451) & (0.6627) & (0.661) & (0.288) & (0.2676) & (1.174) & (1.159) & (0.0551) & (0.0543) & (0.0961) & (0.0946) \\ 
	$Spill(A)^{\tau,dw}_{i,t-1}$ & -0.0424* & -0.0225 & 0.0341 & 0.0382 & 0.0839. & 0.0322 & 0.0474 & 0.0451 & -0.0202 & -0.0286 & 0.2051* & 0.1815. & 0.0043 & 0.0043 & -0.0014 & -0.0041 \\ 
	& (0.0176) & (0.0167) & (0.0659) & (0.066) & (0.0478) & (0.0459) & (0.0556) & (0.0554) & (0.0242) & (0.0224) & (0.0988) & (0.0971) & (0.0046) & (0.0045) & (0.0081) & (0.0079) \\ 
	$TFP^{}_{i,t-1}$ & 0.9444*** & 0.8321*** &  &  &  &  &  &  &  &  &  &  &  &  &  &  \\ 
	& (0.0162) & (0.0246) &  &  &  &  &  &  &  &  &  &  &  &  &  &  \\ 
	$(VA/L)^{}_{i,t-1}$ &  &  & 0.8153*** & 0.6924*** &  &  &  &  &  &  &  &  &  &  &  &  \\ 
	&  &  & (0.0143) & (0.0426) &  &  &  &  &  &  &  &  &  &  &  &  \\ 
	$L^{}_{i,t-1}$ &  &  &  &  & 0.8415*** & 0.96*** &  &  &  &  &  &  &  &  &  &  \\ 
	&  &  &  &  & (0.0145) & (0.0275) &  &  &  &  &  &  &  &  &  &  \\ 
	$W^{}_{i,t-1}$ &  & 0.0579*** &  & 0.1659*** &  & -0.0373* & 0.8332*** & 0.8407*** &  & 0.013 &  & 0.0688. &  & 0.0052** &  & 0.0119*** \\ 
	&  & (0.0091) &  & (0.048) &  & (0.0174) & (0.0129) & (0.0201) &  & (0.0081) &  & (0.0353) &  & (0.0017) &  & (0.0029) \\ 
	$(K/L)^{}_{i,t-1}$ &  & 0.0363*** &  & -0.0544. &  & -0.111*** &  & -0.0604* & 0.8275*** & -0.0434*** &  & -0.2473*** &  & 0.0048* &  & -0.0146*** \\ 
	&  & (0.0077) &  & (0.0297) &  & (0.0209) &  & (0.0249) & (0.016) & (0.0128) &  & (0.0437) &  & (0.002) &  & (0.0036) \\ 
	$(I/L)^{}_{i,t-1}$ &  & -0.0473*** &  & 0.0411. &  & 0.0605*** &  & 0.0718*** &  & 0.0808*** & 0.5464*** & 0.4823*** &  & -0.005*** &  & -3e-04 \\ 
	&  & (0.005) &  & (0.0212) &  & (0.0151) &  & (0.0176) &  & (0.0074) & (0.0208) & (0.0308) &  & (0.0014) &  & (0.0025) \\ 
	$(L^P/L)_{i,t-1}$ &  & 0.1587* &  & 1.114*** &  & 1.336*** &  & 1.127*** &  & -0.7374*** &  & 1.161** & 0.6768*** & 0.6286*** &  & 0.182*** \\ 
	&  & (0.0735) &  & (0.304) &  & (0.209) &  & (0.253) &  & (0.1028) &  & (0.4435) & (0.0193) & (0.0208) &  & (0.0362) \\ 
	$(W^P/W)_{i,t-1}$ &  & 0.0349 &  & -0.2356 &  & -0.4672*** &  & -0.3293* &  & 0.2112*** &  & -0.379 &  & 0.0909*** & 0.5492*** & 0.4784*** \\ 
	&  & (0.0446) &  & (0.1882) &  & (0.1296) &  & (0.1571) &  & (0.0638) &  & (0.2753) &  & (0.0129) & (0.0216) & (0.0225) \\ 
	$(E/L)^{}_{i,t-1}$ &  & -0.048*** &  & -0.0856** &  & -0.0476* &  & -0.0861*** &  & -0.0147 &  & 0.2205*** &  & 0.0039* &  & 0.011*** \\ 
	&  & (0.007) &  & (0.0274) &  & (0.022) &  & (0.0228) &  & (0.0096) &  & (0.0399) &  & (0.0019) &  & (0.0033) \\ 
	$(M/L)^{}_{i,t-1}$ &  & -0.0012 &  & 0.0118 &  & -0.0769*** &  & 0.0225 &  & 0.0297** &  & 0.0139 &  & -0.0062** &  & -0.0043 \\ 
	&  & (0.0073) &  & (0.0294) &  & (0.0206) &  & (0.0247) &  & (0.0102) &  & (0.0432) &  & (0.002) &  & (0.0035) \\ 
	\hline \hline \rule{0pt}{1.075\normalbaselineskip}   Controls &  & Y &  & Y &  & Y &  & Y &  & Y &  & Y &  & Y &  & Y \\ 
	$R^2$ & 0.7438 & 0.7746 & 0.9007 & 0.9024 & 0.929 & 0.9359 & 0.9251 & 0.927 & 0.9135 & 0.927 & 0.8259 & 0.8339 & 0.8778 & 0.8832 & 0.662 & 0.6776 \\ 
	\hline
	\hline
\end{tabular}
\endgroup
\caption{Productivity, labor, capital, and production labor} 
\label{tab:TFP|real_vadd|emp|real_pay|cap|real_invest|share_prode|real_relative_prodw_all_sectors_all_years_FE_weighted_wControls_appendix}
\end{myresizeenv}

\justifying \scriptsize

\noindent
Notes: The table shows the regression results of productivity, labor demand, capital use, and production labor on demand-pull and technology-push effects. The estimation is based on a two-ways weighted fixed-effects (FE) model. The weights used in the regressions are $A^{\mu}_{i,t}$ in the TFP regression and $L_{i,t}$ in all other regressions. 
Each regression include the respective lagged dependent variable as a control. The controls included in all regressions are wages $W_{i,t}$, capital intensity $(K/L)_{i,t}$, investment per capita $(I/L)_{i,t}$, relative wages for production labor $(W^P/W)_{i,t}$, energy intensity $(E/L)^{}_{i,t-1}$, and material inputs per capita $(M/L)^{}_{i,t-1}$.   
Spillovers are calculated on the basis of first-order links. Variables measured in monetary terms are deflated using the industry level price deflators for the value of shipment obtained from the NBER-productivity database \citep{becker2013nber}. To cope with skewness and to obtain tractable coefficients, most variables are pre-processed (taking logs, removing outliers, scaling). Data in logs are patents and output $A^{\alpha }_{i,t}$, centrality $PR^{\alpha, d }_{i,t}$, spillovers $Spill(A)^{\alpha, d }_{i,t}$, employment $L_{i,t}$,  $(K/L)_{i,t}$, $(I/L)_{i,t}$, $W_{i,t}$, $(E/L)^{}_{i,t-1}$, $(M/L)^{}_{i,t-1}$ with $\alpha = \mu, \tau$ and $d = up, dw$. $A^{\alpha }_{i,t}$, $PR^{\alpha, d }_{i,t}$, and $Spill(A)^{\alpha, d }_{i,t}$ are scaled by division by their standard deviation to obtain comparable coefficients across the different network effects. A detailed description of the transformations and descriptive statistics of the regression data before and after the transformations are provided in \ref{app:data}. 
\end{sidewaystable}

\begin{sidewaystable}[H]
\begin{myresizeenv}
\begingroup
\begin{tabular}{|l|cccc|cccc|cccc|cccc|}
	\hline \hline  \rule{0pt}{1.075\normalbaselineskip}  
	& \multicolumn{4}{c|}{ {Productivity}} 
	& \multicolumn{4}{c|}{ {Labor \& Wages}} 
	& \multicolumn{4}{c|}{ {Capital \& Investment}} 
	& \multicolumn{4}{c|}{ {Production labor}} \\
	\hline  \rule{0pt}{1.075\normalbaselineskip}   & $TFP_{i,t}$ & $TFP_{i,t}$ & $(VA/L)_{i,t}$ & $(VA/L)_{i,t}$ & $L_{i,t}$ & $L_{i,t}$ & $W_{i,t}$ & $W_{i,t}$ & $(K/L)_{i,t}$ & $(K/L)_{i,t}$ & $(I/L)_{i,t}$ & $(I/L)_{i,t}$ & $(L^P/L)_{i,t}$ & $(L^P/L)_{i,t}$ & $(W^P/W)_{i,t}$ & $(W^P/W)_{i,t}$ \\ 
	\rule{0pt}{1.075\normalbaselineskip}  & (1) & (2) & (3) & (4) & (5) & (6) & (7) & (8) & (9) & (10) & (11) & (12) & (13) & (14) & (15) & (16) \\ 
	\hline \rule{0pt}{1.075\normalbaselineskip}   $A^{\mu}_{i,t-1}$ & -0.0033 & 0.0358 & -0.359* & -0.0498 & 0.0606 & 0.2302* & -0.2898* & -0.0156 & 0.0782. & 0.0113 & -0.1454 & 0.0202 & 0 & 1e-04 & -0.0171 & -0.0121 \\ 
	& (0.0269) & (0.0256) & (0.1435) & (0.1334) & (0.0875) & (0.0914) & (0.1267) & (0.1125) & (0.0453) & (0.0401) & (0.163) & (0.1756) & (0.0087) & (0.0077) & (0.0147) & (0.0135) \\ 
	$A^{\tau}_{i,t-1}$ & -0.0456 & -0.0725 & -0.4094 & -0.929*** & -0.3931** & -0.5861*** & -0.5227* & -0.6681*** & 0.1251 & 0.1459* & -0.4306 & -0.4793 & -0.0441** & -0.0392* & -5e-04 & 0.0197 \\ 
	& (0.0737) & (0.0535) & (0.254) & (0.2354) & (0.1456) & (0.1568) & (0.2376) & (0.1951) & (0.0848) & (0.0689) & (0.2926) & (0.2939) & (0.0154) & (0.0161) & (0.0283) & (0.0243) \\ 
	$PR^{\mu,up}_{i,t-1}$ & 0.034*** & 0.0199* & 0.1175** & 0.078. & -0.0312 & -0.0077 & 0.0369 & 0.0529 & 0.0066 & 0.0043 & 0.001 & 0.0178 & 0.0029 & 0.0024 & 0.0098. & 0.0106* \\ 
	& (0.0093) & (0.009) & (0.0416) & (0.0401) & (0.0249) & (0.026) & (0.0369) & (0.0383) & (0.0169) & (0.0139) & (0.0571) & (0.0545) & (0.0023) & (0.0022) & (0.005) & (0.0047) \\ 
	$PR^{\mu,dw}_{i,t-1}$ & 0.0299*** & 0.0287*** & 0.1951*** & 0.1946*** & 0.0679*** & 0.0969*** & 0.1466*** & 0.1326*** & -0.0291* & -0.0476*** & 0.1526*** & 0.1777*** & 7e-04 & 0.0013 & 0.0039 & 0.0039 \\ 
	& (0.0086) & (0.0085) & (0.0402) & (0.039) & (0.0198) & (0.0202) & (0.0308) & (0.0291) & (0.0128) & (0.0124) & (0.0405) & (0.0396) & (0.0022) & (0.0023) & (0.0045) & (0.0041) \\ 
	$PR^{\tau,dw}_{i,t-1}$ & 0.1629*** & 0.1313*** & 0.5601*** & 0.4807*** & -0.0242 & 0.0193 & 0.645*** & 0.3945*** & 0.0887* & -0.0065 & 0.2954* & 0.208 & 0.0033 & 0.0088 & -0.0179. & -0.0195. \\ 
	& (0.0461) & (0.0394) & (0.1325) & (0.0957) & (0.0685) & (0.0575) & (0.1311) & (0.0808) & (0.0367) & (0.0257) & (0.126) & (0.1343) & (0.0064) & (0.0058) & (0.0108) & (0.0102) \\ 
	$Spill(A)^{\mu,up}_{i,t-1}$ & -8e-04 & -0.0068 & -0.0355 & -0.0593 & -0.0652 & -0.0675 & -0.1059. & -0.0435 & 0.03 & 0.0551* & -0.1245. & -0.1086 & -0.0014 & -0.0019 & 0.0019 & -0.005 \\ 
	& (0.016) & (0.0141) & (0.069) & (0.0645) & (0.0489) & (0.0481) & (0.0589) & (0.0577) & (0.0277) & (0.0271) & (0.0748) & (0.0782) & (0.0041) & (0.004) & (0.0078) & (0.0071) \\ 
	$Spill(A)^{\mu,dw}_{i,t-1}$ & -0.0121 & -0.0068 & 0.0973. & 0.065 & 0.0698* & 0.0513 & 0.1128** & 0.0667. & -0.0311. & -0.0225 & 0.1198. & 0.1246. & -0.0021 & -0.0018 & 0.0037 & -0.0053 \\ 
	& (0.0108) & (0.0108) & (0.0513) & (0.0512) & (0.0329) & (0.0321) & (0.0433) & (0.0397) & (0.0184) & (0.0163) & (0.0715) & (0.0671) & (0.0034) & (0.0031) & (0.0068) & (0.0062) \\ 
	$Spill(A)^{\tau,up}_{i,t-1}$ & -0.6915 & -0.4913 & -3.865. & -2.116 & -2.37. & -2.604* & -2.985 & -0.6153 & 0.2543 & 0.8965 & -2.12 & -4.18. & -0.0994 & -0.0988 & -0.2208 & -0.1931 \\ 
	& (0.5408) & (0.445) & (2.161) & (2.084) & (1.304) & (1.231) & (1.905) & (1.591) & (0.6839) & (0.5913) & (2.159) & (2.158) & (0.093) & (0.0823) & (0.158) & (0.1577) \\ 
	$Spill(A)^{\tau,dw}_{i,t-1}$ & -0.0858* & -0.0471 & -0.0865 & 0.2426* & 0.2694*** & 0.3773*** & -0.059 & 0.1556. & -0.1474*** & -0.1257*** & 0.1224 & 0.2247 & 0.0226*** & 0.0175* & 0.0105 & 0.0049 \\ 
	& (0.0363) & (0.0331) & (0.1115) & (0.1058) & (0.0727) & (0.0768) & (0.1019) & (0.0873) & (0.0418) & (0.0337) & (0.1416) & (0.1427) & (0.0064) & (0.0071) & (0.0129) & (0.0122) \\ 
	\hline \rule{0pt}{1.075\normalbaselineskip}  AR(1) & 0 & 0 & 0 & 0 & 0 & 0 & 0 & 0 & 0 & 0 & 0 & 0 & 0 & 0 & 0 & 0 \\ 
	AR(2) & 0.0165 & 0.0222 & 0.0032 & 6e-04 & 3e-04 & 0 & 0.0461 & 0.0332 & 9e-04 & 0.0025 & 0.1265 & 0.0301 & 0.6521 & 0.5527 & 0.5203 & 0.3256 \\ 
	Sargan & 0 & 0.0059 & 0 & 0.0035 & 0 & 0.0024 & 0 & 5e-04 & 1e-04 & 9e-04 & 0.0025 & 6e-04 & 0 & 0.0015 & 0.083 & 0.0182 \\ 
	\hline \hline \rule{0pt}{1.075\normalbaselineskip}   Controls &  & Y &  & Y &  & Y &  & Y &  & Y &  & Y &  & Y &  & Y \\ 
	$R^2$ & 0.9554 & 0.9625 & 0.9934 & 0.994 & 0.9877 & 0.9872 & 0.9929 & 0.9943 & 0.9466 & 0.953 & 0.9718 & 0.9724 & 0.9942 & 0.9944 & 0.993 & 0.9933 \\ 
	\hline
	\hline
\end{tabular}
\endgroup
\caption{Productivity, labor, capital, and production labor --- Blundell-Bond estimator.}

\label{tab:TFP|real_vadd|emp|real_pay|cap|real_invest|share_prode|real_relative_prodw_all_sectors_all_years_sys_1step_wControls}
\end{myresizeenv}

\vspace{0.25cm}

\justifying \footnotesize

\noindent
Notes: The table shows the regression results of productivity, labor demand, capital use, and production labor on demand-pull and technology-push effects. The estimation is based on a oneways Blundell-Bond (BB) system GMM model. 
Each regression include the respective lagged dependent variable as a control. The controls included in all regressions are wages $W_{i,t}$, capital intensity $(K/L)_{i,t}$, investment per capita $(I/L)_{i,t}$, relative wages for production labor $(W^P/W)_{i,t}$, energy intensity $(E/L)^{}_{i,t-1}$, and material inputs per capita $(M/L)^{}_{i,t-1}$.   
Spillovers are calculated on the basis of first-order links. Variables measured in monetary terms are deflated using the industry level price deflators for the value of shipment obtained from the NBER-productivity database \citep{becker2013nber}. To cope with skewness and to obtain tractable coefficients, most variables are pre-processed (taking logs, removing outliers, scaling). Data in logs are patents and output $A^{\alpha }_{i,t}$, centrality $PR^{\alpha, d }_{i,t}$, spillovers $Spill(A)^{\alpha, d }_{i,t}$, employment $L_{i,t}$,  $(K/L)_{i,t}$, $(I/L)_{i,t}$, $W_{i,t}$, $(E/L)^{}_{i,t-1}$, $(M/L)^{}_{i,t-1}$ with $\alpha = \mu, \tau$ and $d = up, dw$. $A^{\alpha }_{i,t}$, $PR^{\alpha, d }_{i,t}$, and $Spill(A)^{\alpha, d }_{i,t}$ are scaled by division by their standard deviation to obtain comparable coefficients across the different network effects. A detailed description of the transformations and descriptive statistics of the regression data before and after the transformations are provided in \ref{app:data}. 
The rows AR(1), AR(2), and Sargan show the test statistics of the specification tests, i.e. testing for first- and second-order autocorrelation and the results of a Sargan test for validity of instruments \citep[see][]{roodman2009xtabond2}. 
\end{sidewaystable}

\begin{sidewaystable}[H]
\begin{myresizeenv}
\begingroup
\caption{Productivity, labor, capital, and production labor --- Arellano-Bond estimator.}

\label{tab:TFP|real_vadd|emp|real_pay|cap|real_invest|share_prode|real_relative_prodw_all_sectors_all_years_diff_2step_wControls}
\begin{tabular}{|l|cccc|cccc|cccc|cccc|}
	\hline \hline  \rule{0pt}{1.075\normalbaselineskip}  
	& \multicolumn{4}{c|}{ {Productivity}} 
	& \multicolumn{4}{c|}{ {Labor \& Wages}} 
	& \multicolumn{4}{c|}{ {Capital \& Investment}} 
	& \multicolumn{4}{c|}{ {Production labor}} \\
	\hline  \rule{0pt}{1.075\normalbaselineskip}   & $TFP_{i,t}$ & $TFP_{i,t}$ & $(VA/L)_{i,t}$ & $(VA/L)_{i,t}$ & $L_{i,t}$ & $L_{i,t}$ & $W_{i,t}$ & $W_{i,t}$ & $(K/L)_{i,t}$ & $(K/L)_{i,t}$ & $(I/L)_{i,t}$ & $(I/L)_{i,t}$ & $(L^P/L)_{i,t}$ & $(L^P/L)_{i,t}$ & $(W^P/W)_{i,t}$ & $(W^P/W)_{i,t}$ \\ 
	\rule{0pt}{1.075\normalbaselineskip}  & (1) & (2) & (3) & (4) & (5) & (6) & (7) & (8) & (9) & (10) & (11) & (12) & (13) & (14) & (15) & (16) \\ 
	\hline \rule{0pt}{1.075\normalbaselineskip}   $A^{\mu}_{i,t-1}$ & 0.0311 & -0.005 & -0.1821 & -0.2775 & 0.108 & -0.0092 & 0.0303 & -0.1197 & 0.1205 & -0.0255 & 0.3788 & 0.5065 & -0.0213 & -0.028. & -0.0179 & -0.0567** \\ 
	& (0.0474) & (0.0429) & (0.2127) & (0.2087) & (0.1802) & (0.1727) & (0.1913) & (0.2038) & (0.1109) & (0.0824) & (0.2382) & (0.353) & (0.015) & (0.0146) & (0.0252) & (0.0198) \\ 
	$A^{\tau}_{i,t-1}$ & 0.0193 & -0.0353 & -0.8719 & -0.7764 & -1.201* & -0.8362. & -0.7636 & -0.3618 & 0.1326 & -0.0756 & 0.5511 & 0.6003 & -0.0741. & -0.082 & -0.1396* & -0.0846 \\ 
	& (0.2145) & (0.1451) & (0.7167) & (0.6369) & (0.466) & (0.4727) & (0.5821) & (0.5691) & (0.272) & (0.2861) & (0.9901) & (1.265) & (0.0437) & (0.0505) & (0.0692) & (0.0598) \\ 
	$PR^{\mu,up}_{i,t-1}$ & 0.0064 & 0.0046 & 0.0374 & 0.0287 & 0.0292 & 0.0077 & 0.0246 & 0.0441 & -0.0157 & -0.0148 & -0.0469 & -0.0235 & 0.0076* & 0.0039 & 0.0105* & 0.0125* \\ 
	& (0.0117) & (0.0129) & (0.0526) & (0.0526) & (0.0447) & (0.0443) & (0.0493) & (0.0626) & (0.0204) & (0.0173) & (0.0766) & (0.0947) & (0.0035) & (0.0037) & (0.0049) & (0.0054) \\ 
	$PR^{\mu,dw}_{i,t-1}$ & 0.0084 & 0.0172 & 0.0707 & 0.0395 & 0.0802* & 0.0534 & 0.0482 & 0.0922* & -0.021 & -0.0244 & 0.0782 & 0.0653 & 0.0029 & 4e-04 & 0.0044 & 0.004 \\ 
	& (0.011) & (0.0118) & (0.0494) & (0.0548) & (0.0323) & (0.0339) & (0.0364) & (0.0428) & (0.0209) & (0.0164) & (0.0534) & (0.0825) & (0.0031) & (0.0031) & (0.0052) & (0.0054) \\ 
	$PR^{\tau,dw}_{i,t-1}$ & 0.1399 & -0.0157 & 0.4922 & 0.1913 & 0.299 & -0.0805 & 0.7384. & 0.0626 & 0.1619 & 0.3589* & -1.003. & -0.8996 & -0.005 & -0.0166 & 0.0637 & 0.0939* \\ 
	& (0.1577) & (0.0926) & (0.5016) & (0.4389) & (0.3487) & (0.2779) & (0.4425) & (0.3526) & (0.1698) & (0.1582) & (0.6092) & (0.7414) & (0.0308) & (0.0293) & (0.0579) & (0.0374) \\ 
	$Spill(A)^{\mu,up}_{i,t-1}$ & -0.0065 & -0.0023 & 0.0194 & 0.0535 & -0.085 & -0.1022 & -0.0321 & -0.0941 & 0.0486 & 0.0291 & -0.2236. & -0.1688 & -0.0035 & -0.0018 & -0.0106 & -0.0075 \\ 
	& (0.0228) & (0.0197) & (0.0953) & (0.1056) & (0.073) & (0.0667) & (0.0924) & (0.0812) & (0.0345) & (0.0332) & (0.1156) & (0.1537) & (0.0065) & (0.006) & (0.0098) & (0.0096) \\ 
	$Spill(A)^{\mu,dw}_{i,t-1}$ & -0.0223 & 0.0054 & 0.0348 & 0.0613 & -0.0138 & -0.0043 & 0.006 & 0.028 & -0.0038 & -0.0045 & -0.0617 & -0.0703 & -0.0061 & -0.0039 & -0.008 & -0.0052 \\ 
	& (0.0143) & (0.0139) & (0.0716) & (0.0673) & (0.046) & (0.0513) & (0.0561) & (0.0653) & (0.026) & (0.0259) & (0.094) & (0.1276) & (0.0041) & (0.0047) & (0.0076) & (0.0073) \\ 
	$Spill(A)^{\tau,up}_{i,t-1}$ & -0.1603 & -1 & -1.109 & -2.832 & -3.598 & -5.354. & 1.017 & -6.405. & 1.578 & 2.85* & -3.161 & -4.32 & -0.7326** & -0.3763 & -1.012** & -0.7787. \\ 
	& (0.9856) & (0.8618) & (4.315) & (4.866) & (3.29) & (2.957) & (3.697) & (3.878) & (1.687) & (1.357) & (6.068) & (8.63) & (0.2611) & (0.2475) & (0.3789) & (0.4169) \\ 
	$Spill(A)^{\tau,dw}_{i,t-1}$ & 0.0371 & 0.0349 & 0.4588. & 0.2432 & 0.2026 & 0.0271 & 0.489* & 0.1453 & 0.127 & 0.0485 & 0.1322 & 0.4282 & -0.0097 & -0.0015 & 0.005 & 0.0114 \\ 
	& (0.0652) & (0.0499) & (0.2529) & (0.258) & (0.1997) & (0.1919) & (0.1991) & (0.1978) & (0.0983) & (0.086) & (0.324) & (0.4022) & (0.0146) & (0.0165) & (0.0259) & (0.02) \\ 
	\hline \rule{0pt}{1.075\normalbaselineskip}  AR(1) & 1e-04 & 0 & 0.001 & 1e-04 & 0.2033 & 0.3687 & 5e-04 & 0 & 0.5601 & 0.0526 & 1e-04 & 0.0058 & 0.0032 & 0 & 0.0016 & 0 \\ 
	AR(2) & 0.0048 & 0.0042 & 0.0097 & 0.0202 & 0.0182 & 0.052 & 0.0737 & 0.1603 & 0.0468 & 0.0396 & 0.0468 & 0.0405 & 0.5783 & 0.71 & 0.1537 & 0.0632 \\ 
	Sargan & 4e-04 & 3e-04 & 6e-04 & 3e-04 & 3e-04 & 1e-04 & 1e-04 & 0 & 0.0014 & 0.019 & 0.0086 & 0 & 0.0803 & 0.1101 & 0.08 & 0.0966 \\ 
	\hline \hline \rule{0pt}{1.075\normalbaselineskip}   Controls &  & Y &  & Y &  & Y &  & Y &  & Y &  & Y &  & Y &  & Y \\ 
	$R^2$ & 0.1577 & 0.2142 & 0.2154 & 0.1667 & 0.2476 & 0.2564 & 0.2801 & 0.2852 & 0.1884 & 0.2518 & 0.0432 & 0.0688 & 0.019 & 0.0212 & 0.0028 & 0.0102 \\ 
	\hline
	\hline
\end{tabular}

\endgroup

\end{myresizeenv}

\vspace{0.25cm}

\justifying \footnotesize

\noindent 1977-1992
Notes: The table shows the regression results of productivity, labor demand, capital use, and production labor on demand-pull and technology-push effects. The estimation is based on a two-ways Arellano-Bond (AB) first-difference model. 
Each regression include the respective lagged dependent variable as a control. The controls included in all regressions are wages $W_{i,t}$, capital intensity $(K/L)_{i,t}$, investment per capita $(I/L)_{i,t}$, relative wages for production labor $(W^P/W)_{i,t}$, energy intensity $(E/L)^{}_{i,t-1}$, and material inputs per capita $(M/L)^{}_{i,t-1}$.   
Spillovers are calculated on the basis of first-order links. Variables measured in monetary terms are deflated using the industry level price deflators for the value of shipment obtained from the NBER-productivity database \citep{becker2013nber}. To cope with skewness and to obtain tractable coefficients, most variables are pre-processed (taking logs, removing outliers, scaling). Data in logs are patents and output $A^{\alpha }_{i,t}$, centrality $PR^{\alpha, d }_{i,t}$, spillovers $Spill(A)^{\alpha, d }_{i,t}$, employment $L_{i,t}$,  $(K/L)_{i,t}$, $(I/L)_{i,t}$, $W_{i,t}$, $(E/L)^{}_{i,t-1}$, $(M/L)^{}_{i,t-1}$ with $\alpha = \mu, \tau$ and $d = up, dw$. $A^{\alpha }_{i,t}$, $PR^{\alpha, d }_{i,t}$, and $Spill(A)^{\alpha, d }_{i,t}$ are scaled by division by their standard deviation to obtain comparable coefficients across the different network effects. A detailed description of the transformations and descriptive statistics of the regression data before and after the transformations are provided in \ref{app:data}. 
The rows AR(1), AR(2), and Sargan show the test statistics of the specification tests, i.e. testing for first- and second-order autocorrelation and the results of a Sargan test for validity of instruments \citep[see][]{roodman2009xtabond2}. 

\end{sidewaystable}

\begin{sidewaystable}[H]
\begin{myresizeenv}
\begingroup
\begin{tabular}{|l|cccc|cccc|cccc|cccc|}
	\hline \hline  \rule{0pt}{1.075\normalbaselineskip}  
	& \multicolumn{4}{c|}{ {Productivity}} 
	& \multicolumn{4}{c|}{ {Labor \& Wages}} 
	& \multicolumn{4}{c|}{ {Capital \& Investment}} 
	& \multicolumn{4}{c|}{ {Production labor}} \\
	\hline  \rule{0pt}{1.075\normalbaselineskip}   & $TFP_{i,t}$ & $TFP_{i,t}$ & $(VA/L)_{i,t}$ & $(VA/L)_{i,t}$ & $L_{i,t}$ & $L_{i,t}$ & $W_{i,t}$ & $W_{i,t}$ & $(K/L)_{i,t}$ & $(K/L)_{i,t}$ & $(I/L)_{i,t}$ & $(I/L)_{i,t}$ & $(L^P/L)_{i,t}$ & $(L^P/L)_{i,t}$ & $(W^P/W)_{i,t}$ & $(W^P/W)_{i,t}$ \\ 
	\rule{0pt}{1.075\normalbaselineskip}  & (1) & (2) & (3) & (4) & (5) & (6) & (7) & (8) & (9) & (10) & (11) & (12) & (13) & (14) & (15) & (16) \\ 
	\hline \rule{0pt}{1.075\normalbaselineskip}   $A^{\mu}_{i,t-1}$ & 0.0624* & 0.045. & -0.1046 & -0.18 & -0.2117** & 0.0222 & -0.2794** & -0.2751** & 0.1597*** & -0.0168 & 0.3646* & 0.1793 & -0.0201** & -0.0143. & -0.0236* & -0.0139 \\ 
	& (0.0247) & (0.0258) & (0.1232) & (0.1252) & (0.0795) & (0.0866) & (0.1061) & (0.1062) & (0.0349) & (0.034) & (0.1617) & (0.1757) & (0.0069) & (0.0078) & (0.0118) & (0.0134) \\ 
	$A^{\tau}_{i,t-1}$ & 0.0812 & 0.068 & -0.038 & 0.0484 & -0.8527*** & -0.2916 & -0.3586 & -0.2032 & 0.3294*** & 0.0469 & -0.5414 & -0.2627 & -0.0322. & -0.0273 & -0.0544. & -0.0175 \\ 
	& (0.0568) & (0.0559) & (0.2749) & (0.2792) & (0.1933) & (0.1873) & (0.233) & (0.2369) & (0.0869) & (0.0756) & (0.3925) & (0.3917) & (0.0174) & (0.0175) & (0.0293) & (0.0299) \\ 
	$PR^{\mu,up}_{i,t-1}$ & 7e-04 & -0.0062 & 0.0901. & 0.0806 & 0.0439 & 0.0255 & 0.0538 & 0.0502 & -0.0157 & 4e-04 & 0.1444. & 0.125. & 0.0045 & 0.004 & 0.0071 & 0.0069 \\ 
	& (0.0097) & (0.0093) & (0.0517) & (0.0514) & (0.0366) & (0.0345) & (0.0438) & (0.0436) & (0.0165) & (0.0139) & (0.074) & (0.0721) & (0.0033) & (0.0032) & (0.0056) & (0.0055) \\ 
	$PR^{\mu,dw}_{i,t-1}$ & -0.0101 & -0.0056 & 0.0474 & 0.045 & 0.0655* & 0.0624* & 0.077* & 0.0718* & 0.0123 & -0.0151 & 0.1616** & 0.142** & -0.0051* & -0.0033 & -3e-04 & -7e-04 \\ 
	& (0.0084) & (0.0082) & (0.038) & (0.0385) & (0.0272) & (0.0258) & (0.0322) & (0.0326) & (0.0122) & (0.0105) & (0.0549) & (0.0539) & (0.0024) & (0.0024) & (0.0041) & (0.0041) \\ 
	$PR^{\tau,dw}_{i,t-1}$ & 0.0186 & 0.0183 & 0.344* & 0.4048* & 0.2821* & 0.3528*** & 0.3352* & 0.3485** & -0.0487 & -0.108* & 0.7479*** & 0.6452** & -0.0045 & -0.0073 & 0.0025 & 0.0018 \\ 
	& (0.0321) & (0.0313) & (0.1577) & (0.1597) & (0.112) & (0.1061) & (0.1334) & (0.1346) & (0.0499) & (0.0429) & (0.2236) & (0.2227) & (0.0099) & (0.0099) & (0.0169) & (0.017) \\ 
	$Spill(A)^{\mu,up}_{i,t-1}$ & -0.0106 & 0.0177 & 0.0841 & 0.0908 & -0.0373 & -0.1198 & -0.0062 & -0.0436 & 0.1133. & 0.0512 & 0.3041 & 0.2015 & -0.0055 & -4e-04 & 0.0106 & 0.0031 \\ 
	& (0.0356) & (0.0342) & (0.1851) & (0.1846) & (0.1313) & (0.1233) & (0.1569) & (0.1567) & (0.059) & (0.05) & (0.2654) & (0.2591) & (0.0117) & (0.0116) & (0.0199) & (0.0198) \\ 
	$Spill(A)^{\mu,dw}_{i,t-1}$ & -0.0125 & -0.0015 & 0.068 & 0.0938 & 0.0626 & 0.0656 & 0.1125* & 0.1119* & 0.0284 & -0.0162 & 0.1453 & 0.1259 & -0.0064 & -0.0042 & 0.002 & 0.0021 \\ 
	& (0.0124) & (0.012) & (0.0623) & (0.0622) & (0.0443) & (0.0416) & (0.0527) & (0.0527) & (0.0198) & (0.0168) & (0.0895) & (0.0872) & (0.0039) & (0.0039) & (0.0067) & (0.0067) \\ 
	$Spill(A)^{\tau,up}_{i,t-1}$ & 0.2507 & 0.1165 & 2.111 & 2.194. & 0.4105 & 1.458. & 0.5583 & 0.7734 & -0.2876 & -0.3668 & 5.391** & 5.734** & 0.0741 & 0.082 & -0.2779. & -0.1985 \\ 
	& (0.2681) & (0.2613) & (1.32) & (1.317) & (0.9356) & (0.8823) & (1.119) & (1.117) & (0.4207) & (0.3562) & (1.885) & (1.847) & (0.0836) & (0.0825) & (0.1423) & (0.1409) \\ 
	$Spill(A)^{\tau,dw}_{i,t-1}$ & -0.0296 & -0.022 & -0.0187 & -0.0262 & 0.0554 & -0.0215 & 0.0219 & -0.0128 & -0.0017 & 0.0232 & 0.1863 & 0.14 & -4e-04 & 0.0022 & -0.0042 & -0.0081 \\ 
	& (0.0204) & (0.0195) & (0.1) & (0.0997) & (0.0709) & (0.0666) & (0.0848) & (0.0847) & (0.0319) & (0.027) & (0.1429) & (0.14) & (0.0063) & (0.0063) & (0.0108) & (0.0107) \\ 
	\hline \hline \rule{0pt}{1.075\normalbaselineskip}   Controls &  & Y &  & Y &  & Y &  & Y &  & Y &  & Y &  & Y &  & Y \\ 
	$R^2$ & 0.8645 & 0.8775 & 0.9242 & 0.9264 & 0.9511 & 0.9577 & 0.9442 & 0.9456 & 0.9406 & 0.9583 & 0.8871 & 0.8939 & 0.9317 & 0.9343 & 0.7803 & 0.7885 \\ 
	\hline
	\hline
\end{tabular}
\endgroup
\caption{Productivity, labor, capital, and production labor --- first subperiod.}

\label{tab:TFP|real_vadd|emp|real_pay|cap|real_invest|share_prode|real_relative_prodw_all_sectors_first_subperiod_FE_weighted_wControls}
\end{myresizeenv}

\vspace{0.25cm}

\justifying \footnotesize

\noindent
Notes: The table shows the regression results of productivity, labor demand, capital use, and production labor on demand-pull and technology-push effects. The estimation is based on a two-ways weighted fixed-effects (FE) model. The weights used in the regressions are $A^{\mu}_{i,t}$ in the TFP regression and $L_{i,t}$ in all other regressions. 
Each regression include the respective lagged dependent variable as a control. The controls included in all regressions are wages $W_{i,t}$, capital intensity $(K/L)_{i,t}$, investment per capita $(I/L)_{i,t}$, relative wages for production labor $(W^P/W)_{i,t}$, energy intensity $(E/L)^{}_{i,t-1}$, and material inputs per capita $(M/L)^{}_{i,t-1}$.   
Spillovers are calculated on the basis of first-order links. Variables measured in monetary terms are deflated using the industry level price deflators for the value of shipment obtained from the NBER-productivity database \citep{becker2013nber}. To cope with skewness and to obtain tractable coefficients, most variables are pre-processed (taking logs, removing outliers, scaling). Data in logs are patents and output $A^{\alpha }_{i,t}$, centrality $PR^{\alpha, d }_{i,t}$, spillovers $Spill(A)^{\alpha, d }_{i,t}$, employment $L_{i,t}$,  $(K/L)_{i,t}$, $(I/L)_{i,t}$, $W_{i,t}$, $(E/L)^{}_{i,t-1}$, $(M/L)^{}_{i,t-1}$ with $\alpha = \mu, \tau$ and $d = up, dw$. $A^{\alpha }_{i,t}$, $PR^{\alpha, d }_{i,t}$, and $Spill(A)^{\alpha, d }_{i,t}$ are scaled by division by their standard deviation to obtain comparable coefficients across the different network effects.
A detailed description of the transformations and descriptive statistics of the regression data before and after the transformations are provided in \ref{app:data}. The data includes only the  first subperiod. 
\end{sidewaystable}

\begin{sidewaystable}[H]
\begin{myresizeenv}
\begingroup
\begin{tabular}{|l|cccc|cccc|cccc|cccc|}
	\hline \hline  \rule{0pt}{1.075\normalbaselineskip}  
	& \multicolumn{4}{c|}{ {Productivity}} 
	& \multicolumn{4}{c|}{ {Labor \& Wages}} 
	& \multicolumn{4}{c|}{ {Capital \& Investment}} 
	& \multicolumn{4}{c|}{ {Production labor}} \\
	\hline  \rule{0pt}{1.075\normalbaselineskip}   & $TFP_{i,t}$ & $TFP_{i,t}$ & $(VA/L)_{i,t}$ & $(VA/L)_{i,t}$ & $L_{i,t}$ & $L_{i,t}$ & $W_{i,t}$ & $W_{i,t}$ & $(K/L)_{i,t}$ & $(K/L)_{i,t}$ & $(I/L)_{i,t}$ & $(I/L)_{i,t}$ & $(L^P/L)_{i,t}$ & $(L^P/L)_{i,t}$ & $(W^P/W)_{i,t}$ & $(W^P/W)_{i,t}$ \\ 
	\rule{0pt}{1.075\normalbaselineskip}  & (1) & (2) & (3) & (4) & (5) & (6) & (7) & (8) & (9) & (10) & (11) & (12) & (13) & (14) & (15) & (16) \\ 
	\hline \rule{0pt}{1.075\normalbaselineskip}   $A^{\mu}_{i,t-1}$ & 0.0382. & 0.0363. & -0.0647 & -0.0261 & -0.1156* & -0.053 & -0.1357* & -0.103 & 0.053. & 0.0187 & -0.1321 & -0.1174 & -0.0064 & -0.0113. & -0.0067 & -0.0095 \\ 
	& (0.0214) & (0.0203) & (0.0786) & (0.0783) & (0.0572) & (0.0566) & (0.0655) & (0.0649) & (0.0317) & (0.0306) & (0.1167) & (0.1158) & (0.0058) & (0.0059) & (0.0106) & (0.0104) \\ 
	$A^{\tau}_{i,t-1}$ & 0.0749 & 0.0433 & -0.0348 & 0.0914 & -0.1823 & 0.0413 & 0.2227 & 0.1768 & 0.2483* & 0.171 & -0.3939 & 0.0974 & -0.1102*** & -0.084*** & -0.0569 & -0.015 \\ 
	& (0.0841) & (0.082) & (0.2989) & (0.3077) & (0.2217) & (0.2241) & (0.2496) & (0.2557) & (0.1224) & (0.1206) & (0.4499) & (0.456) & (0.0225) & (0.0231) & (0.041) & (0.0411) \\ 
	$PR^{\mu,up}_{i,t-1}$ & 0.0025 & 0.0043 & 0.026 & 0.021 & 0.0182 & 0.0121 & 0.0452* & 0.0386. & 0.0054 & 0.002 & 0.0228 & 0.0306 & 0.0012 & 0.0012 & 0.0066* & 0.0067* \\ 
	& (0.0066) & (0.0063) & (0.0243) & (0.024) & (0.0179) & (0.0174) & (0.0203) & (0.02) & (0.0099) & (0.0094) & (0.0366) & (0.0356) & (0.0018) & (0.0018) & (0.0033) & (0.0032) \\ 
	$PR^{\mu,dw}_{i,t-1}$ & 0.0043 & 0.0092. & 0.0356. & 0.0453* & 0.0215 & 0.0174 & 0.0238 & 0.026 & -0.0025 & 0 & 0.0672* & 0.0571. & 4e-04 & 0.0011 & -7e-04 & -4e-04 \\ 
	& (0.0057) & (0.0055) & (0.0213) & (0.0213) & (0.0157) & (0.0153) & (0.0178) & (0.0176) & (0.0087) & (0.0083) & (0.032) & (0.0314) & (0.0016) & (0.0016) & (0.0029) & (0.0028) \\ 
	$PR^{\tau,dw}_{i,t-1}$ & 0.0403 & 0.0138 & -0.1792 & -0.1153 & -0.2961*** & -0.1067 & 0.0137 & 0.0439 & 0.2378*** & 0.0961. & -0.2555 & 0.1664 & -0.0062 & -0.0052 & 0.0421** & 0.0589*** \\ 
	& (0.0339) & (0.0338) & (0.1212) & (0.1253) & (0.0875) & (0.0916) & (0.1018) & (0.1043) & (0.0493) & (0.0495) & (0.1795) & (0.1859) & (0.009) & (0.0094) & (0.0161) & (0.0167) \\ 
	$Spill(A)^{\mu,up}_{i,t-1}$ & -0.0074 & -0.0084 & -0.0855. & -0.099* & -0.0605. & -0.051 & -0.0545 & -0.0581 & 0.0375* & 0.0222 & -0.1449* & -0.1296. & -0.0036 & -0.005 & 0.0109. & 0.0111. \\ 
	& (0.0122) & (0.0116) & (0.0467) & (0.0464) & (0.0344) & (0.0334) & (0.0391) & (0.0385) & (0.0191) & (0.0181) & (0.0703) & (0.0687) & (0.0035) & (0.0035) & (0.0064) & (0.0062) \\ 
	$Spill(A)^{\mu,dw}_{i,t-1}$ & -0.0143. & -0.0098 & -0.0135 & -0.0209 & 0.0079 & -0.002 & -0.0026 & -0.0072 & -0.0119 & -0.0065 & 0.0236 & 7e-04 & -0.0011 & -3e-04 & 0.0012 & 0.001 \\ 
	& (0.0086) & (0.0082) & (0.0316) & (0.0313) & (0.0233) & (0.0226) & (0.0264) & (0.026) & (0.0129) & (0.0123) & (0.0474) & (0.0464) & (0.0024) & (0.0024) & (0.0043) & (0.0042) \\ 
	$Spill(A)^{\tau,up}_{i,t-1}$ & -0.8952** & -0.7984* & -3.417** & -3.681** & -2.234* & -2.566** & -3.156** & -3.433*** & 0.928. & 0.903. & -2.214 & -2.735 & -0.3505*** & -0.3291*** & -0.2169 & -0.2245 \\ 
	& (0.3441) & (0.3262) & (1.23) & (1.221) & (0.9085) & (0.8841) & (1.028) & (1.016) & (0.5075) & (0.4811) & (1.853) & (1.812) & (0.0925) & (0.0918) & (0.1675) & (0.1632) \\ 
	$Spill(A)^{\tau,dw}_{i,t-1}$ & -0.0187 & -0.0357 & -0.0887 & -0.0904 & -0.0303 & -0.0211 & -0.0743 & -0.0538 & -0.034 & -0.0081 & 0.0859 & 0.1276 & 0.0082 & 0.0084 & -0.0117 & -0.0099 \\ 
	& (0.0293) & (0.0278) & (0.11) & (0.1088) & (0.0812) & (0.079) & (0.0919) & (0.0904) & (0.045) & (0.0429) & (0.1654) & (0.1612) & (0.0083) & (0.0082) & (0.015) & (0.0145) \\ 
	\hline \hline \rule{0pt}{1.075\normalbaselineskip}   Controls &  & Y &  & Y &  & Y &  & Y &  & Y &  & Y &  & Y &  & Y \\ 
	$R^2$ & 0.7926 & 0.8174 & 0.925 & 0.928 & 0.9423 & 0.9469 & 0.9442 & 0.9468 & 0.9207 & 0.9298 & 0.8616 & 0.8703 & 0.8892 & 0.8933 & 0.6905 & 0.7112 \\ 
	\hline
	\hline
\end{tabular}
\endgroup
\caption{Productivity, labor, capital, and production labor --- second subperiod.}
\label{tab:TFP|real_vadd|emp|real_pay|cap|real_invest|share_prode|real_relative_prodw_all_sectors_second_subperiod_FE_weighted_wControls}
\end{myresizeenv}

\vspace{0.25cm}

\justifying \footnotesize

\noindent
Notes: The table shows the regression results of productivity, labor demand, capital use, and production labor on demand-pull and technology-push effects. The estimation is based on a two-ways weighted fixed-effects (FE) model. The weights used in the regressions are $A^{\mu}_{i,t}$ in the TFP regression and $L_{i,t}$ in all other regressions. 
Each regression include the respective lagged dependent variable as a control. The controls included in all regressions are wages $W_{i,t}$, capital intensity $(K/L)_{i,t}$, investment per capita $(I/L)_{i,t}$, relative wages for production labor $(W^P/W)_{i,t}$, energy intensity $(E/L)^{}_{i,t-1}$, and material inputs per capita $(M/L)^{}_{i,t-1}$.   
Spillovers are calculated on the basis of first-order links. Variables measured in monetary terms are deflated using the industry level price deflators for the value of shipment obtained from the NBER-productivity database \citep{becker2013nber}. To cope with skewness and to obtain tractable coefficients, most variables are pre-processed (taking logs, removing outliers, scaling). Data in logs are patents and output $A^{\alpha }_{i,t}$, centrality $PR^{\alpha, d }_{i,t}$, spillovers $Spill(A)^{\alpha, d }_{i,t}$, employment $L_{i,t}$,  $(K/L)_{i,t}$, $(I/L)_{i,t}$, $W_{i,t}$, $(E/L)^{}_{i,t-1}$, $(M/L)^{}_{i,t-1}$ with $\alpha = \mu, \tau$ and $d = up, dw$. $A^{\alpha }_{i,t}$, $PR^{\alpha, d }_{i,t}$, and $Spill(A)^{\alpha, d }_{i,t}$ are scaled by division by their standard deviation to obtain comparable coefficients across the different network effects.
A detailed description of the transformations and descriptive statistics of the regression data before and after the transformations are provided in \ref{app:data}. The data includes only the  second subperiod. 
\end{sidewaystable}

\FloatBarrier

\newpage
\part*{Supplementary Material}
\FloatBarrier
\appendix
\renewcommand{\appendixname}{Supplementary Material}
\renewcommand{\thesection}{SI.\arabic{section}} \setcounter{section}{0}
\renewcommand{\thefigure}{SI.\arabic{figure}} \setcounter{figure}{0}
\renewcommand{\thetable}{SI.\arabic{table}} \setcounter{table}{0}
\renewcommand{\theequation}{SI.\arabic{table}} \setcounter{equation}{0}
\section{Data processing in detail}
\label{supp:data_processing}
This analysis builds on two distinct sources of data brought into a consistent form that enables the statistical analyses; this is (1) a series of time snapshots of the network of cross-industrial IO flows and patent citations and (2) a panel data set of aggregate statistical indicators at the industry level. 
Industries are identified by 6-digit NAICS codes. The time snapshots cover 5-year intervals from 1977 to 2012. Obtaining these data involved a series of steps of re-formatting and harmonization, which are explained in detail below. 
\subsection{Input-output data}
\label{supp:data_processing_IO}
The IO data is constructed by the composition and harmonization of the historical benchmark tables provided by \href{https://www.bea.gov/industry/historical-benchmark-input-output-tables}{Bureau of Economic Analysis (BEA)}.\footnote{\url{https://www.bea.gov/industry/historical-benchmark-input-output-tables} [accessed on Dec 21, 2020]}
Since 1947, BEA publishes IO tables at the detailed industry level every five years. The data is collected in BEA's quinquennial Survey of Current Business. A detailed manual on BEA's IO data is provided by \citet{BEA2009IO}
\subsubsection{Overview}
The raw data shows monetary transactions between industries. It also covers final demand sectors and public services. For this project, use tables from 1977-2012 are used and harmonized in a series of conversions and processing steps. Over time, industrial classification systems and technical methods of data processing, formatting, and saving have changed. The earliest tables are only available in text format that were edited manually to make it readable for statistical software. A further challenge arises from changes in the classification system, which are most pronounced in the conversion from SIC to NAICS. 

The final data structure is a series of quadratic matrices for each period that show the monetary transactions between industries in NAICS 2002 codes. The data is also used to create a panel of industry level indicators, i.e. outputs, inputs, and growth rates. 
\subsubsection{Processing steps in detail}
Here, the single steps of data processing are explained.
\begin{description}
\item [Step 1:] For each period, the IO data is downloaded separately. Some manual harmonization and data conversions were made to obtain machine-readable, harmonized data tables. For example, the very old data is only available in text format, which is not ready to be read by statistical software. The more recent table are Excel files with many macros and explanations in text. All tables were reformatted individually. The scripts are available in the data publication. 
After this step, all tables have a uniform format, which is a long 3-column table with column (1) as producer ID, column (2) as user ID, and column (3) indicating the monetary value of the goods that flow from producer to user. 

\item [Step 2:] Large quadratic matrices with rows as producers and columns as users are created. The entries $flow^{out,\mu}_{ij,t}$ are flows of goods from $i$ to $j$. Hence, column-wise reading indicates the composition of inputs used by sector $j$ and row-wise reading indicates the composition of customer industries to which industry $i$ delivers. 

\item [Step 3:] This is an intermediate step. All concordance and IO-to-industry conversion tables must be harmonized. Again, some tables are not machine-readable. Moreover, all codes need to be harmonized to obtain a mapping from IO codes for each period to 2002-NAICS codes. Some IO codes map to multiple industries. In this step, tables were created where each row indicates an IO code and all NAICS codes that are attributed to the respective IO.  

\item [Step 4:] NAICS-based IO tables were harmonized and consistency checks were done. For example, it was tested whether the differences in the tables e.g. regarding the sector coverage are negligible. Some normalizations of IO-flows to input (output) shares were made through division by row (column) sum. The full 6-digit list is used as row and column names. 

Not every time snapshot has a full sector coverage. This is a result of reclassification issues, obsolescence, and introduction of new sectors. For example, some finely granulate computer industries were not yet existing in 1977. For these cases, empty vectors are included to present missing sectors to ensure that matrices have the same dimensionality. 

Additional steps of harmonization are done. Rows represent the range of inputs that is used, columns represent customers. After this step, NAICS 6-digit data on IO flows, sector weights (row and column sums), input shares (measured in percentage points), 6-digit distance matrix computed by the input-share dissimilarity are obtained. 

\item [Step 5:] The data was harmonized to quadratic NAICS 2002 matrices. The matrices are 1179 $\times$ 1179 matrices of 6-digit industries.\footnote{473 of 1179 NAICS 6-digit codes are manufacturing and out of these, 307 provide sufficient data to form a balanced panel. These 307 industries are used in the final analyses.} Empty rows and columns are included for industries that are not producing in some $t$, for example if an industry was not yet existing or disappeared over time.  

\item [Step 6:] For each $t$, NAICS $\times$ NAICS matrices were created with flows of goods $flow^{\mu}_{ij,t}$ as entries. 

\end{description}
\subsubsection{Technical and conceptual issues}
\paragraph{General remarks about IO codes, NAICS and SIC}
The original IO data in early years uses IO codes, which are an internal metric of the accounting system used to construct social accounting matrices (SAM). 
These codes are converted into industry codes (SIC and NAICS). 
The classification system has changed over time. Fortunately, the IO-codes in the raw IO tables are largely consistent across time. The accounting codes were converted into SIC and from SIC into NAICS or directly into NAICS if such mapping is available. 
For the data harmonization, all codes were converted into the NAICS-2002 codes as they can be directly mapped to SIC 1987 codes. 
\paragraph{How to deal with accounting codes that are mapped into multiple SIC sectors?}
Some of the accounting codes are associated with multiple SIC sectors, i.e. multiple industries have been aggregated into one accounting position. Information about the strengths of links to each of these subsectors is missing. For reasons of simplification, it is assumed that the accounting position is equally related to all of them. The strength of single links is weighted uniformly by the number of sectors. For example, the IO code 020401 (``Fruits'') is linked to 9 SIC sectors (0171, 0172, 0174, 0175, *0179, *019, *0219, *0259, *029). The links are weighted by factor 1/9. 

\paragraph{How to deal with inconsistencies across time in changing classification systems?}
The accounting codes of 1977 and 1982 data are mapped to SIC 1987. All mappings from accounting positions to SIC are based on the 1987 data, after having ensured that the accounting codes are consistent across time. Also the vast majority of IO-to-SIC-mappings is consistent in 1977 and 1992 data. For 1977 some minor deviations exist, but these can be largely explained by adjustments in the SIC system between 1977 and 1987. 
Some old SIC industries do not exist any longer. A reconstruction is practically not feasible with reasonable effort, given that the expected value added of higher precision is negligibly small, if existing at all. 
The 1977 IO-SIC mapping is only used when 1987-data is not available. 

In the 2002 NAICS file, some IO codes are mapped to a very high number of sub-sectors. This is for example the case for aggregate positions such as retail and wholesale trade and construction. They were kept in the mapping. 
It should be noted that an accounting position that has a link to more than a hundred 6-digit NAICS industries is not necessarily meaningful. This problem was addressed by a series of robustness checks using only a subset of the data, higher levels of aggregation and rounding of IO links that fall below a certain threshold. 

The more recent versions of the classification systems are more detailed. Equal weights were used when one coarse industry maps to several more detailed industry when using another (typically more recent) classification system. Hence, the transaction volume is equally distributed across sub-sectors. 

\subsection{Patent data}
\label{supp:data_processing_patents}
\subsubsection{Overview}
The raw patent data classified by CPC codes are taken from an earlier project. An extensive documentation of the data is provided along with the data, which can be downloaded for reuse under a CC-BY-4.0 license \citep{hotte2023data}.

From the raw data, the CPC classification data, citations among patents with the grant number as ID, and data on the grant year of the patents were used. 
Further, to map patents classified by CPC codes to NAICS 6-digit codes, the concordance tables by \citet{goldschlag2020tracking} was used.\footnote{\url{https://sites.google.com/site/nikolaszolas/PatentCrosswalk} [accessed in Oct 2021]}. 
The \emph{Cooperative Patent Classification (CPC) Crosswalks - Version 1603} file downloaded in October 2021 was used. 

The patents were first reclassified into NAICS taking account of the weighting scheme of \possessivecite{goldschlag2020tracking} concordance. Next, a NAICS-to-NAICS citation network for each 5-year time window was constructed. 

\subsubsection{Processing steps in detail}
To construct a patent citation network among NAICS 6-digit industries, the data were processed in a series of steps. 
\begin{description}
\item[Step 1:] In a first step, patents were mapped to NAICS 2002 codes to create industry level patent stocks as 5-year aggregates covering the period 1973-2012. First, patents were sampled by time window. Then, patents in each time window were aggregated into each 4-digit CPC class, ensuring uniqueness for each entry by patent grant number and CPC 4-digit code. Hence, patents that map to multiple more disaggregate CPC codes that belong to the same 4-digit aggregate were treated as unique entry. The counts at the 4-digit CPC level were subsequently mapped to NAICS 6-digit codes taking account of the weights, i.e. the patent counts are multiplied by the weight whenever one CPC 4-digit class maps to multiple NAICS 6-digit codes. 
\item[Step 2:] Next, the citation data is mapped from citations between individual patents by grant number to citations between NAICS 6-digit industries. The NAICS-to-NAICS edgelist also contains a column with the weight, which indicates the number of citations that flow from one industry to another during each 5-year time window. 
To obtain this edgelist, both the citing and the cited patent were first mapped to CPC, ensuring the uniqueness of patent CPC 4-digit links. Then, they were aggregated and the number of CPC 4-digit-to-CPC 4-digit citations was obtained based on unique pairs of patents. Next, the CPC-to-CPC citations were mapped to NAICS while applying the weight to both the citing and cited NAICS industry. 
\item[Step 3:] The edgelist is used to construct an adjacency matrix with 6-digit industries as row and column names. The data is harmonized with the format of the IO adjacency matrices. Additionally, the adjacency matrices are also created for other levels of aggregation.  

\end{description}
\FloatBarrier
\newpage
\section{Supplementary descriptives}
\label{supp:results_descr}
\subsection{Network plots}

\begin{figure}[h!]
\caption{Upstream networks at the 3-digit level for different periods}
\label{fig:threeDcomp_netw_flow_up}
\hspace{0cm}\textbf{1977-1992}\vspace{0.25cm}

\centering
\begin{subfigure}{0.3\textwidth}		
{\centering
\caption{Input-output}			
\label{fig:threeDcomp_netw_flow_up_noOV_io92}			
\includegraphics[width=\textwidth]{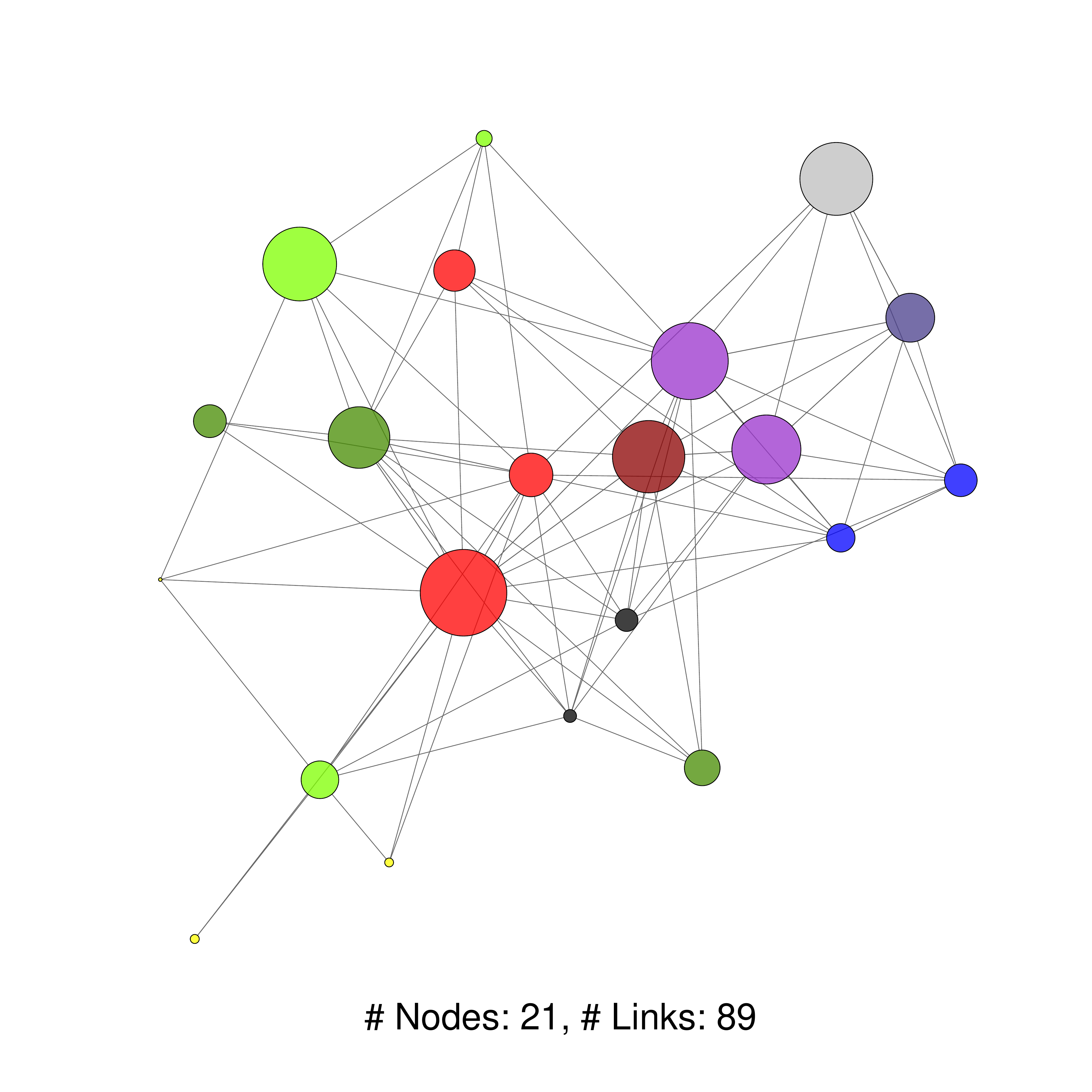}	
}
\end{subfigure}
\begin{subfigure}{0.3\textwidth}		
{\centering
\caption{Patent citations}			
\label{fig:threeDcomp_netw_flow_up_noOV_pat92}			
\includegraphics[width=\textwidth]{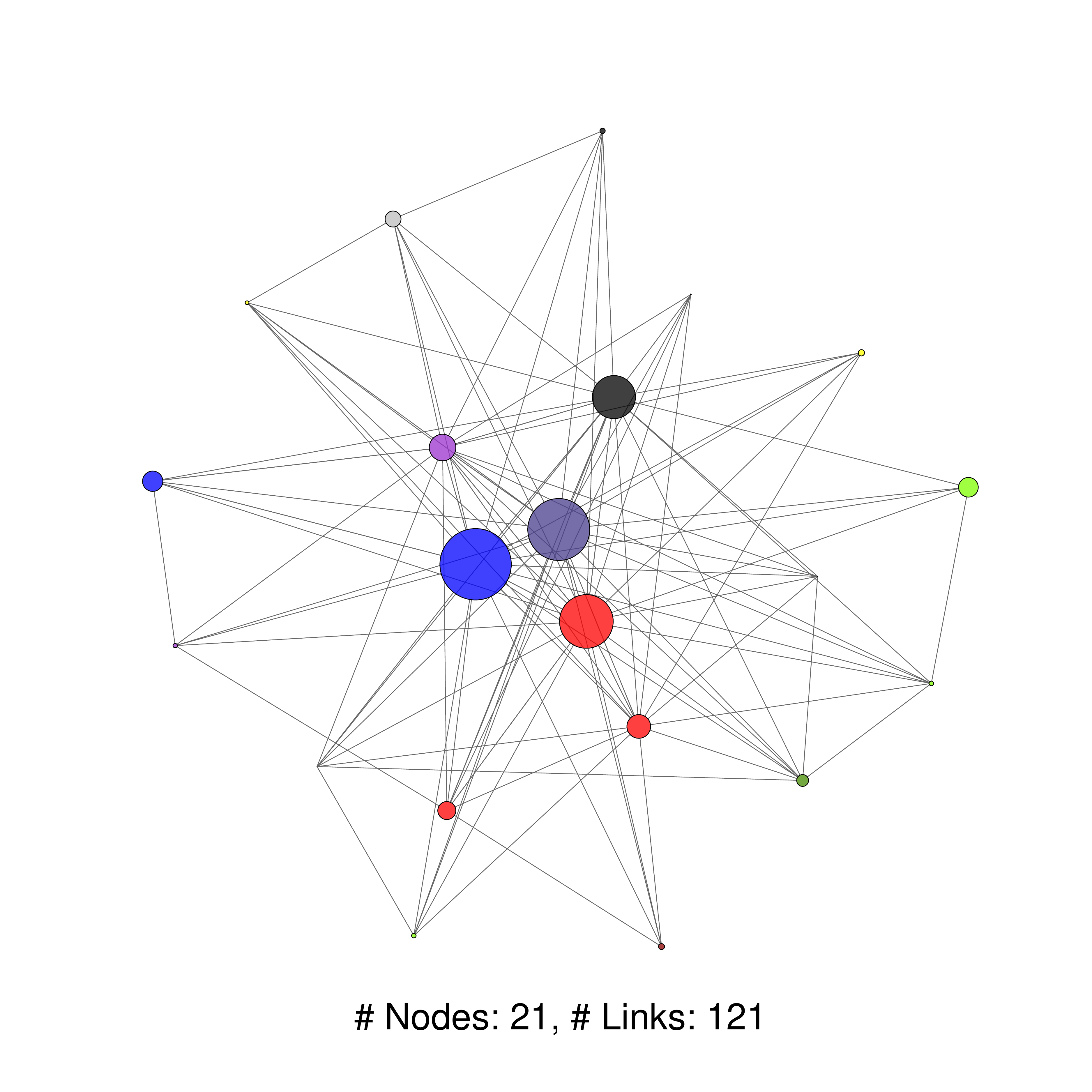}	
}
\end{subfigure}

\hspace{0cm}\textbf{1997-2012}\vspace{0.25cm}

\centering
\begin{subfigure}{0.3\textwidth}		
{\centering
\caption{Input-output}			
\label{fig:threeDcomp_netw_flow_up_noOV_io12}			
\includegraphics[width=\textwidth]{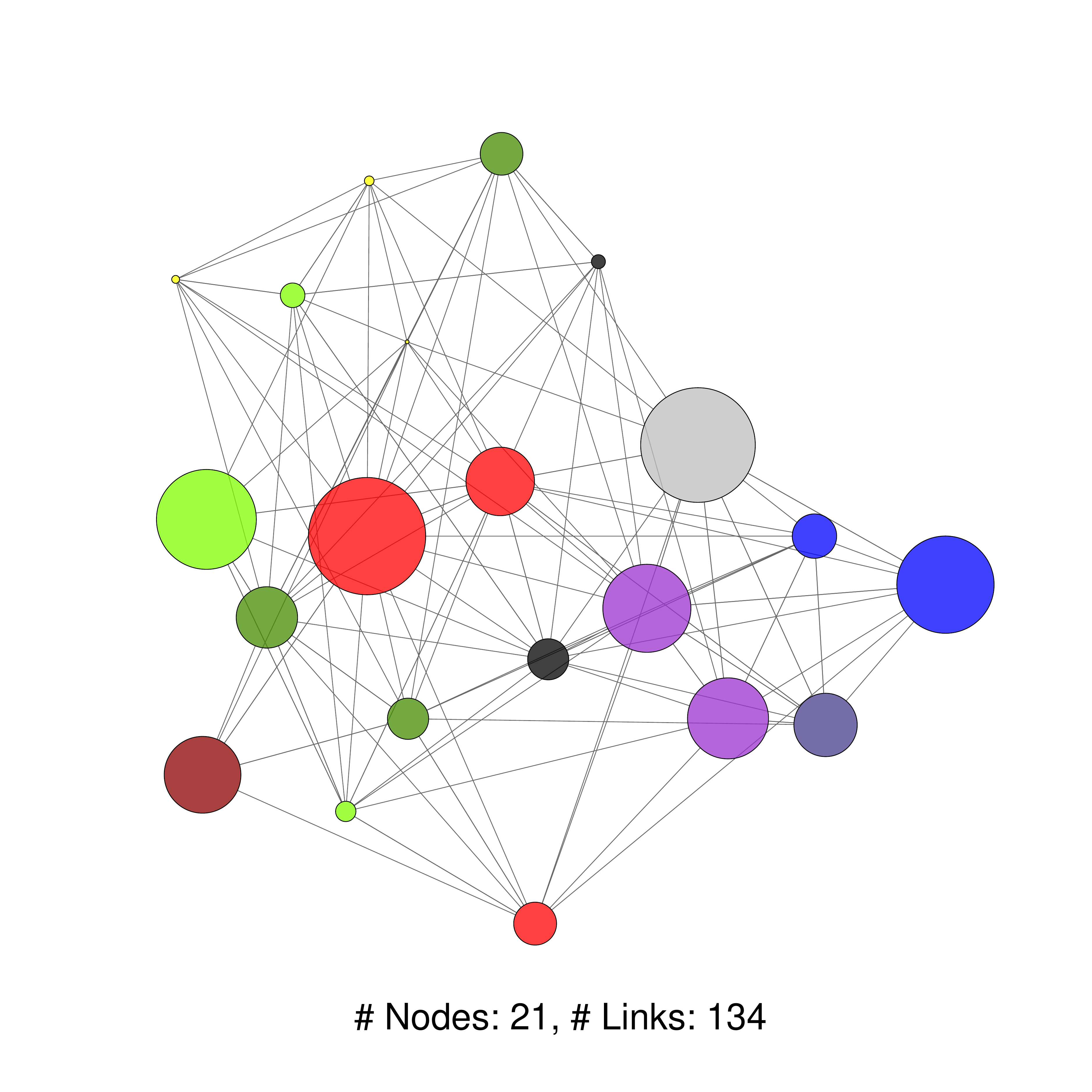}	
}
\end{subfigure}
\begin{subfigure}{0.3\textwidth}		
{\centering
\caption{Patent citations}			
\label{fig:threeDcomp_netw_flow_up_noOV_pat12}			
\includegraphics[width=\textwidth]{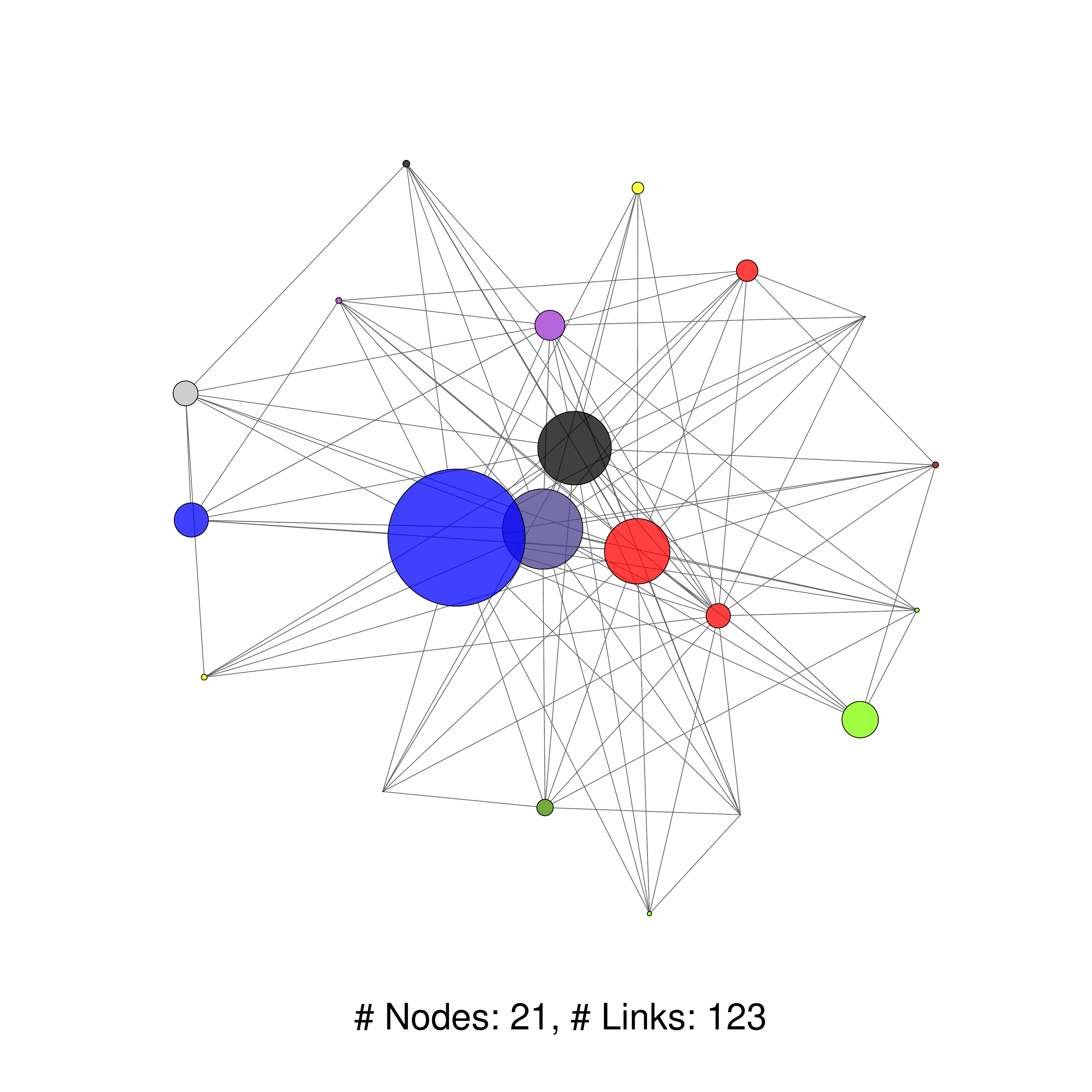}	
}
\end{subfigure}

\includegraphics[width=1\textwidth]{legend_for_nets.pdf}
\footnotesize 
\justifying \noindent
Notes: These figures show the network of upstream links (suppliers) at the 4-digit level for two different periods. A link between a pair of industries $i$ and $j$ is shown if $j$ is a sufficiently important supplier to $i$, i.e. if the average of the weight $w^{in,\alpha}_{ij,t}$ during periods 1977-1992 and 1997-2012 exceeds a threshold level given by the average weight over all industry pairs and all periods plus one standard deviation ($ \text{mean}_{i,j,t} (w^{in, \alpha}_{ij,t}) + \text{sd}_{i,j,t}(w^{in, \alpha}_{ij,t})$). 
The size of the nodes is proportional to the size of an industry $A^{\alpha}_{i,t}$ in the respective layer. 
The figure is generated using the plotting functions of the R-package \emph{igraph}, which makes use of the Fruchtermann-Reingold algorithm to allocate the nodes and edges of the network. 
Self-citations and within-sector IO flows are not shown. The colors indicate broad industrial categories given by groups of 3-digit level industries, i.e. Food (311-312), Textiles (314-316), Fiber (321-323), Petroleum (324), Chemicals (325-327), Metals (331,332), Machinery (333), Electronics (334-335), Transport (336), Other (337-339). 
\end{figure}

\begin{landscape}
\subsection{Industry rankings}
\label{supp:add_descriptives_industry_rankings}
Tables \ref{tab:sixDman_top10_pagerank_in} and \ref{tab:sixDman_top10_pagerank_out} show the Top-10 ranking of industries by up- and downstream centrality as measured by the PageRank.

	\begin{table}[h]
		\caption{Top-10 ranking of industries by the upstream PageRank at the 6-digit level.} 
		\label{tab:sixDman_top10_pagerank_in}
		\begin{myresizeenv}
			\begingroup \scriptsize
			\begin{tabular}{|p{0.15cm}|p{3cm}p{0.7cm}p{0.375cm}p{3cm}p{0.7cm}p{0.375cm}p{3cm}p{0.7cm}p{0.375cm}p{3cm}p{0.7cm}p{0.375cm}|}
				\hline
				\hline \rule{0pt}{1.1\normalbaselineskip}
				& \multicolumn{ 12 }{l|}{\emph{Top-10 industries by Pagerank ($PR^{\mu,up}_{i,t}$) }} \\
				& \multicolumn{3}{c}{ 1977-1982 }&\multicolumn{3}{c}{ 1987-1992 } &\multicolumn{3}{c}{1997-2002} &\multicolumn{3}{c|}{2007-2012} \\
				\hline 
				1 & Petroleum Refineries & 324110 & 0.17 & Petroleum Refineries & 324110 & 0.08 & Petroleum Refineries & 324110 & 0.10 & Copper Refineries & 331411 & 0.05 \\ 
				2 & Copper Refineries & 331411 & 0.05 & Plastics Mat.$\text{ \&}$ Resin  & 325211 & 0.05 & Iron$\text{ \&}$ Steel Mills & 331111 & 0.07 & Iron$\text{ \&}$ Steel Mills & 331111 & 0.05 \\ 
				3 & Plastics Mat.$\text{ \&}$ Resin  & 325211 & 0.04 & Copper Refineries & 331411 & 0.05 & Semiconductor$\text{ \&}$ Device  & 334413 & 0.05 & Automobile Mnft. & 336111 & 0.03 \\ 
				4 & All Petrol.$\text{ \&}$ Coal Prod.  & 324199 & 0.04 & All Petrol.$\text{ \&}$ Coal Prod.  & 324199 & 0.04 & Sawmills & 321113 & 0.04 & Biological Prod.  & 325414 & 0.02 \\ 
				5 & Iron$\text{ \&}$ Steel Mills & 331111 & 0.04 & Chem. Preparations  & 325998 & 0.03 & Plastics Mat.$\text{ \&}$ Resin  & 325211 & 0.04 & Plastics Mat.$\text{ \&}$ Resin  & 325211 & 0.02 \\ 
				6 & Chem. Preparations  & 325998 & 0.02 & Iron$\text{ \&}$ Steel Mills & 331111 & 0.03 & Copper Refineries & 331411 & 0.03 & Ship Building$\text{ \&}$ Repair & 336611 & 0.02 \\ 
				7 & Paperboard Mills & 322130 & 0.02 & Inorganic Dye$\text{ \&}$ Pigm.  & 325131 & 0.03 & Gum$\text{ \&}$ Wood Chem. & 325191 & 0.02 & Aircraft Mnft. & 336411 & 0.02 \\ 
				8 & Organic Chem.  & 325199 & 0.02 & Organic Chem.  & 325199 & 0.03 & Organic Chem.  & 325199 & 0.02 & Dog$\text{ \&}$ Cat Food Mnft. & 311111 & 0.02 \\ 
				9 & Inorganic Dye$\text{ \&}$ Pigm.  & 325131 & 0.01 & Fats$\text{ \&}$ Oils Refin. & 311225 & 0.01 & Machine Shops & 332710 & 0.02 & Petroleum Refineries & 324110 & 0.01 \\ 
				10 & Metal Can Mnft. & 332431 & 0.01 & Nitrogen. Fertl. Mnft. & 325311 & 0.01 & Print Circuit Assembly  & 334418 & 0.02 & Semiconductor$\text{ \&}$ Device  & 334413 & 0.01 \\ 
				\hline \rule{0pt}{1.1\normalbaselineskip}  & \multicolumn{12}{l|}{Quartiles:}\\
				& \multicolumn{3}{c}{ 0.01, 0.01, 0.01 }& \multicolumn{3}{c}{ 0.01, 0.01, 0.01 } & \multicolumn{3}{c}{0.01, 0.01, 0.02}& \multicolumn{3}{c|}{0.01, 0.01, 0.01}  \\
				\hline
				\hline \rule{0pt}{1.1\normalbaselineskip}
				& \multicolumn{ 12 }{l|}{\emph{Top-10 industries by Pagerank ($PR^{\tau,up}_{i,t}$) }} \\
				& \multicolumn{3}{c}{ 1977-1982 }&\multicolumn{3}{c}{ 1987-1992 } &\multicolumn{3}{c}{1997-2002} &\multicolumn{3}{c|}{2007-2012} \\
				\hline 
				1 & Adhesive Mnft. & 325520 & 0.05 & Adhesive Mnft. & 325520 & 0.05 & Semiconductor$\text{ \&}$ Device  & 334413 & 0.05 & Semiconductor$\text{ \&}$ Device  & 334413 & 0.06 \\ 
				2 & Chem. Preparations  & 325998 & 0.05 & Chem. Preparations  & 325998 & 0.05 & Adhesive Mnft. & 325520 & 0.05 & Adhesive Mnft. & 325520 & 0.04 \\ 
				3 & Semiconductor$\text{ \&}$ Device  & 334413 & 0.03 & Semiconductor$\text{ \&}$ Device  & 334413 & 0.04 & Chem. Preparations  & 325998 & 0.04 & Electr. Computer Mnft. & 334111 & 0.04 \\ 
				4 & Power Transm. Equ.  & 333613 & 0.03 & Power Transm. Equ.  & 333613 & 0.03 & Laboratory Apparatus  & 339111 & 0.03 & Chem. Preparations  & 325998 & 0.04 \\ 
				5 & Fastener $\text{ \&}$ Pin  & 339993 & 0.02 & Fastener $\text{ \&}$ Pin  & 339993 & 0.03 & Electr. Computer Mnft. & 334111 & 0.03 & Optical Instrum.$\text{ \&}$ Lens  & 333314 & 0.03 \\ 
				6 & Laboratory Apparatus  & 339111 & 0.02 & Electr. Computer Mnft. & 334111 & 0.03 & Fastener $\text{ \&}$ Pin  & 339993 & 0.03 & Fastener $\text{ \&}$ Pin  & 339993 & 0.03 \\ 
				7 & Speed Changer $\text{ \&}$ Gear  & 333612 & 0.02 & Laboratory Apparatus  & 339111 & 0.02 & Optical Instrum.$\text{ \&}$ Lens  & 333314 & 0.02 & Wireless Communic.  & 334220 & 0.02 \\ 
				8 & Electr. Computer Mnft. & 334111 & 0.02 & Speed Changer $\text{ \&}$ Gear  & 333612 & 0.02 & Power Transm. Equ.  & 333613 & 0.02 & Medical Instrum.  & 339112 & 0.02 \\ 
				9 & Boiler$\text{ \&}$ Heat Exch.  & 332410 & 0.02 & Optical Instrum.$\text{ \&}$ Lens  & 333314 & 0.02 & Speed Changer $\text{ \&}$ Gear  & 333612 & 0.02 & Power Transm. Equ.  & 333613 & 0.02 \\ 
				10 & Optical Instrum.$\text{ \&}$ Lens  & 333314 & 0.02 & Boiler$\text{ \&}$ Heat Exch.  & 332410 & 0.02 & Dental Equ.$\text{ \&}$ Supplies  & 339114 & 0.02 & Elctrmed. Apparatus  & 334510 & 0.02 \\ 
				\hline \rule{0pt}{1.1\normalbaselineskip}  & \multicolumn{12}{l|}{Quartiles:}\\
				& \multicolumn{3}{c}{ 0.01, 0.01, 0.0125 }& \multicolumn{3}{c}{ 0.01, 0.01, 0.01 } & \multicolumn{3}{c}{0.01, 0.01, 0.01}& \multicolumn{3}{c|}{0.01, 0.01, 0.02}  \\
				\hline
				\hline
			\end{tabular}
			\endgroup	
		\end{myresizeenv}
		\vspace{0.25cm}
		\scriptsize \justifying
		
		\noindent
		Notes: Industries are ranked by the PageRank compiled on upstream links $PR^{\alpha,up}_{i,t}$ averaged across the time window indicated in the column header in decreasing order, i.e. showing the largest industries on top. The 6-digit number in the second column of each block shows the NAICS code of the corresponding industry and the third column shows the value of $PR^{\alpha,up}_{i,t}$. 
		The values $PR^{\alpha,up}_{i,t}$ were normalized before through division by the economy-wide average in $t$, i.e. the mean value for each period equals one. 
		The bottom lines of each sub-table show the quartile values as indicators for the skewness of the distribution. 
		Deviations of the median from the average indicate skewness. 
	\end{table}

	\begin{table}[h]
		\caption{Top-10 ranking of industries by the downstream PageRank at the 6-digit level.} 
		\label{tab:sixDman_top10_pagerank_out}
		\begin{myresizeenv}
			\begingroup \scriptsize
			\begin{tabular}{|p{0.15cm}|p{3cm}p{0.7cm}p{0.375cm}p{3cm}p{0.7cm}p{0.375cm}p{3cm}p{0.7cm}p{0.375cm}p{3cm}p{0.7cm}p{0.375cm}|}
				\hline
				\hline \rule{0pt}{1.1\normalbaselineskip}
				& \multicolumn{ 12 }{l|}{\emph{Top-10 industries by Pagerank ($PR^{\mu,dw}_{i,t}$) }} \\
				& \multicolumn{3}{c}{ 1977-1982 }&\multicolumn{3}{c}{ 1987-1992 } &\multicolumn{3}{c}{1997-2002} &\multicolumn{3}{c|}{2007-2012} \\
				\hline  
				1 & Mobile Home Mnft.  & 321991 & 0.07 & Motor Home Mnft. & 336213 & 0.10 & Aircraft Mnft. & 336411 & 0.04 & Petroleum Refineries & 324110 & 0.07 \\ 
				2 & Motor Home Mnft. & 336213 & 0.04 & Ice $\text{ \&}$ Frozen Dessert  & 311520 & 0.04 & Mobile Home Mnft.  & 321991 & 0.04 & Copper Refineries & 331411 & 0.06 \\ 
				3 & Frozen Spec. Food Mnft. & 311412 & 0.03 & Mobile Home Mnft.  & 321991 & 0.04 & Automobile Mnft. & 336111 & 0.04 & Iron$\text{ \&}$ Steel Mills & 331111 & 0.04 \\ 
				4 & Motor Vhcl. Body Mnft. & 336211 & 0.03 & Missile$\text{ \&}$ Space Vhcl.  & 336414 & 0.03 & Motor Home Mnft. & 336213 & 0.03 & Die-Cut$\text{ \&}$ Paper Office  & 322231 & 0.03 \\ 
				5 & Missile$\text{ \&}$ Space Vhcl.  & 336414 & 0.03 & Aircraft Mnft. & 336411 & 0.03 & Ship Building$\text{ \&}$ Repair & 336611 & 0.03 & Plastics Mat.$\text{ \&}$ Resin  & 325211 & 0.02 \\ 
				6 & Ice $\text{ \&}$ Frozen Dessert  & 311520 & 0.03 & Motor Vhcl. Body Mnft. & 336211 & 0.02 & Retail Bakeries & 311811 & 0.03 & Soybean Processing & 311222 & 0.02 \\ 
				7 & Ship Building$\text{ \&}$ Repair & 336611 & 0.02 & Frozen Spec. Food Mnft. & 311412 & 0.02 & Commercial Bakeries & 311812 & 0.03 & All Misc. Electr. Equ.  & 335999 & 0.01 \\ 
				8 & Automobile Mnft. & 336111 & 0.02 & Ship Building$\text{ \&}$ Repair & 336611 & 0.02 & Travel Trailer$\text{ \&}$ Camper  & 336214 & 0.02 & Graphite Prod.  & 335991 & 0.01 \\ 
				9 & Light \& Utility Truck  & 336112 & 0.02 & Travel Trailer$\text{ \&}$ Camper  & 336214 & 0.02 & Oth. Animal Food Mnft. & 311119 & 0.02 & Motor Vhcl. Air-Cond.  & 336391 & 0.01 \\ 
				10 & Heavy Duty Truck Mnft. & 336120 & 0.02 & Electr. Computer Mnft. & 334111 & 0.02 & Dog$\text{ \&}$ Cat Food Mnft. & 311111 & 0.02 & Gum$\text{ \&}$ Wood Chem. & 325191 & 0.01 \\ 
				\hline \rule{0pt}{1.1\normalbaselineskip}  & \multicolumn{12}{l|}{Quartiles:}\\
				& \multicolumn{3}{c}{ 0.01, 0.01, 0.02 }& \multicolumn{3}{c}{ 0.01, 0.01, 0.02 } & \multicolumn{3}{c}{0.01, 0.01, 0.01}& \multicolumn{3}{c|}{0.01, 0.01, 0.01}  \\
				\hline
				\hline \rule{0pt}{1.1\normalbaselineskip}
				& \multicolumn{ 12 }{l|}{\emph{Top-10 industries by Pagerank ($PR^{\tau,dw}_{i,t}$) }} \\
				& \multicolumn{3}{c}{ 1977-1982 }&\multicolumn{3}{c}{ 1987-1992 } &\multicolumn{3}{c}{1997-2002} &\multicolumn{3}{c|}{2007-2012} \\
				\hline 
				1 & Adhesive Mnft. & 325520 & 0.05 & Adhesive Mnft. & 325520 & 0.05 & Semiconductor$\text{ \&}$ Device  & 334413 & 0.05 & Semiconductor$\text{ \&}$ Device  & 334413 & 0.07 \\ 
				2 & Chem. Preparations  & 325998 & 0.05 & Chem. Preparations  & 325998 & 0.05 & Adhesive Mnft. & 325520 & 0.05 & Electr. Computer Mnft. & 334111 & 0.04 \\ 
				3 & Semiconductor$\text{ \&}$ Device  & 334413 & 0.04 & Semiconductor$\text{ \&}$ Device  & 334413 & 0.04 & Chem. Preparations  & 325998 & 0.05 & Adhesive Mnft. & 325520 & 0.04 \\ 
				4 & Power Transm. Equ.  & 333613 & 0.03 & Power Transm. Equ.  & 333613 & 0.03 & Electr. Computer Mnft. & 334111 & 0.03 & Chem. Preparations  & 325998 & 0.04 \\ 
				5 & Fastener $\text{ \&}$ Pin  & 339993 & 0.03 & Fastener $\text{ \&}$ Pin  & 339993 & 0.03 & Fastener $\text{ \&}$ Pin  & 339993 & 0.03 & Optical Instrum.$\text{ \&}$ Lens  & 333314 & 0.03 \\ 
				6 & Electr. Computer Mnft. & 334111 & 0.02 & Electr. Computer Mnft. & 334111 & 0.03 & Optical Instrum.$\text{ \&}$ Lens  & 333314 & 0.02 & Fastener $\text{ \&}$ Pin  & 339993 & 0.03 \\ 
				7 & Speed Changer $\text{ \&}$ Gear  & 333612 & 0.02 & Speed Changer $\text{ \&}$ Gear  & 333612 & 0.02 & Power Transm. Equ.  & 333613 & 0.02 & Power Transm. Equ.  & 333613 & 0.02 \\ 
				8 & Urethane$\text{ \&}$ Foam Prod.  & 326150 & 0.02 & Optical Instrum.$\text{ \&}$ Lens  & 333314 & 0.02 & Speed Changer $\text{ \&}$ Gear  & 333612 & 0.02 & Medical Instrum.  & 339112 & 0.02 \\ 
				9 & Boiler$\text{ \&}$ Heat Exch.  & 332410 & 0.02 & Boiler$\text{ \&}$ Heat Exch.  & 332410 & 0.02 & Urethane$\text{ \&}$ Foam Prod.  & 326150 & 0.01 & Misc. Food Mnft. & 311999 & 0.02 \\ 
				10 & Optical Instrum.$\text{ \&}$ Lens  & 333314 & 0.02 & Urethane$\text{ \&}$ Foam Prod.  & 326150 & 0.02 & Dental Equ.$\text{ \&}$ Supplies  & 339114 & 0.01 & Watch \& Clock Mnft. & 334518 & 0.02 \\ 
				\hline \rule{0pt}{1.1\normalbaselineskip}  & \multicolumn{12}{l|}{Quartiles:}\\
				& \multicolumn{3}{c}{ 0.01, 0.01, 0.01 }& \multicolumn{3}{c}{ 0.01, 0.01, 0.01 } & \multicolumn{3}{c}{0.01, 0.01, 0.01}& \multicolumn{3}{c|}{0.01, 0.01, 0.0175}  \\
				\hline
				\hline
			\end{tabular}
			\endgroup	
		\end{myresizeenv}
		
		\vspace{0.25cm}
		\scriptsize \justifying
		
		\noindent
		Notes: Industries are ranked by the PageRank compiled on downstream links $PR^{\alpha,dw}_{i,t}$ averaged across the time window indicated in the column header in decreasing order, i.e. showing the largest industries on top. The 6-digit number in the second column of each block shows the NAICS code of the corresponding industry and the third column shows the value of $PR^{\alpha,dw}_{i,t}$. 
		The values $PR^{\alpha,dw}_{i,t}$ were normalized before through division by the economy-wide average in $t$, i.e. the mean value for each period equals one. 
		The bottom lines of each sub-table show the quartile values as indicators for the skewness of the distribution. 
		Deviations of the median from the average indicate skewness. 
	\end{table}

\end{landscape}

\FloatBarrier
\section{Disentangled technology-push and demand-pull and the direction of change}
\label{supp:regression_direction_separate_layers}
This section provides additional results for the effects of TP and DP on productivity (TFP and labor productivity), labor demand (employment and wages), capital use (capital intensity and investment per capita), and production labor (share of production workers and relative wage paid for production labor). 

The tables below show the effects from a weighted FE regression, analogous to the extract of the results shown in the main text \ref{subsubsec:direction_of_change}. Additional results relying on dynamic panels methods were used in a comprehensive series of robustness checks. However, as explained above, these models severely suffer from weak instruments. The results should be understood as analysis of conditional correlations without any claims for causality.

\begin{table}[h]
\begin{myresizeenv}
\begingroup
\begin{tabular}{|l|cccccc|cccccc|}
	\hline \hline  \rule{0pt}{1.075\normalbaselineskip}   & \multicolumn{6}{c|}{ \ul{Total factor productivity}} & \multicolumn{6}{c|}{ \ul{Labor productivity}} \\
	\rule{0pt}{1.075\normalbaselineskip}   & \multicolumn{2}{c}{ $\tau \rightarrow TFP_{i,t}$} & 
	\multicolumn{2}{c}{ $\mu \rightarrow TFP_{i,t}$} &
	\multicolumn{2}{c|}{ Both} & 
	\multicolumn{2}{c}{ $\tau \rightarrow (VA/L)_{i,t}$} & 
	\multicolumn{2}{c}{ $\mu \rightarrow (VA/L)_{i,t}$} &
	\multicolumn{2}{c|}{ Both} \\
	
	\hline  \rule{0pt}{1.075\normalbaselineskip}  & $TFP_{i,t}$ & $TFP_{i,t}$ & $TFP_{i,t}$ & $TFP_{i,t}$ & $TFP_{i,t}$ & $TFP_{i,t}$ & $(VA/L)_{i,t}$ & $(VA/L)_{i,t}$ & $(VA/L)_{i,t}$ & $(VA/L)_{i,t}$ & $(VA/L)_{i,t}$ & $(VA/L)_{i,t}$ \\ 
	\rule{0pt}{1.075\normalbaselineskip}  & (1) & (2) & (3) & (4) & (5) & (6) & (7) & (8) & (9) & (10) & (11) & (12) \\ 
	\hline \rule{0pt}{1.075\normalbaselineskip}   $TFP^{}_{i,t-1}$ & 0.9532*** & 0.8322*** & 0.9584*** & 0.8454*** & 0.9444*** & 0.8321*** &  &  &  &  &  &  \\ 
	& (0.016) & (0.0245) & (0.0155) & (0.0244) & (0.0162) & (0.0246) &  &  &  &  &  &  \\ 
	$(VA/L)^{}_{i,t-1}$ &  &  &  &  &  &  & 0.8092*** & 0.6916*** & 0.8066*** & 0.7052*** & 0.8153*** & 0.6924*** \\ 
	&  &  &  &  &  &  & (0.0134) & (0.0421) & (0.0134) & (0.0423) & (0.0143) & (0.0426) \\ 
	$A^{\mu}_{i,t-1}$ & 0.0476** & 0.0278. &  &  & 0.0459** & 0.0275. & -0.1486* & -0.1468* &  &  & -0.1465* & -0.1469* \\ 
	& (0.0166) & (0.016) &  &  & (0.0166) & (0.0159) & (0.0624) & (0.0629) &  &  & (0.0624) & (0.0629) \\ 
	$A^{\tau}_{i,t-1}$ &  &  & 0.098* & 0.0058 & 0.0966* & 0.0096 &  &  & -0.1197 & -0.0679 & -0.1031 & -0.0482 \\ 
	&  &  & (0.0452) & (0.0447) & (0.0452) & (0.0446) &  &  & (0.1588) & (0.1677) & (0.1587) & (0.1675) \\ 
	$PR^{\mu,up}_{i,t-1}$ & 0.0036 & 0.0082. &  &  & 0.0034 & 0.0082. & 0.0414* & 0.0463* &  &  & 0.042* & 0.047* \\ 
	& (0.0051) & (0.0048) &  &  & (0.0051) & (0.0048) & (0.0196) & (0.0196) &  &  & (0.0196) & (0.0196) \\ 
	$PR^{\mu,dw}_{i,t-1}$ & 0.006 & 0.0122** &  &  & 0.0063 & 0.0123** & 0.044** & 0.0557*** &  &  & 0.046** & 0.057*** \\ 
	& (0.0043) & (0.0041) &  &  & (0.0043) & (0.0041) & (0.0161) & (0.0162) &  &  & (0.0162) & (0.0163) \\ 
	$PR^{\tau,dw}_{i,t-1}$ &  &  & 0.0021 & -0.0135 & -0.0041 & -0.0166 &  &  & -0.0782 & -0.0216 & -0.1013 & -0.0463 \\ 
	&  &  & (0.0195) & (0.0192) & (0.0196) & (0.0192) &  &  & (0.073) & (0.0742) & (0.0734) & (0.0745) \\ 
	$Spill(A)^{\mu,up}_{i,t-1}$ & -0.0095 & -0.0083 &  &  & -0.0111 & -0.0117 & -0.0622 & -0.0766. &  &  & -0.0742. & -0.0815* \\ 
	& (0.0103) & (0.0097) &  &  & (0.0103) & (0.0097) & (0.0408) & (0.0408) &  &  & (0.0412) & (0.0411) \\ 
	$Spill(A)^{\mu,dw}_{i,t-1}$ & -0.0167* & -0.0137* &  &  & -0.016* & -0.0135* & -0.0182 & -0.02 &  &  & -0.0161 & -0.0176 \\ 
	& (0.0065) & (0.0062) &  &  & (0.0065) & (0.0062) & (0.0251) & (0.0251) &  &  & (0.0252) & (0.0252) \\ 
	$Spill(A)^{\tau,up}_{i,t-1}$ &  &  & -0.7579*** & -0.7383*** & -0.7711*** & -0.7479*** &  &  & -1.399. & -1.154 & -1.471. & -1.207 \\ 
	&  &  & (0.2119) & (0.2002) & (0.2122) & (0.2002) &  &  & (0.7844) & (0.7871) & (0.7862) & (0.7875) \\ 
	$Spill(A)^{\tau,dw}_{i,t-1}$ &  &  & -0.0418* & -0.0235 & -0.0424* & -0.0225 &  &  & 0.0353 & 0.0355 & 0.0341 & 0.0382 \\ 
	&  &  & (0.0176) & (0.0167) & (0.0176) & (0.0167) &  &  & (0.0658) & (0.066) & (0.0659) & (0.066) \\ 
	$W^{}_{i,t-1}$ &  & 0.059*** &  & 0.0565*** &  & 0.0579*** &  & 0.1687*** &  & 0.1349** &  & 0.1659*** \\ 
	&  & (0.009) &  & (0.009) &  & (0.0091) &  & (0.0477) &  & (0.0477) &  & (0.048) \\ 
	$(K/L)^{}_{i,t-1}$ &  & 0.0349*** &  & 0.0398*** &  & 0.0363*** &  & -0.056. &  & -0.0523. &  & -0.0544. \\ 
	&  & (0.0074) &  & (0.0077) &  & (0.0077) &  & (0.0288) &  & (0.0297) &  & (0.0297) \\ 
	$(L^P/L)_{i,t-1}$ &  & 0.1801* &  & 0.141. &  & 0.1587* &  & 1.17*** &  & 1.109*** &  & 1.114*** \\ 
	&  & (0.0731) &  & (0.0735) &  & (0.0735) &  & (0.3005) &  & (0.3041) &  & (0.304) \\ 
	$(I/L)^{}_{i,t-1}$ &  & -0.0474*** &  & -0.0473*** &  & -0.0473*** &  & 0.0425* &  & 0.0439* &  & 0.0411. \\ 
	&  & (0.005) &  & (0.005) &  & (0.005) &  & (0.0211) &  & (0.0212) &  & (0.0212) \\ 
	$(E/L)^{}_{i,t-1}$ &  & -0.0479*** &  & -0.0454*** &  & -0.048*** &  & -0.0848** &  & -0.0739** &  & -0.0856** \\ 
	&  & (0.0069) &  & (0.0069) &  & (0.007) &  & (0.027) &  & (0.0273) &  & (0.0274) \\ 
	$(M/L)^{}_{i,t-1}$ &  & -0.0019 &  & -0.0021 &  & -0.0012 &  & 0.0069 &  & 0.0115 &  & 0.0118 \\ 
	&  & (0.0072) &  & (0.0073) &  & (0.0073) &  & (0.0287) &  & (0.0294) &  & (0.0294) \\ 
	$(W^P/W)_{i,t-1}$ &  & 0.0369 &  & 0.025 &  & 0.0349 &  & -0.2304 &  & -0.2732 &  & -0.2356 \\ 
	&  & (0.0446) &  & (0.0447) &  & (0.0446) &  & (0.1875) &  & (0.1886) &  & (0.1882) \\ 
	\hline \hline \rule{0pt}{1.075\normalbaselineskip}   Controls &  & Y &  & Y &  & Y &  & Y &  & Y &  & Y \\ 
	$R^2$ & 0.7408 & 0.7726 & 0.7413 & 0.7719 & 0.7438 & 0.7746 & 0.9003 & 0.9022 & 0.8998 & 0.9013 & 0.9007 & 0.9024 \\ 
	\hline
	\hline
\end{tabular}
\endgroup\caption{Productivity effects.}
\label{tab:TFP|real_vadd_all_sectors_all_years_FE_weighted_wControls}
\end{myresizeenv}

\vspace{0.25cm}

\justifying \footnotesize

\noindent
Notes: The table shows the regression results of total factor productivity $TFP$ and labor productivity $(VA/L)_{i,t}$ on demand-pull and technology-push effects. The estimation is based on a two-ways weighted fixed-effects (FE) model. The weights used in the regressions are $A^{\mu}_{i,t}$ in the TFP regression and $L_{i,t}$ in all other regressions.  Spillovers are calculated on the basis of first-order links. Variables measured in monetary terms are deflated using the industry level price deflators for the value of shipment obtained from the NBER-productivity database \citep{becker2013nber}. To cope with skewness and to obtain tractable coefficients, most variables are pre-processed (taking logs, removing outliers, scaling). Data in logs are $TFP$, $(VA/L)_{i,t}$$A^{\alpha }_{i,t}$, $PR^{\alpha, d }_{i,t}$, $Spill(A)^{\alpha, d }_{i,t}$, $L_{i,t}$,  $(K/L)_{i,t}$, $(I/L)_{i,t}$, $W_{i,t}$, $(K/L)^{}_{i,t-1}$, $(E/L)^{}_{i,t-1}$, $(M/L)^{}_{i,t-1}$, $W^{P}_{i,t-1}$ with $\alpha = \mu, \tau$ and $d = up, dw$. $A^{\alpha }_{i,t}$, $PR^{\alpha, d }_{i,t}$, and $Spill(A)^{\alpha, d }_{i,t}$ are scaled by division by their standard deviation to obtain comparable coefficients across the different network effects. A detailed description of the transformations and descriptive statistics of the regression data before and after the transformations are provided in \ref{app:data}. 

\end{table}

\begin{table}[h]
\begin{myresizeenv}
\begingroup
\begin{tabular}{|l|cccccc|cccccc|}
	\hline \hline  \rule{0pt}{1.075\normalbaselineskip}   & \multicolumn{6}{c|}{ \ul{Employment}} & \multicolumn{6}{c|}{ \ul{Wage}} \\
	\rule{0pt}{1.075\normalbaselineskip}   & \multicolumn{2}{c}{ $\tau \rightarrow L_{i,t}$} & 
	\multicolumn{2}{c}{ $\mu \rightarrow L_{i,t}$} &
	\multicolumn{2}{c|}{ Both} & 
	\multicolumn{2}{c}{ $\tau \rightarrow W_{i,t}$} & 
	\multicolumn{2}{c}{ $\mu \rightarrow W_{i,t}$} &
	\multicolumn{2}{c|}{ Both} \\
	
	\hline  \rule{0pt}{1.075\normalbaselineskip}  & $L_{i,t}$ & $L_{i,t}$ & $L_{i,t}$ & $L_{i,t}$ & $L_{i,t}$ & $L_{i,t}$ & $W_{i,t}$ & $W_{i,t}$ & $W_{i,t}$ & $W_{i,t}$ & $W_{i,t}$ & $W_{i,t}$ \\ 
	\rule{0pt}{1.075\normalbaselineskip}  & (1) & (2) & (3) & (4) & (5) & (6) & (7) & (8) & (9) & (10) & (11) & (12) \\ 
	\hline \rule{0pt}{1.075\normalbaselineskip}   $L^{}_{i,t-1}$ & 0.8379*** & 0.9597*** & 0.8334*** & 0.968*** & 0.8415*** & 0.96*** &  &  &  &  &  &  \\ 
	& (0.0145) & (0.027) & (0.0145) & (0.0273) & (0.0145) & (0.0275) &  &  &  &  &  &  \\ 
	$W^{}_{i,t-1}$ &  & -0.0352* &  & -0.0498** &  & -0.0373* & 0.8381*** & 0.8487*** & 0.8171*** & 0.8172*** & 0.8332*** & 0.8407*** \\ 
	&  & (0.0171) &  & (0.0165) &  & (0.0174) & (0.0122) & (0.0199) & (0.0121) & (0.0194) & (0.0129) & (0.0201) \\ 
	$A^{\mu}_{i,t-1}$ & -0.2697*** & -0.09* &  &  & -0.2349*** & -0.0895* & -0.2236*** & -0.2105*** &  &  & -0.2227*** & -0.2082*** \\ 
	& (0.0427) & (0.0438) &  &  & (0.0426) & (0.0438) & (0.0526) & (0.0527) &  &  & (0.0526) & (0.0527) \\ 
	$A^{\tau}_{i,t-1}$ &  &  & -0.4525*** & -0.0185 & -0.4202*** & -0.019 &  &  & -0.0857 & -0.0645 & -0.0747 & -0.0632 \\ 
	&  &  & (0.1158) & (0.117) & (0.1152) & (0.1171) &  &  & (0.1344) & (0.1409) & (0.1337) & (0.1402) \\ 
	$PR^{\mu,up}_{i,t-1}$ & 0.025. & 0.0218 &  &  & 0.0258. & 0.0223. & 0.0579*** & 0.059*** &  &  & 0.0585*** & 0.0596*** \\ 
	& (0.0143) & (0.0135) &  &  & (0.0142) & (0.0136) & (0.0166) & (0.0164) &  &  & (0.0166) & (0.0164) \\ 
	$PR^{\mu,dw}_{i,t-1}$ & 0.0229. & 0.0261* &  &  & 0.024* & 0.0265* & 0.0363** & 0.0399** &  &  & 0.0371** & 0.0398** \\ 
	& (0.0118) & (0.0112) &  &  & (0.0117) & (0.0112) & (0.0136) & (0.0136) &  &  & (0.0136) & (0.0136) \\ 
	$PR^{\tau,dw}_{i,t-1}$ &  &  & -0.1631** & 0.0297 & -0.155** & 0.0136 &  &  & 0.0563 & 0.1111. & 0.0284 & 0.0841 \\ 
	&  &  & (0.0518) & (0.0511) & (0.052) & (0.0514) &  &  & (0.0617) & (0.0622) & (0.0619) & (0.0623) \\ 
	$Spill(A)^{\mu,up}_{i,t-1}$ & -0.0557. & -0.0563* &  &  & -0.0743* & -0.0555. & -0.071* & -0.0705* &  &  & -0.0718* & -0.0672. \\ 
	& (0.0298) & (0.0281) &  &  & (0.0297) & (0.0284) & (0.0345) & (0.0342) &  &  & (0.0348) & (0.0344) \\ 
	$Spill(A)^{\mu,dw}_{i,t-1}$ & 0.0076 & -0.0092 &  &  & 0.0032 & -0.0079 & -0.0072 & -0.0107 &  &  & -0.004 & -0.0085 \\ 
	& (0.0183) & (0.0174) &  &  & (0.0181) & (0.0175) & (0.0212) & (0.0211) &  &  & (0.0212) & (0.0212) \\ 
	$Spill(A)^{\tau,up}_{i,t-1}$ &  &  & -0.653 & -0.434 & -0.754 & -0.4783 &  &  & -1.402* & -1.24. & -1.533* & -1.336* \\ 
	&  &  & (0.5688) & (0.5437) & (0.5666) & (0.5451) &  &  & (0.6643) & (0.6627) & (0.6627) & (0.661) \\ 
	$Spill(A)^{\tau,dw}_{i,t-1}$ &  &  & 0.0938. & 0.0302 & 0.0839. & 0.0322 &  &  & 0.045 & 0.0386 & 0.0474 & 0.0451 \\ 
	&  &  & (0.048) & (0.0459) & (0.0478) & (0.0459) &  &  & (0.0558) & (0.0556) & (0.0556) & (0.0554) \\ 
	$(K/L)^{}_{i,t-1}$ &  & -0.1079*** &  & -0.109*** &  & -0.111*** &  & -0.0502* &  & -0.0602* &  & -0.0604* \\ 
	&  & (0.0205) &  & (0.0209) &  & (0.0209) &  & (0.0242) &  & (0.025) &  & (0.0249) \\ 
	$(L^P/L)_{i,t-1}$ &  & 1.341*** &  & 1.345*** &  & 1.336*** &  & 1.126*** &  & 1.167*** &  & 1.127*** \\ 
	&  & (0.207) &  & (0.209) &  & (0.209) &  & (0.2509) &  & (0.2542) &  & (0.253) \\ 
	$(I/L)^{}_{i,t-1}$ &  & 0.0609*** &  & 0.0614*** &  & 0.0605*** &  & 0.0715*** &  & 0.0772*** &  & 0.0718*** \\ 
	&  & (0.0151) &  & (0.0151) &  & (0.0151) &  & (0.0175) &  & (0.0177) &  & (0.0176) \\ 
	$(E/L)^{}_{i,t-1}$ &  & -0.0483* &  & -0.0458* &  & -0.0476* &  & -0.0885*** &  & -0.0764*** &  & -0.0861*** \\ 
	&  & (0.0219) &  & (0.0219) &  & (0.022) &  & (0.0224) &  & (0.0228) &  & (0.0228) \\ 
	$(M/L)^{}_{i,t-1}$ &  & -0.0779*** &  & -0.0776*** &  & -0.0769*** &  & 0.0195 &  & 0.0236 &  & 0.0225 \\ 
	&  & (0.02) &  & (0.0206) &  & (0.0206) &  & (0.0241) &  & (0.0247) &  & (0.0247) \\ 
	$(W^P/W)_{i,t-1}$ &  & -0.4702*** &  & -0.4825*** &  & -0.4672*** &  & -0.3355* &  & -0.3488* &  & -0.3293* \\ 
	&  & (0.1292) &  & (0.1297) &  & (0.1296) &  & (0.1569) &  & (0.158) &  & (0.1571) \\ 
	\hline \hline \rule{0pt}{1.075\normalbaselineskip}   Controls &  & Y &  & Y &  & Y &  & Y &  & Y &  & Y \\ 
	$R^2$ & 0.9273 & 0.9359 & 0.9276 & 0.9355 & 0.929 & 0.9359 & 0.9248 & 0.9266 & 0.9237 & 0.9257 & 0.9251 & 0.927 \\ 
	\hline
	\hline
\end{tabular}
\endgroup\caption{Labor market outcomes.}
\label{tab:emp|real_pay_all_sectors_all_years_FE_weighted_wControls}
\end{myresizeenv}

\vspace{0.25cm}

\justifying \footnotesize

\noindent
Notes:  The table shows the regression results of employment $L_{i,t}$ and wages $W_{i,t}$ on demand-pull and technology-push effects. The estimation is based on a two-ways weighted fixed-effects (FE) model. The weights used in the regressions are $A^{\mu}_{i,t}$ in the TFP regression and $L_{i,t}$ in all other regressions.  The controls included in all regressions are wages $W_{i,t}$, capital intensity $(K/L)_{i,t}$, investment per capita $(I/L)_{i,t}$, relative wages for production labor $(W^P/W)_{i,t}$, energy intensity $(E/L)^{}_{i,t-1}$, and material inputs per capita $(M/L)^{}_{i,t-1}$.    Spillovers are calculated on the basis of first-order links. Variables measured in monetary terms are deflated using the industry level price deflators for the value of shipment obtained from the NBER-productivity database \citep{becker2013nber}. To cope with skewness and to obtain tractable coefficients, most variables are pre-processed (taking logs, removing outliers, scaling). Data in logs are patents and output $A^{\alpha }_{i,t}$, centrality $PR^{\alpha, d }_{i,t}$, spillovers $Spill(A)^{\alpha, d }_{i,t}$, employment $L_{i,t}$,  $(K/L)_{i,t}$, $(I/L)_{i,t}$, $W_{i,t}$, $(E/L)^{}_{i,t-1}$, $(M/L)^{}_{i,t-1}$ with $\alpha = \mu, \tau$ and $d = up, dw$. $A^{\alpha }_{i,t}$, $PR^{\alpha, d }_{i,t}$, and $Spill(A)^{\alpha, d }_{i,t}$ are scaled by division by their standard deviation to obtain comparable coefficients across the different network effects. A detailed description of the transformations and descriptive statistics of the regression data before and after the transformations are provided in \ref{app:data}. 
\end{table}

\begin{table}[h]
\begin{myresizeenv}
\begingroup
\begin{tabular}{|l|cccccc|cccccc|}
	\hline \hline  \rule{0pt}{1.075\normalbaselineskip}   & \multicolumn{6}{c|}{ \ul{Capital intensity}} & \multicolumn{6}{c|}{ \ul{Investment per capita}} \\
	\rule{0pt}{1.075\normalbaselineskip}   & \multicolumn{2}{c}{ $\tau \rightarrow (K/L)_{i,t}$} & 
	\multicolumn{2}{c}{ $\mu \rightarrow (K/L)_{i,t}$} &
	\multicolumn{2}{c|}{ Both} & 
	\multicolumn{2}{c}{ $\tau \rightarrow (I/L)_{i,t}$} & 
	\multicolumn{2}{c}{ $\mu \rightarrow (I/L)_{i,t}$} &
	\multicolumn{2}{c|}{ Both} \\
	
	\hline  \rule{0pt}{1.075\normalbaselineskip}  & $(K/L)_{i,t}$ & $(K/L)_{i,t}$ & $(K/L)_{i,t}$ & $(K/L)_{i,t}$ & $(K/L)_{i,t}$ & $(K/L)_{i,t}$ & $(I/L)_{i,t}$ & $(I/L)_{i,t}$ & $(I/L)_{i,t}$ & $(I/L)_{i,t}$ & $(I/L)_{i,t}$ & $(I/L)_{i,t}$ \\ 
	\rule{0pt}{1.075\normalbaselineskip}  & (1) & (2) & (3) & (4) & (5) & (6) & (7) & (8) & (9) & (10) & (11) & (12) \\ 
	\hline \rule{0pt}{1.075\normalbaselineskip}   $(K/L)^{}_{i,t-1}$ & 0.69*** & 0.0212 & 0.664*** & 0.0234 & 0.6847*** & 0.0249 &  & -0.325*** &  & -0.3464*** &  & -0.3426*** \\ 
	& (0.0279) & (0.0161) & (0.0277) & (0.0161) & (0.0277) & (0.0162) &  & (0.0693) &  & (0.0699) &  & (0.0698) \\ 
	$(I/L)^{}_{i,t-1}$ &  & 0.0789*** &  & 0.0762*** &  & 0.0765*** & 0.2362*** & 0.1602*** & 0.2912*** & 0.1819*** & 0.23*** & 0.1708*** \\ 
	&  & (0.0076) &  & (0.0076) &  & (0.0077) & (0.0294) & (0.0393) & (0.0278) & (0.0391) & (0.0298) & (0.0393) \\ 
	$A^{\mu}_{i,t-1}$ & 0.1776*** & -0.0122 &  &  & 0.1597*** & -0.0168 & 0.3471* & 0.1329 &  &  & 0.3646* & 0.1793 \\ 
	& (0.0348) & (0.034) &  &  & (0.0349) & (0.034) & (0.1622) & (0.1765) &  &  & (0.1617) & (0.1757) \\ 
	$A^{\tau}_{i,t-1}$ &  &  & 0.3533*** & 0.0565 & 0.3294*** & 0.0469 &  &  & -0.6145 & -0.3806 & -0.5414 & -0.2627 \\ 
	&  &  & (0.0875) & (0.0749) & (0.0869) & (0.0756) &  &  & (0.3941) & (0.3912) & (0.3925) & (0.3917) \\ 
	$PR^{\mu,up}_{i,t-1}$ & -0.0126 & 0.0021 &  &  & -0.0157 & 4e-04 & 0.1389. & 0.1222. &  &  & 0.1444. & 0.125. \\ 
	& (0.0166) & (0.0139) &  &  & (0.0165) & (0.0139) & (0.0743) & (0.0725) &  &  & (0.074) & (0.0721) \\ 
	$PR^{\mu,dw}_{i,t-1}$ & 0.0091 & -0.0157 &  &  & 0.0123 & -0.0151 & 0.164** & 0.1393** &  &  & 0.1616** & 0.142** \\ 
	& (0.0123) & (0.0104) &  &  & (0.0122) & (0.0105) & (0.0548) & (0.0538) &  &  & (0.0549) & (0.0539) \\ 
	$PR^{\tau,dw}_{i,t-1}$ &  &  & -0.0216 & -0.1096* & -0.0487 & -0.108* &  &  & 0.8232*** & 0.673** & 0.7479*** & 0.6452** \\ 
	&  &  & (0.0504) & (0.0427) & (0.0499) & (0.0429) &  &  & (0.2259) & (0.2231) & (0.2236) & (0.2227) \\ 
	$Spill(A)^{\mu,up}_{i,t-1}$ & 0.112. & 0.0492 &  &  & 0.1133. & 0.0512 & 0.351 & 0.2512 &  &  & 0.3041 & 0.2015 \\ 
	& (0.0593) & (0.0499) &  &  & (0.059) & (0.05) & (0.2662) & (0.2599) &  &  & (0.2654) & (0.2591) \\ 
	$Spill(A)^{\mu,dw}_{i,t-1}$ & 0.0292 & -0.0184 &  &  & 0.0284 & -0.0162 & 0.1676. & 0.1421 &  &  & 0.1453 & 0.1259 \\ 
	& (0.0199) & (0.0168) &  &  & (0.0198) & (0.0168) & (0.0897) & (0.0874) &  &  & (0.0895) & (0.0872) \\ 
	$Spill(A)^{\tau,up}_{i,t-1}$ &  &  & -0.4933 & -0.3114 & -0.2876 & -0.3668 &  &  & 4.67* & 5.254** & 5.391** & 5.734** \\ 
	&  &  & (0.4265) & (0.3548) & (0.4207) & (0.3562) &  &  & (1.91) & (1.855) & (1.885) & (1.847) \\ 
	$Spill(A)^{\tau,dw}_{i,t-1}$ &  &  & 0.0024 & 0.0225 & -0.0017 & 0.0232 &  &  & 0.222 & 0.1801 & 0.1863 & 0.14 \\ 
	&  &  & (0.0323) & (0.0269) & (0.0319) & (0.027) &  &  & (0.1444) & (0.1404) & (0.1429) & (0.14) \\ 
	$W^{}_{i,t-1}$ &  & -0.002 &  & -0.0052 &  & -0.0025 &  & 0.0979 &  & 0.1411* &  & 0.1022 \\ 
	&  & (0.0126) &  & (0.0116) &  & (0.0126) &  & (0.0651) &  & (0.0605) &  & (0.0647) \\ 
	$(L^P/L)_{i,t-1}$ &  & -0.3684* &  & -0.3788** &  & -0.3712* &  & 0.867 &  & 1.237 &  & 1.062 \\ 
	&  & (0.1437) &  & (0.1442) &  & (0.1449) &  & (0.7488) &  & (0.7541) &  & (0.751) \\ 
	$(E/L)^{}_{i,t-1}$ &  & 0.0217. &  & 0.0267* &  & 0.0258* &  & 0.1936** &  & 0.1725** &  & 0.1787** \\ 
	&  & (0.0127) &  & (0.0127) &  & (0.0128) &  & (0.0634) &  & (0.0642) &  & (0.0637) \\ 
	$(M/L)^{}_{i,t-1}$ &  & 0.0221 &  & 0.0205 &  & 0.023 &  & 0.1395. &  & 0.1526* &  & 0.1206 \\ 
	&  & (0.015) &  & (0.0149) &  & (0.0151) &  & (0.0773) &  & (0.0767) &  & (0.0773) \\ 
	$(W^P/W)_{i,t-1}$ &  & 0.3604*** &  & 0.3477*** &  & 0.3495*** &  & -1.233** &  & -1.095* &  & -1.123* \\ 
	&  & (0.0869) &  & (0.0869) &  & (0.0869) &  & (0.4468) &  & (0.4485) &  & (0.445) \\ 
	\hline \hline \rule{0pt}{1.075\normalbaselineskip}   Controls &  & Y &  & Y &  & Y &  & Y &  & Y &  & Y \\ 
	$R^2$ & 0.9393 & 0.9579 & 0.938 & 0.958 & 0.9406 & 0.9583 & 0.8849 & 0.892 & 0.8827 & 0.8916 & 0.8871 & 0.8939 \\ 
	\hline
	\hline
\end{tabular}
\endgroup\caption{Effects on capital use.} 
\label{tab:cap|real_invest_all_sectors_all_years_FE_weighted_wControls}
\end{myresizeenv}

\vspace{0.25cm}

\justifying \footnotesize

\noindent
Notes: The table shows the regression results of capital intensity $(K/L)_{i,t}$ and per-capital investment $(I/L)_{i,t}$ on demand-pull and technology-push effects. The estimation is based on a two-ways weighted fixed-effects (FE) model. The weights used in the regressions are $A^{\mu}_{i,t}$ in the TFP regression and $L_{i,t}$ in all other regressions.  The controls included in all regressions are wages $W_{i,t}$, capital intensity $(K/L)_{i,t}$, investment per capita $(I/L)_{i,t}$, relative wages for production labor $(W^P/W)_{i,t}$, energy intensity $(E/L)^{}_{i,t-1}$, and material inputs per capita $(M/L)^{}_{i,t-1}$.    Spillovers are calculated on the basis of first-order links. Variables measured in monetary terms are deflated using the industry level price deflators for the value of shipment obtained from the NBER-productivity database \citep{becker2013nber}. To cope with skewness and to obtain tractable coefficients, most variables are pre-processed (taking logs, removing outliers, scaling). Data in logs are patents and output $A^{\alpha }_{i,t}$, centrality $PR^{\alpha, d }_{i,t}$, spillovers $Spill(A)^{\alpha, d }_{i,t}$, employment $L_{i,t}$,  $(K/L)_{i,t}$, $(I/L)_{i,t}$, $W_{i,t}$, $(E/L)^{}_{i,t-1}$, $(M/L)^{}_{i,t-1}$ with $\alpha = \mu, \tau$ and $d = up, dw$. $A^{\alpha }_{i,t}$, $PR^{\alpha, d }_{i,t}$, and $Spill(A)^{\alpha, d }_{i,t}$ are scaled by division by their standard deviation to obtain comparable coefficients across the different network effects. A detailed description of the transformations and descriptive statistics of the regression data before and after the transformations are provided in \ref{app:data}. 
\end{table}

\begin{table}[h]
\begin{myresizeenv}
\begingroup
\begin{tabular}{|l|cccccc|cccccc|}
	\hline \hline  \rule{0pt}{1.075\normalbaselineskip}   & \multicolumn{6}{c|}{ \ul{Share of production labor}} & \multicolumn{6}{c|}{ \ul{Relative wage for production labor}} \\
	\rule{0pt}{1.075\normalbaselineskip}   & \multicolumn{2}{c}{ $\tau \rightarrow (L^P/L)_{i,t}$} & 
	\multicolumn{2}{c}{ $\mu \rightarrow (L^P/L)_{i,t}$} &
	\multicolumn{2}{c|}{ Both} & 
	\multicolumn{2}{c}{ $\tau \rightarrow (W^P/W)_{i,t}$} & 
	\multicolumn{2}{c}{ $\mu \rightarrow (W^P/W)_{i,t}$} &
	\multicolumn{2}{c|}{ Both} \\
	
	\hline  \rule{0pt}{1.075\normalbaselineskip}  & $(L^P/L)_{i,t}$ & $(L^P/L)_{i,t}$ & $(L^P/L)_{i,t}$ & $(L^P/L)_{i,t}$ & $(L^P/L)_{i,t}$ & $(L^P/L)_{i,t}$ & $(W^P/W)_{i,t}$ & $(W^P/W)_{i,t}$ & $(W^P/W)_{i,t}$ & $(W^P/W)_{i,t}$ & $(W^P/W)_{i,t}$ & $(W^P/W)_{i,t}$ \\ 
	\rule{0pt}{1.075\normalbaselineskip}  & (1) & (2) & (3) & (4) & (5) & (6) & (7) & (8) & (9) & (10) & (11) & (12) \\ 
	\hline \rule{0pt}{1.075\normalbaselineskip}   $(L^P/L)_{i,t-1}$ & 0.499*** & 0.4891*** & 0.4871*** & 0.4796*** & 0.4854*** & 0.4863*** &  & 0.234*** &  & 0.2174*** &  & 0.2212*** \\ 
	& (0.0293) & (0.0333) & (0.0303) & (0.0335) & (0.0301) & (0.0335) &  & (0.0567) &  & (0.057) &  & (0.0573) \\ 
	$(W^P/W)_{i,t-1}$ &  & 0.0557** &  & 0.0537** &  & 0.0539** & 0.4499*** & 0.3903*** & 0.4467*** & 0.3894*** & 0.4417*** & 0.3884*** \\ 
	&  & (0.0198) &  & (0.0199) &  & (0.0199) & (0.033) & (0.0338) & (0.033) & (0.0339) & (0.0331) & (0.034) \\ 
	$A^{\mu}_{i,t-1}$ & -0.0225** & -0.0145. &  &  & -0.0201** & -0.0143. & -0.0249* & -0.0129 &  &  & -0.0236* & -0.0139 \\ 
	& (0.0069) & (0.0078) &  &  & (0.0069) & (0.0078) & (0.0117) & (0.0134) &  &  & (0.0118) & (0.0134) \\ 
	$A^{\tau}_{i,t-1}$ &  &  & -0.0326. & -0.0235 & -0.0322. & -0.0273 &  &  & -0.0582* & -0.0163 & -0.0544. & -0.0175 \\ 
	&  &  & (0.0174) & (0.0174) & (0.0174) & (0.0175) &  &  & (0.029) & (0.0296) & (0.0293) & (0.0299) \\ 
	$PR^{\mu,up}_{i,t-1}$ & 0.0044 & 0.0039 &  &  & 0.0045 & 0.004 & 0.0062 & 0.0064 &  &  & 0.0071 & 0.0069 \\ 
	& (0.0033) & (0.0032) &  &  & (0.0033) & (0.0032) & (0.0056) & (0.0055) &  &  & (0.0056) & (0.0055) \\ 
	$PR^{\mu,dw}_{i,t-1}$ & -0.005* & -0.0031 &  &  & -0.0051* & -0.0033 & 8e-04 & -1e-04 &  &  & -3e-04 & -7e-04 \\ 
	& (0.0024) & (0.0024) &  &  & (0.0024) & (0.0024) & (0.0041) & (0.0041) &  &  & (0.0041) & (0.0041) \\ 
	$PR^{\tau,dw}_{i,t-1}$ &  &  & -0.0094 & -0.0083 & -0.0045 & -0.0073 &  &  & 0.0014 & 0.0019 & 0.0025 & 0.0018 \\ 
	&  &  & (0.0099) & (0.0099) & (0.0099) & (0.0099) &  &  & (0.0167) & (0.0169) & (0.0169) & (0.017) \\ 
	$Spill(A)^{\mu,up}_{i,t-1}$ & -0.0064 & -7e-04 &  &  & -0.0055 & -4e-04 & 0.0106 & 0.002 &  &  & 0.0106 & 0.0031 \\ 
	& (0.0117) & (0.0115) &  &  & (0.0117) & (0.0116) & (0.0199) & (0.0197) &  &  & (0.0199) & (0.0198) \\ 
	$Spill(A)^{\mu,dw}_{i,t-1}$ & -0.007. & -0.0044 &  &  & -0.0064 & -0.0042 & 0.0021 & 0.0023 &  &  & 0.002 & 0.0021 \\ 
	& (0.0039) & (0.0039) &  &  & (0.0039) & (0.0039) & (0.0067) & (0.0066) &  &  & (0.0067) & (0.0067) \\ 
	$Spill(A)^{\tau,up}_{i,t-1}$ &  &  & 0.1114 & 0.1007 & 0.0741 & 0.082 &  &  & -0.253. & -0.1876 & -0.2779. & -0.1985 \\ 
	&  &  & (0.084) & (0.0823) & (0.0836) & (0.0825) &  &  & (0.1416) & (0.1402) & (0.1423) & (0.1409) \\ 
	$Spill(A)^{\tau,dw}_{i,t-1}$ &  &  & -0.0011 & 0.0016 & -4e-04 & 0.0022 &  &  & -0.0028 & -0.0076 & -0.0042 & -0.0081 \\ 
	&  &  & (0.0063) & (0.0062) & (0.0063) & (0.0063) &  &  & (0.0107) & (0.0106) & (0.0108) & (0.0107) \\ 
	$W^{}_{i,t-1}$ &  & -0.0012 &  & -0.0032 &  & -0.0013 &  & -0.0047 &  & -0.0061 &  & -0.0048 \\ 
	&  & (0.0029) &  & (0.0027) &  & (0.0029) &  & (0.0049) &  & (0.0046) &  & (0.0049) \\ 
	$(K/L)^{}_{i,t-1}$ &  & 0.0061* &  & 0.0067* &  & 0.0072* &  & -0.0102. &  & -0.0107* &  & -0.01. \\ 
	&  & (0.0031) &  & (0.0031) &  & (0.0031) &  & (0.0052) &  & (0.0053) &  & (0.0053) \\ 
	$(I/L)^{}_{i,t-1}$ &  & -0.0071*** &  & -0.0077*** &  & -0.0072*** &  & -0.0013 &  & -0.0014 &  & -0.0013 \\ 
	&  & (0.0017) &  & (0.0017) &  & (0.0018) &  & (0.003) &  & (0.003) &  & (0.003) \\ 
	$(E/L)^{}_{i,t-1}$ &  & 0.0053. &  & 0.0058* &  & 0.0058* &  & 0.012* &  & 0.0113* &  & 0.0115* \\ 
	&  & (0.0028) &  & (0.0028) &  & (0.0028) &  & (0.0048) &  & (0.0049) &  & (0.0049) \\ 
	$(M/L)^{}_{i,t-1}$ &  & -0.0039 &  & -0.0033 &  & -0.0034 &  & -3e-04 &  & 0.0016 &  & 7e-04 \\ 
	&  & (0.0034) &  & (0.0034) &  & (0.0035) &  & (0.0059) &  & (0.0058) &  & (0.0059) \\ 
	\hline \hline \rule{0pt}{1.075\normalbaselineskip}   Controls &  & Y &  & Y &  & Y &  & Y &  & Y &  & Y \\ 
	$R^2$ & 0.9311 & 0.9338 & 0.93 & 0.9337 & 0.9317 & 0.9343 & 0.778 & 0.7878 & 0.7791 & 0.7879 & 0.7803 & 0.7885 \\ 
	\hline
	\hline
\end{tabular}
\endgroup
\caption{Effects on the share of production labor.}
\label{tab:share_prode|real_relative_prodw_all_sectors_all_years_FE_weighted_wControls}
\end{myresizeenv}
\vspace{0.25cm}

\justifying \footnotesize

\noindent	
Notes: The table shows the regression results of the share of production labor $(L^P/L)_{i,t}$ and relative wages for production labor $(W^P/W)_{i,t}$ on demand-pull and technology-push effects. The estimation is based on a two-ways weighted fixed-effects (FE) model. The weights used in the regressions are $A^{\mu}_{i,t}$ in the TFP regression and $L_{i,t}$ in all other regressions.  The controls included in all regressions are wages $W_{i,t}$, capital intensity $(K/L)_{i,t}$, investment per capita $(I/L)_{i,t}$, relative wages for production labor $(W^P/W)_{i,t}$, energy intensity $(E/L)^{}_{i,t-1}$, and material inputs per capita $(M/L)^{}_{i,t-1}$. Spillovers are calculated on the basis of first-order links. Variables measured in monetary terms are deflated using the industry level price deflators for the value of shipment obtained from the NBER-productivity database \citep{becker2013nber}. To cope with skewness and to obtain tractable coefficients, most variables are pre-processed (taking logs, removing outliers, scaling). Data in logs are patents and output $A^{\alpha }_{i,t}$, centrality $PR^{\alpha, d }_{i,t}$, spillovers $Spill(A)^{\alpha, d }_{i,t}$, employment $L_{i,t}$,  $(K/L)_{i,t}$, $(I/L)_{i,t}$, $W_{i,t}$, $(E/L)^{}_{i,t-1}$, $(M/L)^{}_{i,t-1}$ with $\alpha = \mu, \tau$ and $d = up, dw$. $A^{\alpha }_{i,t}$, $PR^{\alpha, d }_{i,t}$, and $Spill(A)^{\alpha, d }_{i,t}$ are scaled by division by their standard deviation to obtain comparable coefficients across the different network effects. A detailed description of the transformations and descriptive statistics of the regression data before and after the transformations are provided in \ref{app:data}. 
\end{table}

\FloatBarrier

\section{Sectoral patterns of innovation}
\label{supp:results_sectoral_patterns_of_inno}

\subsection{Map between 6-digit NAICS codes and industry groups}

\begin{center}

\begingroup \scriptsize
\begin{longtable}{| >{\centering\arraybackslash}p{0.75cm}|p{3.75cm}|>{\centering\arraybackslash}p{0.75cm}|>{\centering\arraybackslash}p{0.75cm}|>{\centering\arraybackslash}p{0.75cm}|>{\centering\arraybackslash}p{0.75cm}|>{\centering\arraybackslash}p{0.75cm}|>{\centering\arraybackslash}p{2cm}|}

\caption{Mapping of 6-digit NAICS industries into subsamples}
\label{tab:NAICStoSubsample_map} \\

\hline
\hline
NAICS & Description & Inno intensive & Big industry & Patent central & Market central (up) & Market central (down) & Pavitt sector \\ 
\hline

\endhead 

\hline
\multicolumn{8}{r}{\footnotesize\textit{continue on the next page}}
\endfoot
\hline
\endlastfoot

311111 & Dog \& Cat Food Mnfct & X &  &  &  & X &  \\ 
311119 & Other Animal Food &  & X &  &  & X &  \\ 
311211 & Flour Milling &  & X &  & X &  &  \\ 
311212 & Rice Milling &  &  &  &  &  &  \\ 
311213 & Malt Mnfct &  &  &  &  &  &  \\ 
311222 & Soybean Processing &  & X & X & X & X &  \\ 
311225 & Fats \& Oils Refining &  & X &  & X & X &  \\ 
311230 & Breakfast Cereal Mnfct &  &  &  &  & X &  \\ 
311311 & Sugarcane Mills &  &  &  & X &  &  \\ 
311313 & Beet Sugar Mnfct &  &  &  & X &  &  \\ 
311320 & Chocolate from Cacao Beans &  &  &  &  &  &  \\ 
311330 & Confect from Purch Chocolate & X &  & X &  & X &  \\ 
311340 & Nonchocolate Confectionery & X &  & X &  & X &  \\ 
311411 & Frozen Fruit, Juice \& Veg &  & X &  &  & X &  \\ 
311412 & Frozen Specialty Food & X &  &  &  & X &  \\ 
311421 & Fruit \& Vegetable Canning &  & X &  &  & X &  \\ 
311422 & Specialty Canning &  &  &  &  & X &  \\ 
311423 & Dried \& Dehydrated Food & X &  & X & X & X &  \\ 
311511 & Fluid Milk Mnfct & X & X & X & X & X &  \\ 
311512 & Creamery Butter Mnfct & X &  & X &  & X &  \\ 
311513 & Cheese Mnfct &  & X &  & X & X &  \\ 
311514 & Dry \& Condensed Dairy &  & X &  & X & X &  \\ 
311520 & Ice Cream \& Frozen Dessert &  & X &  &  & X &  \\ 
311611 & Animal Slaughter &  & X &  & X & X &  \\ 
311612 & Meat Proc from Carcasses & X & X & X & X & X &  \\ 
311613 & Meat Byproduct Processing &  & X &  & X & X &  \\ 
311615 & Poultry Processing & X & X & X & X & X &  \\ 
311711 & Seafood Canning &  &  &  &  &  &  \\ 
311811 & Retail Bakeries &  &  &  &  & X &  \\ 
311812 & Commercial Bakeries &  &  &  &  & X &  \\ 
311821 & Cookie \& Cracker & X &  & X &  &  &  \\ 
311822 & Flour Mixes from Purch Flour & X &  & X &  &  &  \\ 
311830 & Tortilla Mnfct & X &  &  &  &  &  \\ 
311911 & Roasted Nuts \& Peanut Butter &  &  &  &  &  &  \\ 
311919 & Other Snack Food Mnfct & X &  & X &  & X &  \\ 
311920 & Coffee \& Tea Mnfct &  & X &  &  & X &  \\ 
311930 & Flavor Syrup \& Concentrate &  & X &  & X &  &  \\ 
311941 & Mayo, Dressing \& Sauce &  &  &  &  & X &  \\ 
311942 & Spice \& Extract & X & X & X & X & X &  \\ 
311991 & Perishable Prepared Food &  &  &  &  &  &  \\ 
311999 & All Other Miscellaneous Food & X & X & X & X & X &  \\ 
312111 & Soft Drink Mnfct & X &  & X &  & X &  \\ 
312112 & Bottled Water Mnfct & X & X &  & X & X &  \\ 
312113 & Ice Mnfct & X &  & X &  & X & Pro \\ 
312120 & Breweries &  & X &  &  & X &  \\ 
312130 & Wineries &  & X & X & X & X &  \\ 
312140 & Distilleries &  & X &  &  & X & Sup \\ 
312210 & Tobacco Stem \& Redrying &  &  &  & X & X &  \\ 
313111 & Yarn Spinning Mills & X & X & X & X &  &  \\ 
313221 & Narrow Fabric Mills &  &  &  & X &  &  \\ 
313230 & Nonwoven Fabric Mills & X & X & X & X &  &  \\ 
313241 & Weft Knit Fabric Mills &  &  &  & X &  &  \\ 
313320 & Fabric Coating Mills & X &  & X & X &  & Pro \\ 
314110 & Carpet \& Rug Mills &  &  &  &  &  &  \\ 
314121 & Curtain \& Drapery Mills &  &  &  &  &  &  \\ 
314991 & Rope, Cordage, \& Twine Mills &  &  &  &  &  &  \\ 
315111 & Sheer Hosiery Mills &  &  &  &  & X &  \\ 
315221 & Men's Under- \& Nightwear & X &  &  &  &  &  \\ 
315231 & Women's Lingerie \& Nightwear &  &  &  &  &  &  \\ 
315292 & Fur \& Leather Apparel &  &  &  &  &  &  \\ 
315991 & Hat, Cap, \& Millinery & X &  & X &  &  &  \\ 
316110 & Leather \& Hide Tann \& Finish &  & X &  & X & X & Sup \\ 
316211 & Rubber \& Plastics Footwear & X &  & X &  & X & Sup \\ 
316991 & Luggage Mnfct & X &  & X &  &  & Sup \\ 
316992 & Women's Handbag \& Purse & X &  &  &  & X & Sup \\ 
321113 & Sawmills &  & X &  & X &  & Sup \\ 
321114 & Wood Preservation &  & X &  & X & X & Sup \\ 
321211 & Hardwood Veneer \& Plywood &  & X &  & X &  & Sup \\ 
321212 & Softwood Veneer \& Plywood &  & X &  & X &  & Sup \\ 
321213 & Engineered Wood Member &  &  &  &  &  & Sup \\ 
321214 & Truss Mnfct &  &  &  &  &  & Sup \\ 
321911 & Wood Window \& Door &  & X & X &  &  & Sup \\ 
321912 & Cut Stock \& Resaw Lumber &  & X & X & X &  & Sup \\ 
321918 & Millwork (including Flooring) &  & X &  & X &  & Sup \\ 
321920 & Wood Container \& Pallet &  & X &  & X &  & Sup \\ 
321991 & Manufact \& Mobile Home &  &  &  &  & X & Sup \\ 
321992 & Prefabricated Wood Building &  & X &  &  & X & Sup \\ 
322110 & Pulp Mills & X & X & X & X &  & Sup \\ 
322121 & Paper (exc Newsprint) Mills & X & X & X & X & X & Pro, Sup\\
322122 & Newsprint Mills &  & X &  & X & X & Sup \\ 
322130 & Paperboard Mills &  & X & X & X & X & Sup \\ 
322211 & Corrugated \& Solid Fiber Box &  & X &  & X & X & Sup \\ 
322212 & Folding Paperboard Box &  & X &  & X &  & Sup \\ 
322213 & Setup Paperboard Box &  & X &  & X &  & Sup \\ 
322221 & Coated Paper \& Plastics Film & X & X & X & X &  & Pro, Sci, Sup\\ 
322231 & Die-Cut \& Paper Office Suppl &  &  &  &  & X & Sup \\ 
322291 & Sanitary Paper Product & X & X & X &  & X & Sup \\ 
323110 & Commercial Lithograph Print &  & X &  & X & X & Sup \\ 
323113 & Commercial Screen Printing &  & X &  & X & X & Sup \\ 
323117 & Books Printing &  & X &  &  & X & Sup \\ 
324110 & Petroleum Refineries &  & X & X & X & X & Sup \\ 
324121 & Asphalt Paving Mixture \& Block & X & X & X &  & X & Pro \\ 
324122 & Asphalt Shle \& Coat Materials &  & X &  &  &  & Pro \\ 
324199 & Petroleum \& Coal Products &  & X &  & X & X & Sup \\ 
325131 & Inorganic Dye \& Pigment &  & X &  & X &  & Sup \\ 
325191 & Gum \& Wood Chemical &  &  &  & X & X & Sup \\ 
325199 & Basic Organic Chemical &  & X &  & X & X & Pro, Sup\\ 
325211 & Plastics Material \& Resin &  & X & X & X & X & Sup \\ 
325212 & Synthetic Rubber Mnfct &  & X &  & X &  & Sup \\ 
325311 & Nitrogenous Fertilizer &  & X & X & X &  & Sup \\ 
325320 & Pesticide \& Agric Chemical & X & X & X &  & X & Sup \\ 
325411 & Medicinal \& Botanical & X & X & X & X & X & Pro \\ 
325414 & Biological Products &  & X &  & X & X & Pro \\ 
325510 & Paint \& Coating & X & X & X & X & X & Sup \\ 
325520 & Adhesive Mnfct & X & X & X & X &  & Sup \\ 
325611 & Soap \& Other Detergent & X & X & X & X & X & Sup \\ 
325612 & Polish \& Other Sanitation & X &  & X & X & X & Sup \\ 
325613 & Surface Active Agent &  & X &  & X &  & Sup \\ 
325910 & Printing Ink Mnfct & X & X & X &  &  & Sup \\ 
325920 & Explosives Mnfct & X &  & X & X &  & Sup \\ 
325991 & Custom Compound of Resins &  & X &  & X &  & Sup \\ 
325992 & Photo Film, Paper \& Chem &  & X & X & X & X & Sup \\ 
325998 & Misc Chem Products & X & X & X & X & X & Pro, Sup\\
326111 & Plastics Bag Mnfct & X & X & X & X &  & Pro \\ 
326112 & Plastics Packag Film \& Sheet &  & X &  & X &  & Pro \\ 
326113 & Unlamin Plastics Film \& Sheet &  & X & X & X &  & Pro \\ 
326121 & Unlamin Plastics Profiles &  & X &  & X &  & Pro \\ 
326122 & Plastics Pipe \& Pipe Fitting & X & X & X & X &  & Pro \\ 
326130 & Lamin Plastics Plate \& Shape & X & X & X & X &  & Pro \\ 
326140 & Polystyrene Foam Product & X & X & X & X &  & Pro \\ 
326150 & Urethane \& Foam & X & X & X & X &  & Pro \\ 
326160 & Plastics Bottle Mnfct &  & X &  & X &  & Pro \\ 
326191 & Plastics Plumbing Fixture &  & X &  & X & X & Pro \\ 
326192 & Resilient Floor Covering &  & X &  & X &  & Pro \\ 
326211 & Tire (exc Retreading) & X & X & X &  &  & Pro \\ 
326212 & Tire Retreading &  & X &  &  & X & Pro \\ 
326220 & Rubber \& Plastics Hoses &  & X &  & X &  & Pro \\ 
326291 & Rubber for Mechanical Use & X &  & X & X &  & Pro \\ 
326299 & All Other Rubber Product & X &  & X & X &  & Pro \\ 
327111 & China Plumbing \& Bathroom &  &  &  &  &  & Pro \\ 
327121 & Brick \& Structural Clay Tile &  &  &  &  &  & Pro \\ 
327211 & Flat Glass Mnfct &  & X & X & X &  & Pro \\ 
327212 & Pressed \& Blown Glass & X & X & X & X &  & Pro \\ 
327213 & Glass Container Mnfct &  & X &  & X &  & Pro \\ 
327215 & Glass Products  (Purch Glass) &  & X & X & X &  & Pro \\ 
327310 & Cement Mnfct & X & X & X & X &  & Pro \\ 
327320 & Ready-Mix Concrete &  & X &  &  & X & Pro \\ 
327331 & Concrete Block \& Brick &  &  &  &  & X & Pro \\ 
327332 & Concrete Pipe Mnfct &  &  &  &  & X & Pro \\ 
327390 & Other Concrete Product & X & X & X &  & X & Pro \\ 
327410 & Lime Mnfct & X &  & X & X &  & Pro \\ 
327420 & Gypsum Product Mnfct & X &  & X & X &  & Pro \\ 
327910 & Abrasive Product Mnfct & X &  & X & X &  & Pro \\ 
327991 & Cut Stone \& Stone Product &  &  &  &  &  & Pro \\ 
327992 & Treated Mineral \& Earth &  &  &  & X &  & Pro \\ 
327993 & Mineral Wool Mnfct &  & X &  &  &  & Pro \\ 
327999 & Misc Nonmetal Mineral Prod &  &  &  & X & X & Pro, Sci\\
331111 & Iron \& Steel Mills &  & X &  & X & X & Sci \\ 
331222 & Steel Wire Drawing &  &  &  & X &  & Pro, Sci\\
331315 & Aluminum Sheet, Plate \& Foil &  & X &  & X &  & Sci \\ 
331411 & Smelting \& Refining of Copper &  & X &  & X & X & Sci \\ 
331421 & Copper Roll, Draw \& Extrud &  & X &  & X &  & Pro, Sci\\
331491 & Nonferr Metal Roll \& Extrud &  & X &  & X &  & Pro, Sci\\ 
331492 & Alloy of Nonferr Metal & X &  & X & X &  & Sci \\ 
332115 & Crown \& Closure &  &  &  & X &  & Pro \\ 
332117 & Powder Metallurgy Part & X &  & X & X &  & Pro \\ 
332211 & Cutlery \& Flatware & X &  & X &  &  & Pro \\ 
332212 & Hand \& Edge Tool & X & X & X & X & X & Pro, Sup\\
332311 & Metal Build \& Component &  & X &  &  & X & Pro \\ 
332313 & Plate Work Mnfct &  & X &  & X &  & Pro, Sci, Sup\\
332321 & Metal Window \& Door &  & X & X &  &  & Pro \\ 
332322 & Sheet Metal Work Mnfct & X & X & X & X &  & Pro \\ 
332323 & Architect Metal Work &  & X &  &  & X & Pro \\ 
332410 & Boiler \& Heat Exchanger & X &  & X &  &  & Pro, Sup\\
332420 & Metal Tank (Heavy Gauge) & X &  & X &  &  & Pro, Sup\\
332431 & Metal Can Mnfct &  & X &  & X &  & Pro \\ 
332439 & Other Metal Container &  & X &  & X &  & Pro \\ 
332510 & Hardware Mnfct & X & X & X & X &  & Pro \\ 
332611 & Spring (Heavy Gauge) &  &  &  &  &  & Pro \\ 
332618 & Other Fabricated Wire Prod &  &  &  & X &  & Pro, Sci\\
332710 & Machine Shops &  & X &  & X & X & Pro \\ 
332721 & Precision Turned Product &  & X &  & X &  & Pro \\ 
332722 & Bolt, Nut \& Screw &  & X & X & X &  & Pro \\ 
332811 & Metal Heat Treating & X & X & X & X &  & Pro \\ 
332812 & Metal Coat, Engrav \& Allied &  & X &  & X &  & Pro \\ 
332813 & Electroplating \& Polishing & X & X & X & X &  & Pro \\ 
332911 & Industrial Valve Mnfct &  & X &  & X &  & Sup \\ 
332912 & Fluid Power Valve \& Hose Fitt &  & X &  & X & X & Sci, Sup\\
332913 & Plumb Fixture Fitt \& Trim &  &  &  &  &  & Pro, Sci, Sup\\ 
332919 & Metal Valve \& Pipe Fitting & X & X & X & X &  & Sci, Sup\\
332991 & Ball \& Roller Bearing & X & X & X & X &  & Sup \\ 
332992 & Small Arms Ammunition & X &  & X &  & X & Pro, Sup\\
332993 & Ammunition (exc Small Arms) & X &  & X &  & X & Pro, Sup\\
332994 & Small Arms Mnfct & X &  & X &  & X & Pro \\ 
332996 & Fabricated Pipe \& Pipe Fitting &  & X &  & X &  & Sci \\ 
333111 & Farm Machinery \& Equipment & X &  & X & X & X & Sup \\ 
333112 & Lawn \& Garden Tractor \& Equ & X &  & X &  & X & Sci, Sup\\
333120 & Construction Machinery &  & X & X &  & X & Pro, Sup\\
333131 & Mining Machinery \& Equ & X &  & X &  & X & Pro, Sci, Sup\\
333132 & Oil \& Gas Field Machinery &  &  &  &  & X & Sup \\ 
333210 & Sawmill \& Woodwork Machine & X &  &  &  &  & Sup \\ 
333220 & Plastics \& Rubber Machinery & X &  & X &  &  & Sup \\ 
333293 & Printing Machinery \& Equ & X &  & X &  &  & Sup \\ 
333294 & Food Product Machinery & X &  & X &  &  & Sup \\ 
333295 & Semiconductor Machinery &  &  &  & X & X & Sup \\ 
333311 & Automatic Vending Machine & X &  & X &  & X & Sup \\ 
333314 & Optical Instrument \& Lens & X &  & X &  &  & Pro \\ 
333315 & Photogr \& Photocopy Equ & X &  & X &  & X & Pro, Sup\\
333414 & Heat Equipment &  &  &  &  & X & Pro, Sup\\
333512 & Machine Tool, Metal Cut & X &  & X &  &  & Sup \\ 
333514 & Die Set, Jig \& Fixture &  &  &  & X &  & Pro, Sup\\
333515 & Cutting \& Machine Tool & X & X & X & X &  & Pro, Sup\\
333516 & Rolling Mill Machinery \& Equ & X &  & X &  & X & Sup \\ 
333611 & Turbine \& Generator Units & X &  & X &  & X & Sup \\ 
333612 & Speed Changer, Drive \& Gear & X & X & X & X &  & Sup \\ 
333613 & Power Transmission Equ & X & X & X & X &  & Sup \\ 
333618 & Other Engine Equipment &  & X &  & X & X & Pro, Sci, Sup\\ 
333911 & Pump \& Pumping Equipment & X & X & X & X & X & Sup \\ 
333912 & Air \& Gas Compressor &  &  &  & X &  & Sup \\ 
333913 & Measuring \& Dispensing Pump & X &  &  &  & X & Sup \\ 
333921 & Elevator \& Moving Stairway & X &  &  &  & X & Sup \\ 
333922 & Conveyor \& Conveying Equ & X &  & X &  & X & Sup \\ 
333923 & Overhead Crane \& Monorail & X &  & X &  & X & Sci, Sup\\
333924 & Truck, Tractor \& Stacker & X &  & X &  &  & Sci, Sup\\ 
333991 & Power-Driven Handtool & X &  & X &  &  & Sup \\ 
333992 & Welding \& Soldering Equ & X &  & X &  &  & Pro, Sup\\
333993 & Packaging Machinery & X &  & X &  &  & Sup \\ 
334111 & Electronic Computer & X & X & X & X & X & Pro \\ 
334112 & Computer Storage Device & X &  & X &  & X & Pro \\ 
334113 & Computer Terminal &  &  &  &  & X & Pro \\ 
334210 & Telephone Apparatus & X & X & X &  & X & Pro \\ 
334220 & Radio, TV, Wirel Communic & X & X & X & X & X & Pro, Sci\\
334310 & Audio \& Video Equipment &  & X &  & X & X & Pro \\ 
334411 & Electron Tube Mnfct & X & X &  & X &  & Pro \\ 
334412 & Bare Printed Circuit Board & X & X & X & X &  & Pro \\ 
334413 & Semiconductor \& Rel Device & X & X & X & X & X & Pro \\ 
334414 & Electronic Capacitor &  & X &  & X &  & Pro \\ 
334417 & Electronic Connector &  & X &  & X &  & Pro \\ 
334418 & Printed Circuit Assembly & X & X & X & X & X & Pro \\ 
334510 & Electromedical \& -therapeutic & X &  & X &  & X & Pro \\ 
334511 & Search, Navig, Nautic System & X &  & X &  & X & Pro \\ 
334512 & Env Ctrl for Resident Use & X &  & X &  &  & Pro \\ 
334513 & Instr for Control  Process & X &  & X & X &  & Pro \\ 
334514 & Fluid Meter \& Count Device & X &  &  &  &  & Pro \\ 
334515 & Measure \& Test Elctr Signals &  &  &  & X &  & Pro \\ 
334517 & Irradiation Apparatus & X &  & X &  &  & Pro \\ 
334518 & Watch, Clock, \& Part & X & X & X & X &  & Pro \\ 
334611 & Software Reproducing & X &  & X &  &  & Sup \\ 
334613 & Magnetic \& Optic Recording & X &  & X &  & X & Pro \\ 
335110 & Electric Lamp Bulb \& Part & X &  & X &  &  & Pro \\ 
335129 & Other Lighting Equipment &  &  &  &  &  & Pro \\ 
335211 & Elctr Housewares \& Home Fan &  &  &  &  &  & Pro \\ 
335221 & Household Cooking Appliance & X &  &  &  &  & Pro \\ 
335222 & Refrigerator \& Freezer & X &  &  &  &  & Pro \\ 
335224 & Household Laundry Equ & X &  &  &  & X & Pro \\ 
335228 & Major Household Appliance &  &  &  &  &  & Pro \\ 
335311 & Power, Distr \& Transformer & X &  & X & X &  & Pro \\ 
335312 & Motor \& Generator & X & X & X & X & X & Pro \\ 
335313 & Switchgear \& Switchboard &  & X & X & X &  & Pro \\ 
335314 & Relay \& Industrial Control &  & X &  & X & X & Pro \\ 
335911 & Storage Battery Mnfct & X &  & X &  &  & Pro \\ 
335912 & Primary Battery Mnfct & X &  &  &  &  & Pro \\ 
335921 & Fiber Optic Cable &  & X &  & X &  & Pro \\ 
335929 & Communic \& Energy Wire &  & X &  & X &  & Pro \\ 
335931 & Current-Carrying Wire Device & X & X & X &  &  & Pro \\ 
335932 & Noncurrent-Carry Wir Device & X & X & X &  &  & Pro \\ 
335991 & Carbon \& Graphite Product & X &  & X & X &  & Pro \\ 
335999 & Miscellaneous Electrical Equ & X &  & X &  & X & Pro \\ 
336111 & Automobile Mnfct & X &  & X & X & X & Pro \\ 
336112 & Light Truck \& Utility Vehicle &  &  &  &  & X & Pro \\ 
336120 & Heavy Duty Truck Mnfct &  &  &  &  & X & Pro \\ 
336211 & Motor Vehicle Body &  & X & X &  & X & Pro, Sup\\
336212 & Truck Trailer Mnfct & X &  &  &  & X & Pro \\ 
336213 & Motor Home Mnfct & X &  &  &  & X & Pro \\ 
336214 & Travel Trailer \& Camper & X &  &  &  & X & Pro \\ 
336311 & Carburetor, Piston \& Valve &  & X &  & X &  & Sup \\ 
336321 & Vehicular Lighting Equipment &  & X & X &  & X & Pro \\ 
336330 & Vhcl Steer \& Suspension Parts &  & X & X &  & X & Pro, Sup\\
336340 & Motor Vehicle Brake System &  & X & X & X & X & Pro, Sup\\ 
336350 & Vhcl Power Train Parts &  & X & X & X & X & Pro \\ 
336360 & Vhcl Seat \& Interior &  & X &  & X & X & Pro, Sci\\
336391 & Motor Vhcl Air-Conditioning &  & X &  & X & X & Sup \\ 
336411 & Aircraft Mnfct & X & X & X & X & X & Sci \\ 
336412 & Aircraft Engine \& Parts &  & X &  & X & X & Sci, Sup\\
336413 & Other Aircraft Parts \& Aux &  & X &  &  & X & Sci \\ 
336414 & Guided Missile \& Space Vhcl & X &  & X &  & X & Sci \\ 
336415 & Missile \& Space Vhcl Propuls &  &  &  &  & X & Sci \\ 
336419 & Missile \& Space Vhcl Parts & X &  &  &  & X & Sci \\ 
336510 & Railroad Rolling Stock & X & X & X &  & X & Pro, Sci\\
336611 & Ship Building \& Repairing & X &  & X &  & X & Sci \\ 
336612 & Boat Building & X &  & X &  & X & Sci \\ 
336991 & Motorcycle, Bicycle, \& Parts & X &  & X &  & X & Sci \\ 
336999 & Transportation Equipment & X &  & X &  & X & Pro, Sci\\
337110 & Wood Kitchen Cabinet &  & X &  &  &  & Pro \\ 
337121 & Upholstered Household Furnit &  &  &  &  & X & Pro \\ 
337122 & Nonupholst Wood Furnit & X &  & X &  &  & Pro \\ 
337124 & Metal Household Furniture & X &  &  &  &  & Pro \\ 
337125 & Other Home Furniture & X &  & X &  &  & Pro \\ 
337211 & Wood Office Furniture &  &  &  &  & X & Pro \\ 
337214 & Office Furniture (exc Wood) & X &  & X &  & X & Pro \\ 
337215 & Showcase, Shelv \& Locker &  & X & X & X & X & Pro, Sup\\
337910 & Mattress Mnfct & X &  & X &  &  & Pro \\ 
337920 & Blind \& Shade Mnfct & X &  &  &  & X & Pro, Sup\\
339111 & Laboratory  \& Furniture & X &  & X &  & X &  \\ 
339112 & Surgical \& Medical Instrument & X & X & X & X & X & Sci \\ 
339113 & Surgical Appliance \& Supplies &  & X &  & X & X & Pro, Sci, Sup\\
339114 & Dental Equipment \& Supplies & X &  & X &  & X & Sci \\ 
339911 & Jewelry (exc Costume) & X &  & X & X &  & Sci \\ 
339920 & Sporting \& Athletic Goods & X &  & X & X & X & Pro, Sci, Sup\\
339931 & Doll \& Stuffed Toy &  &  &  &  & X & Sci \\ 
339941 & Pen \& Mechanical Pencil & X &  &  &  &  & Sci \\ 
339950 & Sign Mnfct & X &  & X &  & X & Pro, Sup\\ 
339992 & Musical Instrument & X &  & X &  & X & Sci \\ 
339993 & Fastener, Button, \& Pin & X &  & X &  &  & Pro, Sic, Sup\\
339994 & Broom, Brush, \& Mop & X &  & X &  &  & Sci \\ 
339995 & Burial Casket Mnfct & X &  & X &  & X & Sci \\ 
339999 & All Other Miscellaneous & X &  & X &  & X & Pro, Sci, Sup \\
\hline
\hline
\end{longtable}
\endgroup

\justifying
\noindent \footnotesize
Notes: This table shows industries are subset into different groups to search for sectoral patterns of innovation. Innovation-intensive is measures as number of citation-weighted patents divided by output ($A^{\tau}/A^{\mu}$). Big industry are industries with above median industry size ($A^{\mu}$). High patent centrality is given $i$ has above median patent PageRank (up- \& downwards) $PR^{\tau,dw}$. High up- and downstream market centrality are given $i$ has above median upstream (downstream) PageRank $PR^{\mu, d}$. Pavitt sectors are identified using the classification proposed by \citep{bogliacino2016pavitt}. The codes are: Pro for production intensive, Sci for Sciene based, Sup for Suppliers dominated. Note that the Pavitt codes are not available for a larger subset of industries belonging to Food processing. 
\end{center}
\FloatBarrier

\subsection{By 2-digit sector}

\subsubsection{Food processing}
\begin{table}[H]
\begin{myresizeenv}
\begingroup

\caption{Demand-pull and technology-push effects in Food processing sectors}			
\label{tab:real_sum_output|citation_weighted_patent_stock_Food_all_years_sys_1step_wControls}
\begin{tabular}{|l|cccc|cccc|cccc|}
\hline \hline  \rule{0pt}{1.075\normalbaselineskip}  
& \multicolumn{4}{c|}{\scriptsize \ul{Type 1}} & \multicolumn{4}{c|}{\scriptsize \ul{Type 2}} & \multicolumn{4}{c|}{\scriptsize \ul{Both}}\\  & \multicolumn{2}{c}{\scriptsize $\tau \rightarrow \mu$} & \multicolumn{2}{c|}{\scriptsize $\mu \rightarrow \tau$}
& \multicolumn{2}{c}{\scriptsize Market} & \multicolumn{2}{c|}{\scriptsize Innovation}
& \multicolumn{4}{c|}{}\\
\hline  \rule{0pt}{1.075\normalbaselineskip}    & $A^{\mu}_{i,t}$ & $A^{\mu}_{i,t}$ & $A^{\tau}_{i,t}$ & $A^{\tau}_{i,t}$ & $A^{\mu}_{i,t}$ & $A^{\mu}_{i,t}$ & $A^{\tau}_{i,t}$ & $A^{\tau}_{i,t}$ & $A^{\mu}_{i,t}$ & $A^{\mu}_{i,t}$ & $A^{\tau}_{i,t}$ & $A^{\tau}_{i,t}$ \\ 
\rule{0pt}{1.075\normalbaselineskip}  & (1) & (2) & (3) & (4) & (5) & (6) & (7) & (8) & (9) & (10) & (11) & (12) \\ 
\hline \rule{0pt}{1.075\normalbaselineskip}   $A^{\mu}_{i,t-1}$ & 0.6983*** & 0.4968*** & 0.0306 & 0.0349 & 0.5831*** & 0.2226 &  &  & 0.6827*** & 0.3424* & 0.067 & 0.0076 \\ 
& (0.0767) & (0.0947) & (0.0391) & (0.0341) & (0.0916) & (0.165) &  &  & (0.1287) & (0.1434) & (0.0473) & (0.0286) \\ 
$A^{\tau}_{i,t-1}$ & -0.3206 & -0.1575 & 1.001*** & 0.9874*** &  &  & 0.9008*** & 0.952*** & -0.328 & -0.2812 & 0.9385*** & 0.9731*** \\ 
& (0.2463) & (0.2534) & (0.0256) & (0.0328) &  &  & (0.1162) & (0.1207) & (0.2739) & (0.2626) & (0.1018) & (0.1088) \\ 
$PR^{\mu,up}_{i,t-1}$ &  &  & 0.0074 & 0.0108 & -0.0592. & 0.0114 &  &  & -0.0897. & 0.0026 & 0.003 & 0.0138 \\ 
&  &  & (0.0114) & (0.0115) & (0.0329) & (0.0519) &  &  & (0.0486) & (0.0524) & (0.013) & (0.0123) \\ 
$PR^{\mu,dw}_{i,t-1}$ &  &  & 0.0041 & 0.0126 & 0.0578 & 0.0944. &  &  & 0.0414 & 0.0759 & 0.0131 & 0.0218* \\ 
&  &  & (0.0069) & (0.0094) & (0.0357) & (0.0531) &  &  & (0.0401) & (0.0479) & (0.0118) & (0.0105) \\ 
$PR^{\tau,dw}_{i,t-1}$ & 0.278 & 0.0677 &  &  &  &  & 0.1238 & 0.0626 & 0.2774 & 0.1359 & 0.1049. & 0.0467 \\ 
& (0.1928) & (0.1103) &  &  &  &  & (0.0798) & (0.0502) & (0.1896) & (0.1282) & (0.0633) & (0.0406) \\ 
$Spill(A)^{\mu,up}_{i,t-1}$ &  &  & 0.0016 & -0.0117 & 0.0229 & 0.07 &  &  & -0.0281 & 0.0401 & -0.0068 & -0.0076 \\ 
&  &  & (0.0243) & (0.0221) & (0.0582) & (0.0804) &  &  & (0.0635) & (0.0804) & (0.0222) & (0.0228) \\ 
$Spill(A)^{\mu,dw}_{i,t-1}$ &  &  & -0.0077 & -0.0095 & 0.1061. & 0.1254 &  &  & 0.0486 & 0.082 & 0.0107 & 0.0235 \\ 
&  &  & (0.0144) & (0.0159) & (0.0617) & (0.0831) &  &  & (0.0679) & (0.0671) & (0.0169) & (0.0167) \\ 
$Spill(A)^{\tau,up}_{i,t-1}$ & 7.858** & 2.557 &  &  &  &  & 2.074* & 0.5733 & 8.253** & 3.865* & 1.495. & 0.2625 \\ 
& (2.687) & (2.012) &  &  &  &  & (0.8828) & (0.8041) & (2.745) & (1.879) & (0.7975) & (0.5631) \\ 
$Spill(A)^{\tau,dw}_{i,t-1}$ & 0.1373 & 0.0379 &  &  &  &  & 0.0174 & -0.004 & 0.1358 & 0.0966 & -0.0094 & -0.0198 \\ 
& (0.1318) & (0.1307) &  &  &  &  & (0.052) & (0.0522) & (0.1447) & (0.1138) & (0.0441) & (0.0402) \\ 
\hline \rule{0pt}{1.075\normalbaselineskip}  AR(1) & 0 & 0 & 7e-04 & 4e-04 & 0 & 2e-04 & 8e-04 & 0.0012 & 0 & 0 & 5e-04 & 5e-04 \\ 
AR(2) & 0.863 & 0.9389 & 0.7458 & 0.7474 & 0.6879 & 0.7258 & 0.6942 & 0.9831 & 0.482 & 0.9749 & 0.8636 & 0.8884 \\ 
Sargan & 2e-04 & 6e-04 & 0.0204 & 0.0152 & 0 & 0.0108 & 0.0118 & 1e-04 & 0.0037 & 0.0018 & 0.0016 & 3e-04 \\ 
Method & 1-step-BB & 1-step-BB & 1-step-BB & 1-step-BB & 1-step-BB & 1-step-BB & 1-step-BB & 1-step-BB & 1-step-BB & 1-step-BB & 1-step-BB & 1-step-BB \\ 
Controls &  & Y &  & Y &  & Y &  & Y &  & Y &  & Y \\ 
$R^2$ & 0.899 & 0.9286 & 0.9949 & 0.9944 & 0.9398 & 0.9168 & 0.9914 & 0.9942 & 0.8947 & 0.9253 & 0.9925 & 0.9937 \\ 
\hline
\hline
\end{tabular}
\endgroup

\end{myresizeenv}

\vspace{0.25cm}

\justifying 
\noindent \scriptsize
Notes: The table shows the regression results of output $A_{i,t}^{\mu}$ and patents $A_{i,t}^{\tau}$ on demand-pull and technology-push effects. The estimation is based on a two-ways Blundell-Bond (BB) system GMM model using a one-step estimation procedure. Spillovers are calculated on the basis of first-order links. Variables measured in monetary terms are deflated using the industry level price deflators for the value of shipment obtained from the NBER-productivity database \citep{becker2013nber}. Instruments are collapsed to avoid instrument proliferation. To cope with skewness and to obtain tractable coefficients, most variables are pre-processed (taking logs, removing outliers, scaling). Data in logs are $A^{\alpha }_{i,t}$, $PR^{\alpha, d }_{i,t}$, $Spill(A)^{\alpha, d }_{i,t}$, $L_{i,t}$,  $(K/L)_{i,t}$, $(I/L)_{i,t}$, $W_{i,t}$, $(K/L)^{}_{ij,t-1}$, $(E/L)^{}_{ij,t-1}$, $(M/L)^{}_{ij,t-1}$, $W^{P}_{ij,t-1}$ with $\alpha = \mu, \tau$ and $d = up, dw$. A detailed description of the transformations and descriptive statistics of the regression data before and after the transformations are provided in \ref{app:data}. 
The rows AR(1), AR(2), and Sargan show the test statistics of the specification tests, i.e. testing for first- and second-order autocorrelation and the results of a Sargan test for validity of instruments \citep[see][]{roodman2009xtabond2}. 
The analysis covers the subset of Food processing industries (NAICS 2-digit code 31). 
\end{table}

\subsubsection{Non-Metallic Manufacturing}
\begin{table}[H]
\begin{myresizeenv}
\begingroup
\caption{Demand-pull and technology-push effects in Non-Metallic manufacturing}

\label{tab:real_sum_output|citation_weighted_patent_stock_NonMetal_all_years_sys_1step_wControls}

\begin{tabular}{|l|cccc|cccc|cccc|}
\hline \hline  \rule{0pt}{1.075\normalbaselineskip}  
& \multicolumn{4}{c|}{\scriptsize \ul{Type 1}} & \multicolumn{4}{c|}{\scriptsize \ul{Type 2}} & \multicolumn{4}{c|}{\scriptsize \ul{Both}}\\  & \multicolumn{2}{c}{\scriptsize $\tau \rightarrow \mu$} & \multicolumn{2}{c|}{\scriptsize $\mu \rightarrow \tau$}
& \multicolumn{2}{c}{\scriptsize Market} & \multicolumn{2}{c|}{\scriptsize Innovation}
& \multicolumn{4}{c|}{}\\
\hline  \rule{0pt}{1.075\normalbaselineskip}    & $A^{\mu}_{i,t}$ & $A^{\mu}_{i,t}$ & $A^{\tau}_{i,t}$ & $A^{\tau}_{i,t}$ & $A^{\mu}_{i,t}$ & $A^{\mu}_{i,t}$ & $A^{\tau}_{i,t}$ & $A^{\tau}_{i,t}$ & $A^{\mu}_{i,t}$ & $A^{\mu}_{i,t}$ & $A^{\tau}_{i,t}$ & $A^{\tau}_{i,t}$ \\ 
\rule{0pt}{1.075\normalbaselineskip}  & (1) & (2) & (3) & (4) & (5) & (6) & (7) & (8) & (9) & (10) & (11) & (12) \\ 
\hline \rule{0pt}{1.075\normalbaselineskip}   $A^{\mu}_{i,t-1}$ & 0.5975*** & 0.6909*** & 0.0484 & 0.0265 & 0.5302*** & 0.5572*** &  &  & 0.4645*** & 0.5605*** & 0.0346 & 0.0327 \\ 
& (0.0984) & (0.0932) & (0.0487) & (0.0507) & (0.117) & (0.0939) &  &  & (0.1369) & (0.097) & (0.0621) & (0.0618) \\ 
$A^{\tau}_{i,t-1}$ & 0.0394 & 0.0849 & 0.994*** & 0.9792*** &  &  & 1.001*** & 0.8991*** & 0.0425 & 0.0981 & 1.027*** & 0.9157*** \\ 
& (0.1719) & (0.2901) & (0.0448) & (0.0476) &  &  & (0.1572) & (0.119) & (0.181) & (0.2866) & (0.1462) & (0.1133) \\ 
$PR^{\mu,up}_{i,t-1}$ &  &  & -0.012 & -0.0125 & -0.0158 & -0.0393 &  &  & -0.0017 & -0.0288 & -0.0037 & -0.0084 \\ 
&  &  & (0.0094) & (0.0112) & (0.0219) & (0.0302) &  &  & (0.024) & (0.0301) & (0.0102) & (0.0124) \\ 
$PR^{\mu,dw}_{i,t-1}$ &  &  & -0.0127 & -0.0097 & 0.0178 & -0.0377 &  &  & 0.022 & -0.0456 & -0.0055 & -0.0078 \\ 
&  &  & (0.0101) & (0.0129) & (0.0153) & (0.0379) &  &  & (0.0188) & (0.0411) & (0.0114) & (0.0136) \\ 
$PR^{\tau,dw}_{i,t-1}$ & 0.0785 & 0.0581 &  &  &  &  & 0.0407 & 0.0661* & 0.1064 & 0.0759 & 0.0294 & 0.0568. \\ 
& (0.0658) & (0.1042) &  &  &  &  & (0.0452) & (0.0319) & (0.0783) & (0.0991) & (0.0445) & (0.0292) \\ 
$Spill(A)^{\mu,up}_{i,t-1}$ &  &  & -0.0041 & -0.0062 & -0.0011 & 0.026 &  &  & 0.0056 & 0.0186 & -0.0221 & -0.0229 \\ 
&  &  & (0.0166) & (0.0158) & (0.0331) & (0.0544) &  &  & (0.0351) & (0.0618) & (0.0187) & (0.0176) \\ 
$Spill(A)^{\mu,dw}_{i,t-1}$ &  &  & 0.0123 & 0.0095 & 0.0391 & 0.0274 &  &  & 0.0372 & 0.025 & -0.0053 & -0.0058 \\ 
&  &  & (0.0137) & (0.0132) & (0.0296) & (0.0395) &  &  & (0.0334) & (0.0429) & (0.0124) & (0.0137) \\ 
$Spill(A)^{\tau,up}_{i,t-1}$ & -0.8038 & 1.117 &  &  &  &  & 1.74*** & 1.664** & -2.134 & -0.3491 & 1.497*** & 1.505** \\ 
& (1.74) & (2.153) &  &  &  &  & (0.5085) & (0.5743) & (1.708) & (2.226) & (0.4398) & (0.4709) \\ 
$Spill(A)^{\tau,dw}_{i,t-1}$ & -0.0462 & -0.0767 &  &  &  &  & -0.0512 & -0.0153 & -0.0767 & -0.1187 & -0.0513 & -0.0175 \\ 
& (0.0707) & (0.086) &  &  &  &  & (0.0373) & (0.0353) & (0.0661) & (0.0904) & (0.0351) & (0.0355) \\ 
\hline \rule{0pt}{1.075\normalbaselineskip}  AR(1) &  & 1e-04 & 0.0015 & 0.0014 &  & 0 & 1e-04 & 0.0063 & 0 & 1e-04 & 0.0022 & 0.0078 \\ 
AR(2) & 0.1328 & 0.345 & 0.2525 & 0.0664 & 0.1232 & 0.5513 & 0.6786 & 0.8131 & 0.2108 & 0.7652 & 0.9401 & 0.9263 \\ 
Sargan & 0.0016 & 0.002 & 0.0014 & 1e-04 & 0 & 1e-04 & 0.0013 & 0.0088 & 0 & 1e-04 & 0.0016 & 0.0041 \\ 
Controls &  & Y &  & Y &  & Y &  & Y &  & Y &  & Y \\ 
$R^2$ & 0.9696 & 0.9462 & 0.9966 & 0.9964 & 0.9716 & 0.9483 & 0.9957 & 0.9952 & 0.9679 & 0.946 & 0.9961 & 0.9955 \\ 
\hline
\hline
\end{tabular}
\endgroup

\end{myresizeenv}

\vspace{0.25cm}

\justifying \footnotesize

\noindent
Notes: The table shows the regression results of output $A_{i,t}^{\mu}$ and patents $A_{i,t}^{\tau}$ on demand-pull and technology-push effects. The estimation is based on a two-ways Blundell-Bond (BB) system GMM model using a one-step estimation procedure. Spillovers are calculated on the basis of first-order links. Variables measured in monetary terms are deflated using the industry level price deflators for the value of shipment obtained from the NBER-productivity database \citep{becker2013nber}. Instruments are collapsed to avoid instrument proliferation. To cope with skewness and to obtain tractable coefficients, most variables are pre-processed (taking logs, removing outliers, scaling). Data in logs are $A^{\alpha }_{i,t}$, $PR^{\alpha, d }_{i,t}$, $Spill(A)^{\alpha, d }_{i,t}$, $L_{i,t}$,  $(K/L)_{i,t}$, $(I/L)_{i,t}$, $W_{i,t}$, $(K/L)^{}_{ij,t-1}$, $(E/L)^{}_{ij,t-1}$, $(M/L)^{}_{ij,t-1}$, $W^{P}_{ij,t-1}$ with $\alpha = \mu, \tau$ and $d = up, dw$. A detailed description of the transformations and descriptive statistics of the regression data before and after the transformations are provided in \ref{app:data}. 
The rows AR(1), AR(2), and Sargan show the test statistics of the specification tests, i.e. testing for first- and second-order autocorrelation and the results of a Sargan test for validity of instruments \citep[see][]{roodman2009xtabond2}. 
The analysis covers the subset of NonMetal industries (NAICS 2-digt code 32).
\end{table}

\subsubsection{Metallic and Machinery manufacturing}
\begin{table}[H]
\begin{myresizeenv}
\begingroup
\caption{Demand-pull and technology-push effects in Metallic and Machinery manufacturing} 

\label{tab:real_sum_output|citation_weighted_patent_stock_Metal_all_years_sys_1step_wControls}

\begin{tabular}{|l|cccc|cccc|cccc|}
\hline \hline  \rule{0pt}{1.075\normalbaselineskip}  
& \multicolumn{4}{c|}{\scriptsize \ul{Type 1}} & \multicolumn{4}{c|}{\scriptsize \ul{Type 2}} & \multicolumn{4}{c|}{\scriptsize \ul{Both}}\\  & \multicolumn{2}{c}{\scriptsize $\tau \rightarrow \mu$} & \multicolumn{2}{c|}{\scriptsize $\mu \rightarrow \tau$}
& \multicolumn{2}{c}{\scriptsize Market} & \multicolumn{2}{c|}{\scriptsize Innovation}
& \multicolumn{4}{c|}{}\\
\hline  \rule{0pt}{1.075\normalbaselineskip}    & $A^{\mu}_{i,t}$ & $A^{\mu}_{i,t}$ & $A^{\tau}_{i,t}$ & $A^{\tau}_{i,t}$ & $A^{\mu}_{i,t}$ & $A^{\mu}_{i,t}$ & $A^{\tau}_{i,t}$ & $A^{\tau}_{i,t}$ & $A^{\mu}_{i,t}$ & $A^{\mu}_{i,t}$ & $A^{\tau}_{i,t}$ & $A^{\tau}_{i,t}$ \\ 
\rule{0pt}{1.075\normalbaselineskip}  & (1) & (2) & (3) & (4) & (5) & (6) & (7) & (8) & (9) & (10) & (11) & (12) \\ 
\hline \rule{0pt}{1.075\normalbaselineskip}   $A^{\mu}_{i,t-1}$ & 0.5624*** & 0.4186*** & 0.0137 & 0.0061 & 0.78*** & 0.7392*** &  &  & 0.7071*** & 0.6474*** & 0.0367. & -0.0015 \\ 
& (0.0373) & (0.059) & (0.0196) & (0.0188) & (0.0689) & (0.0867) &  &  & (0.0903) & (0.0985) & (0.02) & (0.0159) \\ 
$A^{\tau}_{i,t-1}$ & 0.2357 & 0.292 & 1.084*** & 1.089*** &  &  & 1.254*** & 1.182*** & 0.3616 & 0.1782 & 1.246*** & 1.166*** \\ 
& (0.2149) & (0.2619) & (0.0191) & (0.0271) &  &  & (0.1009) & (0.0802) & (0.3341) & (0.3139) & (0.09) & (0.0776) \\ 
$PR^{\mu,up}_{i,t-1}$ &  &  & -0.008 & -0.0039 & -0.0464. & -0.0326 &  &  & -0.0202 & -0.0447 & -0.0091 & -0.008 \\ 
&  &  & (0.007) & (0.0081) & (0.027) & (0.0265) &  &  & (0.028) & (0.031) & (0.0076) & (0.0074) \\ 
$PR^{\mu,dw}_{i,t-1}$ &  &  & 0.0124* & 0.0098. & -0.0193 & 0.0021 &  &  & -0.0308 & -0.0108 & 0.0081 & 0.0059 \\ 
&  &  & (0.005) & (0.0056) & (0.0277) & (0.0252) &  &  & (0.0283) & (0.0252) & (0.0055) & (0.0048) \\ 
$PR^{\tau,dw}_{i,t-1}$ & 0.0222 & 0.0507 &  &  &  &  & -0.0458 & -0.025 & 0.0539 & 0.0986 & -0.0552 & -0.0189 \\ 
& (0.09) & (0.1007) &  &  &  &  & (0.0421) & (0.0257) & (0.15) & (0.1077) & (0.0351) & (0.0258) \\ 
$Spill(A)^{\mu,up}_{i,t-1}$ &  &  & -0.0221* & -0.0106 & 0.0852** & 0.0859** &  &  & 0.098** & 0.0741. & -0.0155 & -0.0091 \\ 
&  &  & (0.009) & (0.01) & (0.0286) & (0.031) &  &  & (0.033) & (0.0392) & (0.0102) & (0.0099) \\ 
$Spill(A)^{\mu,dw}_{i,t-1}$ &  &  & 0.0109 & 0.0103 & -0.1085** & -0.0809* &  &  & -0.1389*** & -0.096** & 0.0023 & 0.0126. \\ 
&  &  & (0.0067) & (0.0075) & (0.0338) & (0.0329) &  &  & (0.0387) & (0.0344) & (0.0067) & (0.0074) \\ 
$Spill(A)^{\tau,up}_{i,t-1}$ & 2.052 & 6.663*** &  &  &  &  & 0.7965 & 0.3382 & 7.514*** & 8.97*** & 0.629 & 0.2332 \\ 
& (1.428) & (1.887) &  &  &  &  & (0.757) & (0.6235) & (1.757) & (2.051) & (0.6482) & (0.5991) \\ 
$Spill(A)^{\tau,dw}_{i,t-1}$ & -0.1443* & -0.246* &  &  &  &  & -0.0858*** & -0.0683** & -0.2378* & -0.2365* & -0.0738** & -0.0615** \\ 
& (0.0678) & (0.0988) &  &  &  &  & (0.0224) & (0.0229) & (0.0963) & (0.1073) & (0.0233) & (0.0224) \\ 
\hline \rule{0pt}{1.075\normalbaselineskip}  AR(1) & 0 & 0 & 0.0269 & 0.0228 & 0 & 0 & 0.0461 & 0.0104 & 0 & 0 & 0.006 & 0.0079 \\ 
AR(2) & 0.4529 & 0.3061 & 0.5856 & 0.5822 & 0.6946 & 0.8913 & 0.3849 & 0.3743 & 0.5486 & 0.4561 & 0.3931 & 0.397 \\ 
Sargan & 0 & 0 & 0 & 0.0029 & 0 & 0 & 0 & 0 & 0 & 0 & 0 & 0 \\ 
Controls &  & Y &  & Y &  & Y &  & Y &  & Y &  & Y \\ 
$R^2$ & 0.9209 & 0.8912 & 0.9957 & 0.9953 & 0.9157 & 0.9124 & 0.9953 & 0.9959 & 0.8972 & 0.8837 & 0.9953 & 0.9959 \\ 
\hline
\hline
\end{tabular}
\endgroup

\end{myresizeenv}

\vspace{0.25cm}

\justifying \footnotesize

\noindent
Notes: The table shows the regression results of output $A_{i,t}^{\mu}$ and patents $A_{i,t}^{\tau}$ on demand-pull and technology-push effects. The estimation is based on a two-ways Blundell-Bond (BB) system GMM model using a one-step estimation procedure. Spillovers are calculated on the basis of first-order links. Variables measured in monetary terms are deflated using the industry level price deflators for the value of shipment obtained from the NBER-productivity database \citep{becker2013nber}. Instruments are collapsed to avoid instrument proliferation. To cope with skewness and to obtain tractable coefficients, most variables are pre-processed (taking logs, removing outliers, scaling). Data in logs are $A^{\alpha }_{i,t}$, $PR^{\alpha, d }_{i,t}$, $Spill(A)^{\alpha, d }_{i,t}$, $L_{i,t}$,  $(K/L)_{i,t}$, $(I/L)_{i,t}$, $W_{i,t}$, $(K/L)^{}_{ij,t-1}$, $(E/L)^{}_{ij,t-1}$, $(M/L)^{}_{ij,t-1}$, $W^{P}_{ij,t-1}$ with $\alpha = \mu, \tau$ and $d = up, dw$. A detailed description of the transformations and descriptive statistics of the regression data before and after the transformations are provided in \ref{app:data}. 
The rows AR(1), AR(2), and Sargan show the test statistics of the specification tests, i.e. testing for first- and second-order autocorrelation and the results of a Sargan test for validity of instruments \citep[see][]{roodman2009xtabond2}. 
The analysis covers the subset of Metal industries (NAICS 2-digit code 33). 
\end{table}

\FloatBarrier
\subsection{Innovation intensity}

\subsubsection{Sectors with a high innovation-intensity}
\begin{table}[H]
\begin{myresizeenv}
\begingroup
\caption{Demand-pull and technology-push effects on Innovation-intensive industries}

\label{tab:real_sum_output|citation_weighted_patent_stock_inno_intensive_TRUE_all_years_sys_1step_wControls}
\begin{tabular}{|l|cccc|cccc|cccc|}
\hline \hline  \rule{0pt}{1.075\normalbaselineskip}  
& \multicolumn{4}{c|}{\scriptsize \ul{Type 1}} & \multicolumn{4}{c|}{\scriptsize \ul{Type 2}} & \multicolumn{4}{c|}{\scriptsize \ul{Both}}\\  & \multicolumn{2}{c}{\scriptsize $\tau \rightarrow \mu$} & \multicolumn{2}{c|}{\scriptsize $\mu \rightarrow \tau$}
& \multicolumn{2}{c}{\scriptsize Market} & \multicolumn{2}{c|}{\scriptsize Innovation}
& \multicolumn{4}{c|}{}\\
\hline  \rule{0pt}{1.075\normalbaselineskip}    & $A^{\mu}_{i,t}$ & $A^{\mu}_{i,t}$ & $A^{\tau}_{i,t}$ & $A^{\tau}_{i,t}$ & $A^{\mu}_{i,t}$ & $A^{\mu}_{i,t}$ & $A^{\tau}_{i,t}$ & $A^{\tau}_{i,t}$ & $A^{\mu}_{i,t}$ & $A^{\mu}_{i,t}$ & $A^{\tau}_{i,t}$ & $A^{\tau}_{i,t}$ \\ 
\rule{0pt}{1.075\normalbaselineskip}  & (1) & (2) & (3) & (4) & (5) & (6) & (7) & (8) & (9) & (10) & (11) & (12) \\ 
\hline \rule{0pt}{1.075\normalbaselineskip}   $A^{\mu}_{i,t-1}$ & 0.5394*** & 0.5587*** & 0.0162 & 0.013 & 0.6693*** & 0.7179*** &  &  & 0.5774*** & 0.7*** & 0.0068 & 0.0026 \\ 
& (0.04) & (0.0416) & (0.0144) & (0.0146) & (0.0669) & (0.0761) &  &  & (0.0784) & (0.0715) & (0.0161) & (0.0162) \\ 
$A^{\tau}_{i,t-1}$ & 0.168 & -0.2289 & 1.083*** & 1.091*** &  &  & 1.225*** & 1.232*** & 4e-04 & -0.3227 & 1.266*** & 1.252*** \\ 
& (0.243) & (0.335) & (0.0153) & (0.0265) &  &  & (0.0702) & (0.0615) & (0.3609) & (0.3554) & (0.0521) & (0.0504) \\ 
$PR^{\mu,up}_{i,t-1}$ &  &  & -0.0014 & -0.0028 & -0.0245 & -0.0202 &  &  & 0.0159 & 0.0078 & -0.0014 & -0.0027 \\ 
&  &  & (0.0041) & (0.0047) & (0.0224) & (0.0224) &  &  & (0.0295) & (0.0307) & (0.0041) & (0.0039) \\ 
$PR^{\mu,dw}_{i,t-1}$ &  &  & 0.0066 & 0.0034 & 0.0384 & 0.0413 &  &  & 0.0775. & 0.0886* & 0.0024 & -4e-04 \\ 
&  &  & (0.0042) & (0.005) & (0.0352) & (0.0323) &  &  & (0.0398) & (0.0356) & (0.0045) & (0.0045) \\ 
$PR^{\tau,dw}_{i,t-1}$ & 0.0854 & 0.2832* &  &  &  &  & -0.008 & -0.0134 & 0.1759 & 0.2504* & -0.0331. & -0.0287* \\ 
& (0.1446) & (0.1117) &  &  &  &  & (0.0275) & (0.0213) & (0.1574) & (0.1062) & (0.0197) & (0.0146) \\ 
$Spill(A)^{\mu,up}_{i,t-1}$ &  &  & -0.0191** & -0.0125. & 0.0494. & 0.0477 &  &  & -0.0156 & -0.0259 & -0.0059 & -0.0051 \\ 
&  &  & (0.0062) & (0.0066) & (0.0283) & (0.0332) &  &  & (0.0333) & (0.0396) & (0.0055) & (0.0057) \\ 
$Spill(A)^{\mu,dw}_{i,t-1}$ &  &  & -5e-04 & 3e-04 & -0.0322 & -0.0239 &  &  & -0.0949** & -0.0871* & 0.0056 & 0.0047 \\ 
&  &  & (0.0048) & (0.0055) & (0.031) & (0.0325) &  &  & (0.0325) & (0.0349) & (0.0048) & (0.0048) \\ 
$Spill(A)^{\tau,up}_{i,t-1}$ & 3.881* & 4.756* &  &  &  &  & 1.78*** & 1.367*** & 9.629*** & 8.218*** & 1.481*** & 1.161*** \\ 
& (1.583) & (2.019) &  &  &  &  & (0.384) & (0.3329) & (2.014) & (2.122) & (0.3058) & (0.277) \\ 
$Spill(A)^{\tau,dw}_{i,t-1}$ & -0.0639 & 0.0259 &  &  &  &  & -0.1079*** & -0.094*** & -0.0014 & 0.1792 & -0.1092*** & -0.1003*** \\ 
& (0.1313) & (0.1377) &  &  &  &  & (0.0207) & (0.0206) & (0.1585) & (0.1504) & (0.0183) & (0.0217) \\ 
\hline \rule{0pt}{1.075\normalbaselineskip}  AR(1) & 0 & 0 & 0 & 0 & 0 & 0 & 0 & 0 & 0 & 0 & 0 & 0 \\ 
AR(2) & 0.9054 & 0.9888 & 0.044 & 0.0938 & 0.9755 & 0.8194 & 0.0434 & 0.0567 & 0.8279 & 0.9325 & 0.0511 & 0.0589 \\ 
Sargan & 0 & 0 & 0 & 0.0016 & 0 & 0 & 0 & 0.0017 & 0 & 0 & 0 & 1e-04 \\ 
Controls &  & Y &  & Y &  & Y &  & Y &  & Y &  & Y \\ 
$R^2$ & 0.916 & 0.8813 & 0.9984 & 0.998 & 0.9181 & 0.9078 & 0.9977 & 0.998 & 0.8858 & 0.8791 & 0.9982 & 0.9984 \\ 
\hline
\hline
\end{tabular}
\endgroup

\end{myresizeenv}

\vspace{0.25cm}

\justifying \footnotesize

\noindent
Notes: The table shows the regression results of output $A_{i,t}^{\mu}$ and patents $A_{i,t}^{\tau}$ on demand-pull and technology-push effects. The estimation is based on a two-ways Blundell-Bond (BB) system GMM model using a one-step estimation procedure. Spillovers are calculated on the basis of first-order links. Variables measured in monetary terms are deflated using the industry level price deflators for the value of shipment obtained from the NBER-productivity database \citep{becker2013nber}. Instruments are collapsed to avoid instrument proliferation. To cope with skewness and to obtain tractable coefficients, most variables are pre-processed (taking logs, removing outliers, scaling). Data in logs are $A^{\alpha }_{i,t}$, $PR^{\alpha, d }_{i,t}$, $Spill(A)^{\alpha, d }_{i,t}$, $L_{i,t}$,  $(K/L)_{i,t}$, $(I/L)_{i,t}$, $W_{i,t}$, $(K/L)^{}_{ij,t-1}$, $(E/L)^{}_{ij,t-1}$, $(M/L)^{}_{ij,t-1}$, $W^{P}_{ij,t-1}$ with $\alpha = \mu, \tau$ and $d = up, dw$. A detailed description of the transformations and descriptive statistics of the regression data before and after the transformations are provided in \ref{app:data}. 
The rows AR(1), AR(2), and Sargan show the test statistics of the specification tests, i.e. testing for first- and second-order autocorrelation and the results of a Sargan test for validity of instruments \citep[see][]{roodman2009xtabond2}. 
The analysis covers the subset of innovation-intensive industries defined by an above median innovation intensity $(A^\tau_{i}/A^\mu_i)$. 
\end{table}

\subsubsection{Sectors with a low innovation-intensity}
\begin{table}[H]
\begin{myresizeenv}
\begingroup
\caption{Demand-pull and technology-push effects in Non-Innovation-intensive industries}						
\label{tab:real_sum_output|citation_weighted_patent_stock_inno_intensive_FALSE_all_years_sys_1step_wControls}

\begin{tabular}{|l|cccc|cccc|cccc|}
\hline \hline  \rule{0pt}{1.075\normalbaselineskip}  
& \multicolumn{4}{c|}{\scriptsize \ul{Type 1}} & \multicolumn{4}{c|}{\scriptsize \ul{Type 2}} & \multicolumn{4}{c|}{\scriptsize \ul{Both}}\\  & \multicolumn{2}{c}{\scriptsize $\tau \rightarrow \mu$} & \multicolumn{2}{c|}{\scriptsize $\mu \rightarrow \tau$}
& \multicolumn{2}{c}{\scriptsize Market} & \multicolumn{2}{c|}{\scriptsize Innovation}
& \multicolumn{4}{c|}{}\\
\hline  \rule{0pt}{1.075\normalbaselineskip}    & $A^{\mu}_{i,t}$ & $A^{\mu}_{i,t}$ & $A^{\tau}_{i,t}$ & $A^{\tau}_{i,t}$ & $A^{\mu}_{i,t}$ & $A^{\mu}_{i,t}$ & $A^{\tau}_{i,t}$ & $A^{\tau}_{i,t}$ & $A^{\mu}_{i,t}$ & $A^{\mu}_{i,t}$ & $A^{\tau}_{i,t}$ & $A^{\tau}_{i,t}$ \\ 
\rule{0pt}{1.075\normalbaselineskip}  & (1) & (2) & (3) & (4) & (5) & (6) & (7) & (8) & (9) & (10) & (11) & (12) \\ 
\hline \rule{0pt}{1.075\normalbaselineskip}   $A^{\mu}_{i,t-1}$ & 0.5031*** & 0.4837*** & 0.0272 & -0.0056 & 0.7667*** & 0.7255*** &  &  & 0.6917*** & 0.7095*** & 0.0506 & -7e-04 \\ 
& (0.0517) & (0.0609) & (0.0379) & (0.042) & (0.0981) & (0.0996) &  &  & (0.1026) & (0.102) & (0.0463) & (0.0457) \\ 
$A^{\tau}_{i,t-1}$ & 0.2706* & 0.3198. & 1.038*** & 1.021*** &  &  & 1.186*** & 1.067*** & 0.1219 & 0.1821 & 1.184*** & 1.098*** \\ 
& (0.1294) & (0.1677) & (0.0313) & (0.0341) &  &  & (0.0864) & (0.0689) & (0.1186) & (0.1288) & (0.0793) & (0.0577) \\ 
$PR^{\mu,up}_{i,t-1}$ &  &  & 0.0038 & 0.0048 & -0.0732*** & -0.0752*** &  &  & -0.069** & -0.0684** & 3e-04 & 0.0033 \\ 
&  &  & (0.0095) & (0.0099) & (0.0216) & (0.0198) &  &  & (0.0231) & (0.0234) & (0.0108) & (0.0099) \\ 
$PR^{\mu,dw}_{i,t-1}$ &  &  & 0.003 & 0.0054 & -0.0341. & -0.0416* &  &  & -0.0389* & -0.0243 & 0.0056 & 0.0067 \\ 
&  &  & (0.006) & (0.0087) & (0.0186) & (0.0188) &  &  & (0.0185) & (0.0226) & (0.0074) & (0.0094) \\ 
$PR^{\tau,dw}_{i,t-1}$ & 0.0542 & 0.2297 &  &  &  &  & -0.1774. & 0.0513 & 0.4148* & 0.4726** & -0.1828. & 0.0132 \\ 
& (0.1453) & (0.1895) &  &  &  &  & (0.1071) & (0.0713) & (0.1701) & (0.1777) & (0.0941) & (0.0641) \\ 
$Spill(A)^{\mu,up}_{i,t-1}$ &  &  & -0.0156 & -0.0052 & 0.0922** & 0.0979** &  &  & 0.0717* & 0.0506 & -0.0225 & -0.009 \\ 
&  &  & (0.0129) & (0.0145) & (0.0318) & (0.0298) &  &  & (0.0311) & (0.0331) & (0.0142) & (0.0159) \\ 
$Spill(A)^{\mu,dw}_{i,t-1}$ &  &  & -0.0033 & 0.0105 & -0.0566 & -0.0359 &  &  & -0.0784 & -0.071 & -0.0086 & 0.0075 \\ 
&  &  & (0.0116) & (0.0114) & (0.0518) & (0.0433) &  &  & (0.0509) & (0.0487) & (0.0126) & (0.0113) \\ 
$Spill(A)^{\tau,up}_{i,t-1}$ & 2.748. & 4.546* &  &  &  &  & 2.408*** & 0.4972 & 1.289 & 3.197. & 2.078*** & 0.8024. \\ 
& (1.47) & (1.92) &  &  &  &  & (0.6854) & (0.7851) & (1.564) & (1.724) & (0.5743) & (0.4467) \\ 
$Spill(A)^{\tau,dw}_{i,t-1}$ & -0.0611 & -0.0415 &  &  &  &  & -0.0322 & -0.0538* & -0.1134. & -0.0983 & -0.0357 & -0.0475* \\ 
& (0.0574) & (0.0679) &  &  &  &  & (0.0303) & (0.0224) & (0.0591) & (0.0625) & (0.029) & (0.0216) \\ 
\hline \rule{0pt}{1.075\normalbaselineskip}  AR(1) & 0 & 0 & 0.0011 & 0.0022 &  & 0 & 2e-04 & 0.0031 &  & 0 & 9e-04 & 0.0029 \\ 
AR(2) & 0.9332 & 0.7449 & 0.6352 & 0.6589 & 0.9511 & 0.9555 & 0.2423 & 0.5757 & 0.9243 & 0.9263 & 0.3571 & 0.496 \\ 
Sargan & 0 & 4e-04 & 1e-04 & 0.0066 & 0 & 0 & 0 & 0 & 0 & 0 & 0 & 0 \\ 
Controls &  & Y &  & Y &  & Y &  & Y &  & Y &  & Y \\ 
$R^2$ & 0.9525 & 0.9232 & 0.9905 & 0.9898 & 0.9491 & 0.9533 & 0.9852 & 0.9879 & 0.9472 & 0.9335 & 0.9859 & 0.9892 \\ 
\hline
\hline
\end{tabular}
\endgroup
\end{myresizeenv}

\vspace{0.25cm}

\justifying \footnotesize

\noindent
Notes: The table shows the regression results of output $A_{i,t}^{\mu}$ and patents $A_{i,t}^{\tau}$ on demand-pull and technology-push effects. The estimation is based on a two-ways Blundell-Bond (BB) system GMM model using a one-step estimation procedure. Spillovers are calculated on the basis of first-order links. Variables measured in monetary terms are deflated using the industry level price deflators for the value of shipment obtained from the NBER-productivity database \citep{becker2013nber}. Instruments are collapsed to avoid instrument proliferation. To cope with skewness and to obtain tractable coefficients, most variables are pre-processed (taking logs, removing outliers, scaling). Data in logs are $A^{\alpha }_{i,t}$, $PR^{\alpha, d }_{i,t}$, $Spill(A)^{\alpha, d }_{i,t}$, $L_{i,t}$,  $(K/L)_{i,t}$, $(I/L)_{i,t}$, $W_{i,t}$, $(K/L)^{}_{ij,t-1}$, $(E/L)^{}_{ij,t-1}$, $(M/L)^{}_{ij,t-1}$, $W^{P}_{ij,t-1}$ with $\alpha = \mu, \tau$ and $d = up, dw$. A detailed description of the transformations and descriptive statistics of the regression data before and after the transformations are provided in \ref{app:data}. 
The rows AR(1), AR(2), and Sargan show the test statistics of the specification tests, i.e. testing for first- and second-order autocorrelation and the results of a Sargan test for validity of instruments \citep[see][]{roodman2009xtabond2}. 
The analysis covers the subset of innovation-intensive industries defined by a below median innovation intensity $(A^\tau_{i}/A^\mu_i)$.
\end{table}

\subsection{Market size}
\subsubsection{Big industries}
\begin{table}[H]
\begin{myresizeenv}

\caption{Demand-pull and technology-push effects in Big industries}
\label{tab:real_sum_output|citation_weighted_patent_stock_big_industry_TRUE_all_years_sys_1step_wControls}
\begingroup
\begin{tabular}{|l|cccc|cccc|cccc|}
\hline \hline  \rule{0pt}{1.075\normalbaselineskip}  
& \multicolumn{4}{c|}{\scriptsize \ul{Type 1}} & \multicolumn{4}{c|}{\scriptsize \ul{Type 2}} & \multicolumn{4}{c|}{\scriptsize \ul{Both}}\\  & \multicolumn{2}{c}{\scriptsize $\tau \rightarrow \mu$} & \multicolumn{2}{c|}{\scriptsize $\mu \rightarrow \tau$}
& \multicolumn{2}{c}{\scriptsize Market} & \multicolumn{2}{c|}{\scriptsize Innovation}
& \multicolumn{4}{c|}{}\\
\hline  \rule{0pt}{1.075\normalbaselineskip}    & $A^{\mu}_{i,t}$ & $A^{\mu}_{i,t}$ & $A^{\tau}_{i,t}$ & $A^{\tau}_{i,t}$ & $A^{\mu}_{i,t}$ & $A^{\mu}_{i,t}$ & $A^{\tau}_{i,t}$ & $A^{\tau}_{i,t}$ & $A^{\mu}_{i,t}$ & $A^{\mu}_{i,t}$ & $A^{\tau}_{i,t}$ & $A^{\tau}_{i,t}$ \\ 
\rule{0pt}{1.075\normalbaselineskip}  & (1) & (2) & (3) & (4) & (5) & (6) & (7) & (8) & (9) & (10) & (11) & (12) \\ 
\hline \rule{0pt}{1.075\normalbaselineskip}   $A^{\mu}_{i,t-1}$ & 0.5116*** & 0.4838*** & -0.0423 & -0.0686. & 0.5763*** & 0.5674*** &  &  & 0.3642* & 0.5871*** & -0.0245 & -0.0573 \\ 
& (0.0619) & (0.0662) & (0.033) & (0.041) & (0.1439) & (0.0975) &  &  & (0.1828) & (0.1029) & (0.0442) & (0.0379) \\ 
$A^{\tau}_{i,t-1}$ & 0.1732 & 0.0343 & 1.058*** & 1.041*** &  &  & 1.11*** & 1.144*** & -0.1267 & 0.0988 & 1.14*** & 1.15*** \\ 
& (0.1068) & (0.0975) & (0.0164) & (0.021) &  &  & (0.1072) & (0.0682) & (0.201) & (0.1008) & (0.0829) & (0.0635) \\ 
$PR^{\mu,up}_{i,t-1}$ &  &  & 0.0133. & 0.0106 & -0.0287 & -0.0361* &  &  & -0.0029 & -0.0319. & 0.0088 & 0.0062 \\ 
&  &  & (0.0073) & (0.0069) & (0.0215) & (0.018) &  &  & (0.03) & (0.0183) & (0.0067) & (0.0064) \\ 
$PR^{\mu,dw}_{i,t-1}$ &  &  & 0.0136*** & 0.0103* & 0.0061 & -0.0069 &  &  & 0.0189 & 0.0026 & 0.0092. & 0.0096. \\ 
&  &  & (0.0038) & (0.005) & (0.0173) & (0.0149) &  &  & (0.027) & (0.0159) & (0.0048) & (0.0056) \\ 
$PR^{\tau,dw}_{i,t-1}$ & 0.0077 & 0.0452 &  &  &  &  & 0.0491 & -0.0061 & 0.254. & 0.0673 & 0.0277 & -0.0013 \\ 
& (0.0604) & (0.0617) &  &  &  &  & (0.0525) & (0.0278) & (0.1462) & (0.0448) & (0.0381) & (0.0262) \\ 
$Spill(A)^{\mu,up}_{i,t-1}$ &  &  & -0.0062 & 0.0056 & 0.0205 & 0.035 &  &  & 0.0102 & 0.0026 & -0.0146 & -0.0034 \\ 
&  &  & (0.012) & (0.0126) & (0.0245) & (0.0259) &  &  & (0.0319) & (0.0268) & (0.0127) & (0.0128) \\ 
$Spill(A)^{\mu,dw}_{i,t-1}$ &  &  & 0.0086 & 0.0177. & -0.0291 & -0.0258 &  &  & -0.0274 & -0.042 & 4e-04 & 0.0132 \\ 
&  &  & (0.0074) & (0.0099) & (0.0223) & (0.0238) &  &  & (0.0249) & (0.0258) & (0.0079) & (0.009) \\ 
$Spill(A)^{\tau,up}_{i,t-1}$ & -1.422 & -0.3816 &  &  &  &  & 1.308** & 0.8623** & -0.1246 & 0.9263 & 1.116** & 0.7302* \\ 
& (0.9352) & (1.095) &  &  &  &  & (0.4987) & (0.3337) & (1.273) & (0.9836) & (0.3849) & (0.3005) \\ 
$Spill(A)^{\tau,dw}_{i,t-1}$ & -0.1058* & -0.0611 &  &  &  &  & -0.0896** & -0.0703** & -0.137. & -0.1369** & -0.09*** & -0.0756*** \\ 
& (0.0436) & (0.0515) &  &  &  &  & (0.0275) & (0.024) & (0.073) & (0.0448) & (0.0246) & (0.0218) \\ 
\hline \rule{0pt}{1.075\normalbaselineskip}  AR(1) & 0 & 0 & 0 & 0 & 0 & 0 & 0.0029 & 0 & 0 & 0 & 0.0011 & 0 \\ 
AR(2) & 0.0196 & 0.024 & 0.0618 & 0.0387 & 0.0209 & 0.006 & 0.1697 & 0.0776 & 0.0228 & 0.0064 & 0.1741 & 0.1719 \\ 
Sargan & 1e-04 & 0 & 0.0233 & 0.1511 & 0 & 0 & 0 & 1e-04 & 0 & 0 & 0 & 0 \\ 
Controls &  & Y &  & Y &  & Y &  & Y &  & Y &  & Y \\ 
$R^2$ & 0.9772 & 0.9725 & 0.9966 & 0.9961 & 0.9768 & 0.977 & 0.9949 & 0.996 & 0.9602 & 0.9737 & 0.9956 & 0.9961 \\ 
\hline
\hline
\end{tabular}
\endgroup
\end{myresizeenv}

\vspace{0.25cm}

\justifying \footnotesize

\noindent
Notes: The table shows the regression results of output $A_{i,t}^{\mu}$ and patents $A_{i,t}^{\tau}$ on demand-pull and technology-push effects. The estimation is based on a two-ways Blundell-Bond (BB) system GMM model using a one-step estimation procedure. Spillovers are calculated on the basis of first-order links. Variables measured in monetary terms are deflated using the industry level price deflators for the value of shipment obtained from the NBER-productivity database \citep{becker2013nber}. Instruments are collapsed to avoid instrument proliferation. To cope with skewness and to obtain tractable coefficients, most variables are pre-processed (taking logs, removing outliers, scaling). Data in logs are $A^{\alpha }_{i,t}$, $PR^{\alpha, d }_{i,t}$, $Spill(A)^{\alpha, d }_{i,t}$, $L_{i,t}$,  $(K/L)_{i,t}$, $(I/L)_{i,t}$, $W_{i,t}$, $(K/L)^{}_{ij,t-1}$, $(E/L)^{}_{ij,t-1}$, $(M/L)^{}_{ij,t-1}$, $W^{P}_{ij,t-1}$ with $\alpha = \mu, \tau$ and $d = up, dw$. A detailed description of the transformations and descriptive statistics of the regression data before and after the transformations are provided in \ref{app:data}. 
The rows AR(1), AR(2), and Sargan show the test statistics of the specification tests, i.e. testing for first- and second-order autocorrelation and the results of a Sargan test for validity of instruments \citep[see][]{roodman2009xtabond2}. 
The analysis covers the subset of big industries defined by an above median industry size $(A^\mu_i)$.
\end{table}

\subsubsection{Small industries}
\begin{table}[H]
\begin{myresizeenv}
\caption{Demand-pull and technology-push effects in Small industries}
\label{tab:real_sum_output|citation_weighted_patent_stock_big_industry_FALSE_all_years_sys_1step_wControls}

\begingroup
\begin{tabular}{|l|cccc|cccc|cccc|}
\hline \hline  \rule{0pt}{1.075\normalbaselineskip}  
& \multicolumn{4}{c|}{\scriptsize \ul{Type 1}} & \multicolumn{4}{c|}{\scriptsize \ul{Type 2}} & \multicolumn{4}{c|}{\scriptsize \ul{Both}}\\  & \multicolumn{2}{c}{\scriptsize $\tau \rightarrow \mu$} & \multicolumn{2}{c|}{\scriptsize $\mu \rightarrow \tau$}
& \multicolumn{2}{c}{\scriptsize Market} & \multicolumn{2}{c|}{\scriptsize Innovation}
& \multicolumn{4}{c|}{}\\
\hline  \rule{0pt}{1.075\normalbaselineskip}    & $A^{\mu}_{i,t}$ & $A^{\mu}_{i,t}$ & $A^{\tau}_{i,t}$ & $A^{\tau}_{i,t}$ & $A^{\mu}_{i,t}$ & $A^{\mu}_{i,t}$ & $A^{\tau}_{i,t}$ & $A^{\tau}_{i,t}$ & $A^{\mu}_{i,t}$ & $A^{\mu}_{i,t}$ & $A^{\tau}_{i,t}$ & $A^{\tau}_{i,t}$ \\ 
\rule{0pt}{1.075\normalbaselineskip}  & (1) & (2) & (3) & (4) & (5) & (6) & (7) & (8) & (9) & (10) & (11) & (12) \\ 
\hline \rule{0pt}{1.075\normalbaselineskip}   $A^{\mu}_{i,t-1}$ & 0.5073*** & 0.5413*** & 0.0229 & 0.02 & 0.7057*** & 0.6489*** &  &  & 0.579*** & 0.6246*** & 0.0158 & -2e-04 \\ 
& (0.047) & (0.042) & (0.0186) & (0.0199) & (0.0571) & (0.0636) &  &  & (0.0774) & (0.0705) & (0.0215) & (0.0206) \\ 
$A^{\tau}_{i,t-1}$ & 0.0663 & -0.0403 & 1.058*** & 1.042*** &  &  & 1.087*** & 1.102*** & 0.139 & 0.0084 & 1.126*** & 1.104*** \\ 
& (0.3422) & (0.2197) & (0.0225) & (0.0221) &  &  & (0.1147) & (0.0707) & (0.4305) & (0.2617) & (0.0914) & (0.0544) \\ 
$PR^{\mu,up}_{i,t-1}$ &  &  & -9e-04 & 1e-04 & -0.0564** & -0.0548** &  &  & -0.0206 & -0.0384. & 0.0017 & 0.001 \\ 
&  &  & (0.0081) & (0.009) & (0.0204) & (0.0197) &  &  & (0.0283) & (0.023) & (0.009) & (0.009) \\ 
$PR^{\mu,dw}_{i,t-1}$ &  &  & 0.0039 & 0.0024 & 0.0322 & 0.0065 &  &  & 0.0731* & 0.0357 & 0.0056 & 0.0027 \\ 
&  &  & (0.0059) & (0.0069) & (0.0351) & (0.0272) &  &  & (0.0371) & (0.0278) & (0.0068) & (0.0065) \\ 
$PR^{\tau,dw}_{i,t-1}$ & 0.1173 & 0.0977 &  &  &  &  & 0.033 & 0.0158 & 0.0895 & 0.07 & 0.0083 & 0.0093 \\ 
& (0.1602) & (0.099) &  &  &  &  & (0.0518) & (0.0362) & (0.1947) & (0.1187) & (0.0377) & (0.0248) \\ 
$Spill(A)^{\mu,up}_{i,t-1}$ &  &  & -0.0161 & -0.0166 & 0.0177 & 0.0184 &  &  & -0.0082 & -0.0046 & -0.0163 & -0.0138 \\ 
&  &  & (0.0103) & (0.0105) & (0.0272) & (0.0303) &  &  & (0.0333) & (0.0342) & (0.0102) & (0.0105) \\ 
$Spill(A)^{\mu,dw}_{i,t-1}$ &  &  & -0.0017 & -0.0023 & -0.0837** & -0.0769* &  &  & -0.0923* & -0.0756* & -0.0038 & 0.0051 \\ 
&  &  & (0.0085) & (0.0083) & (0.0314) & (0.0315) &  &  & (0.0439) & (0.0357) & (0.0086) & (0.0085) \\ 
$Spill(A)^{\tau,up}_{i,t-1}$ & 7.469*** & 5.942** &  &  &  &  & 1.95** & 0.9831 & 9.633*** & 7.002** & 1.542** & 0.7435 \\ 
& (1.958) & (2.071) &  &  &  &  & (0.7049) & (0.7466) & (2.243) & (2.169) & (0.5678) & (0.598) \\ 
$Spill(A)^{\tau,dw}_{i,t-1}$ & -0.0316 & -0.0357 &  &  &  &  & -0.0486 & -0.0553** & -0.0288 & -0.0259 & -0.0547* & -0.0554** \\ 
& (0.0935) & (0.0722) &  &  &  &  & (0.0321) & (0.0211) & (0.111) & (0.0754) & (0.0272) & (0.0188) \\ 
\hline \rule{0pt}{1.075\normalbaselineskip}  AR(1) &  & 0 & 0.0071 & 0.0056 & 0 & 0 & 0.0846 & 0.0059 &  & 0 & 0.0308 & 0.0048 \\ 
AR(2) & 0.9893 & 0.9727 & 0.498 & 0.4887 & 0.6033 & 0.5663 & 0.5482 & 0.3716 & 0.9214 & 0.7744 & 0.4707 & 0.3657 \\ 
Sargan & 1e-04 & 0 & 1e-04 & 2e-04 & 0 & 0 & 0 & 0 & 0 & 0 & 0 & 0 \\ 
Controls &  & Y &  & Y &  & Y &  & Y &  & Y &  & Y \\ 
$R^2$ & 0.8655 & 0.8842 & 0.9948 & 0.9948 & 0.895 & 0.8986 & 0.9929 & 0.9944 & 0.8424 & 0.8825 & 0.9937 & 0.9947 \\ 
\hline
\hline
\end{tabular}
\endgroup
\end{myresizeenv}

\vspace{0.25cm}

\justifying \footnotesize

\noindent
Notes: The table shows the regression results of output $A_{i,t}^{\mu}$ and patents $A_{i,t}^{\tau}$ on demand-pull and technology-push effects. The estimation is based on a two-ways Blundell-Bond (BB) system GMM model using a one-step estimation procedure. Spillovers are calculated on the basis of first-order links. Variables measured in monetary terms are deflated using the industry level price deflators for the value of shipment obtained from the NBER-productivity database \citep{becker2013nber}. Instruments are collapsed to avoid instrument proliferation. To cope with skewness and to obtain tractable coefficients, most variables are pre-processed (taking logs, removing outliers, scaling). Data in logs are $A^{\alpha }_{i,t}$, $PR^{\alpha, d }_{i,t}$, $Spill(A)^{\alpha, d }_{i,t}$, $L_{i,t}$,  $(K/L)_{i,t}$, $(I/L)_{i,t}$, $W_{i,t}$, $(K/L)^{}_{ij,t-1}$, $(E/L)^{}_{ij,t-1}$, $(M/L)^{}_{ij,t-1}$, $W^{P}_{ij,t-1}$ with $\alpha = \mu, \tau$ and $d = up, dw$. A detailed description of the transformations and descriptive statistics of the regression data before and after the transformations are provided in \ref{app:data}. 
The rows AR(1), AR(2), and Sargan show the test statistics of the specification tests, i.e. testing for first- and second-order autocorrelation and the results of a Sargan test for validity of instruments \citep[see][]{roodman2009xtabond2}. 
The analysis covers the subset of small industries defined by a below median industry size $(A^\mu_i)$.  
\end{table}

\subsection{Pavitt sector groups}

\subsubsection{Science-based sectors}
\begin{table}[H]
\begin{myresizeenv}
\begingroup
\begin{tabular}{|l|cccc|cccc|cccc|}
\hline \hline  \rule{0pt}{1.075\normalbaselineskip}  
& \multicolumn{4}{c|}{\scriptsize \ul{Type 1}} & \multicolumn{4}{c|}{\scriptsize \ul{Type 2}} & \multicolumn{4}{c|}{\scriptsize \ul{Both}}\\  & \multicolumn{2}{c}{\scriptsize $\tau \rightarrow \mu$} & \multicolumn{2}{c|}{\scriptsize $\mu \rightarrow \tau$}
& \multicolumn{2}{c}{\scriptsize Market} & \multicolumn{2}{c|}{\scriptsize Innovation}
& \multicolumn{4}{c|}{}\\
\hline  \rule{0pt}{1.075\normalbaselineskip}    & $A^{\mu}_{i,t}$ & $A^{\mu}_{i,t}$ & $A^{\tau}_{i,t}$ & $A^{\tau}_{i,t}$ & $A^{\mu}_{i,t}$ & $A^{\mu}_{i,t}$ & $A^{\tau}_{i,t}$ & $A^{\tau}_{i,t}$ & $A^{\mu}_{i,t}$ & $A^{\mu}_{i,t}$ & $A^{\tau}_{i,t}$ & $A^{\tau}_{i,t}$ \\ 
\rule{0pt}{1.075\normalbaselineskip}  & (1) & (2) & (3) & (4) & (5) & (6) & (7) & (8) & (9) & (10) & (11) & (12) \\ 
\hline \rule{0pt}{1.075\normalbaselineskip}   $A^{\mu}_{i,t-1}$ & 0.5287*** & 0.5553*** & 0.053 & 0.0225 & 0.5106*** & 0.6433*** &  &  & 0.5957*** & 0.642*** & 0.0645 & 0.0443 \\ 
& (0.0779) & (0.1083) & (0.0556) & (0.0426) & (0.1507) & (0.1207) &  &  & (0.179) & (0.1725) & (0.0686) & (0.0391) \\ 
$A^{\tau}_{i,t-1}$ & 0.2759 & 0.3663 & 1.075*** & 1.06*** &  &  & 1.257*** & 1.286*** & 0.0637 & 0.3612 & 1.31*** & 1.246*** \\ 
& (0.3643) & (0.3226) & (0.0247) & (0.047) &  &  & (0.1724) & (0.1029) & (0.5151) & (0.3592) & (0.0984) & (0.0958) \\ 
$PR^{\mu,up}_{i,t-1}$ &  &  & -0.0158 & -0.0119 & 0.0372 & -0.0177 &  &  & 0.0304 & 0.0046 & -0.0093 & -0.0109 \\ 
&  &  & (0.0107) & (0.0102) & (0.0413) & (0.0353) &  &  & (0.0512) & (0.038) & (0.012) & (0.0088) \\ 
$PR^{\mu,dw}_{i,t-1}$ &  &  & 0.0018 & 5e-04 & 0.0249 & 0.0103 &  &  & -0.0159 & -0.0275 & -0.0059 & -0.0039 \\ 
&  &  & (0.0059) & (0.0065) & (0.025) & (0.0213) &  &  & (0.0319) & (0.0266) & (0.0061) & (0.0065) \\ 
$PR^{\tau,dw}_{i,t-1}$ & -0.0979 & -0.0828 &  &  &  &  & 0.0176 & -0.0351 & 0.2431 & 0.0572 & -0.0447 & -0.0311 \\ 
& (0.1809) & (0.1125) &  &  &  &  & (0.0882) & (0.0371) & (0.1901) & (0.1389) & (0.0548) & (0.0352) \\ 
$Spill(A)^{\mu,up}_{i,t-1}$ &  &  & -0.0186 & -0.0144 & 0.0923* & 0.0346 &  &  & 0.1221* & 0.0872 & -0.0052 & -0.0016 \\ 
&  &  & (0.0144) & (0.0163) & (0.0459) & (0.0555) &  &  & (0.0591) & (0.0574) & (0.0133) & (0.0137) \\ 
$Spill(A)^{\mu,dw}_{i,t-1}$ &  &  & 0.0116 & 0.0138 & -0.0604 & -0.0571 &  &  & -0.0918 & -0.0988. & 0.0024 & 0.0101 \\ 
&  &  & (0.0164) & (0.0159) & (0.0557) & (0.0397) &  &  & (0.064) & (0.0525) & (0.0173) & (0.0135) \\ 
$Spill(A)^{\tau,up}_{i,t-1}$ & -4.111 & 1.601 &  &  &  &  & 3.041. & 1.394 & 6.506. & 5.908. & 1.533* & 1.121 \\ 
& (3.27) & (3.121) &  &  &  &  & (1.686) & (0.9076) & (3.857) & (3.244) & (0.7209) & (0.7515) \\ 
$Spill(A)^{\tau,dw}_{i,t-1}$ & -0.177. & -0.0704 &  &  &  &  & -0.1148* & -0.1419*** & -0.2484 & -0.3146* & -0.1134*** & -0.1316*** \\ 
& (0.1058) & (0.1073) &  &  &  &  & (0.0452) & (0.0297) & (0.1984) & (0.1255) & (0.0242) & (0.035) \\ 
\hline \rule{0pt}{1.075\normalbaselineskip}  AR(1) & 7e-04 & 0.0048 & 0.0225 & 0.004 & 0 & 0 &  & 0.0015 & 0 & 1e-04 &  &  \\ 
AR(2) & 0.8111 & 0.8123 & 0.0335 & 0.0074 & 0.8336 & 0.5772 & 0.5283 & 0.4236 & 0.8311 & 0.9539 & 0.1171 & 0.2701 \\ 
Sargan & 0.0024 & 0.0016 & 0.0155 & 0.0552 & 4e-04 & 0.0012 & 0.0015 & 5e-04 & 0.002 & 0.0018 & 0.0012 & 0.0019 \\ 
Controls &  & Y &  & Y &  & Y &  & Y &  & Y &  & Y \\ 
$R^2$ & 0.942 & 0.9467 & 0.9974 & 0.9969 & 0.9555 & 0.958 & 0.9945 & 0.9964 & 0.9304 & 0.9472 & 0.9969 & 0.9969 \\ 
\hline
\hline
\end{tabular}
\endgroup
\caption{Demand-pull and technology-push effects in Science-based sectors.}
\label{tab:real_sum_output|citation_weighted_patent_stock_Science_based_all_years_sys_1step_wControls}
\end{myresizeenv}

\vspace{0.25cm}

\justifying \footnotesize

\noindent 
Notes: The table shows the regression results of output $A_{i,t}^{\mu}$ and patents $A_{i,t}^{\tau}$ on demand-pull and technology-push effects. The estimation is based on a two-ways Blundell-Bond (BB) system GMM model using a one-step estimation procedure. Spillovers are calculated on the basis of first-order links. Variables measured in monetary terms are deflated using the industry level price deflators for the value of shipment obtained from the NBER-productivity database \citep{becker2013nber}. Instruments are collapsed to avoid instrument proliferation. To cope with skewness and to obtain tractable coefficients, most variables are pre-processed (taking logs, removing outliers, scaling). Data in logs are $A^{\alpha }_{i,t}$, $PR^{\alpha, d }_{i,t}$, $Spill(A)^{\alpha, d }_{i,t}$, $L_{i,t}$,  $(K/L)_{i,t}$, $(I/L)_{i,t}$, $W_{i,t}$, $(K/L)^{}_{ij,t-1}$, $(E/L)^{}_{ij,t-1}$, $(M/L)^{}_{ij,t-1}$, $W^{P}_{ij,t-1}$ with $\alpha = \mu, \tau$ and $d = up, dw$. A detailed description of the transformations and descriptive statistics of the regression data before and after the transformations are provided in \ref{app:data}. 
The rows AR(1), AR(2), and Sargan show the test statistics of the specification tests, i.e. testing for first- and second-order autocorrelation and the results of a Sargan test for validity of instruments \citep[see][]{roodman2009xtabond2}. 
The analysis covers the subset of Science-based industries (see \citep{bogliacino2016pavitt}). 
\end{table}

\subsubsection{Suppliers-dominated sectors}
\begin{table}[H]
\begin{myresizeenv}
\caption{Demand-pull and technology-push effects in Supplier dominated sectors.} 
\label{tab:real_sum_output|citation_weighted_patent_stock_Suppliers_dominated_all_years_sys_1step_wControls}
\begingroup
\begin{tabular}{|l|cccc|cccc|cccc|}
\hline \hline  \rule{0pt}{1.075\normalbaselineskip}  
& \multicolumn{4}{c|}{\scriptsize \ul{Type 1}} & \multicolumn{4}{c|}{\scriptsize \ul{Type 2}} & \multicolumn{4}{c|}{\scriptsize \ul{Both}}\\  & \multicolumn{2}{c}{\scriptsize $\tau \rightarrow \mu$} & \multicolumn{2}{c|}{\scriptsize $\mu \rightarrow \tau$}
& \multicolumn{2}{c}{\scriptsize Market} & \multicolumn{2}{c|}{\scriptsize Innovation}
& \multicolumn{4}{c|}{}\\
\hline  \rule{0pt}{1.075\normalbaselineskip}    & $A^{\mu}_{i,t}$ & $A^{\mu}_{i,t}$ & $A^{\tau}_{i,t}$ & $A^{\tau}_{i,t}$ & $A^{\mu}_{i,t}$ & $A^{\mu}_{i,t}$ & $A^{\tau}_{i,t}$ & $A^{\tau}_{i,t}$ & $A^{\mu}_{i,t}$ & $A^{\mu}_{i,t}$ & $A^{\tau}_{i,t}$ & $A^{\tau}_{i,t}$ \\ 
\rule{0pt}{1.075\normalbaselineskip}  & (1) & (2) & (3) & (4) & (5) & (6) & (7) & (8) & (9) & (10) & (11) & (12) \\ 
\hline \rule{0pt}{1.075\normalbaselineskip}   $A^{\mu}_{i,t-1}$ & 0.6488*** & 0.6351*** & 0.0571* & 0.0453* & 0.8444*** & 0.8141*** &  &  & 0.6259*** & 0.6898*** & 0.0431 & 0.0065 \\ 
& (0.0514) & (0.0676) & (0.0256) & (0.0224) & (0.0744) & (0.0882) &  &  & (0.0936) & (0.0942) & (0.0303) & (0.0253) \\ 
$A^{\tau}_{i,t-1}$ & -0.4345 & 0.0346 & 1.044*** & 1.054*** &  &  & 1.187*** & 1.144*** & -0.2396 & -0.0629 & 1.208*** & 1.155*** \\ 
& (0.3649) & (0.3119) & (0.02) & (0.0295) &  &  & (0.0809) & (0.0643) & (0.3478) & (0.3406) & (0.0743) & (0.0705) \\ 
$PR^{\mu,up}_{i,t-1}$ &  &  & -0.013 & -0.0127. & -0.0112 & -0.0362 &  &  & 0.0398 & 0.0144 & -0.0063 & -0.0061 \\ 
&  &  & (0.0086) & (0.0075) & (0.0278) & (0.0268) &  &  & (0.0368) & (0.0325) & (0.0082) & (0.0072) \\ 
$PR^{\mu,dw}_{i,t-1}$ &  &  & -0.0078 & -0.0075 & -0.0079 & -0.0515. &  &  & 0.0323 & 0.01 & -0.0028 & -0.0023 \\ 
&  &  & (0.0062) & (0.0068) & (0.0199) & (0.0308) &  &  & (0.0298) & (0.0353) & (0.0062) & (0.0081) \\ 
$PR^{\tau,dw}_{i,t-1}$ & 0.4023* & 0.1491 &  &  &  &  & 0.0205 & 0.0253 & 0.3914* & 0.2662. & -0.0033 & 0.0308 \\ 
& (0.1655) & (0.1418) &  &  &  &  & (0.0341) & (0.0265) & (0.1672) & (0.1581) & (0.0277) & (0.033) \\ 
$Spill(A)^{\mu,up}_{i,t-1}$ &  &  & -0.0099 & -0.0116 & 0.0358 & 0.0197 &  &  & -0.0153 & -0.0029 & -0.0094 & -0.0134 \\ 
&  &  & (0.0116) & (0.0124) & (0.0263) & (0.0341) &  &  & (0.0438) & (0.0436) & (0.0112) & (0.013) \\ 
$Spill(A)^{\mu,dw}_{i,t-1}$ &  &  & -0.0152. & -0.0158 & -0.0811* & -0.0994* &  &  & -0.0863* & -0.0972* & -0.0118 & -0.013 \\ 
&  &  & (0.009) & (0.0119) & (0.0338) & (0.0436) &  &  & (0.0403) & (0.0422) & (0.0089) & (0.0129) \\ 
$Spill(A)^{\tau,up}_{i,t-1}$ & 5.165. & 5.483* &  &  &  &  & 1.604** & 0.4418 & 5.288. & 5.781* & 1.571** & 0.2136 \\ 
& (2.795) & (2.658) &  &  &  &  & (0.5813) & (0.6069) & (2.698) & (2.91) & (0.5445) & (0.6549) \\ 
$Spill(A)^{\tau,dw}_{i,t-1}$ & -0.1467 & -0.1639 &  &  &  &  & -0.1211*** & -0.0934*** & -0.2471. & -0.204 & -0.1065*** & -0.0985*** \\ 
& (0.1396) & (0.1185) &  &  &  &  & (0.0295) & (0.0272) & (0.1438) & (0.1244) & (0.0286) & (0.0278) \\ 
\hline \rule{0pt}{1.075\normalbaselineskip}  AR(1) &  & 0 & 0 & 0 & 0 & 0 & 0.0015 & 0 &  & 0 & 2e-04 & 0 \\ 
AR(2) & 0.2985 & 0.2953 & 0.0014 & 0.003 & 0.1809 & 0.1426 & 0.1807 & 0.0227 & 0.3221 & 0.2123 & 0.12 & 0.011 \\ 
Sargan & 0 & 0 & 0.0013 & 0.007 & 0 & 0 & 8e-04 & 1e-04 & 0 & 0 & 1e-04 & 0 \\ 
Controls &  & Y &  & Y &  & Y &  & Y &  & Y &  & Y \\ 
$R^2$ & 0.8839 & 0.8934 & 0.9966 & 0.9961 & 0.9347 & 0.9144 & 0.9951 & 0.9958 & 0.8823 & 0.885 & 0.9957 & 0.9949 \\ 
\hline
\hline
\end{tabular}
\endgroup

\end{myresizeenv}

\vspace{0.25cm}

\justifying \footnotesize

\noindent 
Notes: The table shows the regression results of output $A_{i,t}^{\mu}$ and patents $A_{i,t}^{\tau}$ on demand-pull and technology-push effects. The estimation is based on a two-ways Blundell-Bond (BB) system GMM model using a one-step estimation procedure. Spillovers are calculated on the basis of first-order links. Variables measured in monetary terms are deflated using the industry level price deflators for the value of shipment obtained from the NBER-productivity database \citep{becker2013nber}. Instruments are collapsed to avoid instrument proliferation. To cope with skewness and to obtain tractable coefficients, most variables are pre-processed (taking logs, removing outliers, scaling). Data in logs are $A^{\alpha }_{i,t}$, $PR^{\alpha, d }_{i,t}$, $Spill(A)^{\alpha, d }_{i,t}$, $L_{i,t}$,  $(K/L)_{i,t}$, $(I/L)_{i,t}$, $W_{i,t}$, $(K/L)^{}_{ij,t-1}$, $(E/L)^{}_{ij,t-1}$, $(M/L)^{}_{ij,t-1}$, $W^{P}_{ij,t-1}$ with $\alpha = \mu, \tau$ and $d = up, dw$. A detailed description of the transformations and descriptive statistics of the regression data before and after the transformations are provided in \ref{app:data}. 
The rows AR(1), AR(2), and Sargan show the test statistics of the specification tests, i.e. testing for first- and second-order autocorrelation and the results of a Sargan test for validity of instruments \citep[see][]{roodman2009xtabond2}. 
The analysis covers the subset of Suppliers-dominated industries (see \citep{bogliacino2016pavitt}).
\end{table}

\subsubsection{Production-intensive sectors}
\begin{table}[H]
\begin{myresizeenv}
\caption{Demand-pull and technology-push effects in Production-intensive sectors.} 

\label{tab:real_sum_output|citation_weighted_patent_stock_Production_intensive_all_years_sys_1step_wControls}

\begingroup
\begin{tabular}{|l|cccc|cccc|cccc|}
\hline \hline  \rule{0pt}{1.075\normalbaselineskip}  
& \multicolumn{4}{c|}{\scriptsize \ul{Type 1}} & \multicolumn{4}{c|}{\scriptsize \ul{Type 2}} & \multicolumn{4}{c|}{\scriptsize \ul{Both}}\\  & \multicolumn{2}{c}{\scriptsize $\tau \rightarrow \mu$} & \multicolumn{2}{c|}{\scriptsize $\mu \rightarrow \tau$}
& \multicolumn{2}{c}{\scriptsize Market} & \multicolumn{2}{c|}{\scriptsize Innovation}
& \multicolumn{4}{c|}{}\\
\hline  \rule{0pt}{1.075\normalbaselineskip}    & $A^{\mu}_{i,t}$ & $A^{\mu}_{i,t}$ & $A^{\tau}_{i,t}$ & $A^{\tau}_{i,t}$ & $A^{\mu}_{i,t}$ & $A^{\mu}_{i,t}$ & $A^{\tau}_{i,t}$ & $A^{\tau}_{i,t}$ & $A^{\mu}_{i,t}$ & $A^{\mu}_{i,t}$ & $A^{\tau}_{i,t}$ & $A^{\tau}_{i,t}$ \\ 
\rule{0pt}{1.075\normalbaselineskip}  & (1) & (2) & (3) & (4) & (5) & (6) & (7) & (8) & (9) & (10) & (11) & (12) \\ 
\hline \rule{0pt}{1.075\normalbaselineskip}   $A^{\mu}_{i,t-1}$ & 0.5588*** & 0.5395*** & -0.0447. & -0.034 & 0.8319*** & 0.9136*** &  &  & 0.7162*** & 0.713*** & -0.033 & -0.0529. \\ 
& (0.0402) & (0.0583) & (0.0245) & (0.0295) & (0.0773) & (0.0928) &  &  & (0.1022) & (0.095) & (0.0327) & (0.029) \\ 
$A^{\tau}_{i,t-1}$ & 0.4852. & 0.1719 & 1.091*** & 1.105*** &  &  & 1.151*** & 1.154*** & 0.4755 & 0.1159 & 1.167*** & 1.165*** \\ 
& (0.2807) & (0.2333) & (0.027) & (0.0414) &  &  & (0.1501) & (0.0883) & (0.3413) & (0.2732) & (0.1054) & (0.0905) \\ 
$PR^{\mu,up}_{i,t-1}$ &  &  & 0.0815. & 0.0591 & -0.4847** & -0.2179 &  &  & -0.0561. & -0.0466 & 0.008 & 0.0064 \\ 
&  &  & (0.048) & (0.054) & (0.1718) & (0.1765) &  &  & (0.0321) & (0.0346) & (0.0077) & (0.0062) \\ 
$PR^{\mu,dw}_{i,t-1}$ &  &  & 0.1012* & 0.078. & -0.2203 & -0.1363 &  &  & -0.0467 & -0.008 & 0.0112. & 0.0119* \\ 
&  &  & (0.0397) & (0.0421) & (0.2447) & (0.2417) &  &  & (0.0393) & (0.034) & (0.0063) & (0.005) \\ 
$PR^{\tau,dw}_{i,t-1}$ & -1.025 & 0.3851 &  &  &  &  & 0.0969 & -0.0581 & -0.0146 & 0.1327 & -0.002 & -0.0118 \\ 
& (0.9945) & (0.7106) &  &  &  &  & (0.5115) & (0.1969) & (0.1453) & (0.0955) & (0.0479) & (0.0308) \\ 
$Spill(A)^{\mu,up}_{i,t-1}$ &  &  & -0.0503* & -0.0216 & 0.1923* & 0.2629* &  &  & 0.0906. & 0.0757 & -0.0235* & -0.0164 \\ 
&  &  & (0.0215) & (0.0234) & (0.0912) & (0.1053) &  &  & (0.0513) & (0.0564) & (0.01) & (0.0102) \\ 
$Spill(A)^{\mu,dw}_{i,t-1}$ &  &  & 0.0346 & 0.0403 & -0.3194** & -0.3023** &  &  & -0.1326** & -0.0913** & 5e-04 & 0.0105 \\ 
&  &  & (0.0237) & (0.0279) & (0.1073) & (0.1125) &  &  & (0.0419) & (0.0346) & (0.0071) & (0.0079) \\ 
$Spill(A)^{\tau,up}_{i,t-1}$ & 3.149. & 5.418* &  &  &  &  & 2.058* & 1.115 & 8.703*** & 6.739*** & 1.544** & 0.9693. \\ 
& (1.812) & (2.144) &  &  &  &  & (0.9929) & (0.8112) & (2.187) & (2.022) & (0.5982) & (0.5046) \\ 
$Spill(A)^{\tau,dw}_{i,t-1}$ & -0.183. & -0.2227. &  &  &  &  & -0.1257** & -0.0909* & -0.2126* & -0.1512 & -0.0673** & -0.0544* \\ 
& (0.0987) & (0.1289) &  &  &  &  & (0.0442) & (0.0408) & (0.0981) & (0.0937) & (0.0237) & (0.0233) \\ 
\hline \rule{0pt}{1.075\normalbaselineskip}  AR(1) & 0 & 0 & 0.0378 & 0.0344 & 0 & 0 & 0.0416 & 0.0395 &  & 0 & 7e-04 & 0.0363 \\ 
AR(2) & 0.9765 & 0.7431 & 0.5772 & 0.6142 & 0.6127 & 0.3061 & 0.1149 & 0.3658 & 0.9893 & 0.9516 & 0.0072 & 0.3586 \\ 
Sargan & 0 & 0 & 1e-04 & 0.0052 & 0 & 0 & 0 & 0 & 0 & 0 & 0 & 0 \\ 
Controls &  & Y &  & Y &  & Y &  & Y &  & Y &  & Y \\ 
$R^2$ & 0.922 & 0.9124 & 0.9952 & 0.9946 & 0.9216 & 0.9078 & 0.9946 & 0.9952 & 0.8967 & 0.9004 & 0.9948 & 0.995 \\ 
\hline
\hline
\end{tabular}
\endgroup
\end{myresizeenv}

\vspace{0.25cm}

\justifying \footnotesize

\noindent 
Notes: The table shows the regression results of output $A_{i,t}^{\mu}$ and patents $A_{i,t}^{\tau}$ on demand-pull and technology-push effects. The estimation is based on a two-ways Blundell-Bond (BB) system GMM model using a one-step estimation procedure. Spillovers are calculated on the basis of first-order links. Variables measured in monetary terms are deflated using the industry level price deflators for the value of shipment obtained from the NBER-productivity database \citep{becker2013nber}. Instruments are collapsed to avoid instrument proliferation. To cope with skewness and to obtain tractable coefficients, most variables are pre-processed (taking logs, removing outliers, scaling). Data in logs are $A^{\alpha }_{i,t}$, $PR^{\alpha, d }_{i,t}$, $Spill(A)^{\alpha, d }_{i,t}$, $L_{i,t}$,  $(K/L)_{i,t}$, $(I/L)_{i,t}$, $W_{i,t}$, $(K/L)^{}_{ij,t-1}$, $(E/L)^{}_{ij,t-1}$, $(M/L)^{}_{ij,t-1}$, $W^{P}_{ij,t-1}$ with $\alpha = \mu, \tau$ and $d = up, dw$. A detailed description of the transformations and descriptive statistics of the regression data before and after the transformations are provided in \ref{app:data}. 
The rows AR(1), AR(2), and Sargan show the test statistics of the specification tests, i.e. testing for first- and second-order autocorrelation and the results of a Sargan test for validity of instruments \citep[see][]{roodman2009xtabond2}. 
The analysis covers the subset of Production-intensive industries (see \citep{bogliacino2016pavitt}). 
\end{table}

\subsection{Patent centrality}

\subsubsection{Sectors with a high centrality in the innovation layer}
\begin{table}[H]
\begin{myresizeenv}
\caption{Demand-pull and technology-push effects in industries with high patent centrality}

\label{tab:real_sum_output|citation_weighted_patent_stock_patent_central_TRUE_all_years_sys_1step_wControls}
\begingroup
\begin{tabular}{|l|cccc|cccc|cccc|}
\hline \hline  \rule{0pt}{1.075\normalbaselineskip}  
& \multicolumn{4}{c|}{\scriptsize \ul{Type 1}} & \multicolumn{4}{c|}{\scriptsize \ul{Type 2}} & \multicolumn{4}{c|}{\scriptsize \ul{Both}}\\  & \multicolumn{2}{c}{\scriptsize $\tau \rightarrow \mu$} & \multicolumn{2}{c|}{\scriptsize $\mu \rightarrow \tau$}
& \multicolumn{2}{c}{\scriptsize Market} & \multicolumn{2}{c|}{\scriptsize Innovation}
& \multicolumn{4}{c|}{}\\
\hline  \rule{0pt}{1.075\normalbaselineskip}    & $A^{\mu}_{i,t}$ & $A^{\mu}_{i,t}$ & $A^{\tau}_{i,t}$ & $A^{\tau}_{i,t}$ & $A^{\mu}_{i,t}$ & $A^{\mu}_{i,t}$ & $A^{\tau}_{i,t}$ & $A^{\tau}_{i,t}$ & $A^{\mu}_{i,t}$ & $A^{\mu}_{i,t}$ & $A^{\tau}_{i,t}$ & $A^{\tau}_{i,t}$ \\ 
\rule{0pt}{1.075\normalbaselineskip}  & (1) & (2) & (3) & (4) & (5) & (6) & (7) & (8) & (9) & (10) & (11) & (12) \\ 
\hline \rule{0pt}{1.075\normalbaselineskip}   $A^{\mu}_{i,t-1}$ & 0.6268*** & 0.6175*** & -0.0194 & -0.0039 & 0.7156*** & 0.7603*** &  &  & 0.6905*** & 0.7776*** & 0.0103 & -0.0024 \\ 
& (0.0406) & (0.0402) & (0.0176) & (0.017) & (0.0751) & (0.0877) &  &  & (0.096) & (0.0778) & (0.019) & (0.0156) \\ 
$A^{\tau}_{i,t-1}$ & 0.3189. & -0.1059 & 1.134*** & 1.112*** &  &  & 1.288*** & 1.283*** & 0.0176 & -0.3977 & 1.307*** & 1.299*** \\ 
& (0.1758) & (0.2666) & (0.0174) & (0.0342) &  &  & (0.0686) & (0.0668) & (0.3144) & (0.3537) & (0.0573) & (0.0576) \\ 
$PR^{\mu,up}_{i,t-1}$ &  &  & 0.0035 & -0.0034 & -0.0234 & -0.0238 &  &  & -0.0184 & -0.024 & -0.0017 & -0.0025 \\ 
&  &  & (0.0046) & (0.0061) & (0.0255) & (0.0264) &  &  & (0.0327) & (0.0301) & (0.0051) & (0.0041) \\ 
$PR^{\mu,dw}_{i,t-1}$ &  &  & 0.0113** & 0.0067 & 0.0133 & 0.005 &  &  & 0.0159 & 0.0313 & 0.0076. & 0.0058 \\ 
&  &  & (0.0039) & (0.0055) & (0.031) & (0.0331) &  &  & (0.037) & (0.0317) & (0.0044) & (0.0046) \\ 
$PR^{\tau,dw}_{i,t-1}$ & -0.1004 & 0.1441 &  &  &  &  & -0.0354 & -0.0211 & 0.0862 & 0.2605* & -0.0478* & -0.0301 \\ 
& (0.1101) & (0.0994) &  &  &  &  & (0.0261) & (0.023) & (0.1376) & (0.1179) & (0.0242) & (0.0193) \\ 
$Spill(A)^{\mu,up}_{i,t-1}$ &  &  & -0.0123. & -0.0076 & 0.0519. & 0.0781* &  &  & 0.0112 & 0.0232 & -0.0083 & -0.0044 \\ 
&  &  & (0.0064) & (0.0079) & (0.0277) & (0.0347) &  &  & (0.0356) & (0.0377) & (0.0057) & (0.0062) \\ 
$Spill(A)^{\mu,dw}_{i,t-1}$ &  &  & 0.0055 & 0.0081 & -0.0675* & -0.0515 &  &  & -0.0915** & -0.0721* & 0.0105* & 0.012* \\ 
&  &  & (0.005) & (0.0057) & (0.0288) & (0.032) &  &  & (0.0286) & (0.0307) & (0.0048) & (0.005) \\ 
$Spill(A)^{\tau,up}_{i,t-1}$ & 2.676 & 4.904* &  &  &  &  & 2.495*** & 1.82*** & 8.551*** & 7.328** & 2.143*** & 1.562*** \\ 
& (1.682) & (2.118) &  &  &  &  & (0.4845) & (0.4121) & (2.248) & (2.453) & (0.4314) & (0.3562) \\ 
$Spill(A)^{\tau,dw}_{i,t-1}$ & 0.134 & -0.0515 &  &  &  &  & -0.0776* & -0.1248*** & 0.1833 & -0.0208 & -0.0876** & -0.1343*** \\ 
& (0.1849) & (0.1341) &  &  &  &  & (0.0349) & (0.027) & (0.1949) & (0.1495) & (0.0317) & (0.0271) \\ 
\hline \rule{0pt}{1.075\normalbaselineskip}  AR(1) & 0 & 0 & 0 & 0 & 0 & 0 & 0 & 0 &  & 0 & 0 & 0 \\ 
AR(2) & 0.7376 & 0.6504 & 0.0121 & 0.0164 & 0.6387 & 0.4604 & 0.0455 & 0.0676 & 0.6409 & 0.5191 & 0.0782 & 0.105 \\ 
Sargan & 0 & 0 & 0 & 5e-04 & 0 & 0 & 0 & 3e-04 & 0 & 0 & 0 & 0 \\ 
Controls &  & Y &  & Y &  & Y &  & Y &  & Y &  & Y \\ 
$R^2$ & 0.9359 & 0.9221 & 0.9984 & 0.9978 & 0.9357 & 0.9172 & 0.9976 & 0.9981 & 0.9144 & 0.9033 & 0.9981 & 0.9983 \\ 
\hline
\hline
\end{tabular}
\endgroup

\end{myresizeenv}

\vspace{0.25cm}

\justifying \footnotesize

\noindent 
Notes: The table shows the regression results of output $A_{i,t}^{\mu}$ and patents $A_{i,t}^{\tau}$ on demand-pull and technology-push effects. The estimation is based on a two-ways Blundell-Bond (BB) system GMM model using a one-step estimation procedure. Spillovers are calculated on the basis of first-order links. Variables measured in monetary terms are deflated using the industry level price deflators for the value of shipment obtained from the NBER-productivity database \citep{becker2013nber}. Instruments are collapsed to avoid instrument proliferation. To cope with skewness and to obtain tractable coefficients, most variables are pre-processed (taking logs, removing outliers, scaling). Data in logs are $A^{\alpha }_{i,t}$, $PR^{\alpha, d }_{i,t}$, $Spill(A)^{\alpha, d }_{i,t}$, $L_{i,t}$,  $(K/L)_{i,t}$, $(I/L)_{i,t}$, $W_{i,t}$, $(K/L)^{}_{ij,t-1}$, $(E/L)^{}_{ij,t-1}$, $(M/L)^{}_{ij,t-1}$, $W^{P}_{ij,t-1}$ with $\alpha = \mu, \tau$ and $d = up, dw$. A detailed description of the transformations and descriptive statistics of the regression data before and after the transformations are provided in \ref{app:data}. 
The rows AR(1), AR(2), and Sargan show the test statistics of the specification tests, i.e. testing for first- and second-order autocorrelation and the results of a Sargan test for validity of instruments \citep[see][]{roodman2009xtabond2}.
The analysis covers the subset of industries industries with above median $PR^{\tau,dw}_{i,t}$. 
\end{table}

\subsubsection{Sectors with a low centrality in the innovation layer}
\begin{table}[H]
\begin{myresizeenv}
\caption{Demand-pull and technology-push effects in Non-Patent-central sectors.} 

\label{tab:real_sum_output|citation_weighted_patent_stock_patent_central_FALSE_all_years_sys_1step_wControls}

\begingroup
\begin{tabular}{|l|cccc|cccc|cccc|}
\hline \hline  \rule{0pt}{1.075\normalbaselineskip}  
& \multicolumn{4}{c|}{\scriptsize \ul{Type 1}} & \multicolumn{4}{c|}{\scriptsize \ul{Type 2}} & \multicolumn{4}{c|}{\scriptsize \ul{Both}}\\  & \multicolumn{2}{c}{\scriptsize $\tau \rightarrow \mu$} & \multicolumn{2}{c|}{\scriptsize $\mu \rightarrow \tau$}
& \multicolumn{2}{c}{\scriptsize Market} & \multicolumn{2}{c|}{\scriptsize Innovation}
& \multicolumn{4}{c|}{}\\
\hline  \rule{0pt}{1.075\normalbaselineskip}    & $A^{\mu}_{i,t}$ & $A^{\mu}_{i,t}$ & $A^{\tau}_{i,t}$ & $A^{\tau}_{i,t}$ & $A^{\mu}_{i,t}$ & $A^{\mu}_{i,t}$ & $A^{\tau}_{i,t}$ & $A^{\tau}_{i,t}$ & $A^{\mu}_{i,t}$ & $A^{\mu}_{i,t}$ & $A^{\tau}_{i,t}$ & $A^{\tau}_{i,t}$ \\ 
\rule{0pt}{1.075\normalbaselineskip}  & (1) & (2) & (3) & (4) & (5) & (6) & (7) & (8) & (9) & (10) & (11) & (12) \\ 
\hline \rule{0pt}{1.075\normalbaselineskip}   $A^{\mu}_{i,t-1}$ & 0.5915*** & 0.5691*** & 0.0194 & 0.0051 & 0.8507*** & 0.8656*** &  &  & 0.774*** & 0.8428*** & 0.0191 & 0.0084 \\ 
& (0.0488) & (0.0475) & (0.0335) & (0.0339) & (0.0836) & (0.0993) &  &  & (0.0855) & (0.0973) & (0.0351) & (0.0362) \\ 
$A^{\tau}_{i,t-1}$ & 0.3181. & 0.1905 & 1.038*** & 1.006*** &  &  & 1.167*** & 1.093*** & 0.2358 & 0.1918 & 1.157*** & 1.102*** \\ 
& (0.1711) & (0.1636) & (0.0423) & (0.0416) &  &  & (0.0738) & (0.0647) & (0.1623) & (0.1838) & (0.0663) & (0.0579) \\ 
$PR^{\mu,up}_{i,t-1}$ &  &  & 0.004 & 0.0031 & -0.089*** & -0.1136*** &  &  & -0.0752*** & -0.1053*** & 0.0053 & 0.0037 \\ 
&  &  & (0.0085) & (0.0094) & (0.0229) & (0.0252) &  &  & (0.0225) & (0.0267) & (0.0089) & (0.0097) \\ 
$PR^{\mu,dw}_{i,t-1}$ &  &  & 0.0027 & 0.0019 & -0.0425* & -0.0603** &  &  & -0.0292 & -0.0366. & 0.0052 & 0.0034 \\ 
&  &  & (0.0058) & (0.007) & (0.0201) & (0.0222) &  &  & (0.0194) & (0.0217) & (0.0067) & (0.0074) \\ 
$PR^{\tau,dw}_{i,t-1}$ & -0.3511 & -0.6513* &  &  &  &  & -0.2922. & -0.0739 & -0.3931 & -0.7966* & -0.2294. & -0.0897 \\ 
& (0.3047) & (0.2866) &  &  &  &  & (0.15) & (0.1215) & (0.3048) & (0.3953) & (0.1238) & (0.0864) \\ 
$Spill(A)^{\mu,up}_{i,t-1}$ &  &  & -0.0117 & -0.0045 & 0.0981** & 0.0766* &  &  & 0.0799* & 0.0404 & -0.02. & -0.0076 \\ 
&  &  & (0.0115) & (0.012) & (0.0335) & (0.0343) &  &  & (0.0352) & (0.0394) & (0.0121) & (0.0123) \\ 
$Spill(A)^{\mu,dw}_{i,t-1}$ &  &  & -0.003 & 0.0093 & -0.0805 & -0.0973* &  &  & -0.0862. & -0.118* & -0.0116 & 0.0046 \\ 
&  &  & (0.0114) & (0.0112) & (0.0503) & (0.0485) &  &  & (0.0502) & (0.0485) & (0.011) & (0.0109) \\ 
$Spill(A)^{\tau,up}_{i,t-1}$ & 4.489** & 4.793*** &  &  &  &  & 2.047*** & 0.8203 & 3.781** & 4.881** & 1.48*** & 0.7862* \\ 
& (1.45) & (1.382) &  &  &  &  & (0.5552) & (0.6348) & (1.405) & (1.648) & (0.4274) & (0.38) \\ 
$Spill(A)^{\tau,dw}_{i,t-1}$ & -0.0134 & 0.0179 &  &  &  &  & -0.0351 & -0.0544* & 0.0062 & 0.0289 & -0.0446. & -0.0534* \\ 
& (0.0701) & (0.0669) &  &  &  &  & (0.028) & (0.0231) & (0.0675) & (0.0718) & (0.0257) & (0.0222) \\ 
\hline \rule{0pt}{1.075\normalbaselineskip}  AR(1) & 0 & 0 & 6e-04 & 0.0015 & 0 & 0 & 0.0051 & 0.0035 &  & 0 & 0.0077 & 0.0027 \\ 
AR(2) & 0.6202 & 0.5782 & 0.5707 & 0.6415 & 0.6044 & 0.4946 & 0.4388 & 0.5494 & 0.6391 & 0.5207 & 0.4708 & 0.5109 \\ 
Sargan & 0 & 0 & 2e-04 & 0.0058 & 0 & 0 & 0 & 0 & 0 & 0 & 0 & 0 \\ 
Controls &  & Y &  & Y &  & Y &  & Y &  & Y &  & Y \\ 
$R^2$ & 0.9241 & 0.9258 & 0.9893 & 0.9887 & 0.9255 & 0.9077 & 0.9859 & 0.9862 & 0.922 & 0.8927 & 0.9873 & 0.9875 \\ 
\hline
\hline
\end{tabular}
\endgroup

\end{myresizeenv}

\vspace{0.25cm}

\justifying \footnotesize

\noindent 
Notes: The table shows the regression results of output $A_{i,t}^{\mu}$ and patents $A_{i,t}^{\tau}$ on demand-pull and technology-push effects. The estimation is based on a two-ways Blundell-Bond (BB) system GMM model using a one-step estimation procedure. Spillovers are calculated on the basis of first-order links. Variables measured in monetary terms are deflated using the industry level price deflators for the value of shipment obtained from the NBER-productivity database \citep{becker2013nber}. Instruments are collapsed to avoid instrument proliferation. To cope with skewness and to obtain tractable coefficients, most variables are pre-processed (taking logs, removing outliers, scaling). Data in logs are $A^{\alpha }_{i,t}$, $PR^{\alpha, d }_{i,t}$, $Spill(A)^{\alpha, d }_{i,t}$, $L_{i,t}$,  $(K/L)_{i,t}$, $(I/L)_{i,t}$, $W_{i,t}$, $(K/L)^{}_{ij,t-1}$, $(E/L)^{}_{ij,t-1}$, $(M/L)^{}_{ij,t-1}$, $W^{P}_{ij,t-1}$ with $\alpha = \mu, \tau$ and $d = up, dw$. A detailed description of the transformations and descriptive statistics of the regression data before and after the transformations are provided in \ref{app:data}. The rows AR(1), AR(2), and Sargan show the test statistics of the specification tests, i.e. testing for first- and second-order autocorrelation and the results of a Sargan test for validity of instruments \citep[see][]{roodman2009xtabond2}.
The analysis covers the subset of industries industries with below median $PR^{\tau,dw}_{i,t}$. 

\end{table}

\subsection{Upstream centrality in the market}

\subsubsection{Sectors with high upstream centrality in the market}
\begin{table}[H]
\begin{myresizeenv}
\caption{Demand-pull and technology-push effects in sectors with low upstream centrality in the market.}

\label{tab:real_sum_output|citation_weighted_patent_stock_market_central_in_TRUE_all_years_sys_1step_wControls}

\begingroup
\begin{tabular}{|l|cccc|cccc|cccc|}
\hline \hline  \rule{0pt}{1.075\normalbaselineskip}  
& \multicolumn{4}{c|}{\scriptsize \ul{Type 1}} & \multicolumn{4}{c|}{\scriptsize \ul{Type 2}} & \multicolumn{4}{c|}{\scriptsize \ul{Both}}\\  & \multicolumn{2}{c}{\scriptsize $\tau \rightarrow \mu$} & \multicolumn{2}{c|}{\scriptsize $\mu \rightarrow \tau$}
& \multicolumn{2}{c}{\scriptsize Market} & \multicolumn{2}{c|}{\scriptsize Innovation}
& \multicolumn{4}{c|}{}\\
\hline  \rule{0pt}{1.075\normalbaselineskip}    & $A^{\mu}_{i,t}$ & $A^{\mu}_{i,t}$ & $A^{\tau}_{i,t}$ & $A^{\tau}_{i,t}$ & $A^{\mu}_{i,t}$ & $A^{\mu}_{i,t}$ & $A^{\tau}_{i,t}$ & $A^{\tau}_{i,t}$ & $A^{\mu}_{i,t}$ & $A^{\mu}_{i,t}$ & $A^{\tau}_{i,t}$ & $A^{\tau}_{i,t}$ \\ 
\rule{0pt}{1.075\normalbaselineskip}  & (1) & (2) & (3) & (4) & (5) & (6) & (7) & (8) & (9) & (10) & (11) & (12) \\ 
\hline \rule{0pt}{1.075\normalbaselineskip}   $A^{\mu}_{i,t-1}$ & 0.541*** & 0.5399*** & 0.0053 & 1e-04 & 0.5041*** & 0.6168*** &  &  & 0.4558** & 0.6066*** & -0.042 & -0.0139 \\ 
& (0.0527) & (0.0522) & (0.0272) & (0.0233) & (0.1438) & (0.1306) &  &  & (0.1709) & (0.1278) & (0.0397) & (0.0284) \\ 
$A^{\tau}_{i,t-1}$ & 0.1314 & -0.04 & 1.076*** & 1.064*** &  &  & 1.02*** & 1.103*** & -0.2172 & 0.0977 & 1.07*** & 1.102*** \\ 
& (0.177) & (0.1391) & (0.0213) & (0.0303) &  &  & (0.1378) & (0.0906) & (0.208) & (0.1461) & (0.1038) & (0.0881) \\ 
$PR^{\mu,up}_{i,t-1}$ &  &  & 0.0023 & -0.0028 & -0.0175 & -0.0288 &  &  & -0.0107 & -0.029 & 0.0015 & -0.0054 \\ 
&  &  & (0.0072) & (0.0072) & (0.0202) & (0.0193) &  &  & (0.0248) & (0.0185) & (0.0087) & (0.0067) \\ 
$PR^{\mu,dw}_{i,t-1}$ &  &  & 0.0098* & 0.0035 & 0.035. & 0.0175 &  &  & 0.0365 & 0.0188 & 0.0091 & 0.0028 \\ 
&  &  & (0.004) & (0.0055) & (0.0196) & (0.0209) &  &  & (0.023) & (0.0171) & (0.0057) & (0.0061) \\ 
$PR^{\tau,dw}_{i,t-1}$ & -0.0281 & 0.0959 &  &  &  &  & 0.0938 & 0.0129 & 0.169. & 0.024 & 0.0574 & 0.0177 \\ 
& (0.092) & (0.0664) &  &  &  &  & (0.0727) & (0.0392) & (0.0969) & (0.0767) & (0.0497) & (0.0362) \\ 
$Spill(A)^{\mu,up}_{i,t-1}$ &  &  & -0.0083 & 0.0034 & 0.0118 & 0.009 &  &  & 0.0131 & 0.0047 & -0.0128 & 0 \\ 
&  &  & (0.0142) & (0.0133) & (0.0297) & (0.0353) &  &  & (0.034) & (0.0347) & (0.0135) & (0.014) \\ 
$Spill(A)^{\mu,dw}_{i,t-1}$ &  &  & 0.0095 & 0.0185* & -0.0431 & -0.0564. &  &  & -0.0303 & -0.0525. & 0.0132 & 0.0211* \\ 
&  &  & (0.0082) & (0.0083) & (0.0281) & (0.032) &  &  & (0.0284) & (0.0301) & (0.0084) & (0.0085) \\ 
$Spill(A)^{\tau,up}_{i,t-1}$ & -1.394 & 0.8729 &  &  &  &  & 1.201 & 0.7414 & -1.681 & 0.6963 & 0.7794 & 0.4497 \\ 
& (1.345) & (1.449) &  &  &  &  & (0.8165) & (0.6609) & (1.335) & (1.157) & (0.6945) & (0.6744) \\ 
$Spill(A)^{\tau,dw}_{i,t-1}$ & -0.0638 & -0.0453 &  &  &  &  & -0.0679* & -0.0529* & -0.0568 & -0.1018* & -0.0689** & -0.0574** \\ 
& (0.0468) & (0.0502) &  &  &  &  & (0.0304) & (0.0224) & (0.0533) & (0.0449) & (0.0249) & (0.0219) \\ 
\hline \rule{0pt}{1.075\normalbaselineskip}  AR(1) &  & 0 & 0.0173 & 0.0212 & 0 & 0 &  & 0.0105 & 0 &  &  & 0.0056 \\ 
AR(2) & 0.2962 & 0.1914 & 0.4727 & 0.4563 & 0.3395 & 0.2865 &  &  & 0.4521 & 0.3225 &  &  \\ 
Sargan & 1e-04 & 0 & 0.1012 & 0.3343 & 0 & 0 & 0.0013 & 7e-04 & 0 & 0 & 3e-04 & 4e-04 \\ 
Controls &  & Y &  & Y &  & Y &  & Y &  & Y &  & Y \\ 
$R^2$ & 0.9677 & 0.9568 & 0.9955 & 0.9947 & 0.9692 & 0.9641 & 0.9923 & 0.9944 & 0.9632 & 0.9643 & 0.9941 & 0.9947 \\ 
\hline
\hline
\end{tabular}
\endgroup

\end{myresizeenv}

\vspace{0.25cm}

\justifying \footnotesize

\noindent 
Notes: The table shows the regression results of output $A_{i,t}^{\mu}$ and patents $A_{i,t}^{\tau}$ on demand-pull and technology-push effects. The estimation is based on a two-ways Blundell-Bond (BB) system GMM model using a one-step estimation procedure. Spillovers are calculated on the basis of first-order links. Variables measured in monetary terms are deflated using the industry level price deflators for the value of shipment obtained from the NBER-productivity database \citep{becker2013nber}. Instruments are collapsed to avoid instrument proliferation. To cope with skewness and to obtain tractable coefficients, most variables are pre-processed (taking logs, removing outliers, scaling). Data in logs are $A^{\alpha }_{i,t}$, $PR^{\alpha, d }_{i,t}$, $Spill(A)^{\alpha, d }_{i,t}$, $L_{i,t}$,  $(K/L)_{i,t}$, $(I/L)_{i,t}$, $W_{i,t}$, $(K/L)^{}_{ij,t-1}$, $(E/L)^{}_{ij,t-1}$, $(M/L)^{}_{ij,t-1}$, $W^{P}_{ij,t-1}$ with $\alpha = \mu, \tau$ and $d = up, dw$. A detailed description of the transformations and descriptive statistics of the regression data before and after the transformations are provided in \ref{app:data}.
The rows AR(1), AR(2), and Sargan show the test statistics of the specification tests, i.e. testing for first- and second-order autocorrelation and the results of a Sargan test for validity of instruments \citep[see][]{roodman2009xtabond2}.
The analysis covers the subset of industries industries with above median $PR^{\mu,up}_{i,t}$. 
\end{table}

\subsubsection{Sectors with low upstream centrality in the market}
\begin{table}[H]
\begin{myresizeenv}

\caption{Demand-pull and technology-push effects in industries with low upstream centrality in the market.} 		

\label{tab:real_sum_output|citation_weighted_patent_stock_market_central_in_FALSE_all_years_sys_1step_wControls}
\begingroup
\begin{tabular}{|l|cccc|cccc|cccc|}
\hline \hline  \rule{0pt}{1.075\normalbaselineskip}  
& \multicolumn{4}{c|}{\scriptsize \ul{Type 1}} & \multicolumn{4}{c|}{\scriptsize \ul{Type 2}} & \multicolumn{4}{c|}{\scriptsize \ul{Both}}\\  & \multicolumn{2}{c}{\scriptsize $\tau \rightarrow \mu$} & \multicolumn{2}{c|}{\scriptsize $\mu \rightarrow \tau$}
& \multicolumn{2}{c}{\scriptsize Market} & \multicolumn{2}{c|}{\scriptsize Innovation}
& \multicolumn{4}{c|}{}\\
\hline  \rule{0pt}{1.075\normalbaselineskip}    & $A^{\mu}_{i,t}$ & $A^{\mu}_{i,t}$ & $A^{\tau}_{i,t}$ & $A^{\tau}_{i,t}$ & $A^{\mu}_{i,t}$ & $A^{\mu}_{i,t}$ & $A^{\tau}_{i,t}$ & $A^{\tau}_{i,t}$ & $A^{\mu}_{i,t}$ & $A^{\mu}_{i,t}$ & $A^{\tau}_{i,t}$ & $A^{\tau}_{i,t}$ \\ 
\rule{0pt}{1.075\normalbaselineskip}  & (1) & (2) & (3) & (4) & (5) & (6) & (7) & (8) & (9) & (10) & (11) & (12) \\ 
\hline \rule{0pt}{1.075\normalbaselineskip}   $A^{\mu}_{i,t-1}$ & 0.5954*** & 0.6013*** & 0.0168 & 0.0054 & 0.7694*** & 0.7308*** &  &  & 0.687*** & 0.7137*** & 0.0158 & -0.0026 \\ 
& (0.0592) & (0.0594) & (0.0206) & (0.0183) & (0.0481) & (0.0601) &  &  & (0.0647) & (0.0798) & (0.0226) & (0.0215) \\ 
$A^{\tau}_{i,t-1}$ & -0.3968 & -0.3873. & 1.048*** & 1.041*** &  &  & 0.9603*** & 1.024*** & -0.1949 & -0.1691 & 1.026*** & 1.033*** \\ 
& (0.2654) & (0.2182) & (0.0185) & (0.0207) &  &  & (0.1021) & (0.0822) & (0.2142) & (0.2291) & (0.0873) & (0.078) \\ 
$PR^{\mu,up}_{i,t-1}$ &  &  & -0.001 & 8e-04 & -0.0394. & -0.0495* &  &  & 0.0194 & -0.0018 & 0.0096 & 0.0104 \\ 
&  &  & (0.0075) & (0.0095) & (0.021) & (0.0236) &  &  & (0.0316) & (0.0412) & (0.0084) & (0.0101) \\ 
$PR^{\mu,dw}_{i,t-1}$ &  &  & 0.0011 & 0.0026 & 0.027 & -0.0048 &  &  & 0.0771* & 0.0413 & 0.0113. & 0.0111 \\ 
&  &  & (0.0047) & (0.0064) & (0.0216) & (0.0227) &  &  & (0.0304) & (0.0322) & (0.0068) & (0.0076) \\ 
$PR^{\tau,dw}_{i,t-1}$ & 0.457* & 0.3829* &  &  &  &  & 0.129* & 0.083. & 0.3064* & 0.2912. & 0.0794. & 0.0716. \\ 
& (0.1874) & (0.1499) &  &  &  &  & (0.0594) & (0.0463) & (0.151) & (0.1515) & (0.0467) & (0.0404) \\ 
$Spill(A)^{\mu,up}_{i,t-1}$ &  &  & -0.0046 & -0.006 & 0.03 & 0.0415 &  &  & 0.02 & 0.0287 & -0.0079 & -0.0058 \\ 
&  &  & (0.0089) & (0.0098) & (0.0293) & (0.0306) &  &  & (0.0341) & (0.0394) & (0.0094) & (0.0103) \\ 
$Spill(A)^{\mu,dw}_{i,t-1}$ &  &  & -0.0034 & 0.0011 & -0.111*** & -0.1111** &  &  & -0.1291*** & -0.1362*** & -0.0123 & -7e-04 \\ 
&  &  & (0.008) & (0.0078) & (0.0305) & (0.0353) &  &  & (0.0335) & (0.0371) & (0.0081) & (0.0078) \\ 
$Spill(A)^{\tau,up}_{i,t-1}$ & 6.88*** & 7.756** &  &  &  &  & 2.485*** & 1.769* & 6.917*** & 9.159** & 1.994*** & 1.624* \\ 
& (2.08) & (2.613) &  &  &  &  & (0.6966) & (0.7385) & (1.911) & (3.007) & (0.5419) & (0.6353) \\ 
$Spill(A)^{\tau,dw}_{i,t-1}$ & -0.0305 & -0.0296 &  &  &  &  & -0.0513 & -0.0591* & -0.0259 & -0.0964 & -0.0543. & -0.0542* \\ 
& (0.1365) & (0.0991) &  &  &  &  & (0.0395) & (0.0282) & (0.1091) & (0.1012) & (0.0315) & (0.0269) \\ 
\hline \rule{0pt}{1.075\normalbaselineskip}  AR(1) & 0 & 0 & 0 & 0 & 0 & 0 & 0 & 0 & 0 & 0 & 0 & 0 \\ 
AR(2) & 0.5746 & 0.5231 & 0.0435 & 0.0156 & 0.5097 & 0.468 & 0.4398 & 0.1059 & 0.608 & 0.5916 & 0.3685 & 0.1296 \\ 
Sargan & 0 & 0 & 4e-04 & 2e-04 & 0 & 0 & 0 & 0 & 0 & 0 & 0 & 0 \\ 
Controls &  & Y &  & Y &  & Y &  & Y &  & Y &  & Y \\ 
$R^2$ & 0.8405 & 0.8468 & 0.9959 & 0.9952 & 0.9033 & 0.9018 & 0.9906 & 0.9927 & 0.8596 & 0.8411 & 0.9933 & 0.9934 \\ 
\hline
\hline
\end{tabular}
\endgroup
\end{myresizeenv}

\vspace{0.25cm}

\justifying \footnotesize

\noindent 
Notes: The table shows the regression results of output $A_{i,t}^{\mu}$ and patents $A_{i,t}^{\tau}$ on demand-pull and technology-push effects. The estimation is based on a two-ways Blundell-Bond (BB) system GMM model using a one-step estimation procedure. Spillovers are calculated on the basis of first-order links. Variables measured in monetary terms are deflated using the industry level price deflators for the value of shipment obtained from the NBER-productivity database \citep{becker2013nber}. Instruments are collapsed to avoid instrument proliferation. To cope with skewness and to obtain tractable coefficients, most variables are pre-processed (taking logs, removing outliers, scaling). Data in logs are $A^{\alpha }_{i,t}$, $PR^{\alpha, d }_{i,t}$, $Spill(A)^{\alpha, d }_{i,t}$, $L_{i,t}$,  $(K/L)_{i,t}$, $(I/L)_{i,t}$, $W_{i,t}$, $(K/L)^{}_{ij,t-1}$, $(E/L)^{}_{ij,t-1}$, $(M/L)^{}_{ij,t-1}$, $W^{P}_{ij,t-1}$ with $\alpha = \mu, \tau$ and $d = up, dw$. A detailed description of the transformations and descriptive statistics of the regression data before and after the transformations are provided in \ref{app:data}. 
The rows AR(1), AR(2), and Sargan show the test statistics of the specification tests, i.e. testing for first- and second-order autocorrelation and the results of a Sargan test for validity of instruments \citep[see][]{roodman2009xtabond2}.
The analysis covers the subset of industries industries with below median $PR^{\mu,up}_{i,t}$. 
\end{table}

\subsection{Downstream centrality in the market}

\subsubsection{Sectors with a high downstream centrality in the market}
\begin{table}[H]
\begin{myresizeenv}
\begingroup
\begin{tabular}{|l|cccc|cccc|cccc|}
\hline \hline  \rule{0pt}{1.075\normalbaselineskip}  
& \multicolumn{4}{c|}{\scriptsize \ul{Type 1}} & \multicolumn{4}{c|}{\scriptsize \ul{Type 2}} & \multicolumn{4}{c|}{\scriptsize \ul{Both}}\\  & \multicolumn{2}{c}{\scriptsize $\tau \rightarrow \mu$} & \multicolumn{2}{c|}{\scriptsize $\mu \rightarrow \tau$}
& \multicolumn{2}{c}{\scriptsize Market} & \multicolumn{2}{c|}{\scriptsize Innovation}
& \multicolumn{4}{c|}{}\\
\hline  \rule{0pt}{1.075\normalbaselineskip}    & $A^{\mu}_{i,t}$ & $A^{\mu}_{i,t}$ & $A^{\tau}_{i,t}$ & $A^{\tau}_{i,t}$ & $A^{\mu}_{i,t}$ & $A^{\mu}_{i,t}$ & $A^{\tau}_{i,t}$ & $A^{\tau}_{i,t}$ & $A^{\mu}_{i,t}$ & $A^{\mu}_{i,t}$ & $A^{\tau}_{i,t}$ & $A^{\tau}_{i,t}$ \\ 
\rule{0pt}{1.075\normalbaselineskip}  & (1) & (2) & (3) & (4) & (5) & (6) & (7) & (8) & (9) & (10) & (11) & (12) \\ 
\hline \rule{0pt}{1.075\normalbaselineskip}   $A^{\mu}_{i,t-1}$ & 0.6307*** & 0.6515*** & 0.0269 & 0.0041 & 0.8518*** & 0.7987*** &  &  & 0.7486*** & 0.8338*** & -0.0047 & 0.0013 \\ 
& (0.0539) & (0.0451) & (0.0193) & (0.0179) & (0.0688) & (0.0845) &  &  & (0.1088) & (0.079) & (0.0297) & (0.0158) \\ 
$A^{\tau}_{i,t-1}$ & 0.2133 & 0.2429 & 1.049*** & 1.036*** &  &  & 0.9432*** & 1.131*** & 0.1785 & 0.103 & 1.056*** & 1.111*** \\ 
& (0.244) & (0.196) & (0.0154) & (0.0241) &  &  & (0.1314) & (0.0646) & (0.2848) & (0.2163) & (0.0831) & (0.0587) \\ 
$PR^{\mu,up}_{i,t-1}$ &  &  & -0.0033 & -0.0011 & -0.0669** & -0.0535. &  &  & -0.0323 & -0.0469 & 0.0044 & -6e-04 \\ 
&  &  & (0.007) & (0.0082) & (0.0239) & (0.0274) &  &  & (0.0344) & (0.0307) & (0.0097) & (0.0076) \\ 
$PR^{\mu,dw}_{i,t-1}$ &  &  & 0.0082. & 0.0074 & -0.0129 & -0.0213 &  &  & 0.0018 & -0.0052 & 0.0113. & 0.0112* \\ 
&  &  & (0.0043) & (0.0051) & (0.0199) & (0.0232) &  &  & (0.0231) & (0.0206) & (0.006) & (0.0056) \\ 
$PR^{\tau,dw}_{i,t-1}$ & 0.1086 & 0.0517 &  &  &  &  & 0.1528* & 0.0011 & 0.1513 & 0.0864 & 0.0713 & 0.0024 \\ 
& (0.153) & (0.085) &  &  &  &  & (0.0734) & (0.024) & (0.1743) & (0.0773) & (0.0447) & (0.0212) \\ 
$Spill(A)^{\mu,up}_{i,t-1}$ &  &  & -0.0167. & -0.0118 & 0.0643* & 0.0735* &  &  & 0.0733* & 0.0552. & -0.0172. & -0.0144 \\ 
&  &  & (0.0086) & (0.0094) & (0.0279) & (0.0315) &  &  & (0.033) & (0.0319) & (0.0102) & (0.0088) \\ 
$Spill(A)^{\mu,dw}_{i,t-1}$ &  &  & -0.0031 & 7e-04 & -0.0809. & -0.0606 &  &  & -0.084. & -0.086* & -0.0062 & 0.0035 \\ 
&  &  & (0.0087) & (0.0095) & (0.0426) & (0.0427) &  &  & (0.051) & (0.043) & (0.0106) & (0.0083) \\ 
$Spill(A)^{\tau,up}_{i,t-1}$ & 4.756* & 4.962** &  &  &  &  & 3.159** & 0.5732. & 7.422*** & 6.19*** & 1.995*** & 0.6129* \\ 
& (1.932) & (1.767) &  &  &  &  & (0.9771) & (0.3377) & (2.047) & (1.766) & (0.5722) & (0.3068) \\ 
$Spill(A)^{\tau,dw}_{i,t-1}$ & -0.1115 & -0.1055 &  &  &  &  & -0.0438 & -0.0554* & -0.1055 & -0.0564 & -0.0528 & -0.0458. \\ 
& (0.0823) & (0.0784) &  &  &  &  & (0.0478) & (0.0251) & (0.1003) & (0.0856) & (0.0335) & (0.0239) \\ 
\hline \rule{0pt}{1.075\normalbaselineskip}  AR(1) &  & 0 & 0 & 0 & 0 & 0 & 0.0047 & 0 &  & 0 & 0 & 0 \\ 
AR(2) & 0.9448 & 0.8598 & 0.7496 & 0.8759 & 0.9467 & 0.3926 & 0.7496 & 0.9705 & 0.7994 & 0.8165 & 0.8528 & 0.9764 \\ 
Sargan & 0 & 0 & 1e-04 & 0.1108 & 0 & 0 & 6e-04 & 0 & 0 & 0 & 1e-04 & 0 \\ 
Controls &  & Y &  & Y &  & Y &  & Y &  & Y &  & Y \\ 
$R^2$ & 0.903 & 0.9045 & 0.9959 & 0.9949 & 0.9112 & 0.8962 & 0.9848 & 0.9955 & 0.8816 & 0.8953 & 0.9921 & 0.9955 \\ 
\hline
\hline
\end{tabular}
\endgroup
\caption{Demand-pull and technology-push effects in industries with high downstream centrality in the market} 
\label{tab:real_sum_output|citation_weighted_patent_stock_market_central_out_TRUE_all_years_sys_1step_wControls}
\end{myresizeenv}

\vspace{0.25cm}

\justifying \footnotesize

\noindent 
Notes: The table shows the regression results of output $A_{i,t}^{\mu}$ and patents $A_{i,t}^{\tau}$ on demand-pull and technology-push effects. The estimation is based on a two-ways Blundell-Bond (BB) system GMM model using a one-step estimation procedure. Spillovers are calculated on the basis of first-order links. Variables measured in monetary terms are deflated using the industry level price deflators for the value of shipment obtained from the NBER-productivity database \citep{becker2013nber}. Instruments are collapsed to avoid instrument proliferation. To cope with skewness and to obtain tractable coefficients, most variables are pre-processed (taking logs, removing outliers, scaling). Data in logs are $A^{\alpha }_{i,t}$, $PR^{\alpha, d }_{i,t}$, $Spill(A)^{\alpha, d }_{i,t}$, $L_{i,t}$,  $(K/L)_{i,t}$, $(I/L)_{i,t}$, $W_{i,t}$, $(K/L)^{}_{ij,t-1}$, $(E/L)^{}_{ij,t-1}$, $(M/L)^{}_{ij,t-1}$, $W^{P}_{ij,t-1}$ with $\alpha = \mu, \tau$ and $d = up, dw$. A detailed description of the transformations and descriptive statistics of the regression data before and after the transformations are provided in \ref{app:data}. 
The rows AR(1), AR(2), and Sargan show the test statistics of the specification tests, i.e. testing for first- and second-order autocorrelation and the results of a Sargan test for validity of instruments \citep[see][]{roodman2009xtabond2}.
The analysis covers the subset of industries industries with above median $PR^{\mu,dw}_{i,t}$. 
\end{table}

\subsubsection{Sectors with a low downstream centrality in the market}
\begin{table}[H]
\begin{myresizeenv}
\caption{Demand-pull and technology-push effects in sectors with a low downstream centrality in the market.}		

\label{tab:real_sum_output|citation_weighted_patent_stock_market_central_out_FALSE_all_years_sys_1step_wControls}

\begingroup
\begin{tabular}{|l|cccc|cccc|cccc|}
\hline \hline  \rule{0pt}{1.075\normalbaselineskip}  
& \multicolumn{4}{c|}{\scriptsize \ul{Type 1}} & \multicolumn{4}{c|}{\scriptsize \ul{Type 2}} & \multicolumn{4}{c|}{\scriptsize \ul{Both}}\\  & \multicolumn{2}{c}{\scriptsize $\tau \rightarrow \mu$} & \multicolumn{2}{c|}{\scriptsize $\mu \rightarrow \tau$}
& \multicolumn{2}{c}{\scriptsize Market} & \multicolumn{2}{c|}{\scriptsize Innovation}
& \multicolumn{4}{c|}{}\\
\hline  \rule{0pt}{1.075\normalbaselineskip}    & $A^{\mu}_{i,t}$ & $A^{\mu}_{i,t}$ & $A^{\tau}_{i,t}$ & $A^{\tau}_{i,t}$ & $A^{\mu}_{i,t}$ & $A^{\mu}_{i,t}$ & $A^{\tau}_{i,t}$ & $A^{\tau}_{i,t}$ & $A^{\mu}_{i,t}$ & $A^{\mu}_{i,t}$ & $A^{\tau}_{i,t}$ & $A^{\tau}_{i,t}$ \\ 
\rule{0pt}{1.075\normalbaselineskip}  & (1) & (2) & (3) & (4) & (5) & (6) & (7) & (8) & (9) & (10) & (11) & (12) \\ 
\hline \rule{0pt}{1.075\normalbaselineskip}   $A^{\mu}_{i,t-1}$ & 0.594*** & 0.5005*** & -0.0221 & -0.0253 & 0.472*** & 0.4261*** &  &  & 0.4649*** & 0.4814*** & -0.0388 & -0.0224 \\ 
& (0.0555) & (0.0527) & (0.0355) & (0.0362) & (0.0735) & (0.0899) &  &  & (0.0829) & (0.0896) & (0.0376) & (0.0378) \\ 
$A^{\tau}_{i,t-1}$ & -0.1408 & -0.3558. & 1.066*** & 1.052*** &  &  & 1.094*** & 1.058*** & 0.1376 & -0.4097 & 1.098*** & 1.061*** \\ 
& (0.2237) & (0.2154) & (0.0315) & (0.035) &  &  & (0.1424) & (0.0733) & (0.3105) & (0.2783) & (0.1309) & (0.0793) \\ 
$PR^{\mu,up}_{i,t-1}$ &  &  & 0.0144. & 0.0122. & -0.0013 & -0.0163 &  &  & 0.0078 & -0.0216 & 0.0074 & 0.0071 \\ 
&  &  & (0.0084) & (0.0069) & (0.0263) & (0.0238) &  &  & (0.0306) & (0.0243) & (0.008) & (0.0071) \\ 
$PR^{\mu,dw}_{i,t-1}$ &  &  & -0.017* & -0.017* & 0.0484 & 0.033 &  &  & 0.0519 & 0.0201 & -0.0193** & -0.0226** \\ 
&  &  & (0.0081) & (0.0083) & (0.0306) & (0.0302) &  &  & (0.0357) & (0.0322) & (0.0067) & (0.0081) \\ 
$PR^{\tau,dw}_{i,t-1}$ & 0.0222 & 0.1383 &  &  &  &  & 0.0141 & 0.0304 & -0.2027 & 0.0798 & 0.0038 & 0.0297 \\ 
& (0.1298) & (0.1215) &  &  &  &  & (0.0704) & (0.0451) & (0.1915) & (0.1464) & (0.067) & (0.0405) \\ 
$Spill(A)^{\mu,up}_{i,t-1}$ &  &  & 0.0022 & 0.0125 & 0.0865** & -0.0156 &  &  & 0.0424 & -0.0166 & 0.0011 & 0.0177 \\ 
&  &  & (0.0117) & (0.0123) & (0.0324) & (0.0414) &  &  & (0.0404) & (0.0462) & (0.011) & (0.0136) \\ 
$Spill(A)^{\mu,dw}_{i,t-1}$ &  &  & 0.0011 & 0.0083 & 0.0317 & -0.0342 &  &  & 0.0048 & -0.0439 & -0.0016 & 0.0077 \\ 
&  &  & (0.0086) & (0.0089) & (0.0281) & (0.0313) &  &  & (0.0314) & (0.0323) & (0.0087) & (0.0085) \\ 
$Spill(A)^{\tau,up}_{i,t-1}$ & 4.632** & 1.952 &  &  &  &  & 1.839*** & 0.7189. & 5.329* & 1.241 & 1.564** & 0.5707 \\ 
& (1.697) & (1.528) &  &  &  &  & (0.4848) & (0.4341) & (2.219) & (1.608) & (0.5233) & (0.4725) \\ 
$Spill(A)^{\tau,dw}_{i,t-1}$ & 0.07 & 0.024 &  &  &  &  & -0.061* & -0.0465* & 0.0797 & 0.061 & -0.0563* & -0.0491* \\ 
& (0.0516) & (0.0532) &  &  &  &  & (0.0279) & (0.0203) & (0.0604) & (0.0629) & (0.0275) & (0.0231) \\ 
\hline \rule{0pt}{1.075\normalbaselineskip}  AR(1) & 0 & 0 & 0.0365 & 0.0269 & 0 & 0 &  & 0.0163 &  & 0 &  & 0.0188 \\ 
AR(2) & 0.873 & 0.9963 & 0.8164 & 0.7997 & 0.9125 & 0.8953 &  & 0.5992 & 0.6816 & 0.9007 &  & 0.7343 \\ 
Sargan & 0 & 0 & 0 & 0 & 0 & 0 & 0 & 0 & 0 & 0 & 0 & 1e-04 \\ 
Controls &  & Y &  & Y &  & Y &  & Y &  & Y &  & Y \\ 
$R^2$ & 0.9516 & 0.9495 & 0.9955 & 0.9955 & 0.9597 & 0.9405 & 0.9952 & 0.995 & 0.9417 & 0.9396 & 0.9955 & 0.995 \\ 
\hline
\hline
\end{tabular}
\endgroup
\end{myresizeenv}
\vspace{0.25cm}

\justifying \footnotesize

\noindent 
Notes: The table shows the regression results of output $A_{i,t}^{\mu}$ and patents $A_{i,t}^{\tau}$ on demand-pull and technology-push effects. The estimation is based on a two-ways Blundell-Bond (BB) system GMM model using a one-step estimation procedure. Spillovers are calculated on the basis of first-order links. Variables measured in monetary terms are deflated using the industry level price deflators for the value of shipment obtained from the NBER-productivity database \citep{becker2013nber}. Instruments are collapsed to avoid instrument proliferation. To cope with skewness and to obtain tractable coefficients, most variables are pre-processed (taking logs, removing outliers, scaling). Data in logs are $A^{\alpha }_{i,t}$, $PR^{\alpha, d }_{i,t}$, $Spill(A)^{\alpha, d }_{i,t}$, $L_{i,t}$,  $(K/L)_{i,t}$, $(I/L)_{i,t}$, $W_{i,t}$, $(K/L)^{}_{ij,t-1}$, $(E/L)^{}_{ij,t-1}$, $(M/L)^{}_{ij,t-1}$, $W^{P}_{ij,t-1}$ with $\alpha = \mu, \tau$ and $d = up, dw$. A detailed description of the transformations and descriptive statistics of the regression data before and after the transformations are provided in \ref{app:data}. 
The rows AR(1), AR(2), and Sargan show the test statistics of the specification tests, i.e. testing for first- and second-order autocorrelation and the results of a Sargan test for validity of instruments \citep[see][]{roodman2009xtabond2}.
The analysis covers the subset of industries industries with below median $PR^{\mu,dw}_{i,t}$. 
\end{table}

\end{document}